\documentclass[11pt,a4paper]{article}
\usepackage[english]{babel}
\usepackage[mathscr]{eucal}

\usepackage[affil-it]{authblk}

\usepackage{amsmath,amsbsy,amsthm,amsfonts,amssymb,makeidx,enumerate,bm,wasysym}
\usepackage{courier,vmargin,graphicx,wrapfig,float}
\usepackage{caption,subcaption}
\usepackage{epstopdf}
\usepackage{theorem}
\usepackage{verbatim}

\numberwithin{equation}{subsection}

\setmarginsrb{2.2 cm}{1.3 cm}
             {2.2 cm}{1.4 cm}
             {0.9 cm}{0.9 cm}
             {0.8 cm}{0.8 cm}

\newcommand{\boma}[1]{{\mbox{\boldmath $#1$} }}

\newcommand{\Hs}[1]{\bb{H}^{{#1}}_{\ss}}

\def\quadr{a)\,}
\def\tria{b)\,}
\def\spetria{c)\,}
\def\separ{d)\,}
\def\holom{e)\,}

\def\Le{\mathbb{L}}
\def\Ti{\mathbb{T}}
\def\Ma{\mathbb{M}}

\def\complessi{{\mathbf{C}}}
\def\complessic{\complessi_{\!\times}}

\def\reali{{\mathbf{R}}}

\def\einsu{\mathfrak{A}}
\def\einsd{\mathfrak{E}}
\def\einsf{\mathfrak{F}}

\def\AA{{\mathcal A}}
\def\al{\alpha}
\def\be{\beta}
\def\BB{{\mathcal B}}
\def\BBB{ {\mathscr B} }
\def\Bm{\mathcal{P}}
\def\Bp{\mathcal{Q}}
\def\bx{{\bf x}}
\def\conm{\mathfrak{E}^{\mbox{{\tiny $(m)$}}}}
\def\cu{\mathsf{k}}
\def\Dd{{\mathcal D}}
\def\drom{\dot{\rho}^{\mbox{{\tiny $(m)$}}}}
\def\ds{n}
\def\Ec{\mathcal{E}}
\def\EE{\mathscr{E}}
\def\eep{\zeta}
\def\elm{\ell}
\def\elp{L}

\def\Fc{\mathcal{F}}
\def\FF{\mathscr{F}}
\def\Fre{\mbox{{\scriptsize \cite{Fre}}}}
\def\f{\varphi}
\def\fs{\f_{\ast}}
\def\ft{\dot{\varphi}}
\def\ftt{\ddot{\varphi}}

\def\ga{\gamma}
\def\Ga{\Gamma}
\def\Hs{H_{*}}
\def\LL{{\mathscr L}}
\def\LLL{{\mathcal L}}
\def\lp{\eta}
\def\Mel{M_{\elm,\elp}}
\def\MK{\MM^\ds_{\cu}}
\def\MM{{\mathcal M}}
\def\om{\omega}
\def\Omm{\Omega^{\mbox{{\tiny $(m)$}}}}
\def\Omf{\Omega^{\mbox{{\tiny $(\phi)$}}}}
\def\Omk{\Omega^{\mbox{{\tiny $(k)$}}}}
\def\OmL{\Omega^{\mbox{{\tiny $(\Lambda)$}}}}
\def\Oms{\Omm_{*}}
\def\pfi{p^{\mbox{{\tiny $(\phi)$}}}}
\def\ppm{p^{\mbox{{\tiny $(m)$}}}}
\def\qt{\dot{q}}
\def\rofi{\rho^{\mbox{{\tiny $(\phi)$}}}}
\def\rom{\rho^{\mbox{{\tiny $(m)$}}}}
\def\roms{\rom_{*}}
\def\sec{\mbox{sec}}
\def\sgn{\mbox{sgn}}
\def\st{s_{\ast}}
\def\tc{\tau}
\def\tcin{\tc_{in}}
\def\tcp{\tc'}
\def\tcs{\tc_{*}}

\def\tcz{\tc_{0}}
\def\te{\theta}
\def\Tf{{T^{\mbox{{\tiny $(\phi)$}}}}}
\def\tm{t_{min}}
\def\Tm{{T^{\mbox{{\tiny $(m)$}}}}}
\def\tp{t_{max}}
\def\ts{t_{*}}
\def\tu{v}
\def\V{{\mathscr V}}

\def\VV{{\mathscr V}}
\def\wF{w^{\mbox{{\tiny $(\phi)$}}}}
\def\xt{\dot{x}}
\def\xtt{\ddot{x}}
\def\yt{\dot{y}}
\def\ytt{\ddot{y}}
\def\zt{\dot{z}}

\def\dd{\displaystyle}

\def\leqs{\leqslant}
\def\geqs{\geqslant}
\def\parn{\par\noindent}
\def\salto{\vskip 0.2truecm \noindent}

\def\beq{\begin{equation}}
\def\feq{\end{equation}}
\def\barray{\begin{array}}
\def\farray{\end{array}}

\title{\huge{Integrable scalar cosmologies\\ with matter and curvature}\vspace{0.5cm}}

\author[1,2,$\ast$]{Davide Fermi}
\author[1,$\dagger$]{Massimo Gengo}
\author[1,2,$\ddagger$,(a)]{Livio Pizzocchero}
\affil[1]{Dipartimento di Matematica, Universit\`a degli Studi di Milano,\protect\\
Via Saldini 50, I-20133 Milano, Italy\vspace{0.2cm}}
\affil[2]{Istituto Nazionale di Fisica Nucleare, Sezione di Milano,\protect\\
Via Celoria 16, I-20133 Milano, Italy\vspace{0.3cm}}
\affil[$\ast$]{davide.fermi@unimi.it\vspace{0.1cm}}
\affil[$\dagger$]{massimo.gengo@unimi.it\vspace{0.1cm}}
\affil[$\ddagger$]{livio.pizzocchero@unimi.it\vspace{0.3cm}}
\affil[(a)]{corresponding author}

\date{}

\begin{document}
\begin{titlepage}
\maketitle
\thispagestyle{empty}

\begin{abstract}
We show that several integrable (i.e., exactly solvable) scalar cosmologies considered by
Fr\'e, Sagnotti and Sorin (Nuclear Physics \textbf{B 877}(3) (2013), 1028--1106)
can be generalized to include cases where the spatial curvature is not zero and,
besides a scalar field, matter or radiation are present with an equation
of state $\ppm = w\, \rom$; depending on the specific form of the self-interaction potential
for the field, the constant $w$ can be arbitrary or must be fixed suitably.
\end{abstract}
\vspace{0.2cm} \noindent
\textbf{Keywords}:
Scalar field cosmologies, exact solutions, early universe, inflation, dark energy.
\vspace{0.15cm}\\
\textbf{MSC\! 2010}:
83-XX,  
83C15,  
83F05,  
70H06.  
\vspace{0.15cm}\\
\textbf{PACS\! 2010}:
04.40.Nr, 
04.20.Jb, 
98.80.Cq, 
95.36.+x. 
\end{titlepage}

\makeatother

\tableofcontents

\vfill\eject\noindent
$\phantom{a}$\vspace{-0.9cm}

\section{Introduction}
\textbf{Scalar fields in cosmology.}
The consideration of scalar fields in cosmological
models has a long history, and arises from
different motivations.
On one hand the inflaton, i.e.,
the entity driving inflation, is often
modeled as a scalar field. This approach originates from
the work of some scholars at the beginning
of the 1980's: let us mention,
in particular, Linde \cite{Lind},
Madsen and Coles \cite{Mads}.
On the other hand, a scalar field can be used as a model for dark energy.
This idea seems to have appeared
in a 1988 paper by Ratra and Peebles \cite{Peeb};
Caldwell, Dave and Steinhardt \cite{Cald}
are credited for introducing, ten years later, the
term ``quintessence'' to indicate
a scalar model of dark energy. \parn
It is hardly the case
to recall that the notion of dark energy
was experimentally consolidated during the
same years, following the publication (between 1998 and 1999) of
the observational results by
the High-Z Supernova Search Team \cite{Riess} and
the  Supernova Cosmology Project \cite{Perl}. \parn
To our knowledge, Saini, Raychaudhury, Sahni and Starobinsky \cite{Starob}
were the first to set up a strictly quantitative
connection between a scalar field model of dark energy
and the observational data of \cite{Perl,Riess}.
More precisely, the paper \cite{Starob} determines the most probable shape
of the self-interaction potential for the dark energy scalar field,
fitting the data of \cite{Perl,Riess} on luminosity distance and redshift
for the epoch ranging from present time to the time when the scale factor
was half of its present value.
\parn
Most of the papers cited before and in the sequel,
as well as the present work, rely on a paradigm
in which the universe is homogeneous and isotropic at each
time, and the scalar field (modeling the inflaton or dark energy)
is treated classically
({\footnote{An alternative approach treats the scalar field
as a quantum object, and replaces the deterministic values
of the related classical observables with the expectation
values arising from the underlying quantum theory.
It is a well known fact that the computation of these expectation values
is typically plagued by the occurrence of divergences, which must be treated
with some kind of renormalization procedure. In this connection, let us mention
that zeta-function regularization provides a very elegant approach,
allowing to cure the said divergences by pure analytic continuation
(see, e.g., \cite{CFP,FPWS,FPpoint} and the references cited therein).
We will not discuss these issues any further in this work.\vspace{-0.2cm}}}).
Due to the assumptions of homogeneity and isotropy,
the spacetime metric has the form
of Friedmann-Lema\^{i}tre-Robertson-Walker (FLRW),
possibly with non zero spatial curvature. For the same
reasons, it is assumed that the scalar field only depends on time.
These cosmological models might involve, besides
the scalar field, some other form of matter described as a perfect
fluid. Here and in the sequel the term ``matter'' is used in a
broad sense, and includes the case of a radiation gas.
The presence of matter fluids is typical of models where
the scalar field represents dark energy; these often encompass
the whole history of the universe, except for the very
early stages, and include epochs in which the matter contribution
is dominant with respect to that of dark energy. On the contrary,
models for the very early, inflationary stage of the universe
typically ignore the role of matter and just focus
the attention on the inflaton scalar field.
\parn
All the above models give rise to systems of ODEs,
describing the time evolution of the main actors, which include,
especially, the scale factor in the FLRW metric and the scalar field.
\parn
Let us point out some additional features
of these cosmological models. It is commonly assumed
that the scalar field is minimally coupled to gravity,
and that it does not interact directly with matter,
if the latter is present;
correspondingly, the stress energy tensors
of the scalar field and of the matter fluid are separately conserved.
In absence of different indications, all papers
cited in the sequel fit into the scheme just outlined (spatial homogeneity
and isotropy, minimal coupling of the scalar field with gravity,
no direct interaction between the field and matter).
\vspace{0.1cm}\\
\textbf{Integrable scalar cosmologies.}
Since the late 1980's, the rising physical interest for
cosmologies with scalar fields stimulated the search
for \textsl{integrable} models, in which the evolution
equations can be solved
explicitly. It turned out that this is possible
for models with certain features, like a special
functional form for the self-interaction potential of the
scalar field.
Of course, the availability of exact solutions is a significant
advantage with respect to numerical integration, since it
allows to identify details and conceptual aspects that could
be missed otherwise.
\parn
\newpage\noindent
Since the very beginning of these investigations,
it was understood that exact solutions can be
obtained assuming an exponential form for the
self-interaction potential of the scalar field.
Let us describe these results with the normalizations
of the present work, which we borrow from \cite{Fre} where
a suitable dimensionless version $\f$ of the scalar field is introduced
(for the precise definition see subsection \ref{subsec:2.3}, especially
Eq.\,\eqref{fiad}). \parn
In 1987 Barrow \cite{Barr2} assumed a potential of the form
$\V(\f) =$ const.\,$e^{-\lambda \f}$ (with $\lambda $ another arbitrary constant),
and a vanishing spatial curvature; he presented
a particular exact solution of the evolution
equations (but \textsl{not} the general solution)
for the case of a scalar field alone.
In the previously mentioned paper \cite{Peeb} of 1988, Ratra and Peebles
considered the same exponential potential
as in \cite{Barr2} (with
$\lambda >0$) and zero spatial curvature; they presented
some particular exact solutions of the evolution
equations, both for the
case of a scalar field alone and for a model
with a scalar field and pressure-less matter (dust)
({\footnote{In this connection, let us recall that the solution derived in presence of matter
was regarded by the authors as too peculiar to be physically relevant.\vspace{-0.2cm}}}).
In the same year, Burd and Barrow \cite{Burd} considered again the potential
$\V(\f) = $ const.\,$e^{-\lambda \f}$ (with $\lambda >0$), with possibly non-zero
spatial curvature in arbitrary spacetime dimension $n+1$; they proposed a detailed
stability analysis of the model and presented some new exact solutions exhibiting
the transition to power-law inflation at late times.
\parn
In 1990 de Ritis, Marmo, Platania, Rubano, Scudellaro and Stornaiolo
\cite{Mar1} investigated a cosmology with a scalar
field (and no matter) in the case of zero spatial curvature,
suggesting its use as an inflationary model.
To analyze the evolution equations, these authors
proposed a systematic use of
the Lagrangian viewpoint. In this way
they proved that the only potentials giving
rise to a Noether symmetry
for the system have the form (with the normalizations
of the present work) $\V(\f) = $
const.\,$e^{\f}$\! + const.\,$e^{-\f}$\!
+ const.\,. Moreover, they constructed
the general solution of the evolution equations
for this class of potentials. The same authors
extended these results to the case of a field
with non-minimal coupling to gravity in \cite{Mar2}.
\parn
In 1996 Zhuk \emph{et al.} \cite{KasZh1,Zhuk} (see also \cite{KasZh2})
examined cosmological models where the spacetime was the product
of a time line with an arbitrary number of Einstein spaces of arbitrary dimensions,
and the content of the universe was described by a family of scalar fields.
The scalar fields were assumed to be minimally-coupled to gravity,
self-interacting with a potential to be specified (yet, with no mutual interactions),
and to fulfill each one a prescribed
equation of state. Under additional constraints, some classical
(and quantum) integrable cases were obtained, deducing a posteriori
the corresponding self-interaction potentials for the fields.
The resulting solutions described inflationary cosmologies in some
cases, and wormholes in other cases.
\parn
In 1998 Chimento \cite{Chim} analyzed some cosmological models driven by two scalar
fields, one of them self-interacting with an exponential potential of the form
$\V(\f) = $ const.\,$e^{-\lambda \f}$ (as in \cite{Barr2,Barr,Peeb})
and the other one free and with non-zero mass.
Exact general solutions were obtained and examined in detail;
remarkably, these solutions show a transition
from expansion dominated by the free scalar field to
that dominated by the self-interacting field, yielding a power-law inflation. \parn
The potential $\V(\f) = $ const.\,$e^{\f}$\! + const.\,$e^{-\f}$\!
+ const. was reconsidered in 2002 by Rubano and Scudellaro \cite{Rub},
and in 2012 by Piedipalumbo, Scudellaro, Esposito and Rubano \cite{Piedipalumbo},
again for zero spatial curvature.
These authors showed that the solvability of the evolution
equations is preserved by the addition of dust;
they proposed this model for describing dark energy
and matter up to the present time.
\parn
With the notable exceptions of \cite{Burd,KasZh1,Zhuk} (and \cite{KasZh2}),
all papers \cite{Barr2,Chim,Mar1,Mar2,Piedipalumbo,Peeb,Rub}
mentioned above deal with spacetimes of dimension $3+1$.
\vspace{0.1cm}\\
\textbf{The integrable cosmologies of Fr\`e, Sagnotti and Sorin \cite{Fre}.}
The cited paper (published in 2013) considers the FLRW cosmologies with a self-interacting scalar field,
no matter and zero spatial curvature, in arbitrary spacetime dimension $n+1$.
The analysis of these models is based on the Lagrangian formalism,
and on the possibility of using gauge transformations
for the time coordinate.
More precisely, the approach of \cite{Fre}
describes a cosmology of the above type as a Lagrangian
system with \hfill two \hfill degrees \hfill of \hfill freedom \hfill plus \hfill the \hfill
constraint \hfill of \hfill zero \hfill energy;
\hfill the \hfill Lagrangian \hfill coordinates \hfill are,
\vfill \eject \noindent
{~}
\vskip -0.5cm \noindent
basically, the instantaneous values
of the scale factor and of the scalar field. The Lagrangian depends on the scalar field self-potential
and on a gauge function (describing
the choice of the time coordinate), to be
specified according to convenience in the investigation
of integrable cases.
\parn
For nine classes of self-potentials individuated in
\cite{Fre} (see, in particular, Table 1 on page 1048 of this work)
the Lagrange equations are solvable by quadratures, for arbitrary
initial data.
The reason of solvability is that, after a convenient choice of the gauge function
and a suitable change of the Lagrangian coordinates,
one of the following features (i-iv) occurs:
\vspace{0.05cm}\\
i) the Lagrangian is quadratic, so it gives rise
to linear evolution equations;
\parn
ii) the Lagrange equations have a triangular structure,
which essentially means that one of the equations
involves only one of the (new) Lagrangian coordinates;
\parn
iii) the Lagrangian is separable, i.e., it is the sum
of a Lagrangian depending only on the first coordinate
and a Lagrangian depending only on the second one.
In this case there are two independent subsystems, each one with just
one degree of freedom and a conserved energy, which can be used to
reduce to quadratures the corresponding evolution equation. \parn
iv) the Lagrangian is a function of a complex coordinate
(equivalent to a pair of real coordinates), with
a suitable ``holomorphic structure''; this fact ensures
the conservation of a complex valued ``energy'' function,
which allows to solve
by quadratures the evolution equations.
\vspace{0.05cm}\\
With the notations of \cite{Fre} and of the present
work, the first one of the nine potential classes
is formed by functions of the form $\V(\f) = $ const.$\,e^{\f}$ + const.$\, e^{-\f}$
+ const.; this case is solvable by the linearity of the Lagrange
equations (see (i)), and this result extends to any spacetime dimension
the previous results of \cite{Piedipalumbo,Rub} on
these potentials in dimension $3+1$. \parn
The second potential class of \cite{Fre}
is formed by potentials of the form
$\V(\f) = $ const.\,$e^{2 \,\gamma \,\f}$ + const.$\,e^{(1 + \gamma) \f}$,
with $\gamma$ another arbitrary constant; this case is solvable
due to the triangular structure of the Lagrange equations (see the previous
item (ii)).
\parn
It is not the case to illustrate now the remaining seven classes of potentials
described by \cite{Fre}; we will meet each one of them in the sequel of this paper.
Here we only highlight that such potentials are built using the exponential
and some functions closely related to it (namely, hyperbolic or trigonometric functions),
together with their inverses. \parn
Paper \cite{Fre} subsequently passes from the Lagrangian to the
Hamiltonian formalism and investigates the Liouville integrable cases, i.e.,
the cases in which there is a second constant of motion besides
the Hamiltonian; this second constant of motion (when its exists)
is automatically in involution with the Hamiltonian and, since the system under
analysis has two degrees of freedom, the standard theories of
Liouville and Hamilton-Jacobi allow to solve Hamilton's equations
by quadratures. In this investigation, the authors of \cite{Fre}
benefit from the existing literature on Hamiltonian
systems with two degrees of freedom which possess a second constant
of motion. They extract from the previous literature $26$ ``sporadic''
classes of Hamiltonian systems with such features; these correspond to cosmological
models with $26$ classes of self-interaction potentials $\V(\f)$,
which are referred to as the ``sporadic potentials''.
\parn
Remarkably, the last two sporadic potentials considered in \cite{Fre}
have elementary trigonometric forms, and the associated cosmological models
are related via suitable transformations to two Toda-type lattices.
At least one of these two trigonometric potentials has a close relation with the
integrable models of class 9 mentioned before, a fact already noticed in \cite{Fre}.
It was later demonstrated by Sokolov and Sorin \cite{SokSor}
that all the ``sporadic potentials'' can actually be regarded as
particular or limit cases of the nine non-sporadic integrable classes.
\vspace{0.03cm}\\
\textbf{The present paper: adding matter or curvature to the Fr\`e-Sagnotti-Sorin
integrable cosmologies.} Gravity and the self-interacting scalar field
are the only actors in the cosmologies of \cite{Fre}.
It is natural to wonder if the integrable models of \cite{Fre}
can be generalized adding a (homogeneous) matter fluid and/or
removing the assumptions of
zero spatial curvature. This is the subject addressed in the present work;
here we extend a more limited analysis of the same issue performed in the PhD thesis
of one of us (M.G.) \cite{thesis}, that was supervised by the other
authors of the present work (D.F. and L.P.).
\newpage\noindent
As for the matter fluid, we admit a standard equation of state of the form
$\ppm = w \,\rom$, where $\ppm$ is the pressure,
$\rom$ is the density and $w$ is a constant; moreover, we consider
an unspecified value $\cu$ for the spatial curvature.
\parn
We assume no direct interaction between the matter fluid and
the scalar field so that, as already observed, there are separate
conservation laws for the corresponding stress-energy tensors.
The conservation equation for the matter fluid can be directly
integrated, yielding the explicit dependence of the density $\rom$ on the
scale factor. This information, as well as the presence of spatial curvature,
can be implemented in the Lagrangian formalism; in the end,
any cosmology of the type outlined above is described
as a Lagrangian system with two degrees of freedom,
in which the basic coordinates are (again) the instantaneous
values of the scale factor
and of the scalar field. \parn
The Lagrangian derived in this way contains, as
in \cite{Fre}, an unspecified
``gauge function'' related to the choice of the time
coordinate; in comparison with the cited work, our
Lagrangian has two additional terms depending
on the scale factor and on the gauge function,
corresponding to the matter fluid and to the spatial curvature.
({\footnote{The matter term also depends
on an unspecified constant, related to
the matter density at some reference time.\vspace{-0.cm}}}) \parn
The next step in this construction is the reconsideration
of the nine potential classes of \cite{Fre}, with
the related choices of the gauge function and
of new Lagrangian coordinates. It is natural to
wonder if such choices, allowing to solve
the evolution equations in the purely scalar models
of \cite{Fre}, do in fact ensure solvability also in presence of matter and/or spatial curvature,
for one of the reasons (i-iv) listed in the previous paragraph. \parn
This problem is addressed in the PhD thesis \cite{thesis}
for the first two potential classes of \cite{Fre}. Concerning
the class 1 potentials
$\V(\f) = $ const.$\,e^{\f}$ + const.$\, e^{-\f}$ + const.,
it is found that the model is still solvable
with zero spatial curvature and the
addition of matter with $w=0$ (dust), due to the linearity
of the evolution equation; indeed, the linearizability
of this cosmological model in spacetime dimension $3+1$
had already been established in
\cite{Piedipalumbo}, so we are just
extending the result of the cited papers
to an arbitrary dimension $n+1$. \parn
Concerning the class 2 potentials $\V(\f) = $ const.$\,e^{2 \ga \f}$ + const.$\, e^{(1 + \ga) \f}$,
the thesis \cite{thesis} finds that the system maintains
its integrability features (of the type indicated in (ii))
for suitable values of the parameter $\ga$
in presence of zero spatial curvature and matter with
arbitrary $w$, or in presence of arbitrary spatial curvature
and matter with $w=1/n$ or $w=2/n-1$ ($n+1$ is again
the spacetime dimension). It should be noted
that the case $w=1/n$, in which the stress-energy tensor
of matter has zero trace, can be interpreted as a radiation
gas. In all these cases, the value of $\ga$ must be appropriately
fixed (for example, $\ga=w$ in the case with $w$ arbitrary). \parn
The present work reports the above results from the thesis \cite{thesis},
completes the analysis of class 2 potentials finding
further integrable cases and then discusses the remaining
seven classes of potentials listed by \cite{Fre} showing
that, for each one of these classes, there are several integrable
extensions of the model with spatial curvature and/or matter.
\vspace{0.13cm}\\
\textbf{Detailed discussion of the solutions.}
Of course, after discovering the mechanism ensuring the
integrability of a cosmological model it is essential
to write explicitly its solution and
to analyze it from a qualitative and quantitative viewpoint,
so as to answer questions like the following:
Does the model exhibit a Big Bang?
Is there a Big Crunch, or does the universe exist forever (in terms of the
cosmic time)? What about the asymptotic behavior of the scale factor and
of the energy densities
of matter and of the scalar field near the Big Bang, near the Big Crunch,
or in the infinitely far future?
Which type of energy is dominating
in these limits? Is there a particle horizon associated to the Big Bang?
Is the model realistic for the whole history of universe, for most of it
or al least for some stage? If so,
can one fix the free parameters and/or the constants of integration
in the solution of the model so as to fit the available observational data? \parn
The integrable cases found in this paper
adding matter or spatial curvature to models from \cite{Fre} are a lot,
so it is not possible to treat explicitly the above issues for all of them.
Therefore, we perform the above mentioned discussion just for some case studies.
\vfill\eject\noindent
$\phantom{a}$\vspace{-0.9cm}\\
Firstly, we consider the case of self-interaction potential $\V(\f)$
belonging to class 1, with dust and no spatial curvature.
If one identifies dust with ordinary matter and the scalar field with dark energy,
this model describes with a good approximation the content of the universe
for most of its history, from the end of the radiation dominated era to the very far future.
The analysis presented here follows the thesis \cite{thesis}, and somehow
refines the investigation of the authors who discovered this integrable
case \cite{Piedipalumbo,Rub}.
The behavior of the solution
of this model depends on the parameters in the potential $\V(\f)$ and on the
integration constants; we show how to choose them so that
the early universe is dominated by matter, the late universe by
dark energy and the (dimensionless) energy densities of
these two entities at present time have the values suggested
by observational evidence (of course, in this computation
we assume that the spacetime dimension is $3+1$). \parn
Next, we consider a scalar field with a class 2 potential and the addition of
matter and curvature.
Among the many integrable cases of this model, listed elsewhere in the paper, we
choose the one with zero spatial curvature and matter with
arbitrary equation of state parameter $w$
(and with the parameter $\ga$ in $\V(\f)$ fixed by the previously mentioned prescription $\ga=w$,
which ensures the triangular structure of the Lagrange equations). Again, there are
many subcases of this model: we choose one with $0 <w<1$ and suitable
signs of the coefficients in the potential $\V(\f)$, which exhibits
a Big Bang and no Big Crunch. The asymptotic behavior of the relevant observables
near the Big Bang and in the very far future is determined for arbitrary $w \in (0,1)$.
Sticking to this subcase, we subsequently
fix the spacetime dimension to be $3+1$ and set $w=1/3$ (radiation gas). With
these choices, we individuate a solution of the Lagrange equations that,
although being entirely built with elementary functions, has a rather
complicated structure implying a stage in which the scale factor
grows exponentially with the cosmic time, preceded and followed by
epochs in which the scale factor behaves like a power of the cosmic time.
We show that the free parameters of the model and the constants of
integration appearing in this solution can be adjusted so that the exponential
growth occurs in the very early universe and the scale factor is increased,
say, by a factor $3 \times 10^{20}$ in a time interval between $0.5 \times 10^{-34}$
seconds and $1.5 \times 10^{-34}$ seconds after the Big Bang. This is the behavior
postulated by inflationary theories: we think it can be of some
interest to obtain such a behavior from an exact solution of the
Einstein equations with the simultaneous presence of radiation
and of a scalar field; clearly, the latter ought to be interpreted as
the inflaton in this model. \parn
The last case study considered in this paper is associated to class 7 potentials;
the spatial curvature is zero and a type of matter is present with $w= (\ell-1)/(\ell+1)$,
where $\ell \geqs 2$ is an integer. This case is discussed since
it provides a rather interesting example of separable system (see item (iii)
in the second last paragraph). Indeed, upon introducing a suitable pair $(x,y)$
of Lagrangian coordinates, the Lagrangian is found to be the sum of two Lagrangians
depending separately on $x,y$ (and their time derivatives). The first
Lagrangian describes a non-linear repulsor with potential
energy proportional to $-x^{2 \ell}$; the second one
described a non-linear oscillator with potential energy
proportional to $y^{2 \ell}$. Using the conservation of energy for these
separate subsystems, we derive quadrature formulas for their motions and
then return to the original variables of the model, i.e., the scale factor
and the scalar field, ultimately performing a qualitative and quantitative analysis
of their behavior. In this way we find, for example, that
the system exhibits a Big Bang and an exponential growth of
the scale factor (as a function of cosmic time) in the very
far future; at intermediate times, there is a competition
between the behaviors associated to the previously mentioned
repulsor and oscillator, whose effects depend on the
parameters in the potential $\V(\f)$ and on the values
assumed for the constants of integration.
\vspace{0.03cm}\\
\textbf{Organization of the paper.} Section \ref{bafre} and the related Appendix \ref{appa}
present  some general facts on cosmologies with a scalar field minimally
coupled to gravity and with a matter fluid (not interacting directly with
the scalar field, with a given equation of state $\ppm =\ppm(\rom)$). After
some generalities about the action functional and the stress-energy tensors
of the field and of the matter fluid,
we focus the attention on the homogeneous and isotropic case, in which
the spacetime metric has the FLRW form, and the equation of state for
matter is assumed to have the form $\ppm = w\,\rom$; this yields the Lagrangian
setting with two degrees of freedom mentioned in the previous paragraphs.
\newpage\noindent
Section \ref{admat} considers the nine potential classes of
Fr\`e, Sagnotti and Sorin, and lists the integrable cases that we have
obtained adding matter or curvature. Section
\ref{sec:Qual} and the related Appendices \ref{appsol},
\ref{appEsttc}, \ref{appclass7} present the explicit solutions for some integrable
cases of Section \ref{admat}, accompanied by a qualitative and quantitative
analysis. Here we present the results mentioned in the previous paragraph, i.e.:
a review of the Rubano-Scudellaro-Pedipalumbo-Esposito model with dust
\cite{Piedipalumbo,Rub} (subsection \ref{piedipal}), with a somehow refined
qualitative and quantitative analysis; a general discussion
of class 2 potentials with the addition of matter (subsection \ref{class2}),
that includes the previously mentioned  model for inflation
(paragraph \ref{Models2}); an analysis of
an integrable case with a class 7 potential and matter, yielding
the previously mentioned model with a nonlinear repulsor and
a nonlinear oscillator (subsection \ref{subclass7}).
\vspace{0.2cm}\\
\textbf{Final remarks.} (a) One could wonder if the present
integrability results (or those of \cite{Fre}) could be extended to the case
of non minimal coupling between gravity and the scalar field; we refer mainly
to the case of a standard curvature coupling,
in which the action functional for the system contains
a term proportional to $R\, \f^2$ ($R$ is the scalar curvature).
This problem certainly deserves further investigation. There is some
hope to obtain such extensions for the purely scalar models
of \cite{Fre}, using some formal transformations proposed
in the literature \cite{Fut,Maed} to connect minimally coupled theories to systems
with curvature coupling. However, the cited transformations refer to systems with
no type of matter fluid, so they cannot be used for the cosmologies with matter
of this work.
\parn
(b) We already pointed out that no direct interaction between the matter
fluid and the scalar field is ever considered in this paper. However,
let us mention that some integrable FLRW cosmological models with such an interaction
have previously appeared in the literature; we refer in particular, to the very recent
work of Piedipalumbo, De Laurentis and Capozziello
\cite{PLC}, where the scalar field represents dark energy and a possible
interaction with (dark) matter is considered (see also the references
cited therein).
\parn
(c) In most of the integrable cases presented in this work, a particle horizon appears;
this fact can be checked by hand noting that the reciprocal of the scale factor, viewed as a function
of cosmic time, diverges in a non-integrable way
at the Big Bang. In the case of non-positive spatial curvature,
the deep reason for this fact is explained
in \cite{phantom}; therein it is shown that
a particle horizon occurs in all homogeneous and
isotropic cosmologies with non positive spatial curvature,
a self-interacting scalar field minimally coupled to gravity
and a matter fluid
with equation of state $\ppm = w\,\rom$, fulfilling the
strong energy condition.
As shown in \cite{phantom}, the particle horizon is absent
if, instead of a canonical scalar field, one considers a
\textsl{phantom} field whose action functional contains
an anomalous term corresponding to a \textsl{negative} kinetic energy.
It would be of some interest to search for FLRW integrable cosmologies
with a phantom scalar field and matter; this subject is
left to future investigations.
\parn
(d) In the present work, in \cite{Fre} and in most of
the other previously cited papers, the attention is focused
on a ``direct problem'':
find for \textsl{arbitrary} initial data the solution of a cosmological model with
a pre-assigned potential for the
scalar field and, possibly, with matter having a suitable
equation of state. On the other hand, there is also an ``inverse problem":
find the scalar field self-potential producing
a time evolution with a prescribed feature
in a FLRW cosmologies with a purely scalar content, or including a matter fluid.
To our knowledge, the first examples of this inverse approach
date back to 1980's and 1990's: we will mention, in particular,
the papers by Lucchin and Matarrese \cite{Luc}, Barrow \cite{Bar90}, Ellis and
Madsen \cite{Ema},
Eashter \cite{East}. More recently, nice ``inverse'' results have
been obtained by Dimakis, Karagiorgos, Zampeli, Paliathanasis,
Christodoulakis and Terzis
\cite{Dima}, and by
Barrow and Paliathanasis
\cite{Barr}; the same approach is also partly employed in \cite{phantom},
for the case of a phantom field. The feature
specified in the cited papers to determine the scalar field potential is,
for example, the dependence on cosmic time of one of the following
observables: the scale factor, the Hubble parameter,
the ratio between the pressure and the density
produced by the scalar field alone, or jointly by scalar field
and matter.
The distinction between the ``direct'' and ``inverse''
problems outlined above is essential to understand
the difference between the present work and
the ones we have just mentioned.
\newpage\noindent

\section{The reference framework}\label{bafre}

\subsection{A general cosmological model with matter and a scalar field}
Throughout this paper we employ units in which the speed of light and
the reduced Planck's constant are $c=1$ and $\hbar=1$. As a consequence,
indicating with $\Le, \Ti$ and $\Ma$ the spaces of lengths, times
and masses we have $\Le=\Ti=\Ma^{-1}$. \parn
Let us introduce a cosmological model living in a spacetime of dimension
\beq d = \ds + 1\,, \quad \mbox{with\, $\ds = 2,3,4,...$} \feq
(of course, $\ds$ stands for the spatial dimension). Spacetime coordinates are
typically indicated with $(x^\mu)$, and the line element reads $d s^2 = g_{\mu \nu}\, d x^\mu d x^\nu$.
The metric $(g_{\mu\nu})$ has signature $(-,+,...,+)$ and the corresponding covariant
derivative, Ricci tensor and scalar curvature are respectively denoted with
$\nabla_{\!\mu}, R_{\mu \nu}$ and $R$.
\parn
We assume that the content of the universe consists of:\vspace{-0.2cm}
\begin{enumerate}
\item[(i)] a scalar field $\phi$ (of dimension $\Le^{-(\ds - 1)/2}$), minimally coupled to
gravity and self-interacting with potential $V(\phi)$ (of dimension $\Le^{-(\ds+1)})$;\vspace{-0.2cm}
\item[(ii)] some kind of matter which can be described as a perfect fluid with mass-energy $\rom$ and
pressure $\ppm$, fulfilling an assigned equation of state $\ppm = \ppm(\rom)$.
Such matter does not interact directly with the scalar field.
Let us also remark that here and in the sequel the term ``matter'' is used
in a very generic sense (e.g., it possibly refers to a radiation gas).
\end{enumerate}
The action functional $\mathcal{S}$ for the above model depends on the spacetime metric,
on the scalar field history and on the matter history (defined as in \cite{Hawk}) with the law
\beq \mathcal{S} :=\!
\int\! d^{\ds+1}x\, \sqrt{|g|}
\left[ {R \over 2 \kappa^2_\ds} - {1 \over 2}\, g^{\mu \nu} \partial_{\mu} \phi\, \partial_{\nu} \phi
- V(\phi) - \rom \right] , \label{Functional} \feq
where $g := \det (g_{\mu \nu})$ and $\kappa_\ds$ (of dimension $\Le^{(\ds-1)/2}$)
is, up to a numerical factor, the square root of the universal gravitational constant.
Note that $\mathcal{S}$ is dimensionless in our units with $\hbar=1$.
\parn
Demanding $\mathcal{S}$ to be stationary under variations of the metric $(g_{\mu\nu})$
entails the Einstein equations
\beq R_{\mu \nu} - {1 \over 2}\, g_{\mu \nu} R =
\kappa^2_\ds \big(\Tf_{\!\mu \nu} + \Tm_{\!\mu \nu}\big)~, \label{eqs1} \feq
where the r.h.s. contains the stress-energy tensors of
the scalar field and of the matter fluid:
\begin{gather}
\Tf_{\mu \nu} := \partial_{\mu} \phi\, \partial_{\nu} \phi - {1 \over 2}\, g_{\mu \nu}\,
\partial_{\alpha} \phi\, \partial^{\,\alpha} \phi - g_{\mu \nu}\, V(\phi)~; \label{tifield0} \\
\Tm_{\mu \nu} := \big(\ppm\!+\!\rom\big) U_\mu  U_\nu + \ppm g_{\mu \nu}~, \label{Tm}
\end{gather}
with $U^\mu$ indicating the $(\ds+1)$-velocity of the fluid.\parn
The stationary condition for $\mathcal{S}$ with respect to variations of the field $\phi$
gives the Klein-Gordon-type equation
\beq \Box\, \phi = V'(\phi) ~, \label{eqsX} \feq
where $\Box\, \phi := \nabla_{\!\mu} \nabla^\mu \phi = {1 \over \sqrt{|g|}}\,
\partial_\mu \big(\sqrt{|g|}\, g^{\mu\nu}\, \partial_\nu \phi\big)$
(recall that $g := \det (g_{\mu\nu})$).
\parn
Finally, the stationarity of $\mathcal{S}$ under variations of the matter history gives the conservation law
for the stress-energy tensor of the matter fluid, namely
\beq \nabla_{\!\mu}\, \Tm^{\mu}_{\nu} = 0~. \label{eqs2} \feq
Of course the Einstein equations \eqref{eqs1}, along with the Bianchi identity,
imply the conservation of the total stress-energy tensor $\Tf_{\!\mu \nu} + \Tm_{\!\mu \nu}$.
Combined with Eq.\,\eqref{eqs2}, this implies
\beq \nabla_{\!\mu}\, \Tf^{\mu}_{\nu} = 0 ~. \label{eqs3} \feq
On the other hand, from the explicit expression \eqref{tifield0} of $\Tf_{\!\mu\nu}$ one gets
\beq \nabla_{\!\mu}\, \Tf^{\mu}_{\nu} =
\big(\Box \, \phi - V'(\phi)\big)\,\partial_\nu\phi ~. \label{eqsY} \feq
Thus, Eqs.\,\eqref{eqs1}, \eqref{eqsX} and \eqref{eqs2} are not independent:
in fact, one has the chain of implications $\big($\eqref{eqs1} and \eqref{eqs2}$\big)$ $\Rightarrow$ \eqref{eqs3} $\Rightarrow$
\eqref{eqsX} (at points where $\partial_{\nu} \phi \neq 0$). These considerations
on Eqs.\,\eqref{eqs1} \eqref{eqsX} \eqref{eqs2} have partial converses
which are easily described
for special geometries, such as a FLRW spacetime:
in this case, to be addressed in the following, Eqs.\,\eqref{eqsX} \eqref{eqs2}
imply that some of the Einstein equations \eqref{eqs1} are actually constraints,
holding at all times if and only if they are fulfilled at a particular time.\parn
From here to the end of the paper, we assume that the
equation of state for the matter fluid reads
\beq \ppm = w\,\rom\,, \label{pwro} \feq
for some suitable real constant $w$, in principle arbitrary.
When $w=0$, the fluid is a \textsl{dust}; if $w=1/\ds$ the trace
$\Tm^{\mu}_{\mu}$ vanishes, as typical of a \textsl{radiation gas};
for $w = 1$ one speaks of \textsl{stiff matter}; if $w = -1$ matter
behaves like a \textsl{cosmological constant} (see the forthcoming Eqs.\,\eqref{wm1cosm}-\eqref{eqsma1bis} and
the related discussion).
Besides, let us mention that the weak, dominant and strong energy conditions
for $ \Tm_{\!\mu \nu}$ are respectively equivalent to (see, e.g., \cite{Hawk,Ma18})
\begin{gather}
\rom \geqs 0 ~, \quad w \geqs -1 ~, \label{C1} \\
\rom \geqs 0~, \quad -1 \leqs w \leqs 1 ~, \label{C2} \\
\rom \geqs 0~, \quad w \geqs {2 \over \ds} - 1 ~. \label{C3}
\end{gather}

\paragraph{Comparison with \cite{Fre}.}
As already stressed, \cite{Fre} considers a scalar field
as the only content of the universe; thus, any statement
of the present paper involving the matter fluid has no counterpart
in the cited work. \parn
Here and in the sequel, we employ notations
as close as possible to those of \cite{Fre}; however there are a few
minor differences, to be pointed out step by step.
For the moment, let us mention that our convention
$(-,+,+,...,+)$ for the metric signature is opposite to the convention $(+,-,-,...)$
employed in \cite{Fre}. Following \cite{Fre}, we
indicate with $d$ the spacetime dimension; however,
differently from \cite{Fre} we often refer to
the space dimension $\ds$ and write $d=\ds+1$.
In particular, our constant $\kappa_{\ds}$ coincides
with the quantity $k_{d}$ of \cite{Fre} (with $d=\ds+1$).
\parn
In the sequel, as in \cite{Fre} we restrict our
attention to the case of a FLRW
geometry that we describe using similar
notations, apart from the symbol $\tc$ for cosmic time
replacing the notation $t_{c}$ of \cite{Fre}. In addition,
let us stress that we admit arbitrary
values for the constant, spatial sectional curvature,
while \cite{Fre} discusses only the case of zero curvature.

\subsection{The homogeneous and isotropic case}\label{subsec:2.3}
From here to the end of the paper, the general model of the previous subsection
is specialized to the case of a spatially homogeneous and isotropic universe.
\vspace{-0.4cm}

\paragraph{Spacetime and its metric.}
To implement the above assumptions
we consider a FLRW spacetime, given by the product of the time line
and of a (simply connected) Riemannian manifold $\MK $ of
constant sectional curvature $\cu$ (of dimension $\Le^{-2}$).
Using the cosmic time $\tc$ and any system of coordinates
$\bx = (x^i)_{i=1,...,d}$ for $\MK$, we have
\beq d s^2 = - \,d \tc^2 + a^2(\tc)\,d \ell^2 =
- \,d \tc^2 + a^2(\tc)\,h_{i j}(\bx)\,d x^i d x^j\,, \label{robw0} \feq
where $d \ell^2 = h_{i j}(\bx)\, d x^i d x^j$ is the line element of $\MK$
and $a(\tc) > 0$ is the dimensionless scale factor; typically, the latter
is fixed so that $ a(\tcs)=1 $ at some reference time $\tcs$.
\parn
For our purposes it is convenient to use in place of $\tc$ a dimensionless
``time'' coordinate $t$, implicitly defined by
\beq d \tc = \te\,e^{\BB(t)} \, d t\,, \label{taut} \feq
where $\BB(t)$ is a dimensionless ``gauge function'', to be fixed according
to convenience, and $\te$ is a constant of dimension $\Ti \equiv \Le$.
This re-parametrization of time is suggested in \cite{Fre} where, however, the
analogue of Eq.\,\eqref{taut} contains no dimensional constant
$\te$ and reads $d \tc = e^{\BB(t)} dt$; due to this, the coordinate
$t$ of \cite{Fre} has dimension $\Ti$.
\parn
Having $\te$ at our disposal, we re-express the scalar curvature $\cu$
in terms of a dimensionless coefficient $k$, setting
\beq \cu = {k \over \te^2} ~; \feq
to compare with \cite{Fre}, let us recall that
$\cu=0$ therein. \parn
Having introduced the time coordinate $t$, we can regard the scale factor
as a function of it, i.e., $a = a(t)$; inspired again by \cite{Fre}, we write
\beq a(t) = e^{\AA(t)/\ds}\,, \label{aeaa} \feq
where $\AA$ is a dimensionless function. Substituting
Eqs.\,\eqref{taut} and \eqref{aeaa} into Eq.\,\eqref{robw0},
we obtain for the spacetime metric the representation
\beq d s^2 = -\,\te^2\, e^{2 \BB(t)} d t^2 + e^{2 \AA(t)/\ds} d \ell^2
\equiv -\,\te^2\, e^{2 \BB(t)} d t^2 + e^{2 \AA(t)/\ds} \, h_{i j}(\bx)\, d x^i d x^j ~, \label{robw} \feq
which coincides with the one given in \cite{Fre} apart from
the presence of the constant $\te$ and from the extension
to non-zero values for the curvature $\cu$ of $d \ell^2$.\parn
In the sequel we always use the spacetime coordinate system
\beq (x^\mu) \equiv (x^0, x^i):= (t, \bx) \qquad (\mu = 0,...,\ds;\, i=1,...,\ds)~, \label{sc0i} \feq
where, as above, $\bx = (x^i)$ are coordinates on $\MK$; Greek indexes
always range from $0$ to $\ds$, Latin indexes from $1$ to $\ds$.
Moreover, we indicate derivatives with respect to $t$ with dots, namely,\vspace{-0.1cm}
\beq \dot{~} \, \equiv {d \over d t}~. \feq
In Appendix \ref{appa} we report the explicit
expressions of the Ricci tensor $R_{\mu \nu}$
and of the scalar curvature $R$ for the metric
\eqref{robw}. \parn
Let us indicate with $U^\mu$ the $(\ds+1)$-velocity of the
FLRW frame (i.e., the future-directed, timelike vector field
tangent to the lines with $\bx=$ const., normalized so that $U^{\mu} U_{\mu} = -1$);
we have
\beq U^\mu = \te^{-1}\,e^{-\BB(t)} \;\delta^{\mu}_{~ 0}~,
\qquad  U_\mu = -\,\te\, e^{\BB(t)} \,\delta_{\mu 0}~.
\label{umurob} \feq

\paragraph{Scalar field and matter content.}
Let us now introduce the dimensionless rescaled versions $\f,\VV$
of the field and of the potential, defined so that
\beq \phi = \sqrt{{\ds-1 \over \ds}}~ {\f \over \kappa_\ds}~,
\qquad V(\phi) = {\ds-1 \over \ds}\; {\VV(\f) \over \kappa^2_\ds \,\te^2}~. \label{fiad} \feq
In the sequel, the terms ``scalar field'' and ``potential'' will be
frequently employed to indicate these rescaled quantities.
Let us also remark that in \cite{Fre} there are similar rescaled objects
$\f_{\Fre} = \f$ and $\VV_{\Fre}(\f_{\Fre}) = \VV(\f)/\theta^2$. \parn
To comply with the hypothesis of spatial homogeneity, we assume that the field
and the matter density depend only on time:
\beq \f = \f(t)~, \qquad \rom = \rom(t)~. \label{firomt} \feq
In addition, we assume the matter fluid to be at rest
in the FLRW frame, which entails that its $(\ds+1)$-velocity $U^\mu$
is fixed as in Eq.\,\eqref{umurob}. Let us also recall that we are considering
an equation of state of the form $\ppm = w \,\rom$ (see Eq.\,\eqref{pwro}).
\parn
In Appendix \ref{appa} we compute the stress-energy
tensors of the scalar field and matter fluid
starting from the general expressions \eqref{tifield0} \eqref{Tm}
and implementing the assumptions \eqref{pwro} \eqref{fiad}
\eqref{firomt}. The conclusion is that $\Tf_{\mu \nu}$ has the form
of the stress-energy tensor for a perfect fluid with the $(\ds+1)$-velocity
$U^\mu$ of the FLRW frame (see Eq.\,\eqref{umurob}), and
with appropriate density and pressure; more precisely,
\beq
\begin{array}{c}
 \dd{\Tf_{\mu \nu} = \big(\pfi +\rofi\big)\, U_\mu\, U_\nu + \pfi\, g_{\mu \nu}~,}
  \vspace{0.2cm} \\
\dd{\rofi \!:= {1 \over \kappa^2_\ds\, \te^2}\, {\ds-1 \over \ds}
\left(e^{- 2 \BB(t)}\, {\dot{\f}^2 \over 2} + \VV(\f)\right), \qquad
\pfi \!:= {1 \over \kappa^2_\ds\, \te^2}\, {\ds-1 \over \ds}
\left(e^{- 2 \BB(t)}\, {\dot{\f}^2 \over 2} - \VV(\f)\right)} .
\end{array}
\label{rofi}
\feq
In the sequel, we often refer to the ``equation of state coefficient''
\beq
\wF := \frac{\pfi}{\rofi}~, \label{defwfi}
\feq
depending on $ t $ and defined whenever $ \rofi(t) \neq 0$.
\vspace{-0.1cm}

\paragraph{Evolution equations.}
We refer to Appendix \ref{appa} for all the statements reported in this paragraph.
Let us first notice that the conservation law \eqref{eqs2} for the stress-energy tensor
of the matter fluid
is fulfilled if and only if $\rom(t) = \roms\, e^{-(w+1) \AA(t)}$, where
$\roms$ is an integration constant with the dimension of $\rom$, i.e.,
$\Ma/\Le^{\ds} = \Le^{-(\ds + 1)}$. For future convenience we set
$\roms \!= \ds(\ds-1)\, \Oms/(2\, \kappa^2_\ds\, \te^2)$, with $\Oms$
a dimensionless constant, so that
\beq \rom = {\ds (\ds-1)\, \Oms \over 2\,\kappa^2_\ds\, \te^2}\; e^{-(w+1) \AA}~. \label{pmrom1} \feq
Note that $ \mbox{sgn}(\rom) = \mbox{sgn}\big(\Oms\big)$ at all times;
unless otherwise stated, in the sequel we will typically assume $\Oms \geq 0$.
\parn
Next, let us consider the Einstein equations \eqref{eqs1}, that we re-write using
the above expression for $\rom$ and the related expression for $\ppm = w\,\rom$.
There are just two independent equations, respectively corresponding to the group of indexes
$(\mu,\nu) \!=\! (i,j) \!\in\! \{1,...,\ds\}^2$ and $(\mu,\nu) \!=\! (0,0)$ in Eq.\! \eqref{eqs1}:
\beq \begin{array}{c}
\dd{\einsu = 0 ~,} \vspace{0.1cm}\\
\dd{\einsu := \ddot{\AA} + {\dot{\AA}^2 \over 2} - \dot{\AA}\, \dot{\BB}
+ {\dot{\f}^2 \over 2} - e^{2 \BB}\, \VV(\f)
+ {\ds^2\, \Oms\, w \over 2} \, e^{2 \BB -(w+1) \AA}
+ {\ds (\ds-2)\, k \over 2} \, e^{2\BB - 2 \AA/\ds} \,;}
\end{array} \label{EE1} \feq
\beq \begin{array}{c}
\dd{\einsd =0 ~,} \vspace{0.1cm}\\
\dd{\einsd :=  {\dot{\AA}^2 \over 2} - {\ft^2 \over 2} - e^{2 \BB}\, \VV(\f)
- {\ds^2\, \Oms \over 2} \, e^{2 \BB -(w+1) \AA} + {\ds^2\, k \over 2} \, e^{2 \BB - 2 \AA/\ds}\,.}
\end{array}\label{EE2} \feq
Finally, note that the field equation \eqref{eqsX} becomes (with $\VV' := d \VV/d \f$)
\beq \einsf = 0~, \qquad \einsf :=
\ftt + (\dot{\AA} - \dot{\BB}) \, \ft + e^{2 \BB} \, \VV'(\f)~. \label{ensca} \feq
Before proceeding, let us point out that a triple equivalent to the above set of equations
\eqref{EE1} \eqref{EE2} \eqref{ensca}
is obviously given by $\einsu - \einsd=0$, $2 \einsd=0$, $\einsf=0$;
for $\Oms=0$ and $k=0$, the latter triple coincides with
that reported in Eq.\,(2.12) of \cite{Fre}.
\vspace{-0.1cm}

\paragraph{An anticipation.}
For the moment, $\BB$ is treated as an unspecified function of $t$
(the same viewpoint is assumed in Appendix \ref{appa}). Starting from the forthcoming
subsection \ref{secgaug} to the end of the paper, following \cite{Fre} we will assume
$\BB(t) = \BBB(\AA(t),\f(t))$ for some assigned
function $\BBB$, and refer to this procedure as
a \textsl{gauge fixing}. Thus, $\AA$ and $\f$ will
be ultimately recognized as the true degrees of
freedom of the model.
\vfill\eject\noindent
$\phantom{a}$\vspace{-1.4cm}

\paragraph{Independence considerations.}
Regardless of the previously mentioned gauge fixing, Eqs.\,\eqref{EE1} \eqref{EE2} \eqref{ensca}
are not independent, as illustrated in the forthcoming items (i)(ii).\vspace{-0.2cm}
\begin{enumerate}
\item[(i)] We already pointed out that, in view of Eqs.\,\eqref{eqs3}\eqref{eqsY},
the field equation $\einsf = 0$ (equivalent to Eq.\,\eqref{eqsX})
is in fact a consequence of the other evolution equations $\einsu = 0$, $\einsd = 0$
(equivalent to Eqs.\,\eqref{eqs1}\eqref{eqs2}) in the spacetime
region where the scalar field is non-constant.
As a matter of fact, in the present setting it can be checked by direct computations that
$\ft\,\einsf = \dot{\AA} \, \einsu  - \,\dot{\einsd} -
(\dot{\AA} - 2 \dot{\BB})\, \einsd$, yielding
\beq \einsu =0,~ \einsd=0 \quad \Rightarrow \quad \einsf=0~~\mbox{when}~~
\ft \neq 0\,. \label{230} \vspace{-0.2cm}\feq
\item[(ii)] As a partial converse, let us consider the relations $\einsu =0$,
$\einsf=0$ supplemented with the initial condition $\einsd(t_0) = 0$
(requiring $\einsd$ to vanish at some given time $t_0$); we claim that
\beq
\einsu =0,~ \einsf=0,~ \einsd(t_0) = 0 \quad \Rightarrow \quad \einsd=0~~ \mbox{at all times}.
\feq
To prove this, let us reconsider the identity in the text line before Eq.
\eqref{230}. If $\einsu =0$, $\einsf=0$ (at all times),
this implies $\dot{\einsd} + \,(\dot{\AA} - 2 \dot{\BB} ) \, \einsd = 0$ whence
$(d/ d t) (e^{\AA - 2 \BB} \einsd)=0$; the latter relation, supplemented with
the initial condition $\einsd(t_0)=0$, gives $\einsd=0$ at all times.\vspace{-0.2cm}
\end{enumerate}
In the sequel we stick to the viewpoint expressed in item (ii): we regard
$\einsu=0$ and $\einsf=0$ as the authentic evolution equations for the
model, and $\einsd=0$ as a \textsl{constraint} that is fulfilled at
all times as soon as it is fulfilled by the initial data at some fixed time $t_0$.
\vspace{-0.4cm}

\paragraph{Solutions with maximal domain; Big Bang and Big Crunch.}\label{subsec:2.5}
Of course, each solution $(\AA(t),\BB(t), \f(t))$ of the system
$\einsu=0$, $\einsf=0$ (and $\einsd=0$)
is well defined for $ t $ in a suitable interval $I \subset \reali$.
From now on, when we speak of a solution we always assume $ I $ to be maximal
(i.e., that the solution cannot be extended to a larger interval).
Let $ I = (t_{in}, t_{fin}) $, where $ -\infty \leqs t_{in} < t_{fin} \leqs +\infty $.
Recall that $a(t) \equiv e^{\AA(t)/\ds}$ is the scale factor and that
$t$, $ \tc $ are related by Eq.\,\eqref{taut}, which is equivalent to\vspace{-0.05cm}
\beq \tc(t) \,=\, \te \!\int_{t_{r}}^ t\! dt'\; e^{\BB(t')} \label{BIGBANG} \feq
(here $ t_{r}$ is chosen arbitrarily).
If $ a(t) \rightarrow 0 $  (i.e., $\AA(t) \to - \infty$) for $ t \rightarrow t_{in}^{+} $
and $e^{\BB(t)}$ is integrable in a right neighborhood of $t_{in}$
(initial singularity at a finite cosmic time),
we say that the model has a  \textit{Big Bang} at $ t = t_{in}$.
If $a(t) = e^{\AA(t)/\ds}$ vanishes and $e^{\BB(t)}$ is integrable
for $ t \rightarrow t_{fin}^{-}$ (final singularity at a finite
cosmic time),
we say that the model has a  \textsl{Big Crunch} at $ t = t_{fin}$.
\vspace{-0.4cm}

\paragraph{Particle horizon.}
Suppose the model has a Big Bang at $\tcin = \tc(t_{in})$.
The lapse of conformal time that has passed from the Big Bang
to any cosmic time $\tc = \tc(t)$ is\vspace{-0.1cm}
\beq \Theta(\tc) \,:= \int_{\tcin}^{\tc}\! \frac{d \tcp}{a(\tcp)} \;=\;
\te \int_{t_{in}}^{t}\!\! dt'\; e^{\BB(t') - \AA(t')/n} ~. \label{horizon} \feq
The above integral can be finite or infinite.
The interpretation of $\Theta(\tc)$ is well known,
and can be summarized as follows, writing $\bf{p}_0, \bf{p}$, etc. for
the points of $\MK$ and $\mbox{dist}$ for the distance on $\MK$ related
to the metric $d \ell^2$ (see Eq.\,\eqref{robw0}): for each $\bf{p} \in \MK$,
the ball $\mathcal{B}({\bf{p}},\tc) :=
\{ {\bf{p}}_0 \in \MK~|~\mbox{dist}({\bf{p}}_0, {\bf{p}}) \leq \Theta(\tc)\}$
is the subset of $\MK$ formed by the points $\bf{p}_0$ which had
the time to interact causally with $\bf{p}$ from the Big Bang up to $\tc$
({\footnote{In fact, it can be shown that there exists a causal curve starting
from $\bf{p}_0$ at a time $\tcz \in (\tcin,\tc)$ and
ending at $\bf{p}$ at time $\tc$ if and only if
$\mbox{dist}({\bf{p}_0}, {\bf{p}}) \leq \Theta(\tc)$.\vspace{-0.4cm}}}).
This subset is the whole $\MK$ if and only if $\Theta(\tc) \geqs \delta_{\cu}$,
where $\delta_{\cu} := \,\sup \{ \mbox{dist}({\bf{p}}_0, {\bf{p}})~|~{\bf{p}}_0 \in \MK \}$
is the diameter of $\MK$, in fact independent of $\bf{p}$.
One has $\delta_{\cu} = + \infty$ if $\cu \leqs 0$ and $\delta_{\cu} = \pi/\sqrt{\cu} =
\theta \pi/\sqrt{k}$ if $\cu >0$.
\parn
Of course the situation where $\Theta(\tc) \geqs \delta_{\cu}$ is
of special interest, since it allows to explain the homogeneity of the universe
at time $\tc$ by standard thermodynamical arguments. In the opposite case
$\Theta(\tc) < \delta_{\cu}$, we say that \textsl{there is a particle horizon}
at time $\tc$; when $k \leqs 0$
the condition for a particle horizon reads $\Theta(\tc) < +\infty$,
and this happens at some time $\tc$ if and only if it happens at all
times $\tc$ after the Big Bang. \parn
Many FLRW cosmologies present
particle horizons; it was shown in \cite{phantom} that
any FLRW cosmology with $k \leqs 0$,
a (minimally coupled) scalar field and a matter fluid
with equation of state $\ppm = w \rom$ has a particle
horizon ($\Theta(\tc) < +\infty$
for all $\tc$ after the Big Bang) if the matter fluid fulfills in
the  strict sense the strong energy condition
(i.e., if the inequalities for $\rom, w$ in Eq.\,\eqref{C3} hold
strictly, with $\geqs$ replaced by $>$).
\vspace{-0.3cm}

\paragraph{Cosmological constant behavior for matter.} Let us specialize our considerations
to the case where the parameter in the equation of state \eqref{pwro} for matter is
\beq\label{wm1cosm}
w=-1~.
\feq
With this position, the equation of state itself and Eq.\,\eqref{pmrom1} reduce to
\beq \ppm = \, \mbox{const.} = - \rom \,, \qquad
\rom = \, \mbox{const.} = {\ds (\ds-1)\, \Oms \over 2\,\kappa^2_\ds\, \te^2}~, \label{pmrom1wm1} \feq
which entail for the matter stress-energy tensor the expression
\beq \Tm_{\!\mu \nu} := - {\ds (\ds-1)\, \Oms \over 2\,\kappa^2_\ds\, \te^2}\; g_{\mu \nu}~. \label{Tmwm1} \feq
Moving this term from the left-hand side to the right-hand side of the Einstein equations \eqref{eqs1} we get
\beq R_{\mu \nu} - {1 \over 2}\, g_{\mu \nu} R +
{\ds(\ds-1)\, \Oms \over 2\, \te^2}\; g_{\mu \nu} =
\kappa^2_{\ds} \, \Tf_{\mu \nu}~,  \label{eqsma1bis} \feq
corresponding to a model with cosmological constant
$\Lambda = \ds (\ds-1)\, \Oms/(2\, \te^2)$ (of dimension $\Le^{-2}$).\vspace{-0.2cm}

\paragraph{Cosmological constant behavior for the field.} \label{subsec:2.7}
Let us now search for a solution with
\beq \f(t) = \mbox{const.} \equiv \f_{0} ~. \label{cosm1} \feq
Eq.\,\eqref{ensca} entails that the above condition can be fulfilled if and only if
\beq \VV'(\f_0) = 0 ~. \label{cosm2}\feq
In this case, the field stress-energy tensor becomes (recall Eq.\,\eqref{rofi})
\beq \Tf_{\mu \nu} = - \,{\ds - 1 \over \ds\,\kappa^2_{\ds}\,\te^2}\; \VV(\f_0)
\, g_{\mu \nu} ~. \label{tifield2} \feq
Bringing this term to the right-hand side of the Einstein equations \eqref{eqs1} we obtain
\beq R_{\mu \nu} - {1 \over 2}\, g_{\mu \nu} R + {\ds(\ds -1) \, \OmL \over 2\,\te^2} \, g_{\mu \nu} =
\kappa^2_{\ds} \, \Tm_{\mu \nu}~, \qquad \OmL := {2 \over \ds^2} \,\VV(\f_{0})~, \label{eqs1bis} \feq
corresponding to a model with a cosmological constant $\Lambda = \ds(\ds-1)\,\OmL/(2 \, \te^2)$
(note that $ \OmL$ is dimensionless while $\Lambda$ has dimension $\Le^{-2}$, as expected).
\parn
Let us also mention that, according to Eqs.\,\eqref{rofi} \eqref{defwfi}
\beq \f =~\mbox{const.}~= \f_0 \quad \Leftrightarrow \quad \pfi = - \rofi \quad \Leftrightarrow \quad
\wF = - 1\,, \label{wficos}
\feq
where the second equivalence holds under the complementary assumption
$\rofi \neq 0$, or $\VV(\f_0) \neq 0$.
\vspace{-0.1cm}

\paragraph{Hubble parameter.}
The time-dependent Hubble parameter is given by
\beq H := {1 \over a}\,{d a \over d\tc} = {e^{-\BB} \dot{\AA} \over \ds\,\te}~. \label{H} \feq
Here the first equality is the standard definition in terms of the
scale factor $a$ and cosmic time $\tc$, while the second equality follows
from Eqs.\,\eqref{taut} \eqref{aeaa}.
\vfill\eject\noindent

\paragraph{The dimensionless density parameters.}
These are the time-dependent quantities
\beq \Omm := {2\,\kappa^2_{\ds} \over \ds (\ds -1)}\, {\rom \over H^2}~, \qquad\;
\Omf := {2\,\kappa^2_{\ds} \over \ds (\ds -1)}\, {\rofi \over H^2}~, \qquad\;
\Omk := -\,{k \over \te^2 H^2\, a^2}~. \label{Omegadef} \feq
From Eqs.\,\eqref{aeaa}, \eqref{rofi}, \eqref{pmrom1} and \eqref{H} we get
\beq \Omm = {\ds^2\, \Oms \, e^{2 \BB - (w+1) \AA} \over \dot{\AA}^{2}}\, , \qquad
\Omf = {\ft^2 + 2\,e^{2 \BB} \,\VV(\f) \over \dot{\AA}^{2}} \, , \qquad
\Omk = -\,{ \ds^2 \, k\; e^{2 \BB - 2 \AA/\ds} \over \dot{\AA}^{2}}\, . \label{Omega} \feq
By comparison with Eq.\,\eqref{EE2}, we see that
\beq \einsd =0 \quad \Leftrightarrow \quad \Omm +\Omf + \Omk = 1~.
\label{OmegaSum}
\feq
The parameters $\Omm$ and $\Omk$ are standard objects in cosmology (see, e.g., \cite{Wei}).
$\Omf$ plays a role similar to the dimensionless parameter $\OmL : = 2\,\Lambda/(\ds (\ds-1) H^2)$
usually considered when a cosmological term $\Lambda\, g_{\mu \nu}$ is present
in the Einstein equations. \parn
In agreement with the remark after Eq.\,\eqref{robw0}, in the sequel we often set
$a(t_{*})=1$, i.e. $\AA(t_{*}) = 0$,
at some reference time $t_{*}$. Moreover, fixing $\te := 1/|H(t_{*})|$\,,
from Eq.\,\eqref{H} we obtain $|\dot{\AA}(t_{*})| = \ds \, e^{\BB(t_{*})}$\,. By comparison
with the first relation in Eq.\,\eqref{Omega}, these facts entail the identity
\beq
\Omm(t_{*})= \Oms ~. \label{OmsId}
\feq

\subsection{Lagrangian viewpoint}
Let us return to the general expression \eqref{Functional} for the action functional,
and evaluate it on a history of the type considered in subsection \ref{subsec:2.3}.
A computation sketched in Appendix \ref{appa} yields
\beq \mathcal{S} = {1 \over \kappa^2_\ds \, \te}\!
\int\! d^\ds \bx\, \sqrt{h(\bx)} \int\! d t
\left[ {\ds - 1 \over \ds}\, \LLL(\AA, \f, \BB,\dot{\AA}, \ft) +
{d \over dt}\! \left( e^{\AA - \BB} \dot{\AA} \right) \right] ,\!
\label{esse} \feq
where $h(\bx) := \det(h_{i j}(\bx))$ and
\beq
\LLL(\AA, \f, \BB,\dot{\AA}, \ft) := e^{\AA - \BB} \left( -\,{\dot{\AA}^2 \over 2} + {\dot{\f}^2 \over 2} \right)
- e^{\AA + \BB} \, \VV(\f)
- {\ds^2\, \Oms \over 2}\; e^{-w \AA + \BB}
+ {\ds^2\,k \over 2} \; e^{{\ds - 2 \over \ds} \AA + \BB}\,. \label{eLL}
\feq
In Eq.\,\eqref{esse}, the integral $\int\! d^n \bx\,\sqrt{h(\bx)}$
is an irrelevant multiplicative factor (although infinite if $k \!\leqs\! 0 $);
the total $t$-derivative in the integral is also irrelevant.
In conclusion, $\mathcal{S}$ is related to the (dimensionless) Lagrangian function $\LLL$
written in Eq.\,\eqref{eLL}, which is degenerate since it does not depend on $\dot{\BB}$. \parn
Independently of the previous considerations, it can be checked by direct computations
that the Lagrange equations induced by $\LLL$ are equivalent to the evolution
equations of the model under analysis. In fact, the Lagrangian derivatives
\beq {\delta \LLL \over \delta q} := -\,{d \over d t}\left( {\partial \LLL \over \partial \qt} \right)
+\, {\partial \LLL \over \partial q} \qquad (q=\AA,\f,\BB) \label{LagDer}\feq
are such that
\beq {\delta \LLL \over \delta \AA} = e^{\AA - \BB} \, \einsu~, \qquad
{\delta \LLL \over \delta \f}= -\,e^{\AA - \BB} \, \einsf ~, \qquad
{\delta \LLL \over \delta \BB} = e^{\AA - \BB} \, \einsd~, \feq
which ensures the equivalence between the Lagrange equations
$\delta \LLL/\delta q =0~ (q=\AA,\f,\BB)$ and the evolution equations
$\einsu=0$, $\einsf=0$, $\einsd=0$ (see Eqs.\,\eqref{EE1} \eqref{EE2} \eqref{ensca}).
We already noted that such evolution equations are not independent;
from the present Lagrangian viewpoint, this is a consequence
of the degeneracy of $\LLL$.\parn
Finally, let us mention that for $\Oms=0$ and $k=0$ the Lagrangian \eqref{eLL}
coincides with the one appearing in Eq. (2.11) of \cite{Fre}.

\subsection{Gauge fixing and the energy constraint}
\label{secgaug}
From here to the end of this work we assume that
\beq \BB = \BBB(\AA,\f)~, \label{bB} \feq
where $\BBB$ is a suitable function, referred to as the \textsl{gauge function} in the sequel
(the same attitude is proposed in \cite{Fre} for the special case $\Oms=k=0$).\parn
Of course, the evolution equations are still $\einsu=0$, $\einsf=0$, $\einsd=0$.
Besides, the results of the previous paragraphs continue to hold, with $\BB$ fixed
according to Eq.\,\eqref{bB} and
\beq \dot{\BB} = \partial_\AA \BBB(\AA,\f)\,\dot{\AA} +
\partial_\f \BBB(\AA,\f)\, \ft~.
\label{bBt} \feq
Under the same gauge fixing, the Lagrangian $\LLL$ of Eq.
\eqref{eLL} becomes
\begin{gather}
\LL(\AA,\f, \dot{\AA},\ft) := \label{eL}  \\
 e^{\AA - \BBB(\AA,\f)} \!\left(\! -\,{\dot{\AA}^2 \over 2} + {\dot{\f}^2 \over 2} \right)
- e^{\AA + \BBB(\AA,\f)} \, \VV(\f)
- \,{\ds^2\, \Oms \over 2}\; e^{-w \AA + \BBB(\AA,\f)}
+ \,{\ds^2\, k \over 2} \; e^{{\ds - 2 \over \ds} \AA + \BBB(\AA,\f)} \,.  \nonumber
\end{gather}
Note that $\LL$ is a non-degenerate Lagrangian of mechanical type,
whose kinetic part is induced by a metric of signature $(-,+)$ on
the $(\AA,\f)$ configuration space.
Let us introduce the Lagrangian derivatives (cf. Eq.\,\eqref{LagDer})
\beq {\delta \LL \over \delta q} := -\,{d \over d t}\left( {\partial \LL \over \partial \qt} \right)
+ {\partial \LL \over \partial q} \qquad (q=\AA,\f) \feq
and the energy function
\begin{gather}
\Ec := \sum_{q\,=\,\AA,\,\f} \qt\, {\partial \LL \over \partial \qt} - \LL \label{efun} \\
= e^{\AA - \BBB(\AA,\f)}\! \left( -\, {\dot{\AA}^2 \over 2} + {\dot{\f}^2 \over 2} \right)
+ \,e^{\AA + \BBB(\AA,\f)} \, \VV(\f)
+ \,{\ds^2\, \Oms \over 2}\, e^{-w \AA + \BBB(\AA,\f)}
- \,{\ds^2\, k \over 2} \, e^{{\ds - 2 \over \ds} \AA + \BBB(\AA,\f)} \,. \nonumber
\end{gather}
Of course, $\Ec$ is a constant of motion for the Lagrange equations
$\delta \LL/\delta q =0~ (q=\AA,\f)$. Moreover, it can be easily checked that
\beq \left( \barray{c} \delta \LL/\delta \AA \\ \delta \LL/\delta \f \\ \Ec
\farray \right) = e^{\AA - \BBB}
\left( \barray{ccc} 1 & 0 & \partial_\AA \BBB \\
 0 & - 1 & \partial_\f \BBB \\
 0 & 0 & - 1 \farray \right)\!
\left( \barray{c} \einsu \\ \einsf \\ \einsd \farray \right) , \label{system3} \feq
\beq \left( \barray{c} \einsu \\ \einsf \\ \einsd \farray \right) = e^{\BBB -\AA}
\left( \barray{ccc} 1 & 0 & \partial_\AA \BBB \\
 0 & -1 & - \partial_\f\BBB \\
 0 & 0 & - 1 \farray \right) \!\left( \barray{c} \delta \LL/\delta \AA \\ \delta \LL/\delta \f \\ \Ec
\farray \right) , \label{system4} \feq
where $\einsu,\einsf,\einsd$ are evaluated with $\BB,\dot{\BB}$ as in Eqs.\,\eqref{bB} \eqref{bBt}.
\parn
Summing up: \textsl{after gauge fixing, the evolution equations
$\einsu=0$, $\einsf=0$, $\einsd=0$ are equivalent to the
Lagrange equations $\delta \LL/\delta q =0~ (q=\AA,\f)$, supplemented
with the condition $\Ec=0$} (the latter condition is satisfied at all
times if and only if it is fulfilled by the initial datum
$\big(\AA(t_0), \dot{\AA}(t_0), \f(t_0), \ft(t_0)\big)$). \parn
From now on, to analyze the dynamics of our cosmological model we systematically refer
to the Lagrangian $\LL$ of Eq.\,\eqref{eL} and to the \textsl{energy constraint} $\Ec=0$.
Whenever we speak of a solution of (one or all) these equations, we always tacitly
assume that the interval of definition is maximal;
this convention is consistent with the domain prescriptions of subsection \ref{subsec:2.5},
and will be applied also to the solutions obtained using Lagrangian coordinates different
from $(\AA,\f)$ (say, the coordinates $ (x,y) $ of the next sections).
The plan for the sequel is to consider specific choices for $ \VV $, allowing to solve
explicitly the corresponding Lagrange equations.
\vfill\eject\noindent

\section{Adding matter and curvature to the integrable models of Fr\'e, Sa\-gnot\-ti and Sorin}\label{admat}
Let us repeat once more that \cite{Fre} considers purely scalar, spatially
flat cosmologies, i.e., models with no matter content and zero spatial
curvature. Referring to this framework, Fr\'e, Sagnotti and Sorin identified
nine classes of self-interaction potentials $\VV(\f)$ for the scalar field that,
after an appropriate gauge fixing $\BB = \BBB(\AA,\f)$ and a suitable coordinate
transformation for the Lagrangian $\LL(\AA,\f, \dot{\AA}, \dot{\f})$, produce solvable
Lagrange equations.
The gauge function $\BBB(\AA,\f)$ and the coordinate transformation just mentioned
are given explicitly in \cite{Fre} (together with the energy constraint) for each
one of the nine potential classes; these results are summarized in \cite{Fre}.
\parn
In this section we show that, for all classes of potentials in the cited
paper \ref{Fre}, extended cosmological models including matter and possibly curvature
can be introduced and solved explicitly using the same coordinate transformations
employed in \cite{Fre} for the corresponding, purely scalar cosmologies.
In these extended cosmologies the matter fluid has an equation of state of the form
$\ppm = w\, \rom$ (see Eq.\,\eqref{pwro}), where the coefficient $w$ either has a fixed specific value or remains
arbitrary. In the cases with arbitrary $w$ (occurring for three of the nine potential classes),
some free parameter $\gamma$ labeling the potentials becomes a prescribed function of $w$.
\parn
To the best of our knowledge, the possibility to build integrable
extensions with matter or curvature was previously unknown for
all the cosmologies in \cite{Fre}, with the notable exception
of class 1 potentials which was analyzed in
\cite{Piedipalumbo} a short time before the publication of \cite{Fre}
in the case of matter with $w=0$ (dust), zero curvature and space dimension $\ds=3$.
\parn
The following subsections \ref{subsclass1}-\ref{subsclass9} present extended
cosmologies for the nine potential classes in \cite{Fre},
starting from the case of \cite{Piedipalumbo} (here generalized to
an arbitrary space dimension). In each subsection we indicate how the Lagrangian
function can be reduced by a proper gauge fixing and a suitable coordinates transformation
to one of the canonical forms analyzed in the forthcoming paragraph.
Following the strategies outlined in the said paragraph, the Lagrange equations
can be systematically reduced to quadratures in all cases of interest;
in particular, explicit expressions for the corresponding solutions can always be derived.
These expressions can be used to investigate the chief qualitative features
of each specific model: presence of a Big Bang and corresponding asymptotic behavior;
presence of a Big Crunch or, in absence of it, long time evolution of the system;
behavior of the density parameters. We will exemplify these issues for some
subcases of the nine classes in Section \ref{sec:Qual}.
\vspace{-0.3cm}

\paragraph{Solvable Lagrangian systems arising in the analysis of the nine potential classes.}
In the subsequent subsections \ref{subsclass1}--\ref{subsclass9}
we will replace the Lagrangian coordinates $\AA,\f$ with either
a new pair of real coordinates or with a complex one, with the rationale of obtaining
simple canonical forms for the Lagrange equations.
Under these coordinate transformations, the Lagrangian function $\LL$
assumes one of the forms described below (which are actually the same
forms occurring in \cite{Fre} in the case of zero spatial curvature and no matter content).
\parn
Let us point out a fact that will never be mentioned again in
the sequel: like the quantities $\AA, \f, \VV(\f)$, the new Lagrangian coordinates
$x, y, \xi, \eta$, etc. introduced in the sequel are all dimensionless.
\vspace{0.1cm}\\
\textsl{\quadr Quadratic Lagrangian.}
Assume that there exist two real Lagrangian coordinates $x,y$ such that $\LL(x,y,\xt,\yt)$
is the difference between two quadratic functions of the variables $(\xt,\yt)$ and
$(x,y)$. In this case the Lagrange equations
are linear and can be decoupled via additional
linear coordinate transformations. It is unnecessary to
give further details on this elementary
case, that will appear in subsection \ref{subsclass1}.
\vspace{0.1cm}\\
\textsl{\tria Triangular Lagrangian.}
Assume that there exist two real Lagrangian coordinates $x,y$ such that
\beq \LL(x,y,\xt,\yt) = - \mu \,\xt\, \yt + u(x)\, y - h(x)~, \label{triagx} \feq
for some $\mu \in \reali \backslash \{0 \}$ and some pair of smooth functions $u,h$.
The corresponding energy function is
\beq \Ec(x,y,\xt,\yt) = - \mu \,\xt\, \yt - u(x)\, y + h(x)~, \label{energ} \feq
while the Lagrange equations $\delta \LL/\delta y=0$ and $\delta \LL/\delta x=0$ are, respectively,
\begin{gather}
\mu\, \xtt + u(x) = 0~, \label{lagg2} \\
\mu\, \ytt + u'(x)\, y = h'(x) \label{lagg1}
\end{gather}
($u', h'$ are the derivatives of $u, h$).
The system \eqref{lagg2} \eqref{lagg1} is clearly triangular, since
Eq.\eqref{lagg2} involves only the unknown function $x(t)$; the system  can be reduced to
quadratures, following the procedure described hereafter. Firstly, note that Eq.\,\eqref{lagg2}
describes a one-dimensional conservative system,
admitting as a constant of motion the energy
\beq \Fc(x, \xt) := {1 \over 2}\, \mu\, \xt^2 + U(x)~, \qquad \mbox{with\, $U$ \,s.t.\, $U'= u$}~. \label{effe} \feq
Any solution $t \!\mapsto\! x(t)$ of Eq.\,\eqref{lagg2} with energy $\Fc(x(t),\xt(t)) \equiv \Fc$
fulfills $(2/\mu) (\Fc - U(x(t)) =$ $\xt^2(t) \geqs 0$, and thus it takes values within a connected
component of the region $\{ x ~|~(2/\mu) (\Fc - U(x)) \geqs 0 \}$.
For any such solution, let $t_0 < t_1$ be fixed instants in its domain of definition and assume that
({\footnote{Throughout the paper, $\sgn$ indicates the sign function. This is
such that: $\sgn(z) \!=\! -1$ for $z \!<\! 0$, $\sgn(z) \!=\! 0$ for $z \!=\! 0$
and $\sgn(z) \!=\! +1$ for $z \!>\! 0$.}})
\beq x(t_0) = x_0, \qquad \sgn\, \xt(t) = \mbox{const.} \equiv \sigma \in \{\pm 1\}~
\,\mbox{for all\, $t \in (t_0, t_1)$}~. \label{anal1} \feq
Then,
\beq \xt(t) = \sigma\, \sqrt{{2 \over \mu} \Big(\Fc - U\big(x(t)\big)\Big)} \quad
\mbox{for all\, $t \in (t_0, t_1)$}~, \label{quax} \feq
which entails
\beq \sqrt{\mu \over 2} \int_{x_0}^{x(t)}\!\! {d x \over \sqrt{\Fc - U(x)}} = \sigma\, (t - t_0)
\quad \mbox{for all\, $t \in [t_0, t_1]$}~. \label{anal3} \feq
Next, let $t \mapsto y(t)$ be a map forming, together with the previous function $t \mapsto x(t)$,
a solution of the system \eqref{lagg2} \eqref{lagg1} and consider the total energy $\Ec(x(t),y(t),\xt(t),\yt(t)) \equiv \Ec$.
Since $\xt(t)$ does not vanish and has constant sign for $t \in (t_0, t_1)$, there exists a smooth function
\beq Y : J \to \reali~, \qquad J := \big\{ x(t)~\big|~t \in (t_0, t_1) \big\}\,,  \feq
such that
\beq y(t) = Y\big(x(t)\big) \qquad \mbox{for all\, $t \in (t_0, t_1)$}~. \label{Yy} \feq
Correspondingly we have $\dot{y} = Y'(x)\, \xt$, whence $\xt\, \yt =  Y'(x)\,\xt^2
= (2/\mu) \big(\Fc - U(x)\big) Y'(x)$. Inserting the latter expression for $\xt\, \yt$
and the relation \eqref{Yy} for $y$ into Eq.\,\eqref{energ}, for $x = x(t) \in J$ we obtain
\beq \Ec = - 2 \,\big(\Fc - U(x)\big)\, Y'(x) - u(x)\, Y(x) + h(x)~; \label{enery} \feq
equivalently, recalling that $u = U'$, we have
\beq Y'(x) =  - \,{U'(x) \over 2 \big(\Fc - U(x)\big)}\, Y(x) - {\Ec - h(x) \over 2 \big(\Fc - U(x)\big)}
\qquad \mbox{for~ $x \in J$}~. \label{ypr} \feq
Noting that the latter is a linear inhomogeneous ODE for $Y$, by elementary arguments we get
\beq \begin{array}{c}
\dd{Y(x) = \sqrt{\Fc - U(x)}\, \big(P(\Ec,\Fc;x) + K\big) \qquad \mbox{for $x \in J$}~;} \\
\dd{K \in \reali~, \qquad
P(\Ec,\Fc;\,\cdot\,) \quad\mbox{s.t.}\quad \partial_x P(\Ec,\Fc;x) = - \,{\Ec - h(x) \over 2 \big(\Fc - U(x)\big)^{3/2}}~.} \label{quay}
\end{array} \feq
Eqs.\,\eqref{quax} \eqref{Yy} \eqref{quay} give the desired reduction to quadratures of
the system \eqref{lagg2} \eqref{lagg1} (on any time interval where $\xt$ has constant sign).
\parn
Of course, a Lagrangian of the form
\beq \LL(x,y,\xt,\yt) = - \mu\, \xt\, \yt + u(y)\, x - h(y) \label{triagy} \feq
can be treated in a similar way, interchanging the roles of $x$ and $y$.
\vfill\eject\noindent
\textsl{\spetria Harmonic triangular Lagrangian}. A very simple subcase of the previous framework
occurs if the Lagrangian has the triangular form \eqref{triagx} with $u(x) = \lambda\,x - \nu$
for some $\lambda,\nu \in \reali$, so that
\beq \LL(x,y,\xt,\yt) = - \mu\, \xt\, \yt + (\lambda\, x - \nu)\, y - h(x)~, \label{triax} \feq
where $\mu \!\in\! \reali \backslash \{0\}$ and $h$ is a smooth function.
The Lagrange equations \eqref{lagg2} \eqref{lagg1} become
\begin{gather}
\mu\, \xtt + \lambda\, x = \nu~, \label{lag2} \\
\mu\, \ytt + \lambda\, y = h'(x)~. \label{lag1}
\end{gather}
The above system is again triangular, but in this case both equations \eqref{lag2} \eqref{lag1} are elementary.
Depending on the sign of $\lambda/\mu$, Eq.\,\eqref{lag2} describes a harmonic oscillator,
a ``free particle" or a harmonic repulsor, with a constant external force $\nu$.
In the sequel, for the sake of brevity we will use the term
``harmonic system'' to indicate a system of any one of the three kinds just mentioned.
Regarding $x = x(t)$ as a known function, Eq.\,\eqref{lag1} describes
another harmonic system with a time-dependent external force $h'(x(t))$. \parn
As an example, assume that $\lambda/\mu > 0$ and set
\beq \om := \sqrt{\lambda/\mu}~; \feq
then, up to a time shift $t \mapsto t + \mbox{const.}$, the general solution of Eq.\,\eqref{lag2}
is of the form
\beq x(t) = A\, \sin(\om t) + {\nu \over \lambda}~, \label{xA} \feq
where $A \in \reali$ is an arbitrary constant.
Let $t_0 \in \reali$ and $J \subset \reali$ be any open real interval such that $t_0 \in J$
and the integral appearing in the forthcoming Eq.\,\eqref{yInt} exists for all $t \in J$;
then, the general solution on $J$ of the evolution equation obtained inserting the expression \eqref{xA}
for $x(t)$ into Eq.\,\eqref{lag1} is given by
\beq y(t) = B \cos(\om t) + C \sin(\om t)
+ {1 \over \mu\, \om} \int_{t_0}^t d s\; \sin\!\big(\om(t-s)\big)\; h'\Big(A\,\sin(\om s) + {\nu \over \lambda}\Big)\,, \label{yInt}\feq
where $B,C \in \reali$ are arbitrary constants. \parn
Of course, we obtain a system with similar solvability features
interchanging the roles of $x$ and $y$ in the Lagrangian \eqref{triax}.
\salto
\textsl{\separ Separable Lagrangian.}
Assume there exist two real Lagrangian coordinates $x_1,x_2$ such that
({\footnote{\label{pafoot} Additive constants appearing in a Lagrangian,
like the term $-C$ in Eq.\,\eqref{lagsep}, are usually regarded as irrelevant.
However, in the applications considered in this paper
we will always be interested in solutions of the Lagrange equations fulfilling the
energy constraint $\Ec= 0$ (see subsection \ref{secgaug}); in this connection,
additive constants appearing in the definition of the Lagrangian $\LL(x_1,x_2,\xt_1,\xt_2)$
cannot be neglected, since they contribute to the energy function
$\Ec := \sum_{i=1}^2 \dot{x_i} (\partial \LL/\partial \dot{x}_i) - \LL$
(see, e.g., Eq.\,\eqref{elagsep}).\vspace{-0.3cm}}})
\beq \LL(x_1,x_2,\xt_1,\xt_2) = \LL_1(x_1,\xt_1) + \LL_2(x_2,\xt_2) - C~,~~
\LL_i(x_i,\xt_i) := {1 \over 2}\, \mu_i\, \xt_i^2 - U_i(x_i) \label{lagsep} \feq
where $C \!\in\! \reali$, $\mu_i \!\in\! \reali \backslash \{0 \}$ and $U_i$ is a smooth
real function for $i=1,2$. The Lagrange equations $0 = \delta \LL/\delta x_i$
($i=1,2$) describe two decoupled subsystems admitting as constants of motion the energies
\beq \Ec_i(x_i, \xt_i) := {1 \over 2}\, \mu_i\, \xt_i^2 + U_i(x_i)~. \feq
Of course, the total energy corresponding to the Lagrangian \eqref{lagsep} is given by
\beq \Ec(x_1, x_2, \xt_1, \xt_2) = \Ec_1(x_1, \xt_1) + \Ec_2(x_2, \xt_2) + C~. \label{elagsep} \feq
Any pair of motions $x_i(t)$ ($i = 1,2$) of the separate subsystems with corresponding energies $\Ec_i$
are confined within connected regions where $\sgn(\mu_i) (\Ec_i - U(x_i)) \geqs 0$,
and can be reduced to quadratures via the relations
\beq \sqrt{\mu_i \over 2}\int_{x_i(t_0)}^{x_i(t)} {d x_i \over \sqrt{\Ec_i - U_i(x_i)}} = \sigma_i (t-t_0)~, \label{quasep} \feq
for any $t$ such that $\sigma_i := \sgn \, \xt_i(t) = \mbox{const.} \in \{\pm 1\}$ on $(t_0, t)$.
\salto
\textsl{\holom One-dimensional, holomorphic conservative system.}
From here to the end of the paper, $\Re z, \Im z$ and $\overline{z}$ indicate
the real part, the imaginary part and the conjugate of any complex number $z$.
Assume that there exist an open subset $\Dd \!\subset\! \complessi$ and a complex
Lagrangian coordinate $z \!\in\! \Dd$ such that
({\footnote{Also in this case, the constant $C$ in Eq.\,\eqref{lagcomp} is not irrelevant for our purposes:
see footnote \ref{pafoot} on page \pageref{pafoot}.}})
\beq \LL(z, \dot{z}) = - \Im\!\left( {1 \over 2}\,\mu\, \dot{z}^2  - U(z)\right) - C
\qquad (z \in \Dd, \dot{z} \in \complessi)\,, \label{lagcomp} \feq
where $\mu \!\in\! \complessi \backslash \{0 \}$, $C \!\in\! \reali$ and
$U : \Dd \to \complessi$ is an holomorphic function.
Let us point out that
$\LL(z, \dot{z}) = - {1 \over 2 i}\! \left( {1 \over 2}\,\mu\, \dot{z}^2  - U(z)\right) +
{1 \over 2 i}\! \left( {1 \over 2}\,\bar{\mu}\, \dot{\bar{z}}^2  - \overline{U(z)}\right)$
(where $\dot{\overline{z}} := \overline{\dot{z}\,}$);
so the Lagrange equations $\delta \LL/\delta z=0$ and $\delta \LL/\delta \overline{z} =0$
respectively read
\begin{gather}
\mu \, \ddot{z} = -\,U'(z)~, \label{fodeiib1} \\
\bar{\mu} \, \ddot{\overline{z}}= -\,\overline{U'(z)} \label{fodeiib2}
\end{gather}
($U'$ is the complex derivative of $U$)
({\footnote{Note that $\partial_z\, \overline{U(z)} = 0$ and $\partial_{\bar{z}}\,\overline{U(z)} = \overline{U'(z)}$.\vspace{-0.4cm}}}).
It appears that Eq.\,\eqref{fodeiib2} is the complex conjugate of Eq.\,\eqref{fodeiib1},
therefore the cited equations are fully equivalent.
\parn
From now on we fix the attention on Eq.\,\eqref{fodeiib1}.
This possesses as a constant of motion the ``complexified'' energy
\beq \mathfrak{E}(z, \dot{z}) = {1 \over 2}\,\mu\, \dot{z}^2 + U(z) ~\in~ \complessi~. \label{encomp} \feq
Starting from here, we derive a quadrature formula by a natural adaptation
to the complex framework of the approach usually employed for
real, one dimensional conservative systems.
More precisely, consider an open set $\mathscr{P}
\subset \complessi \backslash \{0 \}$ such that the map
$p \mapsto p^2$ is biholomorphic between $\mathscr{P}$ and
$\mathscr{P}^2 := \{ p^2~|~p \in \mathscr{P} \}$;
denote with $\sqrt{\phantom{X}} : \mathscr{P}^2 \to \mathscr{P}$ the inverse of this map.
Let $z$ be any solution of Eq.\,\eqref{fodeiib1} (hence, of Eq.\,\eqref{fodeiib2})
with complex energy $\mathfrak{E}(z(t), \dot{z}(t)) \equiv \mathfrak{E}$;
moreover, let $\mathscr{O} \subset \complessi$ be an open, simply connected subset
such that $(2/\mu)\,(\mathfrak{E} - U)(\mathscr{O}) \subset \mathscr{P}^2$
and assume that $z(t') \in \mathscr{O}$, $\dot{z}(t') \in \mathscr{P}$ for all $t' \in [t_0, t]$.
Then, we have
\beq \sqrt{\mu \over 2} \int_{z(t_0)}^{z(t)} {d z \over \sqrt{\mathfrak{E} - U(z)}}\,=\, t - t_0~,
\label{quafc} \feq
where $\int_{z(t_0)}^{z(t)}$ indicates the integration along
any path in $\mathscr{O}$ with initial point $z(t_0)$ and
final point $z(t)$ (the integral is independent
of the chosen path). \parn
To proceed, let us remark that the usual energy function
$\Ec := \dot z \, {\partial \LL/\partial \dot{z}} +
\dot{\overline{z}} \, {\partial \LL/\partial \dot{\overline{z}}} - \LL$
associated to the Lagrangian \eqref{lagcomp} is given by
\beq
\Ec(z, \dot{z}) = -\, \Im\!\left( {1 \over 2}\,\mu\, \dot{z}^2  + U(z)\right) + C
= - \,\Im \mathfrak{E}(z, \dot{z})+ C~.
\feq
Of course, there are subcases in which Eqs.\,\eqref{fodeiib1} \eqref{fodeiib2}
can be integrated by elementary means
without even referring to Eq.\,\eqref{quafc}. In particular, if
\beq U(z) = {1 \over 2}\;\varsigma\, z^2 \qquad (\varsigma \in \complessi)~, \feq
Eq.\,\eqref{fodeiib1} has the elementary form
\beq \mu\, \ddot{z} + \varsigma\, z = 0~; \feq
we refer to this system as a ``complex harmonic system''.
\vfill\eject\noindent
$\phantom{a}$\vspace{-1.3cm}

\subsection{Class 1 potentials}
\label{subsclass1}
The first class of potentials in \cite{Fre} has the form
\beq \VV(\f) := V_1\, e^{\f} + V_2\, e^{-\f} +  2\, V_0 \qquad (V_0,V_1, V_2 \in \reali)~. \label{vclass1} \feq
For these potentials, the cited reference suggests to introduce the gauge function $\BBB$
and the coordinates $x,y$, defined by
\begin{gather}
\BBB(\AA,\f) := 0~; \label{gau1}\\
\AA = \log(x\, y)~, \qquad \f = \log(x/y) \qquad (x,y\!>\!0) ~. \label{cor1}
\end{gather}
In the case of no matter and zero curvature ($\Oms\!=0$, $k=0$),
the above positions give rise to a quadratic Lagrangian and,
consequently, to linear evolution equations for $x$ and $y$.
Let us implement the same positions in our framework with
matter and curvature, searching for additional cases
with a quadratic Lagrangian.
Eqs.\,\eqref{vclass1}, \eqref{gau1} and \eqref{cor1}
yield the following expressions for the Lagrangian \eqref{eL}
and the energy function \eqref{efun}:
\begin{gather}
\LL(x,y,\xt,\yt) = -\, 2 \,\xt\,\yt - V_1\, x^2 - V_2\, y^2 - 2\, V_ 0 \, x\, y
- \,{\ds^2\,\Oms \over 2}\; (x\, y)^{-w}
+ \,{\ds^2 k \over 2} \; (x\, y)^{\ds - 2 \over \ds} ~ ; \label{eL1} \\
\Ec(x,y,\xt,\yt) = - \,2 \,\xt\,\yt + V_1\, x^2 + V_2\, y^2 + 2\,V_ 0 \, x\, y
+ \,{\ds^2\,\Oms \over 2}\; (x\, y)^{-w}
- \,{\ds^2 k \over 2} \; (x\, y)^{\ds - 2 \over \ds} ~. \label{efun1}
\end{gather}
The Lagrangian \eqref{eL1} is still quadratic, up to additive constants, in
the following cases with matter or curvature (i.e., with $(\Oms,k)$ not
constrained to be $(0,0)$):
\vskip 0.1cm \noindent
\textbf{i) $\boma{k=0}$, $\boma{w=0}$ (dust)}; \parn
\textbf{ii) $\boma{k=0}$, $\boma{w=-1}$ (cosmological constant)}; \parn
\textbf{iii) $\boma{\ds = 2}$, $\boma{w=0}$ (dust)}; \parn
\textbf{iv) $\boma{\ds = 2}$, $\boma{w=-1}$ (cosmological constant)}; \parn
\textbf{v) $\boma{\Oms=0}$, $\boma{\ds = 2}$}.
\vskip 0.1cm \noindent
In each one of these cases, the Lagrange equations in the coordinates
$x,y$ form a linear system, and can be decoupled via further linear coordinate
changes.
Let us stress that the admissible solutions must fulfill the energy constraint
$\Ec=0$, as well as the conditions $x(t),y(t)>0$ (cf. Eq.\,\eqref{cor1}).
\parn
As an example, let us consider case (i), providing (at least if $\ds=3$)
a rather realistic model of the universe for most of its history; this is
just the case considered (for $\ds =3$) in \cite{Piedipalumbo}.
The Lagrangian \eqref{eL1} and the energy \eqref{efun1} reduce, respectively, to
\begin{gather}
\LL(x,y,\xt,\yt) = - \,2\, \xt\,\yt
- V_1\, x^2 - V_2\, y^2 - 2\, V_ 0 \, x\, y - {\ds^2\, \Oms\over 2} ~, \label{eL10} \\
\Ec(x,y,\xt,\yt) = - \,2\, \xt\, \yt + V_1\, x^2 + V_2\, y^2 + 2\, V_ 0 \, x\, y
+ {\ds^2\, \Oms \over 2}~.  \label{efun10}
\end{gather}
The Lagrange equations decouple under a further, linear
change of coordinates $(x,y) \mapsto (u,v)$. For example, if $V_1 V_2 >0$
we can put
\beq x = {1 \over 2} \left({V_2 \over V_1}\right)^{\!\!1/4} (u - v)~, \qquad
y = {1 \over 2} \left({V_1 \over V_2}\right)^{\!\!1/4} (u + v)~, \label{xyuv} \feq
which transforms the Lagrangian \eqref{eL10} and the energy \eqref{efun10} into
\begin{gather}
\begin{array}{c}
\dd{\LL(u,v, \dot{u}, \dot{v}) = \LL_1(u, \dot{u}) + \LL_2(v, \dot{v}) - {\ds^2 \over 2}\; \Oms ~,}\vspace{0.1cm} \\
\dd{\LL_1(u, \dot{u}) := -\,{1 \over 2}\,\dot{u}^2 - {\sqrt{V_1 V_2} + V_0 \over 2}\; u^2\,, \qquad
\LL_2(v, \dot{v}) := {1 \over 2}\,\dot{v}^2 - {\sqrt{V_1 V_2} - V_0 \over 2}\; v^2\,;}
\end{array} \vspace{-0.1cm} \label{lasep}\\
\begin{array}{c}
\dd{\Ec(u,v, \dot{u}, \dot{v}) = \Ec_1(u, \dot{u}) + \Ec_2(v, \dot{v}) + {\ds^2 \over 2}\; \Oms ~,} \\
\dd{\Ec_1(u, \dot{u}) := -\,{1 \over 2}\,\dot{u}^2 + {\sqrt{V_1 V_2} + V_0 \over 2}\; u^2\,, \qquad
\Ec_2(v, \dot{v}) := {1 \over 2}\,\dot{v}^2 + {\sqrt{V_1 V_2} - V_0 \over 2}\; v^2 \,.} \label{Euv}
\end{array}
\end{gather}
The separable Lagrangian \eqref{lasep} gives rise to the system of uncoupled equations
\beq \ddot{u} - \big(\sqrt{V_1 V_2} + V_0\big)\, u=0~, \qquad
\ddot{v} + \big(\sqrt{V_1 V_2} - V_0\big)\, v=0~, \label{uvLeq} \feq
whose solutions can be determined by elementary means.
Let us point out that in the present case, the admissible solutions are those
fulfilling $u(t) > |v(t)|$ (which is equivalent to $x(t),y(t) \!>\! 0$).
More details about the qualitative behavior of such solutions will be given
in subsection \ref{piedipal}.

\subsection{Class 2 potentials}\label{subsec2}
The second class of potentials in \cite{Fre} is formed by the
functions
\beq \VV(\f) := V_1\, e^{2\, \gamma\, \f} + V_{2}\, e^{(1 + \gamma)\, \f}
\qquad \big(V_1, V_2 \in \reali\,,\; \gamma \in \reali \backslash\{\pm 1\}\big)~. \label{vclass2} \feq
Ref.\,\cite{Fre} suggests to study these potentials fixing the gauge function $\BBB$
and introducing new coordinates $x,y$ as follows:
\begin{gather}
\BBB(\AA,\f) := - \,\ga\, \f~; \vspace{-0.1cm} \label{gau2} \\
\AA = \log\!\big(x^{1 \over 1 +\ga}\, y^{1 \over 1 - \ga}\big)~, \qquad
\f = \log\!\big(x^{1 \over 1 +\ga}\, y^{-{1 \over 1 - \ga}}\big) \qquad (x,y >0)~. \label{coor2}
\end{gather}
In the case of no matter and zero curvature, the Lagrangian obtained via these
prescriptions has the harmonic triangular form \eqref{triax}.
\parn
Let us now apply the same prescriptions \eqref{gau2} \eqref{coor2} in our
framework with matter and curvature, and search for additional
triangular cases. The Lagrangian \eqref{eL} and the energy \eqref{efun} become, respectively,
\begin{gather}
\LL(x,y,\xt,\yt) = \nonumber \\
- \,{2\, \xt\, \yt\, \over 1 - \ga^2} - V_1\, x\, y - V_2\, x^{2 \over 1 + \ga}
- {\ds^2\, \Oms \over 2}\; x^{-{w + \ga \over 1 + \ga}}\, y^{-{w - \ga \over 1 -\ga}}
+ {\ds^2 k \over 2}\; x^{\ds (1 - \ga) - 2 \over \ds (1 + \ga)}\, y^{\ds (1 + \ga) - 2 \over \ds (1 - \ga)}\,, \label{el2}\\
\Ec(x,y,\xt,\yt) = \nonumber \\
- \,{2\, \xt\, \yt\, \over 1 - \ga^2} + V_1\, x\, y + V_2\, x^{2 \over 1 + \ga}
+ {\ds^2\, \Oms \over 2}\; x^{-{w + \ga \over 1 + \ga}}\, y^{-{w - \ga \over 1 -\ga}}
- {\ds^2 k \over 2}\; x^{\ds (1 - \ga) - 2 \over \ds (1 + \ga)}\, y^{\ds (1 + \ga) - 2 \over \ds (1 - \ga)}\,. \label{efun2}
\end{gather}
The Lagrangian \eqref{el2} has a triangular structure
in the cases with matter or curvature listed below.
In each one of these cases, the admissible solutions
are those fulfilling the energy constraint $\Ec=0$
and the conditions $x(t),y(t)>0$ (cf. Eq.\,\eqref{coor2}).
\vspace{-0.4cm}

\paragraph{i) $\boma{k=0}$, $\boma{\ga = w \neq \pm 1}$\,.} The Lagrangian \eqref{el2}
and the energy \eqref{efun2} become
\begin{gather}
\LL(x,\xt,y,\yt) = - \,{2 \over 1 \!-\! w^2}\; \xt \, \yt - V_1 \, x \, y - V_2 \, x^{{2 \over 1 + w}}
- {\ds^2\,\Oms \over 2} \; x^{- {2 w \over 1 + w}}~, \label{L21}\\
\Ec(x,\xt,y,\yt) = - \,{2 \over 1 \!-\! w^2}\; \xt \, \yt + V_1 \, x \, y + V_2 \, x^{{2 \over 1 + w}}
+ {\ds^2\,\Oms \over 2}\; x^{- {2 w \over 1 + w}}~. \label{E21}
\end{gather}
The Lagrangian \eqref{L21} has the harmonic triangular structure \eqref{triax}
and the related equations $\delta \LL/\delta y=0$, $\delta \LL/\delta x=0$ read, respectively,
\begin{equation}\begin{array}{c}
\dd{\xtt - {(1 - w^2)\,V_1 \over 2}\; x = 0 ~,} \\
\dd{\ytt - {(1 - w^2)\,V_1 \over 2}\; y = (1 - w)\, V_2\, x^{1 - w \over 1 + w}
- {w\,(1 - w)\,\ds^2\,\Oms \over 2}\; x^{-{1 + 3 w \over 1 + w}} ~.}\label{l11}
\end{array} \end{equation}
The equation in the first line of \eqref{l11} describes a harmonic system
({\footnote{In the generalized sense stipulated for this term in the discussion
after Eqs.\,\eqref{lag2} \eqref{lag1}; the specific kind of this harmonic system depends
on the sign of $(1- w^2)\, V_1$. Similar explanations will never be repeated in the sequel.}});
when its general solution is substituted into the equation in the second line of
\eqref{l11}, the latter can be interpreted in terms of a forced harmonic system.
Notably, the system \eqref{l11} can be treated by elementary means.
\vfill\eject\noindent
$\phantom{a}$\vspace{-1.3cm}

\paragraph{ii) $\boma{\ga = w = {1 \over \ds}}$ (radiation gas).}
The Lagrangian \eqref{el2} and the energy \eqref{efun2} reduce to
\begin{gather}
\LL(x,\xt,y,\yt) = -\,{2 \, \ds^2 \over \ds^2 \!-\! 1}\; \xt \, \yt
- \left(V_1\, x  - { \ds^2 k \over 2}\; x^{{\ds-3 \over n + 1}} \right) y
- \left( V_2\, x^{2 n \over n + 1} + {\ds^2\, \Oms \over 2}\; x^{- {2  \over n + 1}} \right), \label{L2ii}\\
\Ec(x,\xt,y,\yt) = -\,{2 \, \ds^2 \over \ds^2 \!-\! 1}\; \xt \, \yt +
\left(V_1 \, x  - {\ds^2 k \over 2}\, x^{{\ds-3 \over n + 1}} \right) y
+ \left( V_2\, x^{2 n \over n + 1} + {\ds^2\, \Oms\over 2} \; x^{- {2  \over n + 1}} \right).
\end{gather}
The present Lagrangian has the triangular form \eqref{triagx}, and can be treated
with the methods described below the cited equation; in particular, note that Eqs.\,\eqref{triagx} \eqref{effe}
are fulfilled in the present case with
$u(x) = - V_1 \, x  + { \ds^2 \over 2}\,k\, x^{{\ds-3 \over n + 1}}$ and
$U(x) = - {V_1 \over 2}\, x^2 + {\ds^2 (\ds+1)\, k \over 4 (\ds-1)}\, x^{{2 (\ds-1) \over n + 1}}$.
\vspace{-0.4cm}

\paragraph{ii$_0$) $\boma{\ds = 3}$, $\boma{\ga = w = {1 \over 3}}$\,.}
In this particular subcase of case (ii), the Lagrangian \eqref{el2}
and the energy function \eqref{efun2} read
\begin{gather}
\LL(x,\xt,y,\yt) = - \,{9 \over 4} \, \xt \, \yt - \left(V_1\, x - {9\,k \over 2}\right) y - V_2 \, x^{{3/2}}
- {9\, \Oms \over 2}\; x^{- {1/2}}~, \label{L2ii0}\\
\Ec(x,\xt,y,\yt) = - \,{9 \over 4} \, \xt \, \yt + \left(V_1\,x - {9\,k \over 2}\right) y + V_2 \, x^{{3/2}}
+ {9\, \Oms \over 2}\; x^{- {1/2}}~.
\end{gather}
The Lagrangian \eqref{L2ii0} is of the harmonic triangular form \eqref{triax}
and the related equations $\delta \LL/\delta y=0$, $\delta \LL/\delta x=0$ read
\begin{equation}
\xtt - {4\,V_1 \over 9}\; x = -\,2\,k~, \qquad
\ytt - {4\,V_1 \over 9}\; y = {2\,V_2 \over 3}\; x^{1/2} - \,\Oms\, x^{-{3/2}}\,. \label{e3d}
\end{equation}
The first equation in \eqref{e3d} describes a harmonic system with a constant ``curvature force'';
once $x(t)$ has been determined, the second equation in \eqref{e3d} describes another forced harmonic system.
Also in this case, we have a pair of equations which can be solved by elementary means.
\vspace{-0.2cm}

\paragraph{iii) $\boma{\ga = w = {2 \over \ds} - 1}$\,.}\!\!
({\footnote{The position $w = 2/\ds - 1$ makes this case perhaps less interesting than
the previous ones, since it gives $ w < 0 $ for $ \ds \geqs 3 $
(for $\ds =2$ one has $w = 0$, typical of a dust fluid).
Nonetheless, if $ w = 2/\ds - 1 $ and we assume $\Oms \geqs 0$ in Eq.\,\eqref{pmrom1}
(non-negative matter density), the requirements \eqref{C1} \eqref{C2} 
corresponding to the weak and dominant energy conditions are both fulfilled
for any $\ds \geqs 2$ (in fact, for $\ds \geqs 1$).\vspace{-0.2cm}}})
The Lagrangian \eqref{el2} and the energy \eqref{efun2} take the form
\vspace{-0.2cm}
\begin{gather}
\LL(x,\xt,y,\yt) = -\, {\ds^2  \over 2 (n\!-\!1)} \; \xt \, \yt - V_1\, x \, y - V_2 \, x^{n}
- {\ds^2 (\Oms \!- k) \over 2}\; x^{\ds-2}~, \label{L2iii}\\
\Ec(x,\xt,y,\yt) = -\, {\ds^2  \over 2 (n\!-\!1)} \; \xt \, \yt + V_1\, x \, y + V_2 \, x^{n}
+ {\ds^2 (\Oms \!- k) \over 2}\; x^{\ds-2}~.
\end{gather}
The Lagrangian \eqref{L2iii} has the harmonic triangular form \eqref{triax}
and the related equations $\delta \LL/\delta y=0$, $\delta \LL/\delta x=0$ give
\vspace{-0.2cm}
\begin{equation}\begin{array}{c}
\dd{\xtt - {2\,(\ds-1)\, V_1 \over \ds^2} \; x = 0 ~,} \vspace{0.1cm}\\
\dd{\ytt  - {2\, (\ds-1)\, V_1 \over \ds^2} \; y = {2\,(\ds-1)\, V_2 \over \ds}\; x^{\ds-1}
+ (\ds-2)(\ds-1)\, \big( \Oms -k \big)\, x^{\ds-3}~.}  \label{ex2}
\end{array}\end{equation}
Again, the equation for $x(t)$ describes a harmonic system and,
once this function has been determined, the equation for $y(t)$ describes a forced harmonic system.
\vspace{-0.4cm}

\paragraph{iv) $\boma{\ga = {2 \over \ds} - 1}$, $\boma{w = {4 \over \ds} - 3}$\,.}
The Lagrangian \eqref{el2} and the energy \eqref{efun2} are, respectively,
\vspace{-0.2cm}
\begin{gather}
\LL(x,\xt,y,\yt) = -\, {\ds^2 \over 2 (n \!-\! 1)}\; \xt \, \yt -
\left (V_1 \, x  + {\ds^2\,\Oms \over 2} \; x^{2 n - 3} \right) y
- \left( V_2\, x^{n} - {\ds^2 k \over 2}\; x^{- 2 n} \right) , \label{L2iv}\\
\Ec(x,\xt,y,\yt) = -\, {\ds^2 \over 2 (n \!-\! 1)}\; \xt \, \yt +
\left (V_1 \, x  + {\ds^2\,\Oms \over 2}\; x^{2 n - 3} \right) y
+ \left( V_2\, x^{n} - {\ds^2 k \over 2}\; x^{- 2 n} \right) .
\end{gather}
The Lagrangian \eqref{L2iv} has the triangular structure \eqref{triagx}, an can be
treated with the corresponding methods; in particular, Eqs.\,\eqref{triagx} \eqref{effe}
are fulfilled in the present case with
$u(x) = - V_1 \, x  - { \ds^2 \,\Oms \over 2}\, x^{2 n - 3}$ and
$U(x) = - {V_1 \over 2}\, x^2 - { \ds^2\, \Oms \over 4 (\ds-1)}\,x^{2 n - 2}$.
\vfill\eject\noindent

\paragraph{iv$_0$) $\boma{n = 2}$, $\boma{\ga = 0}$, $\boma{w = -1}$ (cosmological constant).}
This is the subcase of case (iv) corresponding to $n=2$.
The Lagrangian \eqref{el2} and the energy function \eqref{efun2} reduce to
\begin{gather}
\LL(x,\xt,y,\yt) = -\, 2  \, \xt \, \yt - \big(V_1  + 2\, \Oms\big) \, x \, y - V_2 \, x^{2} + 2\, k~, \label{L2iv0} \\
\Ec(x,\xt,y,\yt) = -\, 2  \, \xt \, \yt + \big(V_1  + 2\, \Oms\big) \, x \, y + V_2 \, x^{2} - 2\, k~.
\vspace{-0.1cm}
\end{gather}
The Lagrangian \eqref{L2iv0} has the harmonic triangular form \eqref{triax}.
The related equations $\delta \LL/\delta y=0$, $\delta \LL/\delta x=0$ read
\begin{equation}
\xtt - \left({V_1 \over 2} + \Oms\! \right) x = 0 ~, \qquad
\ytt - \left({V_1 \over 2} + \Oms\!\right) y = V_2\, x ~; \label{ey4d}
\vspace{-0.1cm}
\end{equation}
they describe, respectively, a harmonic system and another harmonic system with an external
force proportional to $x(t)$.
Up to a constant, the Lagrangian \eqref{L2iv0} also belongs to
the class of the quadratic Lagrangians.
\vspace{-0.3cm}

\paragraph{v) $\boma{k = 0}$, $\boma{\ga = {w+1 \over 2} \neq \pm 1}$\,.}
The Lagrangian \eqref{el2} and the energy \eqref{efun2} become
\begin{gather}
\LL(x,\xt,y,\yt) = - \,{8 \over (1\!-\!w)(3 \!+\! w)}\; \xt \, \yt -
\left (V_1 \, x  + { \ds^2\,\Oms \over 2}\; x^{- {1 + 3 w \over 3 + w}} \right) y
- V_2\, x^{{4 \over 3 + w}} ~, \label{L2v} \\
\Ec(x,\xt,y,\yt) = - \,{8 \over (1\!-\!w)(3 \!+\! w)}\; \xt \, \yt +
\left (V_1 \, x  + { \ds^2\, \Oms \over 2}\; x^{- {1 + 3 w \over 3 + w}} \right) y
+ V_2\, x^{{4 \over 3 + w}} ~.
\end{gather}
The Lagrangian \eqref{L2v} has the triangular form \eqref{triagx}, an can be treated
with the corresponding methods; in particular, Eqs.\,\eqref{triagx} \eqref{effe}
are fulfilled in the present case setting
$u(x) = - V_1 \, x  - { \ds^2\,\Oms \over 2}\, x^{- {1 + 3 w \over 3 + w}}$ and
$U(x) = - {V_1 \over 2}\, x^2 - {(3+w)\,\ds^2 \, \Oms \over 4(1-w)}\, x^{{2 (1-w) \over 3 + w}}$.
\vspace{-0.3cm}

\paragraph{v$_0$) $\boma{k=0}$, $\boma{\ga = {1 \over 3}}$, $\boma{w = - {1 \over 3}}$\,.}
This is a subcase of case (v), corresponding to $w=-1/3$.
The Lagrangian \eqref{el2} and the energy \eqref{efun2} reduce to
\begin{gather}
\LL(x,\xt,y,\yt) = - \,{9 \over 4}  \, \xt \, \yt - \left(V_1\, x + {\ds^2\,\Oms \over 2} \right) y
- V_2 \, x^{{3/2}}~, \label{L2v0} \\
\Ec(x,\xt,y,\yt) = - \,{9 \over 4}  \, \xt \, \yt + \left(V_1\, x + {\ds^2\,\Oms \over 2} \right) y
+ V_2 \, x^{{3/2}}~.
\end{gather}
The Lagrangian \eqref{L2v0} has the harmonic triangular form \eqref{triax}.
The equations $\delta \LL/\delta y=0$, $\delta \LL/\delta x=0$ read
\begin{gather}
\xtt - {4\, V_1 \over 9}\; x = {2\, \ds^2\,\Oms \over 9} ~, \qquad
\ytt - {4\, V_1 \over 9}\; y = {2\,V_2 \over 3}\, x^{1/2} \label{ey5c}
\end{gather}
and describe two forced harmonic systems, the first one with a constant
external force and the second one with an external force depending on $x(t)$.
\vspace{-0.3cm}

\paragraph{vi) $\boma{\ga = {1 \over \ds}}$, $\boma{w = {2 \over \ds} - 1}$\,.}
The Lagrangian \eqref{el2} and the energy \eqref{efun2} read
\begin{gather}
\LL(x,\xt,y,\yt) = - {2 \, \ds^2 \over \ds^2 \!-\! 1}\; \xt \, \yt -
\left (V_1 \, x  + { \ds^2 (\Oms\!-k) \over 2}\; x^{{\ds-3 \over \ds+1}} \right) y
- V_2\, x^{{2 n \over \ds+1}} ~, \label{L2vi} \\
\Ec(x,\xt,y,\yt) = - {2 \, \ds^2 \over \ds^2 \!-\! 1}\; \xt \, \yt +
\left (V_1 \, x  + { \ds^2 (\Oms\!-k) \over 2}\; x^{{\ds-3 \over \ds+1}} \right) y
+ V_2\, x^{{2 n \over \ds+1}} ~.
\end{gather}
The Lagrangian \eqref{L2vi} has the triangular structure \eqref{triagx}, and can be treated
with the corresponding methods; in particular, Eqs.\,\eqref{triagx} \eqref{effe}
are fulfilled in the present case with
$u(x) = - V_1 \, x  - { \ds^2 (\Oms -k) \over 2}\, x^{{\ds-3 \over \ds+1}}$ and
$U(x) = - {V_1 \over 2}\, x^2 - {(\ds+1)\,\ds^2 (\Oms -k) \over 4 (\ds-1)}\, x^{{2 (\ds-1) \over \ds+1}}$.
\vfill\eject\noindent

\paragraph{vi$_0$) $\boma{\ds = 3}$, $\boma{\ga = {1 \over 3}}$, $\boma{w = - {1 \over 3}}$\,.}
This is the subcase of case (vi) corresponding to $\ds=3$.
The Lagrangian \eqref{el2} and the energy \eqref{efun2} reduce to
\begin{gather}
\LL(x,\xt,y,\yt) = - \,{9 \over 4}  \, \xt \, \yt - \left(V_1\, x + {9\,(\Oms\!-k) \over 2} \right) y
- V_2 \, x^{{3/2}}~,\label{L2vi0}\\
\Ec(x,\xt,y,\yt) = - \,{9 \over 4}  \, \xt \, \yt + \left(V_1 \,x + {9\,(\Oms\!-k) \over 2} \right) y
+ V_2 \, x^{{3/2}}~.
\end{gather}
The Lagrangian \eqref{L2vi0} has the harmonic triangular form \eqref{triax}
and the equations $\delta \LL/\delta y=0$, $\delta \LL/\delta x=0$ give
\begin{gather}
\xtt - {4\, V_1 \over 9}\; x = 2\,\big(\Oms\!-k\big) ~,  \qquad
\ytt - {4\, V_1 \over 9}\; y = {2\, V_2 \over 3}\; x^{1/2} ~. \label{ey6d}
\end{gather}
Again, we have a harmonic system with a constant external force and another harmonic system
with an external force depending on $x(t)$.

\subsection{Class 3 potentials}
Let us now consider potentials of the form
\beq \VV(\f) = V_{1}\, e^{2\, \f} + V_{2} \qquad (V_1, V_2 \in \reali)~. \label{VCase3} \feq
Ref.\,\cite{Fre} treats these potentials fixing the gauge function $\BBB$
and introducing a pair of Lagrangian coordinates $x,y$, in the following way:
\begin{gather}
\BBB(\AA,\f) := -\,\f ~, \label{gau3} \\
\AA = {1 \over 2}\,(\log x) + y~; \qquad
\f = {1 \over 2}\,(\log x) - y \qquad (x \!>\!0\,, ~y \!\in\! \reali)~. \label{coor3}
\end{gather}
In the case of no matter and zero curvature, the Lagrangian obtained with the above positions
has the harmonic triangular form \eqref{triax}.
Let us make the same positions in our framework with matter and curvature,
and search for other triangular cases. Eqs.\,\eqref{VCase3}\eqref{gau3}\eqref{coor3}
yield for the Lagrangian \eqref{eL} and for the energy \eqref{efun} the following expressions:
\begin{gather}
\LL(x,\xt,y,\yt) = -\,\xt\,\yt - V_{1}\, x - V_{2}\,e^{2 y}
-\,{\ds^2\,\Oms \over 2}\;x^{-{1+w \over 2}}\,e^{(1-w)y}
+ {\ds^2 k \over 2}\; x^{-{1 \over \ds}}\,e^{{2(\ds-1) \over \ds}\,y} ~; \label{l3} \\
\Ec(x,\xt,y,\yt) = -\,\xt\,\yt + V_{1}\, x + V_{2}\,e^{2 y}
+\,{\ds^2\,\Oms \over 2}\; x^{-{1+w \over 2}}\,e^{(1-w)y}
- {\ds^2 k \over 2}\; x^{-{1 \over \ds}}\,e^{{2(\ds-1) \over \ds}\,y} ~. \label{eqen}
\end{gather}
It is evident that the Lagrangian \eqref{l3} cannot have a triangular structure when $k \neq 0$;
thus, we set $k=0$ and search for triangular cases with matter, i.e., with $\Oms \neq 0$.
Below we give a list of such cases; in each one of these cases, the admissible solutions
are those fulfilling the energy constraint $\Ec=0$ and the condition $x(t) >0$ (see Eq.\,\eqref{coor3}).
\vspace{-0.3cm}

\paragraph{i) $\boma{k = 0}$, $\boma{w = -1}$ (cosmological constant).}
The Lagrangian \eqref{l3} and the energy \eqref{eqen} read
\begin{gather}
\LL(x,y,\xt,\yt) = -\, \xt \, \yt - V_1\, x - \left(V_2 + {\ds^2\,\Oms \over 2}\right) e^{2 y}~, \label{L3i}\\
\Ec(x,y,\xt,\yt) = -\, \xt \, \yt + V_1\, x + \left(V_2 + {\ds^2\,\Oms \over 2}\right) e^{2 y}~. \label{E3i}
\end{gather}
The Lagrangian \eqref{L3i} has the harmonic triangular form \eqref{triagy}.
The related equations $\delta \LL/\delta x=0$, $\delta \LL/\delta y=0$ read
\beq \ytt = V_1 ~, \qquad \xtt = 2\left(V_{2} + {\ds^2\, \Oms \over 2}\right) e^{2\,y} ~; \feq
their general solution is
\beq \begin{array}{c}
\dd{x(t) = \left(V_{2} + {\ds^2\, \Oms \over 2}\right) {e^{2 \be - {\al^2 \over V_1}}\! \over V_1}
\left({\sqrt{\pi}\,(V_1\, t \!+\! \al) \over \sqrt{V_1}}\; \mbox{Erfi}\Big({V_1\, t \!+\! \al \over \sqrt{V_1}}\Big)
\!- e^{{(V_1 t + \al)^2 \over V_1}} \right)\! + \ga\,t + \delta ~,} \vspace{0.1cm}\\
\dd{y(t) = {V_1 \over 2} \; t^2 + \al\,t + \be ~,} \label{solxy}
\end{array} \feq
where $\al,\be,\ga,\delta$ are arbitrary integration constants and $\mbox{Erfi}$ is the
imaginary error function.\parn
Eqs.\,\eqref{E3i} and \eqref{solxy} imply
$\Ec = V_1\,\delta - \alpha\,\gamma$, so the energy constraint $\Ec = 0$ holds
if and only if $V_1\,\delta = \alpha \, \gamma$.
The issue of finding a maximal interval where $x(t)>0$
is strictly related to the choice of the integration constants in Eq.\,\eqref{solxy}.
\vspace{-0.4cm}

\paragraph{ii) $\boma{k=0}$, $\boma{w= - 3}$\,.}
The Lagrangian \eqref{l3} and the energy \eqref{eqen} become, respectively,
\begin{gather}
\LL(x,y,\xt,\yt) = - \,\xt \, \yt - \left(V_1  + {\ds^2\,\Oms \over 2}\;e^{4 y}\right) x - V_2\, e^{2 y}~, \label{L3ii}\\
\Ec(x,y,\xt,\yt) = - \,\xt \, \yt + \left(V_1  + {\ds^2\,\Oms \over 2}\; e^{4 y}\right) x + V_2\, e^{2 y}~.
\end{gather}
The Lagrangian \eqref{L3ii} has the triangular structure \eqref{triagy}, an can be treated
with the corresponding methods; in particular, Eq.\,\eqref{triagy} and the analogue of Eq.\,\eqref{effe}
are fulfilled in the present case with
$u(y) = - V_1  - {\ds^2\,\Oms \over 2}\, e^{4 y}$ and
$U(y) = - V_1\, y - {\ds^2\,\Oms \over 8}\, e^{4 y}$\,.
\vspace{-0.4cm}

\paragraph{iii) $\boma{k = 0}$, $\boma{V_2 = 0}$, $\boma{w = 1}$ (stiff matter).}
In this case the potential does also belong to the class 2
discussed in subsection \ref{subsec2}
(since $\VV(\f)$ is of the form \eqref{vclass2} with $V_2 = 0$ and $\gamma = 1$). \parn
The Lagrangian \eqref{l3} and the energy \eqref{eqen} read, respectively,
\begin{gather}
\LL(x,y,\xt,\yt) = -\, \xt \, \yt - V_1\, x - {\ds^2\,\Oms \over 2}\; x^{-1}~, \\
\Ec(x,y,\xt,\yt) = - \,\xt \, \yt + V_1\, x + {\ds^2\,\Oms \over 2}\; x^{-1}~. \label{En3iii}
\end{gather}
We have a harmonic triangular Lagrangian, of the form \eqref{triax} with $\lambda = \sigma=0$.
The corresponding equations $\delta \LL/\delta y=0$, $\delta \LL/\delta x=0$ are
\beq \xtt = 0 ~, \qquad \ytt = V_1 - {\ds^2\,\Oms \over 2} \;x^{-2} ~. \feq
Again, the general solution can be expressed in terms of elementary functions and reads
\begin{equation}
x(t) = \al\, t + \be ~, \qquad
y(t) = {1 \over 2\,\al^2}\, \Big(V_1\,(\al\, t + \be)^2 + \ds^2\, \Oms \log(\al\, t + \be) \Big)
+ \ga\,t + \delta ~, \label{solxy3iii}
\end{equation}
where $\al,\be,\ga,\delta$ are integration constants.
Eqs.\,\eqref{En3iii} \eqref{solxy3iii} imply $\Ec = - \al\,\ga$, showing that
the energy constraint $\Ec = 0$ holds only if $\al\,\ga = 0$.
Finding a maximal interval where $x(t)>0$ is a trivial task, once the
integration constants in Eq.\,\eqref{solxy3iii} have been assigned.

\subsection{Class 4 potentials}
Let us consider the potential
\beq \VV(\f) = V\,\f\; e^{2\, \f} \quad~ (V \!\in \reali) ~. \label{VCase4} \feq
Ref.\,\cite{Fre} suggests to treat these potentials using the gauge function $\BBB$
and the coordinates $x,y$, defined by
\begin{gather}
\BBB(\AA,\f) := -\,(\AA + 2 \f) ~, \label{gau4} \\
\AA = {1 \over 4}\,(\log x) + y ~, \qquad \f = {1 \over 4}\,(\log x) - y
\qquad (x \!>\!0\,, ~y \!\in\! \reali)~. \label{cor3}
\end{gather}
In the case with no matter and zero curvature, these prescriptions yield
again a harmonic triangular Lagrangian of the form \eqref{triax}.
\vfill\eject\noindent
$\phantom{a}$\vspace{-1.cm}\\
Also in this case, we extend the treatment of \cite{Fre} to our framework with
matter and curvature and search for additional triangular cases. With the positions
\eqref{VCase4}\eqref{gau4}\eqref{cor3}, the Lagrangian \eqref{eL} and the energy \eqref{efun}
become, respectively,
\begin{gather}
\LL(x,\xt,y,\yt) =  -\,{1 \over 2}\,\xt\,\yt - V\! \left({1 \over 4}\log x - y\right)
- {\ds^2\,\Oms \over 2} \;x^{-{3+w \over 4}} e^{(1-w)y}
+ {\ds^2 k \over 2}\; x^{-{\ds+1 \over 2\ds}} e^{{2(\ds-1) \over \ds}\,y} ~, \label{l4}\\
\Ec(x,\xt,y,\yt) = -\,{1 \over 2}\,\xt\,\yt + V\! \left({1 \over 4}\log x - y\right)
+ {\ds^2\,\Oms \over 2} \; x^{-{3+w \over 4}} e^{(1-w)y}
- {\ds^2 k \over 2}\; x^{-{\ds+1 \over 2\ds}} e^{{2(\ds-1) \over \ds}\,y} ~. \label{aren}
\end{gather}
In presence of matter ($\Oms \!>\! 0$), the only case where
the Lagrangian \eqref{l4} has a triangular structure is the following.
\vspace{-0.4cm}

\paragraph{i) $\boma{k = 0}$, $\boma{w = 1}$ (stiff matter).}
The Lagrangian \eqref{l4} and the energy \eqref{aren} reduce to
\begin{gather}
\LL(x,y,\xt,\yt) = - \,{1 \over 2}\, \xt \, \yt + V\, y - \left({V \over 4}\, \log x + {\ds^2\,\Oms \over 2}\; x^{-1} \right), \label{L4i}\\
\Ec(x,y,\xt,\yt) = - \,{1 \over 2}\, \xt \, \yt - V\, y + \left({V \over 4}\, \log x + {\ds^2\,\Oms \over 2}\; x^{-1} \right) .
\end{gather}
The Lagrangian \eqref{L4i} has the harmonic triangular form \eqref{triax}.
The related equations $\delta \LL/\delta y=0$, $\delta \LL/\delta x=0$ entail
\beq \xtt = - 2\,V ~, \qquad \ytt = {V \over 2}\,x^{-1} - \ds^2\,\Oms\,x^{-2} ~, \feq
and their general solution is given by
\begin{gather}
x(t) = - \,V\, t^2 + \al\,t + \be ~, \label{solxy4} \\
y(t) = \ga\,t + \delta + {1 \over 4}\, \log\!\big(\!- V t^2 \!+\! \al\, t \!+\! \be\big)
- \,{(4\,\ds^2\, \Oms\! - \Delta)\,(\al \!-\! 2 V t)\, \over 4\Delta^{3/2}}
\,\log\!\left({\sqrt{\Delta} + \al - 2 V t \over \sqrt{\Delta} - \al + 2 V t}\right) , \nonumber
\end{gather}
where $\al,\be,\ga,\delta$ are integration constants and $\Delta := \al^2 + 4 V \be$.
From Eqs.\,\eqref{aren} \eqref{solxy4} it follows that
$\Ec = [4\,\ds^2\,\Oms V -\Delta\,(\alpha\,\gamma + 2\,\delta\,V)]/(2\,\Delta)$,
which shows that the energy constraint $\Ec = 0$ holds if and only if
$\alpha\,\gamma = 2\,V\,(2\,\ds^2\,\Oms\! - \delta\,\Delta)/\Delta$.
Again, finding a (maximal) interval where $x(t)>0$ is an elementary task,
once the integration constants in Eq.\,\eqref{solxy4} have been assigned.

\subsection{Class 5 potentials}\label{class5}
This class is formed by potentials of the form
\beq \VV(\f) = V_1\,\log(\coth \f) + V_2 \qquad (V_1,V_2 \!\in\! \reali) ~. \label{VCase5} \feq
It should be noted that $\VV(\f)$ is well defined only for $\f \in (0,\infty)$
(apart from the trivial case where $V_1=0$, entailing $\VV(\f) = $ constant $=V_2$).
Ref.\,\cite{Fre} suggests to analyze these potentials by means of the gauge function $\BBB$
and of the new coordinates $x,y$ defined by
\begin{gather}
\BBB(\AA,\f) := - \AA\,,  \label{gau5} \\
\AA = {1 \over 2}\,\log\!\left({x^2\! - y^2 \over 2}\right) , \qquad
\f = {1 \over 2}\, \log\!\left({x + y \over x - y}\right)
\qquad (x \!>\! y \!>\! 0)\,. \label{cor5}
\end{gather}
In absence of matter and curvature, the Lagrangian $\LL(x,y,\xt,\yt)$ obtained
via these prescriptions is separable.
Following our general approach, let now us implement the prescriptions
of \cite{Fre} in our framework with matter and curvature,
and search for additional separable cases.
Eqs.\,\eqref{VCase5} \eqref{gau5} \eqref{cor5} yield for the Lagrangian \eqref{eL} and
for the energy \eqref{efun} the expressions
\begin{gather}
\LL(x,\xt,y,\yt) = \nonumber \\
{\yt^2\! - \xt^2 \over 4} - V_1 \,(\log x - \log y) - V_2
- {\ds^2\, \Oms \over 2} \left({x^2\! - \!y^2 \over 2}\right)^{\!\!-{1 + w \over 2}} \!
+ {\ds^2 k \over 2}\left({x^2\! -\! y^2 \over 2}\right)^{\!\!- {1 \over \ds}} \!, \label{LCase5} 
\end{gather}
\vfill\eject\noindent
$\phantom{a}$\vspace{-1.cm}
\begin{gather}
\Ec(x,\xt,y,\yt) = \nonumber \\
{\yt^2\! - \xt^2 \over 4} + V_1 \,(\log x - \log y) + V_2
+ {\ds^2\, \Oms \over 2} \left({x^2\! - \!y^2 \over 2}\right)^{\!\!-{1 + w \over 2}} \!
- {\ds^2 k \over 2} \left({x^2\! -\! y^2 \over 2}\right)^{\!\!- {1 \over \ds}} \!. \label{ECase5}
\end{gather}
The Lagrangian \eqref{LCase5} is given by the sum of two functions depending separately on $(x,\xt)$
and $(y,\yt)$, plus two additional terms proportional to $\Oms$ and $k$, respectively,
both consisting of suitable powers of $x^2 - y^2$.
Besides the case with $\Oms = 0$ and $k = 0$,
the only cases where the latter additional terms are themselves separable or disappear,
yielding again a separable Lagrangian of the form \eqref{lagsep},
are those where $k=0$ and the exponent $-(1+w)/2$
equals $0$ or $1$. We discuss these two cases in the sequel, keeping in mind that the
corresponding Lagrange equations can be reduced to quadratures as indicated in Eq.\,\eqref{quasep};
let us also repeat that admissible solutions must also fulfill
the energy constraint $\Ec=0$ and the conditions $x(t) \!>\! y(t) \!>\! 0$
(cf. Eq.\,\eqref{cor5}).
\vspace{-0.3cm}

\paragraph{i) $\boma{k = 0}$, $\boma{w = -1}$ (cosmological constant).}
The Lagrangian \eqref{LCase5} and the energy \eqref{ECase5} become, respectively,
\begin{gather}
\begin{array}{c}
\dd{\LL(x,y,\xt,\yt) = \LL_1(x,\xt) + \LL_2(y,\yt) - V_2  + 2\, \ds^2\,\Oms~,} \vspace{0.1cm}\\
\dd{\LL_1(x,\xt) = -\, {1 \over 4}\,\xt^2 - V_1 \log x~, \qquad \LL_2(y,\yt) = {1 \over 4}\,\yt^2 + V_1 \log y~;}
\end{array} \label{L5i}\\
\begin{array}{c}
\dd{\Ec(x,y,\xt,\yt) = \Ec_1(x,\xt) + \Ec_2(y,\yt) + V_2  - 2 \,\ds^2\, \Oms~,} \vspace{0.1cm}\\
\dd{\Ec_1(x,\xt) = - \,{1 \over 4}\,\xt^2 + V_1 \log x~, \qquad \Ec_2(y,\yt) = {1 \over 4}\,\yt^2 - V_1 \log y~.}
\end{array}\vspace{-0.cm}
\end{gather}

\paragraph{ii) $\boma{k = 0}$, $\boma{w = -3}$\,.}
The Lagrangian \eqref{LCase5} and the energy \eqref{ECase5} become
\begin{gather}
\LL(x,y,\xt,\yt) = \LL_1(x,\xt) + \LL_2(y,\yt) - V_2~, \label{L5ii} \\
\LL_1(x,\xt) = - \,{1 \over 4}\,\xt^2 - V_1 \log x + 4 \,\ds^2\, \Oms\, x^2~, \qquad
\LL_2(y,\yt) = {1 \over 4}\,\yt^2 + V_1 \log y - 4 \,\ds^2\, \Oms\, y^2~; \nonumber \\
\Ec(x,y,\xt,\yt) = \Ec_1(x,\xt) + \Ec_2(y,\yt) + V_2~, \\
\Ec_1(x,\xt) = -\, {1 \over 4}\,\xt^2 + V_1 \log x - 4 \,\ds^2\, \Oms\, x^2~, \qquad
\Ec_2(y,\yt) = {1 \over 4}\,\yt^2 - V_1 \log y + 4 \,\ds^2\, \Oms\, y^2~. \nonumber
\end{gather}

\subsection{Class 6 potentials}
\label{subclass6}
Let us consider potentials of the form
\beq \VV(\f) = V_1\,\arctan e^{-2\,\f} + V_2 \qquad (V_1,V_2 \!\in \!\reali) ~. \label{VCase6} \feq
In connection with this class, \cite{Fre} suggests to consider the gauge function
\beq \BBB(\AA,\f) := -\,\AA ~ \label{gau6} \feq
and to replace the Lagrangian coordinates $(\AA,\f)$ with a conveniently defined, complex variable $z$.
Regarding this complex setting, it can be useful to specify the following conventions,
somehow implicit in the cited reference:
\vspace{-0.1cm}
\begin{itemize}
\item[$\centerdot$] $\complessic := \complessi \backslash (-\infty, 0]$
is the open region in the complex plane $\complessi$ obtained removing the
negative real semi-axis. Correspondingly, we consider the determination of the
argument function given by
\beq \arg: \complessic \backslash (-\infty, 0] \to (-\pi, \pi)~, \quad z \mapsto \arg z~. \label{argdef}\feq
This is such that $\arg z = 0$ for $z \in (0,\infty)$ and
$\arg(\bar{z}) = - \arg z$ for all $z \in \complessic$.
\item[$\centerdot$] The usual, natural logarithm $\log: (0,\infty) \to \reali$ possesses the extension
\beq \log : \complessic \to \reali + i (-\pi, \pi), \quad
z \mapsto \log z := \log |z| + i \arg z~, \label{logdef}\feq
with $\arg$ as in Eq.\,\eqref{argdef}. Such an extension is a holomorphic function fulfilling
$e^{\log z} = z$ and $\log \bar{z} = \log |z| - i \arg z = \overline{\log z}$ for all $z \in \complessic$.
\end{itemize}
\vfill\eject\noindent
$\phantom{a}$\\
Keeping these premises in mind, the complex formalism of \cite{Fre} can be described as follows:
the coordinates $(\AA,\f) \in \reali^2$ are replaced by a complex coordinate $z \in \Dd$ with
\beq \Dd := \{ z \in \complessi~|~\Re z, \Im z > 0 \} \subset \complessic~, \feq
related to $(\AA, \f)$ by
\begin{gather}
\AA = {1 \over 2} \log\!\left({z^2 - \bar{z}^2 \over 2i}\right)
= {1 \over 2} \log \big(2 \, \Re z \Im z\big)\,, \label{traz} \qquad
\f = {1 \over 2} \log \!\left(i\, {z + \bar{z} \over z - \bar{z}}\right)
= {1 \over 2} \log\!\left({\Re z \over \Im z}\right) .
\end{gather}
The correspondence $z \mapsto (\AA,\f)$ defined above is one-to-one between
$\Dd$ and $\reali^2$, with inverse
\beq z = {1 \over \sqrt{2}} \left( e^{\AA + \f} + i \, e^{\AA - \f} \right). \feq
Let us note that the second relation in Eq.\,\eqref{traz} implies $e^{- 2 \f} = \Im z/\Re z$,
which allows to express the potential \eqref{VCase6} as
\beq \VV(\f) =
V_1 \, \arctan {\Im z \over \Re z} + V_2 = V_1 \arg z + V_2 = V_1\, \Im \log z + V_2~. \feq
In absence of matter and curvature, \cite{Fre} expresses the Lagrangian function
associated to a potential of the form \eqref{VCase6} fixing the gauge and introducing
a complex coordinate $z$ as above; the Lagrangian $\mathscr{L}(z,\zt)$ obtained in this
way is of the holomorphic type \eqref{lagcomp}.
Following our general mindset, hereafter we try to generalize the results of \cite{Fre}
to cases with matter or curvature. To this purpose, let us first note that the prescriptions
\eqref{gau6} \eqref{traz} yield the following expressions for the Lagrangian \eqref{eL}
and for the energy \eqref{efun}:
\begin{gather}
\LL(z,\dot{z}) = -\, \Im\! \left({1 \over 2}\,\dot{z}^2 + V_1 \log z \!\right)
- {\ds^2\,\Oms \over 2}\; \big(\Im z^2\big)^{-{1+w \over 2}}\!
+ {\ds^2 k \over 2}\; \big(\Im z^2\big)^{-{1 \over \ds}} - V_2~ ,\label{LCase6} \\
\Ec(z,\dot{z}) = -\, \Im\! \left({1 \over 2}\,\dot{z}^2 - V_1 \log z \!\right)
+ {\ds^2\,\Oms \over 2}\; \big(\Im z^2\big)^{-{1+w \over 2}}\!
- {\ds^2 k \over 2}\; \big(\Im z^2\big)^{-{1 \over \ds}} + V_2 \label{ECase6}
\end{gather}
(here and in the sequel, $\Im z^2$ stands for $\Im (z^2)$).
Let us point out that Eqs.\,\eqref{LCase6} \eqref{ECase6} have the same
structure as Eqs.\,\eqref{LCase5}\eqref{ECase5} in the previous section,
with the replacement $(x,y) \!\to\! (z,\bar{z})$.
\parn
We are interested in cases in which the Lagrangian \eqref{LCase6} maintains the holomorphic structure
\eqref{lagcomp} even in presence of matter or curvature.
Clearly, this occurs only if $k=0$ and the exponent
$-(1+w)/2$ equals $0$ or $1$, which yields the cases discussed below.
Let us recall that admissible solutions must fulfill the energy constraint $\Ec = 0$,
besides the condition $z(t) \!\in\! \Dd$.
\vspace{-0.cm}

\paragraph{i) $\boma{k = 0}$, $\boma{w = -1}$ (cosmological constant).}
The Lagrangian \eqref{LCase6} and the energy \eqref{ECase6} become, respectively,
\begin{gather}
\LL(z,\dot{z}) =  -\, \Im\! \left({1 \over 2}\,\dot{z}^2 + V_1 \log z \right)
- {\ds^2\,\Oms \over 2} - V_2~, \\
\Ec(z,\dot{z}) =  -\, \Im\! \left({1 \over 2}\,\dot{z}^2 - V_1 \log z \right)
+ {\ds^2\,\Oms \over 2} + V_2~.
\end{gather}
One can apply the methods described below Eq.\,\eqref{lagcomp}
to solve the corresponding Lagrange equations by quadratures;
in the present case the holomorphic function $U$ and the complexified
energy $\mathfrak{E}$ of Eqs.\,\eqref{lagcomp} \eqref{encomp} are
$U(z) = - V_1 \log z$\,, $\mathfrak{E}(z, \dot{z}) = {1 \over 2}\,\dot{z}^2 - V_1 \log z$.
\vspace{-0.4cm}

\paragraph{ii) $\boma{k = 0}$, $\boma{w = -3}$\,.}
The Lagrangian \eqref{LCase6} and the energy \eqref{ECase6} reduce to\vspace{-0.1cm}
\begin{gather}
\LL(z,\dot{z}) =  -\, \Im\! \left({\dot{z}^2 \over 2} + V_1 \log z + {\ds^2\,\Oms \over 2}\; z^2 \right) - V_2~, \\
\Ec(z,\dot{z}) =  -\, \Im\! \left({\dot{z}^2 \over 2} - V_1 \log z - {\ds^2\,\Oms \over 2}\; z^2 \right) + V_2~.
\vspace{-0.1cm}
\end{gather}
Again, one should refer to the methods reported below Eq.\,\eqref{lagcomp};
in the present case the holomorphic function $U$ and the complexified
energy function $\mathfrak{E}$ of Eqs.\,\eqref{lagcomp} \eqref{encomp} are given by
$U(z) = - V_1 \log z - {\ds^2\,\Oms \over 2} \, z^2$\,,
$\mathfrak{E}(z, \dot{z}) = {1 \over 2}\,\dot{z}^2 - V_1 \log z - {\ds^2\,\Oms \over 2} \, z^2$.

\subsection{Class 7 potentials}
Let us consider a potential of the form\vspace{-0.1cm}
\beq
\VV(\f) = V_1\,\big(\!\cosh(\ga\,\f)\big)^{{2 \over \ga} - 2}
+ V_2\,\big(\!\sinh(\ga\,\f)\big)^{{2 \over \ga} - 2} \qquad
\big(V_1,V_2 \!\in\! \reali\,,~\ga \!\in\! \reali \backslash\{0\}\big) ~; \label{VCase7} \vspace{-0.1cm}
\feq
here and in the sequel we assume
\beq \f \in I_{\ga, V_2}~, \quad~
I_{\ga, V_2} := \left\{\!\begin{array}{ll}
\dd{(-\infty,+\infty)}          &   \dd{\mbox{if}~~ {2 \over \ga} - 2 \in \{0,1,2,...\}~~\mbox{or}~~V_2=0 \,,} \vspace{0.2cm}\\
\dd{ \sgn(\ga)\, (0,+\infty)}    &   \dd{\mbox{if}~~ {2 \over \ga} - 2 \not \in \{0,1,2,...\}~~\mbox{and}~~ V_2 \neq 0\,,}
\end{array}
\right.
\label{igavd}
\feq
where $\sgn(\ga) \, (0,+\infty) :=
\big\{ \sgn(\ga) \psi~|~\psi \!\in\! (0,+\infty) \big\}$
(this set equals $(0,+\infty)$ if $\ga \!>\! 0$, and $(-\infty,0)$ if $\ga \!<\! 0$).
In any case, $I_{\ga, V_2}$ is a maximal interval where the function
$\VV$ in Eq.\,\eqref{VCase7} is well defined and smooth.
Let us also note that
\beq {2 \over \ga} - 2 = h \in \{0,1,2,...\} \qquad \Leftrightarrow \qquad
\ga = {2 \over h + 2}~, \quad h \in \{0,1,2,...\}~.
\feq
To treat a potential of the above form, \cite{Fre} suggests to use the gauge function
$\BBB$ and the real coordinates $x,y$, determined as follows:\vspace{-0.1cm}
\begin{gather}
\BBB(\AA,\f) = (1-2\ga)\,\AA ~; \label{gau7}\\
\AA = {1 \over 2\,\ga}\,\log(x^2\! - y^2) ~, \qquad
\f = {1 \over 2\,\ga}\,\log\!\left({x+y \over x-y}\right)~,
\qquad (x,y) \in \mathcal{D}_{\ga, V_2} \subset \reali^2 . \label{cor7}\vspace{-0.1cm}
\end{gather}
The domain $\mathcal{D}_{\ga, V_2}$ is not indicated explicitly
in \cite{Fre}, but it is evident that
\beq \mathcal{D}_{\ga, V_2} := \left\{\!\!\begin{array}{ll}
\dd{\big\{ (x,y) \!\in\! \reali^2 ~\big|~ x>0\,,\; - x < y < x \big\}}
& \dd{\mbox{if}~~ {2 \over \ga} - 2 \in \{0,1,2,...\}~~\mbox{or}~~V_2=0 \,,} \vspace{0.2cm} \\
\dd{\big\{ (x,y) \!\in\! \reali^2 ~\big|~ x> y > 0 \big\} }
& \dd{\mbox{if}~~{2 \over \ga} - 2 \not\in \{0,1,2,...\} ~~\mbox{and}~~ V_2 \neq 0\,.}
\end{array}
\right.
\label{cor7bis} \feq
In fact, it can be readily checked that the map $(x,y) \mapsto (\AA,\f)$ described
in Eq.\,\eqref{cor7} is a smooth diffeomorphism between the open sets
$\mathcal{D}_{\ga, V_2}$ and $\reali \times I_{\ga, V_2}$,
with inverse function given by
\beq x = {1 \over 2} \left( e^{\ga (\AA + \f)} + e^{\ga (\AA - \f)} \right) , \qquad
y = {1 \over 2} \left( e^{\ga (\AA + \f)} - e^{\ga (\AA - \f)} \right) . \feq
In the case of no matter and zero curvature, the Lagrangian $\LL(x,y,\xt,\yt)$
obtained with the above prescriptions is separable.
As usual, let us try to use the same prescriptions adding
matter and curvature. Eqs.\,\eqref{VCase7}\eqref{gau7}\eqref{cor7}
allow to express the Lagrangian \eqref{eL} and the energy \eqref{efun}, respectively, as\vspace{-0.2cm}
\begin{gather}
\LL(x,\xt,y,\yt) = \nonumber \\
{\yt^2\!- \!\xt^2 \over 2\,\ga^2}
- V_1\, x^{{2 \over \ga} - 2}\! - V_2 \, y^{{2 \over \ga} - 2}
- {\ds^2\, \Oms \over 2}\; (x^2\!-y^2)^{{1 - 2 \ga - w\over 2\ga}}\!
+ {\ds^2 k\over 2}\; (x^2\!-y^2)^{{\ds(1-\ga) - 1} \over \ds\, \ga} , \label{LCase7} \\
\Ec(x,\xt,y,\yt) = \nonumber \\
{\yt^2\!- \!\xt^2 \over 2\,\ga^2}
+ V_1\, x^{{2 \over \ga} - 2}\! + V_2 \, y^{{2 \over \ga} - 2}
+ {\ds^2\, \Oms \over 2}\; (x^2\!-y^2)^{{1 - 2 \ga - w\over 2\ga}}\!
- {\ds^2 k \over 2}\; (x^2\!-y^2)^{{\ds(1-\ga) - 1} \over \ds\, \ga} .\label{ECase7}
\end{gather}
Similarly to the case of class 5 potentials dealt with in subsection \ref{class5},
the Lagrangian \eqref{LCase7} is given by the sum of two functions
depending separately on $(x, \xt)$ and on $(y, \yt)$,
plus two extra terms proportional to $\Oms$ and $k$, respectively,
which consist of powers of $x^2 - y^2$ with exponents ${1 - 2 \ga - w\over 2\ga}$
and ${\ds(1-\ga) - 1 \over \ds\, \ga}$.
Besides the case with $\Oms = 0$ and $k = 0$,
the only situations in which these extra terms are themselves separable or disappear,
yielding a separable Lagrangian of the form \eqref{lagsep}, are listed in the following.
The resulting (decoupled) Lagrange equations for $x(t), y(t)$ can be reduced to quadratures
as indicated in Eq.\,\eqref{quasep}; the admissible solutions must also fulfill
the condition $(x(t), y(t)) \in \mathcal{D}_{\ga, V_2}$ (cf. Eqs.\,\eqref{cor7}\eqref{cor7bis})
and the energy constraint $\Ec = 0$.
\vspace{-0.4cm}

\paragraph{i) $\boma{k = 0}$, $\boma{\ga = {1-w \over 2} \neq 0}$\,.}
The Lagrangian \eqref{LCase7} and the energy \eqref{ECase7} take the form\vspace{-0.1cm}
\begin{gather}
\begin{array}{c}
\dd{\LL(x,y,\xt,\yt) = \LL_1(x,\xt) + \LL_2(y,\yt) - {\ds^2\,\Oms \over 2} ~,}\vspace{0.1cm}\\
\dd{\LL_1(x,\xt) = - \,{2 \over (1\!-\!w)^2}\,\xt^2 - V_1\, x^{{2 (1 + w) \over 1 - w}}~, \qquad
\LL_2(y,\yt) = {2 \over (1\!-\!w)^2}\,\yt^2 - V_2\, y^{{2 (1 + w) \over 1 - w}}~;} \label{L7i}
\end{array} \\
\begin{array}{c}
\dd{\Ec(x,y,\xt,\yt) = \Ec_1(x,\xt) + \Ec_2(y,\yt) + {\ds^2\,\Oms \over 2}~,} \vspace{0.1cm} \\
\dd{\Ec_1(x,\xt) = - \,{2 \over (1\!-\!w)^2}\,\xt^2 + V_1\, x^{{2 (1 + w) \over 1 - w}}~, \qquad
\Ec_2(y,\yt) = {2 \over (1\!-\!w)^2}\,\yt^2 + V_2\, y^{{2 (1 + w) \over 1 - w}}~ .} \label{E7i}
\end{array}
\end{gather}
Let us mention that in the subcases $w=-1$ (cosmological constant), $w=-1/3$ and $w=0$ (dust), the
exponent of $x$ in $\LL_1(x,\xt)$ and of $y$ in $\LL_2(y,\yt)$ becomes,
respectively, $0,1$ and $2$, so that the Lagrange equations are elementary.
\vspace{-0.3cm}

\paragraph{ii) $\boma{k=0}$, $\boma{\ga = {1-w \over 4} \neq 0}$\,.}
The Lagrangian \eqref{LCase7} and the energy \eqref{ECase7} become
\begin{gather}
\LL(x,y,\xt,\yt) = \LL_1(x,\xt) + \LL_2(y,\yt)~, \\
\LL_1(x,\xt) = - \,{8 \over (1\!-\!w)^2}\,\xt^2 - V_1\, x^{{2 (3 + w) \over 1 - w}}\! - {\ds^2\,\Oms \over 2}\; x^2\,, \nonumber \\
\LL_2(y,\yt) = {8 \over (1\!-\!w)^2}\,\yt^2 - V_2\, y^{{2 (3 + w) \over 1 - w}}\! - {\ds^2\,\Oms \over 2}\; y^2 \,; \nonumber\vspace{-0.3cm}
\end{gather}
\begin{gather}
\Ec(x,y,\xt,\yt) = \Ec_1(x,\xt) + \Ec_2(y,\yt)~, \\
\Ec_1(x,\xt) = - \,{8 \over (1\!-\!w)^2}\,\xt^2 + V_1\, x^{{2 (3 + w) \over 1 - w}} + {\ds^2\,\Oms \over 2}\; x^2 \,, \nonumber \\
\Ec_2(y,\yt) = {8 \over (1\!-\!w)^2}\,\yt^2 + V_2\, y^{{2 (3 + w) \over 1 - w}} + {\ds^2\,\Oms \over 2}\; y^2 \,. \nonumber
\end{gather}
The subcases $w=-3, -5/3, -1$ are elementary, since $x$ and $y$
appear in $\LL_1(x,\xt)$ and $\LL_2(y,\yt)$ with exponent $2$, or
with exponents $1$ and $2$.
\vspace{-0.3cm}

\paragraph{iii) $\boma{\Oms=0}$, $\boma{\ga = {\ds-1 \over \ds}}$\,.}
The Lagrangian \eqref{LCase7} and the energy \eqref{ECase7} reduce to
\begin{gather}
\begin{array}{c}
\dd{\LL(x,y,\xt,\yt) = \LL_1(x,\xt) + \LL_2(y,\yt) + {\ds^2 k \over 2}~,} \vspace{0.1cm} \\
\dd{\LL_1(x,\xt) = -\, {\ds^2 \over 2 (\ds\!-\!1)^2}\,\xt^2 - V_1\, x^{{2 \over \ds-1}} ~, \qquad
\LL_2(y,\yt) = {\ds^2 \over (\ds\!-\!1)^2}\,\yt^2 - V_2\, y^{{2 \over \ds-1}} ~;}
\end{array}\label{L7iii}
\end{gather}
\begin{gather}
\begin{array}{c}
\dd{\Ec(x,y,\xt,\yt) = \Ec_1(x,\xt) + \Ec_2(y,\yt) - {\ds^2 k \over 2} ~,} \vspace{0.1cm} \\
\dd{\Ec_1(x,\xt) = -\, {\ds^2 \over 2 (\ds\!-\!1)^2}\,\xt^2 + V_1\, x^{{2 \over \ds-1}} ~, \qquad
\Ec_2(y,\yt) = {\ds^2 \over (\ds\!-\!1)^2}\,\yt^2 + V_2\, y^{{2 \over \ds-1}} ~.}
\end{array}
\end{gather}
In the subcases $n=2$ and $n=3$ the
exponent of $x$ in $\LL_1(x,\xt)$ and of $y$ in $\LL_2(y,\yt)$ becomes, respectively,
$2$ and $1$, so the Lagrange equations are elementary.
\vspace{-0.4cm}

\paragraph{iv) $\boma{\Oms=0}$, $\boma{\ga = {\ds-1 \over 2 \ds}}$\,.}
The Lagrangian \eqref{LCase7} and the energy \eqref{ECase7} read
\begin{gather}
\LL(x,y,\xt,\yt) = \LL_1(x,\xt) + \LL_2(y,\yt)~, \label{L7iv2a} \\
\LL_1(x,\xt) = - \,{2 \, \ds^2 \over 2 (\ds\!-\!1)^2}\,\xt^2
- V_1\, x^{{2 (\ds+1) \over \ds-1}} + {\ds^2 k \over 2}\; x^2\,, \!\qquad
\LL_2(y,\yt) = {2 \ds^2 \over 2 (\ds\!-\!1)^2}\, \yt^2
- V_2\, y^{{2 (\ds+1) \over \ds-1}} - {\ds^2 k \over 2}\; y^2\,; \nonumber \\
\Ec(x,y,\xt,\yt) = \Ec_1(x,\xt) + \Ec_2(y,\yt)~, \\
\Ec_1(x,\xt) = - {2 \, \ds^2 \over 2 (\ds\!-\!1)^2}\,\xt^2
+ V_1\, x^{{2 (\ds+1) \over \ds-1}} - {\ds^2 k \over 2}\; x^2\,, \!\qquad
\Ec_2(y,\yt) = {2 \ds^2 \over 2 (\ds\!-\!1)^2}\, \yt^2
+ V_2\, y^{{2 (\ds+1) \over \ds-1}} + {\ds^2 k \over 2}\; y^2 \,. \nonumber
\vspace{-0.2cm}
\end{gather}
\vspace{-0.3cm}

\paragraph{v) $\boma{\ga = - {\ds-1 \over \ds}}$, $\boma{w = - {\ds-2 \over \ds}}$\,.}
The Lagrangian \eqref{LCase7} and the energy \eqref{ECase7} take the forms
\begin{gather}
\begin{array}{c}
\dd{\LL(x,y,\xt,\yt) = \LL_1(x,\xt) + \LL_2(y,\yt) - {\ds^2\, (\Oms\! - k) \over 2}~,} \vspace{0.1cm} \\
\dd{\LL_1(x,\xt) = - \,{\ds^2 \over 2 (\ds\!-\!1)^2}\,\xt^2 - V_1\, x^{{2 \over \ds-1}} \,, \qquad
\LL_2(y,\yt) = {\ds^2 \over (\ds\!-\!1)^2}\,\yt^2 - V_2\, y^{{2 \over \ds-1}} \,;}
\end{array} \\
\begin{array}{c}
\dd{\Ec(x,y,\xt,\yt) = \Ec_1(x,\xt) + \Ec_2(y,\yt) + {\ds^2\,(\Oms\! - k) \over 2}~,} \vspace{0.1cm}\\
\dd{\Ec_1(x,\xt) = - \,{\ds^2 \over 2 (\ds\!-\!1)^2}\,\xt^2 + V_1\, x^{{2 \over \ds-1}} \,, \qquad
\Ec_2(y,\yt) = {\ds^2 \over (\ds\!-\!1)^2}\,\yt^2 + V_2\, y^{{2 \over \ds-1}} \,. }
\end{array}
\end{gather}
Let us note the close similarities with case (iii):
$\LL_1$ and $\LL_2$ coincide with the homonymous functions in \eqref{L7iii},
and $\LL$ is like the homonymous Lagrangian in the same equation with $k$
replaced by $k - \Oms$. The subcases
$n=2,3$ are elementary, for the same reasons indicated in subcase (iii).
\vspace{-0.3cm}

\paragraph{vi) $\boma{\ga = {\ds-1 \over \ds}}$, $\boma{w = - {3\ds - 4  \over \ds}}$\,.}
The Lagrangian \eqref{LCase7} and the energy \eqref{ECase7} become
\begin{gather}
\LL(x,y,\xt,\yt) = \LL_1(x,\xt) + \LL_2(y,\yt) +  {\ds^2 k \over 2} ~, \\
\LL_1(x,\xt) = - \,{\ds^2 \over 2 (\ds\!-\!1)^2}\,\xt^2
- V_1\, x^{{2 \over \ds-1}} - {\ds^2\,\Oms \over 2}\; x^2 ~, \nonumber \\
\LL_2(y,\yt) = {\ds^2 \over (\ds\!-\!1)^2}\,\yt^2
- V_2\, y^{{2 \over \ds-1}} + {\ds^2\,\Oms \over 2}\; y^2 ~; \nonumber
\end{gather}
\begin{gather}
\Ec(x,y,\xt,\yt) = \Ec_1(x,\xt) + \Ec_2(y,\yt) - {\ds^2 k \over 2} ~, \\
\Ec_1(x,\xt) = - \,{\ds^2 \over 2 (\ds\!-\!1)^2}\,\xt^2
+ V_1\, x^{{2 \over \ds-1}} + {\ds^2\,\Oms \over 2}\; x^2 ~, \nonumber \\
\Ec_2(y,\yt) = {\ds^2 \over (\ds\!-\!1)^2}\,\yt^2
+ V_2\, y^{{2 \over \ds-1}} - {\ds^2\,\Oms \over 2}\; y^2 ~. \nonumber
\end{gather}
The subcases $n\!=\!2,3$ are elementary, since $x,y$ appear in $\LL_1(x,\xt),\LL_2(y,\yt)$
with exponent $2$, or with exponents $1,2$.\vspace{-0.2cm}
\vfill\eject\noindent

\paragraph{vii) $\boma{\ga = - {\ds-1 \over 2 \ds}}$, $\boma{w = {1 \over \ds}}$ (radiation gas).}
The Lagrangian \eqref{LCase7} and the energy \eqref{ECase7} read
\begin{gather}
\LL(x,y,\xt,\yt) = \LL_1(x,\xt) + \LL_2(y,\yt) - {\ds^2\,\Oms \over 2}~, \\
\LL_1(x,\xt) = - \,{2 \, \ds^2 \over 2 (\ds\!-\!1)^2}\,\xt^2\!
- V_1\, x^{{2 (\ds+1) \over \ds-1}}\! + {\ds^2 \over 2}\,k\, x^2\,, \qquad
\LL_2(y,\yt) = {2\, \ds^2 \over 2 (\ds\!-\!1)^2}\, \yt^2
- V_2\, y^{{2 (\ds+1) \over \ds-1}}\! - {\ds^2 k \over 2}\; y^2 \,; \nonumber
\end{gather}
\begin{gather}
\Ec(x,y,\xt,\yt) = \Ec_1(x,\xt) + \Ec_2(y,\yt) + {\ds^2\,\Oms \over 2}~, \\
\Ec_1(x,\xt) = - \,{2 \, \ds^2 \over 2 (\ds\!-\!1)^2}\,\xt^2\!
+ V_1\, x^{{2 (\ds+1) \over \ds-1}}\! - {\ds^2 k \over 2}\; x^2\,, \quad
\Ec_2(y,\yt) = {2\, \ds^2 \over 2 (\ds\!-\!1)^2}\, \yt^2
+ V_2\, y^{{2 (\ds+1) \over \ds-1}}\! + {\ds^2 k \over 2}\; y^2 \,. \nonumber
\end{gather}
Note the strong similarities with case (iv):
$\LL_1$ and $\LL_2$ are as in Eq.\,\eqref{L7iv2a}, while $\LL$ is like the homonymous
Lagrangian in the same equation with the additional constant $- {\ds^2 \over 2}\,\Oms$.
\vspace{-0.4cm}

\paragraph{viii) $\boma{\ga = {\ds-1 \over 2 \ds}}$, $\boma{w = - {\ds-2 \over \ds}}$\,.}
The Lagrangian \eqref{LCase7} and the energy \eqref{ECase7} reduce to
\begin{gather}
\LL(x,y,\xt,\yt) = \LL_1(x,\xt) + \LL_2(y,\yt)~, \\
\LL_1(x,\xt) = -\,{2 \, \ds^2 \over 2 (\ds\!-\!1)^2}\,\xt^2\!
- V_1\, x^{{2 (\ds+1) \over \ds-1}}\! - {\ds^2\,(\Oms\!-\!k) \over 2}\; x^2\,, \nonumber\\
\LL_2(y,\yt) = {2\, \ds^2 \over 2 (\ds\!-\!1)^2}\, \yt^2\!
- V_2\, y^{{2 (\ds+1) \over \ds-1}}\! + {\ds^2\,(\Oms\!-\!k) \over 2}\; y^2 \,; \nonumber
\end{gather}
\begin{gather}
\Ec(x,y,\xt,\yt) = \Ec_1(x,\xt) + \Ec_2(y,\yt)~, \\
\Ec_1(x,\xt) = -\,{2 \, \ds^2 \over 2 (\ds\!-\!1)^2}\,\xt^2\!
+ V_1\, x^{{2 (\ds+1) \over \ds-1}}\! + {\ds^2\,(\Oms\!-\!k) \over 2}\; x^2\,, \nonumber\\
\Ec_2(y,\yt) = {2\, \ds^2 \over 2 (\ds\!-\!1)^2}\, \yt^2\!
+ V_2\, y^{{2 (\ds+1) \over \ds-1}}\! - {\ds^2\,(\Oms\!-\!k) \over 2}\; y^2 \,; \nonumber
\end{gather}
Note that $\LL_1$, $\LL_2$ and $\LL$ are like the homonymous Lagrangians in Eq.\,\eqref{L7iv2a},
with $k$ replaced by $k - \Oms$.

\subsection{Class 8 potentials}
Let us consider potentials of the form
\beq \begin{array}{c}
\dd{\VV(\f) = C \;\big(\!\cosh (2 \,\ga\,\f)\big)^{{1 \over \ga}-1}\;
\sin\!\left[\Big({1 \over \ga} - 1\Big) \arctan\!\Big(\,{1 \over \sinh (2\, \ga\, \f)}\Big) + \vartheta \right]}
\vspace{0.1cm}\\
\dd{\big(C \in [0,+\infty),\vartheta \in [0, 2 \pi)\,,\;\ga \in \reali \backslash\{0\}\big) ~.}\label{VCase8}
\end{array} \feq
These make sense for $\f \in \reali \backslash \{0\}$. However it is natural
to require that $\f$ ranges within a connected domain; so, we assume
$\f  \!\in\! \sgn(\ga) \,(0,+\infty)$ (see the explanation
after Eq. \eqref{igavd}; this means
$\f \!\in\! (0,+\infty)$ if $\gamma \!>\!0$ and $\f \!\in\! (-\infty,0)$
if $\gamma \!<\!0$). The alternative choice $\f \!\in\! \sgn(\ga)\,(-\infty,0)$,
of obvious meaning, could be treated similarly.\parn
Like the class 6 potentials addressed in subsection \ref{subclass6}, the present class 8
can be treated using a complex formalism. To this purpose, let $\complessic$, $\arg$ and $\log$
be defined as in Eqs.\,\eqref{argdef} \eqref{logdef} of subsection \ref{subclass6}
(see also the related comments); in addition, let us put
\beq z^{\lambda} := e^{\lambda \log z} \qquad
\mbox{for $z \!\in\! \complessic$, $\lambda \!\in\! \reali$}. \feq
For any $z,\lambda$ as above, the map $z \mapsto z^{\lambda}$ is holomorphic on $\complessic$
and we have $z^{\lambda} = |z|^{\lambda}\, e^{i \lambda\, arg z}$,
$\overline{z^{\lambda}} = {\bar{z}}^{\lambda}$. \parn
Potentials of the form \eqref{VCase8} were treated in \cite{Fre} fixing the gauge function $\BBB$
and replacing the Lagrangian coordinates $(\AA,\f)$ with a complex variable $z$, setting
\begin{gather}
\BBB(\AA,\f) = (1-2\ga)\,\AA ~; \label{gau8}\\
\AA = {1 \over 2 \ga} \log\!\left({z^2 \!-\! \bar{z}^2 \over 2i}\right)
= {1 \over 2 \ga} \log \big(2 \, \Re z \Im z\big) ~, \label{cor8} \qquad
\f = {1 \over 2 \ga} \log \!\left(i\, {z + \bar{z} \over z - \bar{z}}\right)
= {1 \over 2 \ga} \log\! \left({\Re z \over \Im z}\right)
\end{gather}
(here and in the sequel $\Re z^2, \Im z^2$ stand for $\Re(z^2), \Im(z^2)$).
In the above, $z$ is assumed to belong to a suitable domain $\Dd \subset \complessi$,
which is not described explicitly in \cite{Fre} and that we take as follows:
\beq \Dd := \big\{ z \!\in\! \complessi~\big|~\Re z \!>\! \Im z \!>\! 0 \big\}
= \big\{ z \!\in\! \complessic ~\big|~0 \!<\! \arg z \!<\! {\pi/ 4} \big\}~;
\label{defdi} \feq
the coordinate transformation \eqref{cor8} is one-to-one between the domain $\Dd$ and the set
$\{ (\AA,\f)~|~\AA \!\in\! \reali,\,  \f \in \sgn(\ga)\,(0,+\infty) \}$.
\parn
From here to the end of this subsection, we stick to the position \eqref{defdi}.
The inverse of the map $(\AA,\f) \mapsto z \in \Dd$
described in Eq.\,\eqref{cor8} is given by
\beq z = {1 \over \sqrt{2}} \left( e^{\ga(\AA + \f)} + i \, e^{\ga(\AA - \f)} \right) . \feq
To proceed, we claim that Eqs.\,\eqref{VCase8} \eqref{cor8} entail the following identities
({\footnote{The first identity in Eq.\,\eqref{vvf} follows trivially from
the expression for $\AA$ in Eq.\,\eqref{cor8} and from the basic equality $(z^2-\bar{z}^2)/(2i) = \Im z^2$.
To obtain the second identity in Eq.\,\eqref{vvf}, first notice that the expression
for $\f$ in Eq.\,\eqref{cor8} entails $e^{2 \ga \f} = (\Re z)/(\Im z)$;
writing $\cosh$ and $\sinh$ in terms of exponentials, from the latter identity we infer
$$ \cosh(2\, \ga\, \f) = {(\Re z)^2 + (\Im z)^2 \over 2 \,\Re z \Im z} = {|z|^2 \over \Im z^2}~, \qquad
\sinh(2 \,\ga\, \f) = {(\Re z)^2 - (\Im z)^2 \over 2 \,\Re z \Im z} = {\Re z^2 \over \Im z^2}~. $$
In view of the above relations, starting from Eq.\,\eqref{VCase8} and setting $C := V\,e^{-i\theta}$
we infer the following chain of identities:
\begin{align*}
\VV(\f) & = \big(\!\cosh (2\,\ga\,\f)\big)^{{1 \over \ga}-1}\;
\Im \!\left(\!V \, e^{i \left({1 \over \ga} - 1\right)\arctan\left({1 \over \sinh (2 \ga\, \f)}\right)}\!\right)
= {|z|^{{2 \over \ga}-2} \over (\Im z^2)^{{1 \over \ga} - 1}}\;
\Im\!\left(\!V \, e^{i \left({1 \over \ga} - 1\right)\arctan\left(\!{\Im z^2 \over \Re z^2}\!\right)}\!\right)\vspace{-0.2cm}\\
& = {|z|^{{2 \over \ga}-2} \over (\Im z^2)^{{1 \over \ga} - 1}}\;
\Im \!\left(V \, e^{i \left({1 \over \ga} - 1\right) \arg z^2} \right)
= {\Im \!\left(V |z|^{{2 \over \ga}-2} \; e^{i \left({2 \over \ga} - 2\right) \arg z} \right)
\over (\Im z^2)^{{1 \over \ga} - 1}} = {\Im \!\left(V\,z^{{2 \over \ga}-2} \right) \over (\Im z^2)^{{1 \over \ga} - 1}}~.
\end{align*}
\vspace{-0.3cm}
}}):
\begin{gather}
e^{\AA} = (\Im z^2)^{1 \over 2 \ga}~; \qquad \quad
\VV(\f) = {\Im \!\left(V z^{{2 \over \ga}-2} \right) \over (\Im z^2)^{{1 \over \ga} - 1}}
\quad \mbox{with} \quad V := C\, e^{i \theta} \,. \label{vvf}
\end{gather}
The main result of \cite{Fre} about potentials of the form \eqref{VCase8}
with no matter and zero curvature is that the Lagrangian $\LL(z,\zt)$ arising
from the gauge choice \eqref{gau8} and the coordinate change \eqref{cor8}
is of the holomorphic type \eqref{lagcomp}. As usual,
we try to generalize this result using the prescriptions of \cite{Fre}
in presence of matter and curvature. Using Eqs.\,\eqref{VCase8}\eqref{gau8}\eqref{cor8}
and some related identities, especially Eq.\,\eqref{vvf}),
the Lagrangian \eqref{eL} and the corresponding energy \eqref{efun} become
\begin{gather}
\LL(z,\dot{z}) = -\,\Im\! \left({1 \over 2 \ga^2}\,\dot{z}^2 + V\,z^{{2 \over \ga} - 2} \right)
- {\ds^2\,\Oms \over 2}\; (\Im z^2)^{{1 - 2\ga - w \over 2\ga}}
+ {\ds^2 k \over 2}\; (\Im z^2)^{{\ds(1-\ga)-1 \over \ds\,\ga}} \,, \label{LCase8} \\
\Ec(z,\dot{z}) = -\,\Im\! \left({1 \over 2 \ga^2}\,\dot{z}^2 - V\,z^{{2 \over \ga} - 2} \right)
+ {\ds^2\,\Oms \over 2}\; (\Im z^2)^{{1 - 2\ga - w \over 2\ga}}
- {\ds^2 k \over 2}\; (\Im z^2)^{{\ds(1-\ga)-1 \over \ds\,\ga}} \,. \label{ECase8}
\end{gather}
Let us note the close analogies between the present Lagrangian \eqref{LCase8}
and the Lagrangian \eqref{LCase7}; in fact, writing $\Im z^2 = (z^2 - \bar{z}^2)/(2 i)$
and using similar relations (in particular for $\Im \zt^2$) we see
that the role played in Eq.\,\eqref{LCase8} by the complex pair $(z, \bar{z})$
is similar to the role played in Eq.\,\eqref{LCase7} by the real pair
$(x,y)$. \parn
We now search for cases with matter or curvature, in which the Lagrangian \eqref{LCase8} has the
holomorphic structure \eqref{lagcomp}. The problem is similar to that of finding the
separable cases for the Lagrangian \eqref{LCase7}; it can be treated fixing the attention
on the terms in Eq.\,\eqref{LCase8} containing $\Im z^2$, which have coefficients proportional
to $\Oms$ or $k$ and exponents ${1 - 2 \ga - w\over 2\ga}$ and ${\ds(1-\ga) - 1 \over \ds\, \ga}$.
It appears that the Lagrangian \eqref{LCase8} has the holomorphic structure \eqref{lagcomp}
if $k=0$ and ${1 - 2 \ga - w\over 2\ga} \in \{0,1\}$,
or $\Oms = 0$ and ${\ds(1-\ga) - 1 \over \ds\, \ga} \in \{0,1\}$, or
${1 - 2 \ga - w\over 2\ga} \in \{0,1\}$ and ${\ds(1-\ga) - 1 \over \ds\, \ga} \in \{0,1\}$.
This yields the following 8 cases, which can be all reduced to quadratures following the
prescriptions below Eq.\,\eqref{lagcomp}.
\vspace{-0.4cm}

\paragraph{i) $\boma{k = 0}$, $\boma{\ga = {1-w \over 2} \neq 0}$\,.}
The Lagrangian \eqref{LCase8} and the energy \eqref{ECase8} reduce to
\begin{gather}
\LL(z,\zt) = -\, \Im\!\left({2 \over (1 \!-\! w)^2}\,\zt^2 + V\, z^{{2 (1 + w) \over 1 - w}} \right)
- {\ds^2\,\Oms \over 2} ~, \label{L8i}\\
\Ec(z,\zt) = - \,\Im\!\left({2 \over (1 \!-\! w)^2}\,\zt^2 - V\, z^{{2 (1 + w) \over 1 - w}} \right)
+ {\ds^2\,\Oms \over 2} ~.
\end{gather}
Let us mention that in the subcases $w=-1$ (cosmological constant), $w=-1/3$ and $w=0$ (dust),
the exponent of $z$ in $\LL$ becomes,
respectively, $0,1$ and $2$, so that the Lagrange equations are elementary.
For example, if $w=0$ we have $\LL(z,\zt) = - \Im(2\, \zt^2 + V\,z^2) - {\ds^2 \over 2}\,\Oms$
and the Lagrange equations reduce to $\ddot{z} = {V \over 2}\, z$, thus describing
a complex harmonic system.
\vspace{-0.4cm}

\paragraph{ii) $\boma{k=0}$, $\boma{\ga = {1-w \over 4} \neq 0}$\,.}
The Lagrangian \eqref{LCase8} and the energy \eqref{ECase8} become
\begin{gather}
\LL(z,\zt) = -\, \Im \left( {8 \over (1 \!-\! w)^2}\,\zt^2 + V\, z^{{2 (3 + w) \over 1 - w}}
+ {\ds^2\,\Oms \over 2}\; z^2 \right) , \\
\Ec(z,\zt) = - \,\Im \left( {8 \over (1 \!-\! w)^2}\,\zt^2 - V\, z^{{2 (3 + w) \over 1 - w}}
- {\ds^2\,\Oms \over 2}\; z^2 \right) .
\end{gather}
The subcases $w=-3, -5/3, -1$ are elementary, since $z$
appears in $\LL(z,\zt)$ with exponents $0,1,2$.
\vspace{-0.4cm}

\paragraph{iii) $\boma{\Oms=0}$, $\boma{\ga = {\ds-1 \over \ds}}$\,.}
The Lagrangian \eqref{LCase8} and the energy \eqref{ECase8} take the forms
\begin{gather}
\LL(z,\zt) = -\, \Im \left( {\ds^2 \over 2 (\ds\!-\!1)^2}\,\zt^2 + V\, z^{{2 \over \ds-1}} \right)
+ {\ds^2 k \over 2} ~,\label{L8iii}\\
\Ec(z,\zt) = - \,\Im \left( {\ds^2 \over 2 (\ds\!-\!1)^2}\,\zt^2 - V\, z^{{2 \over \ds-1}} \right)
- {\ds^2 k \over 2} ~.
\end{gather}
The subcases $n\!=\!2$ and $n\!=\!3$ are elementary, since $z$ appears in $\LL(z,\zt)$
with exponent $2$ and $1$, respectively.
\vspace{-0.4cm}

\paragraph{iv) $\boma{\Oms=0}$, $\boma{\ga = {\ds-1 \over 2 \ds}}$\,.}
The Lagrangian \eqref{LCase8} and the energy \eqref{ECase8} reduce to
\begin{gather}
\LL(z,\zt) = -\, \Im \left( {2 \ds^2 \over (\ds-1)^2}\,\zt^2
+ V\, z^{{2 (\ds+1) \over \ds-1}} - {\ds^2 k \over 2}\; z^2 \right) , \label{L8iv} \\
\Ec(z,\zt) = -\, \Im \left( {2 \ds^2 \over (\ds-1)^2}\,\zt^2
- V\, z^{{2 (\ds+1) \over \ds-1}} + {\ds^2 k \over 2}\; z^2 \right) .\vspace{-0.4cm}
\end{gather}
$\phantom{a}$\vspace{-0.8cm}

\paragraph{v) $\boma{\ga = -{\ds-1 \over \ds}}$, $\boma{w = - {\ds-2 \over \ds}}$\,.}
The Lagrangian \eqref{LCase8} and the energy \eqref{ECase8} become
\begin{gather}
\LL(z,\zt)= -\, \Im \left( {\ds^2 \over 2 (\ds\!-\!1)^2}\, \zt^2
+ V\, z^{{2 \over \ds-1}} \right) - {\ds^2\,(\Oms \!-\! k) \over 2} ~, \label{L8v}\\
\Ec(z,\zt)= - \Im \left( {\ds^2 \over 2 (\ds\!-\!1)^2}\,\zt^2
- V\, z^{{2 \over \ds-1}} \right) + {\ds^2\,(\Oms \!-\! k) \over 2} ~.
\end{gather}
The Lagrangian \eqref{L8v} is like the Lagrangian \eqref{L8iii},
with $k$ replaced by $k - \Oms$.
Besides, let us mention that the subcases $n=2,3$ are elementary,
for the same reasons indicated in (iii).
\vfill\eject\noindent

\paragraph{vi) $\boma{\ga = {\ds-1 \over \ds}}$, $\boma{w = - {3 \ds-4 \over \ds}}$\,.}
The Lagrangian \eqref{LCase8} and the energy \eqref{ECase8} take the forms
\begin{gather}
\LL(z,\zt) = -\, \Im \left( {\ds^2 \over 2 (\ds\!-\!1)^2}\,\zt^2
+ V\, z^{{2 \over \ds-1}} + {\ds^2\,\Oms \over 2}\; z^2 \right)
+ {\ds^2 k \over 2} ~, \\
\Ec(z,\zt) = -\, \Im \left( {\ds^2 \over 2 (\ds-1)^2}\,\zt^2
- V\, z^{{2 \over \ds-1}} - {\ds^2\,\Oms \over 2}\; z^2 \right)
- {\ds^2 k \over 2} ~.
\end{gather}
The subcases $n=2$ and $n = 3$ are elementary, since $z$ appears
in $\LL(z,\zt)$ with exponents $2$ and $1$.
\vspace{-0.4cm}

\paragraph{vii) $\boma{\ga = - {\ds-1 \over 2 \ds}}$, $\boma{w = {1 \over \ds}}$ (radiation gas).}
The Lagrangian \eqref{LCase8} and the energy \eqref{ECase8} read
\begin{gather}
\LL(z,\zt) = - \Im \left( {2\, \ds^2 \over (\ds\!-\!1)^2}\,\zt^2
+ V\, z^{{2 (\ds+1) \over \ds-1}} - {\ds^2 k \over 2}\; z^2 \right)
- {\ds^2\,\Oms \over 2}~, \label{L8vii}\\
\Ec(z,\zt) = - \Im \left( {2\,\ds^2 \over (\ds\!-\!1)^2}\,\zt^2
+ V\, z^{{2 (\ds+1) \over \ds-1}} - {\ds^2 k \over 2}\; z^2 \right)
+ {\ds^2\,\Oms \over 2} ~.
\end{gather}
The Lagrangian \eqref{L8vii} is like the Lagrangian \eqref{L8iv}
with the additional constant $- {\ds^2 \over 2}\,\Oms$.
\vspace{-0.4cm}

\paragraph{viii) $\boma{\ga = {\ds-1 \over 2\ds}}$, $\boma{w = - {\ds-2 \over \ds}}$\,.}
The Lagrangian \eqref{LCase8} and the energy \eqref{ECase8} become
\begin{gather}
\LL(z,\zt) = -\,\Im \left( {2\,\ds^2 \over (\ds\!-\!1)^2}\,\zt^2
+ V\, z^{{2 (\ds+1) \over \ds-1}}
- {\ds^2 \,(\Oms \!-\! k) \over 2}\; z^2 \right) , \label{L8viii} \\
\Ec(z,\zt) = -\,\Im\left( {2 \ds^2 \over (\ds\!-\!1)^2}\,\zt^2
- V\, z^{{2 (\ds+1) \over \ds-1}}
+ {\ds^2\,(\Oms \!-\! k) \over 2}\; z^2 \right) ,
\end{gather}
Note that the Lagrangian \eqref{L8viii} is like the Lagrangian \eqref{L8iv},
with $k$ replaced by $k - \Oms$.

\subsection{Class 9 potentials}
\label{subsclass9}
Let us finally consider potentials of the form
\beq \VV(\f) = V_1\,e^{2\,\ga\,\f} + V_2\, e^{{2 \over \ga}\,\f}
\qquad \big(V_1,V_2 \!\in\! \reali\,,\; \ga \!\in\! (-1,1) \backslash \{0\}\big) ~. \label{VCase9} \feq
Ref.\,\cite{Fre} treats this class of potentials fixing the gauge function $\BBB$
and introducing new Lagrangian coordinates $(x,y)$ related to $(\AA,\f)$ by a
``Lorentz transformation'', in the following way:
\begin{gather}
\BBB(\AA,\f) = \AA ~; \label{gau9} \\
\AA = {x - \ga\, y \over \sqrt{1-\ga^2}} ~, \qquad
\f = {y - \ga\,x \over \sqrt{1- \ga^2}} \qquad (x,y \!\in\! \reali)~. \label{cor9}
\end{gather}
In absence of matter and curvature, the Lagrangian $\LL(x,y,\xt,\yt)$ obtained in this way
in \cite{Fre} is separable.
We now add matter and curvature, and use again the above prescriptions
trying to find new separable cases. Eqs.\,\eqref{VCase9}\eqref{gau9}\eqref{cor9}
yield the following expressions for the Lagrangian \eqref{eL} and the energy \eqref{efun}:
\begin{gather}
\LL(x,\xt,y,\yt) = \nonumber \\
{\dot{y}^2 \!-\! \dot{x}^2 \over 2}
- V_1\,e^{2 \sqrt{1-\ga^2}\,x}\! - V_2 \,e^{{2 \sqrt{1 - \ga^2} \over \ga}\,y}\!
- {\ds^2\,\Oms \over 2}\; e^{{1-w \over \sqrt{1- \ga^2}}\,(x-\ga\,y)}\!\!
+ {\ds^2 k \over 2}\; e^{{2(\ds-1) \over \ds\,\sqrt{1- \ga^2}}\,(x- \ga\,y)} ;\!\! \label{LCase9} \\
\Ec(x,\xt,y,\yt) = \nonumber \\
{\dot{y}^2 \!-\! \dot{x}^2 \over 2}
+ V_1\,e^{2 \sqrt{1-\ga^2}\,x}\! + V_2 \,e^{{2 \sqrt{1 - \ga^2} \over \ga}\,y}\!
+ {\ds^2\,\Oms \over 2}\; e^{{1-w \over \sqrt{1- \ga^2}}\,(x-\ga\,y)}\!\!
- {\ds^2 k \over 2}\; e^{{2(\ds-1) \over \ds\,\sqrt{1- \ga^2}}\,(x- \ga\,y)} .\!\! \label{ECase9}
\end{gather}
The only situation with matter or curvature where the Lagrangian \eqref{LCase9} is separable,
is the one described hereafter.
\vfill\eject\noindent

\paragraph{i) $\boma{k= 0}$, $\boma{w = 1}$ (stiff matter).}
The Lagrangian \eqref{LCase9} and the energy \eqref{ECase9} reduce to
\begin{gather}
\begin{array}{c}
\dd{\LL(x,y,\xt,\yt) = \LL_1(x,\xt) + \LL_2(y,\yt) - {\ds^2\,\Oms \over 2} ~,} \\
\dd{\LL_1(x,\xt) := -\,{1 \over 2}\,\xt^2 - V_1\,e^{2 \sqrt{1-\ga^2}\,x}~, \qquad
\LL_2(y,\yt) := {1 \over 2}\,\yt^2 - V_2\,e^{2 \sqrt{1-\ga^2}\,y} ~;}
\end{array}\\
\begin{array}{c}
\dd{\Ec(x,y,\xt,\yt) = \Ec_1(x,\xt) + \Ec_2(y,\yt) + {\ds^2\,\Oms \over 2} ~,} \\
\dd{\Ec_1(x,\xt) := -\,{1 \over 2}\,\xt^2 + V_1\,e^{2 \sqrt{1-\ga^2}\,x}~, \qquad
\Ec_2(y,\yt) := {1 \over 2}\,\yt^2 + V_2\,e^{2 \sqrt{1-\ga^2}\,y} ~.}
\end{array}
\end{gather}

\section{Explicit form and detailed analysis of some spatially flat solutions}\label{sec:Qual}
In the previous section, we provided lists of integrable cases with matter or space curvature associated
to the nine potential classes of Fr\`e-Sagnotti-Sorin. Let us recall that each one
of these cases is solvable for one of the reasons (a--e) indicated at the beginning of
section \ref{admat} (linearity of the Lagrange equations, triangular or harmonic triangular Lagrangian,
separable Lagrangian, one-dimensional holomorphic and conservative system).
\parn
Of course, after indicating a reason for the solvability of the Lagrange equations
one should derive the explicit form of the (general) solution and analyze it both
qualitatively and quantitatively. In particular, one should investigate the occurrence of an initial Big Bang
singularity and the related presence of a particle horizon (see Eqs.\,\eqref{BIGBANG} \eqref{horizon}
and the associated comments), as well as the possible development of a Big Crunch or,
in absence of it, the evolution of the system for long times.
In connection with these issues, it is essential to determine the asymptotic behavior of the main elements
of the model: the scale factor $a$, the scalar field $\f$ and the
related equation of state parameter $\wF$ (see Eqs.\,\eqref{rofi}\eqref{defwfi}),
together with the density parameters $\Omm$, $\Omf$, $\Omk$.
If the model turns out to be physically plausible, at least for some epoch
in the evolution of the universe, one should also
make sensible choices for the parameters in the potential $\V(\f)$ and for the constants of integration
of the solution, so as to make contact with the available observational data.
In the forthcoming subsections \ref{piedipal}-\ref{subclass7} we discuss the above issues (or some
of them) for some specific cases, taken as examples.
\parn
All the cases to be analyzed in the sequel have vanishing scalar curvature, i.e.,
\beq k = 0 ~, \label{k0}\feq
and possess a Big Bang (to which we devote most of our attention) at
\beq t = t_{in} = 0 ~. \label{BigBangt0}\feq
Following Eq.\,\eqref{BIGBANG}, we define the cosmic time as
\beq
\tc(t) \,:=\, \theta \int_0^t dt'\;e^{\BB(t')}~, \label{tauti}
\feq
keeping in mind that the integrability of $e^{\BB}$ in a right neighborhood
of $t = 0$ is required by the very definition of Big Bang. Of course,
$\tc(t) \to 0^{+}$ for $t \to 0^{+}$ and we can speak of the inverse
function $t = t(\tc)$.
\parn
In the sequel we always require for matter a positive density:
$\Omm >0$ at all times, which happens if and only if
\beq
\Oms > 0 ~. \label{Omspos}
\feq
\parn
Let us note that the assumption \eqref{k0} and
Eq.\,\eqref{Omega} give $\Omk = 0$;
from here and from \eqref{OmegaSum} we infer that, at all times,
\beq
\Omm \!+ \Omf = 1 ~. \label{OmmOmf}
\feq
Making reference to the above relation, we will say that \textit{matter dominates at the Big Bang}
if $\Omm(t) \to 1$ (or equivalently, $\Omf(t) \to 0$) for $t \to 0^{+}$; conversely,
we will say that \textit{the scalar field} \textsl{dominates at the
Big Bang} if $\Omf(t) \to 0$ (or equivalently, $\Omm(t) \to 1$)
in the same limit. The cases where \textit{matter} or
\textsl{the scalar field dominate at the Big Crunch}, if this exists,
can be defined similarly.
\parn
In general, the discussion on the notion of  particle horizon requires to consider,
at any time $\tc = \tc(t)$, the integral
\beq
\Theta(\tc) \,:= \int_{0}^{\tc}\!\! \frac{d \tc'}{a(\tc')} \,=\,
\te \int_{0}^{t}\! dt'\; e^{\BB(t')-\AA(t')/\ds} ~; \label{horizonk}
\feq
see Eq.\,\eqref{horizon} and the related comments. The above integral
converges at some time $\tau >0$ if and only if it converges
at all times, and in this case there is a particle horizon.
According to the result of \cite{phantom} mentioned
after Eq. \eqref{horizon}, a particle horizon occurs if
the strong energy condition \eqref{C3} is fulfilled strictly
by the matter fluid; we have already assumed
a positive density for this fluid, so \eqref{C3}
holds strictly if and only if
\beq
w > {2 \over \ds} - 1 ~. \label{stEC}
\feq
\subsection{Solutions for class 1 potentials with dust}\label{piedipal}
Let us consider an $(\ds+1)$-dimensional, spatially flat cosmology
with matter content described by a dust fluid; accordingly,
besides Eq.\,\eqref{k0} we posit
\beq
w=0 ~. \label{w0}
\feq
Moreover, we assume that the self-interaction potential for the field is given by
\beq
\VV(\f) = V_1\, e^{\f} + V_2\, e^{-\f}~, \qquad \mbox{with~ $V_1,V_2 > 0$}~; \label{vclass1sol}
\feq
for notational convenience, in the sequel we shall put
\beq
V := \sqrt{V_1\,V_2} \,>\, 0 ~. \label{VV1V2}
\feq
The model depicted above was previously studied by Rubano and Scudellaro \cite{Rub},
and by Piedipalumbo, Scudellaro, Esposito and Rubano \cite{Piedipalumbo},
in the physically most relevant case with spatial dimension $\ds = 3$.
Hereafter, we review within our framework the results of \cite{Piedipalumbo,Rub},
generalizing them to the case of arbitrary $\ds \geqs 2$; in addition, we discuss
the asymptotic behavior of the density parameters $\Omm,\Omf$ near the Big Bang.
\salto
To begin with, let us notice that the potential \eqref{vclass1sol} is clearly of the form \eqref{vclass1}
(with $V_0 \!=\! 0$ and the conditions stated above on $V_1,V_2$);
to be more precise, as a consequence of Eqs.\,\eqref{k0}\eqref{w0}, the cosmological model
under analysis belongs to the integrable subcase (i) of class 1, discussed previously
in subsection \ref{subsclass1}.
In this connection, let us recall that it is convenient to fix the gauge function
$\BBB(\AA,\f)$ as in Eq.\,\eqref{gau1}, which gives $\BB = 0$. In view of Eq.\,\eqref{taut},
this implies that the cosmic time $\tc$ and the coordinate time $t$ are linearly related:
\beq
t = \tc/\te ~. \label{tctlin}
\feq
According to subsection \ref{subsclass1}, the Lagrangian function for
the model that we are considering is separable and can be reduced to quadratures by
introducing a new pair of coordinates $u,v$, related to $\AA,\f$ via
(cf. Eqs.\,\eqref{cor1} \eqref{xyuv})
\beq
\AA = \log \left({u^2 - v^2 \over 4}\right) , \qquad
\f = \log \left(\!\sqrt{{V_2 \over V_1}}\;{u-v \over u + v}\right) . \label{Afuv}
\feq
The Lagrange equations have the form \eqref{uvLeq} that, with
the previous assumptions, reduces to
\beq \ddot{u} - V u=0~, \qquad
\ddot{v} + V v=0~; \label{uvLeqRed} \feq
the corresponding solutions are readily found to be
\beq
u(t) = A\, \cosh\big(\sqrt{V}\, t\big) + B\, \sinh\big(\sqrt{V}\, t\big)~, \qquad
v(t) = C\, \cos\big(\sqrt{V}\, t\big) + D\, \sin\big(\sqrt{V}\, t\big)~, \label{uvsol}
\feq
where $A,B,C,D \in \reali$ are suitable integration constants.
\parn
From Eqs.\,\eqref{Euv}\eqref{uvsol}, by elementary computations we infer
the following expression for the energy $\Ec \equiv \Ec(u,v, \dot{u}, \dot{v})$ of the system:
\beq
\Ec = {V \over 2}\,\big(A^2 - B^2 + C^2 + D^2\big) + {\ds^2\, \Oms \over 2} ~.
\feq
Taking the above relation into account, to fulfill the energy constraint $\Ec = 0$ we set
\beq
\Oms = {V \over \ds^2}\,\big(B^2 - A^2 - C^2 - D^2\big) ~. \label{E0Oms}
\feq
Furthermore, for fixed values of the parameters we take as a domain
for the solutions \eqref{uvsol} the maximal interval $I \subset \reali$
such that (see the comment at the end of subsection \ref{subsclass1})
\beq
t_{in} \equiv  0 \in I \qquad \mbox{and} \qquad u(t) > |v(t)| ~~ \mbox{for all $t \in I$}\,.
\label{Idef}
\feq
Finally let us mention that Eqs.\,\eqref{rofi} \eqref{defwfi} \eqref{Omega} with
$\BB = 0$ and Eqs.\,\eqref{vclass1sol} \eqref{Afuv} give the following representations
for the coefficient $\wF$ in the field equation of state and for the matter
density parameter $\Omm$:
\begin{gather}
\wF = {(\dot{u}\,v - u\,\dot{v})^2 - V (u^4\!-v^4) \over (\dot{u}\,v - u\,\dot{v})^2 + V (u^4\!-v^4)}~, \label{wfipal} \\
\Omm = {\ds^2\, \Oms \,(u^2 - v^2)  \over (u\,\dot{u} - v\, \dot{v})^{2}}~. \label{Ommpal}
\end{gather}
\vspace{0.1cm}

\subsubsection{Big Bang analysis}\label{subBigPP}
Let us wonder under which conditions the solution \eqref{uvsol} produces a Big Bang
at some instant, that we conventionally choose as the time origin $t=0$
(cf. Eq.\,\eqref{BigBangt0}).
It is required that $a(t) \to 0$ (i.e., $\AA(t) \to -\infty$) for $t \to 0^{+}$
and, even prior to this, that $a(t)$ (hence, $\AA(t)$) is well defined
in a right neighborhood of $t=0$; in terms of the functions $u(t),v(t)$, this amounts to demand
\beq
u^2(t) - v^2(t) \to 0 \quad~ \mbox{and} \quad~ u(t) > |v(t)|
\qquad \mbox{for $t \to 0^{+}$} \label{ask}
\feq
(here and in the sequel, an expression of the form ``$f(t) >0$ for $t \to 0^{+}$''
means that there exits some $\epsilon > 0$ such that $f(t) >0$ for all $t \in (0,\epsilon)$).
\parn
On the other hand, from the explicit expressions written in Eq.\,\eqref{uvsol},
it follows straightforwardly that
\beq u(t) = A + B \sqrt{V}\,t + {1 \over 2}\,V\! A\,t^2\! + O(t^3)\,, ~~~
v(t) = C + D \sqrt{V}\,t - {1 \over 2}\,V C\,t^2\! + O(t^3) \quad ~
\mbox{for $t \to 0^{+}$}. \label{uvasy}
\feq
The above relations show that the first condition in Eq.\,\eqref{ask} is fulfilled
if and only if $A^2\! - C^2 = 0$, while it is necessary to assume that $A \geqs 0$
in order to satisfy the second condition in the same equation; thus, we require
\beq A = |C| \geqs 0 ~. \label{acz} \feq
In the sequel we will distinguish three subcases fulfilling the latter constraint.
\vspace{-0.3cm}

\paragraph{i) $\boma{A = C > 0}$.}
In this case the second condition in Eq.\,\eqref{ask} holds if and only if $B \!\geq\! D$.
It should be noticed that for $B \!=\! D$ the energy constraint written in the
form \eqref{E0Oms} entails $\Oms \!= -\,{2 \over \ds^2}\,V A^2 \!<\! 0$; since
this contradicts our general assumption \eqref{Omspos}, from now we assume
\beq B > D ~. \feq
To go on, we note that the previously mentioned expressions for $\AA$, $\f$, $\wF$ and $\Omm$
(see Eqs.\,\eqref{Afuv} \eqref{E0Oms} \eqref{wfipal} \eqref{Ommpal}) imply the following, for $t \to 0^{+}$:
\begin{gather}
\AA(t) = \log t + \log\!\left({A\,(B - D)\,\sqrt{V} \over 2}\right) +\, O(t)~; \label{AAtasy}\\
\f(t) = \log t + \log\!\left({(B - D)\,V_2^{3/4} \over 2A\, V_1^{1/4}}\right) +\, O(t)~; \\
\wF(t) = 1 - {8\,A\, \sqrt{V} \over B - D}\; t + O(t^2)~; \\
\Omm(t) = {2\,(B^2 - D^2 - 2A^2)\,\sqrt{V} \over A(B-D)}\;t + O(t^2)~. \label{Omegabis2}
\end{gather}
Of course, similar expansions in terms of the cosmic time $\tc$ can be obtained
simply recalling that $t = \tc/\theta$ (see Eq.\,\eqref{tctlin}); in particular,
from Eqs.\,\eqref{aeaa}\eqref{AAtasy} we obtain
\beq
a(\tc) = \left({A\,(B - D)\,\sqrt{V} \over 2}\right)^{\!\!{1 /\ds}} (\tc/\theta)^{{1 /\ds}}
+ O\!\left((\tc/\te)^{({1/\ds}) + 1}\right) \qquad \mbox{for $\tc/\te \to 0^+$}\,. \label{a(t)sez5}
\feq
From Eqs.\,\eqref{horizonk}\eqref{a(t)sez5}, noting that $1/\ds < 1$ for $\ds \geqs 2$
we infer the existence of a particle horizon
({\footnote{Making reference to the comments related to Eq.\,\eqref{stEC},
let us remark that in the present case the strong energy condition is in fact fulfilled
as an equality for $n = 2$ and as a strict inequality for $n \geqs 3$ (since $w = 0 > {2 \over n} - 1$).}}).
Eq.\,\eqref{Omegabis2} shows that $\Omm(t) \to 0$, indicating that the scalar
field dominates close to the Big Bang.
\vspace{-0.3cm}

\paragraph{ii) $\boma{A = -\,C > 0}$.}
This is qualitatively very similar to the previous case. The second condition
in Eq.\,\eqref{ask} holds only if $B \!\geq\! -D$; yet, for $B \!=\! -D$ the energy
constraint \eqref{E0Oms} yields a negative matter density $\Oms < 0$, thus violating
our hypothesis \eqref{Omspos}. So, we assume
\beq B > - D ~. \feq
Then, for $t \to 0^{+}$ we have
\begin{gather}
\AA(t) = \log t + \log\!\left({A\,(B + D)\,\sqrt{V} \over 2}\right) +\, O(t)~; \label{AtAsy}\\
\f(t) = -\,\log t -\, \log\!\left({(B + D)\,V_1^{3/4} \over 2A\,V_2^{1/4}}\right) +\, O(t)~; \\
\wF(t) = 1 - {8\,A\, \sqrt{V} \over B + D}\; t + O(t^2)~; \\
\Omm(t) = {2\,(B^2 - D^2 - 2A^2)\,\sqrt{V} \over A(B+D)}\;t + O(t^2)~. \label{OmmtAC}
\end{gather}
Correspondingly, from Eqs.\,\eqref{aeaa}\eqref{tctlin}\eqref{AtAsy} we get
\beq
a(\tc) = \left({A\,(B + D)\,\sqrt{V} \over 2}\right)^{\!\!{1 /\ds}} (\tc/\theta)^{{1 /\ds}}
+ O\!\left((\tc/\te)^{({1/\ds}) + 1}\right) \qquad \mbox{for $\tc/\te \to 0^+$}\,,
\feq
which allows us to infer that there is a particle horizon.
Moreover, Eq.\,\eqref{OmmtAC} shows that the scalar field dominates at the Big Bang.
\vfill\eject\noindent

\paragraph{iii) $\boma{A = C = 0}$.}
In this setting the second condition in Eq.\,\eqref{ask} holds if and only if
\beq B > |D| \geqs 0~, \qquad D \in \reali ~. \label{BD} \feq
As a consequence, we have a strictly positive matter density
$\Oms \!= (B^2 - D^2)\,V/\ds^2 \!>\! 0$ (see Eq.\,\eqref{E0Oms}),
in agreement with the general condition stated in Eq.\,\eqref{Omspos}.
\parn
Regarding the asymptotic behavior of the system near the Big Bang,
note that for $t \to 0^+$ we have
\begin{gather}
\AA(t) = 2\,\log t + \log\!\left({(B^2 - D^2)\,V \over 4}\right) +\, O(t^2)~; \label{AtAsy2}\\
\f(t) = \log\!\left({B - D \over B + D}\sqrt{{V_2 \over V_1}}\right) +\, O(t^2)~; \\
\wF(t) = - 1 + {8\,B^2 D^2\, V \over 9\,(B^4 - D^4)}\; t^2 + O(t^4)~; \label{cosmoc}\\
\Omm(t) = 1 - {(B^2 + D^2)\,V \over B^2 - D^2}\;t^2 + O(t^4) ~. \label{OmmtAC0}
\end{gather}
Furthermore, Eqs.\,\eqref{aeaa}\eqref{tctlin}\eqref{AtAsy2} entail
\beq
a(\tc) = e^{\AA(\tc/\theta)/\ds} = \left({(B^2 - D^2)\,V \over 4}\right)^{\!\!{1 /\ds}} (\tc/\theta)^{{2/\ds}}
+ O\!\left((\tc/\te)^{({2/\ds}) + 2}\right) \qquad \mbox{for $\tc/\te \to 0^+$}\,. \label{a(t)sez5AC0}
\feq
From Eqs.\,\eqref{horizonk}\eqref{a(t)sez5AC0} we infer that a particle horizon
is present if $\ds \geqs 3$, and absent if $\ds = 2$ (in the latter case the integral
in Eq.\,\eqref{horizonk} diverges logaritmically)
({\footnote{Since we are assuming $w = 0$, for $\ds = 2$ the strong energy condition
is only fulfilled as an equality (see Eq.\,\eqref{stEC}).
Due to this, the hypothesis in \cite[Eq.\,(34)]{phantom} is not satisfied, which explains
why the general conclusions of \cite[Prop.\,1]{phantom} do not hold in this case.
This suggests that the hypotheses underlying \cite[Prop.\,1]{phantom} are somehow optimal.}}).
Eq.\,\eqref{cosmoc} indicates a field equation of state close to that of a
cosmological constant (recall Eq.\,\eqref{wficos}). On the other hand, Eq.\,\eqref{OmmtAC0}
shows that $\Omm(t) \rightarrow 1$; thus, differently from the
previous cases, here matter dominates near the Big Bang.

\subsubsection{Far future analysis}
First of all, let us remark that the bare solutions written in Eq.\,\eqref{uvsol} make sense
for any $t \in \reali$. However, one should not forget that the second condition in
Eq.\,\eqref{Idef} puts severe restrictions on the maximal admissible domain $I \subset \reali$
for such solutions.
In presence of a Big Bang at $t = 0$, the most enticing scenarios are those corresponding
to an endless evolution of the universe, namely,
\beq
I = (0,+\infty) ~. \label{IInf}
\feq
In the sequel we restrict the attention to cases where the integration constants $A,B,C,D$
characterizing the solutions \eqref{uvsol} are such that the condition \eqref{IInf} is verified
({\footnote{As a matter of fact, there do exist such admissible choices of $A,B,C,D$.
For example, making reference to the cases analyzed in the previous subsection \ref{subBigPP},
it can be checked by direct inspection that Eq.\,\eqref{IInf} certainly holds if
$$
A = C > 0\,, ~B > D > 0 \qquad \mbox{or} \qquad
A = - C > 0\,, ~B > -D > 0 \qquad \mbox{or} \qquad
A = C = 0\,, ~B > |D| > 0 ~.
$$
In all the cases mentioned above, noting that $\cosh(z) \!>\! |\cos(z)|$ and $\sinh(z) \!>\! |\sin(z)|$
for any $z \!>\! 0$, it can be checked by direct inspection that $u(t) > |v(t)|$ for all $t \in (0,\infty)$.
\vspace{-0.3cm}}}),
and proceed to investigate the asymptotic behavior of the corresponding
cosmological model for $t \to +\infty$.
\parn
From the explicit expressions for $u(t),v(t)$ written in Eq.\,\eqref{uvsol} we easily infer the following:
$u(t) = {A + B \over 2}\,e^{\sqrt{V}\,t} + O\big(e^{-\sqrt{V}\,t}\big)$ for $t \to +\infty$;
$v(t)$ is an oscillatory motion, with $|v(t)| \leqs |C| + |D|$ for all $t \in (0,+\infty)$.
Thus, we see a posteriori that the second condition in Eq.\,\eqref{Idef} is fulfilled
in a neighborhood of $+\infty$ if and only if
\beq
A + B > 0 ~,
\feq
which we assume from now on. With this assumption the solutions \eqref{uvsol} are
admissible, at least, in neighborhood of infinity and we have the following asymptotics
for $t \to + \infty$ (recall Eqs.\,\eqref{Afuv} \eqref{E0Oms} \eqref{wfipal} \eqref{Ommpal}):
\begin{gather}
\AA(t) = 2\sqrt{V}\;t + 2 \log\!\left({A+B \over 4}\right) + O\big(e^{-2\sqrt{V}\,t}\big) ~; \label{AAtasyInf}\\
\f(t) = \log\sqrt{{V_2 \over V_1}} + O\big(e^{-\sqrt{V}\,t}\big) ~; \\
\wF(t) = - 1 + O\big(e^{-2\sqrt{V}\,t}\big)~; \\
\Omm(t) = {4\,\big(B^2 - A^2 - C^2 - D^2\big)V \over (A+B)^2}\;e^{-2\sqrt{V}\,t}
+ O\big(e^{-4\sqrt{V}\,t}\big) ~.
\end{gather}
It is straightforward to derive similar expansions in terms of the cosmic time $\tc$,
recalling that Eq.\,\eqref{tctlin} gives $t = \tc/\theta$.
From Eqs.\,\eqref{aeaa}\eqref{AAtasyInf} we deduce
\beq
a(\tc) = \left({A + B \over 4}\right)^{\!\!{2/\ds}} e^{(2/\ds)\sqrt{V}\,(\tc/\theta)}
+ O\!\left(e^{-(\ds-1)(2/\ds)\sqrt{V}\,(\tc/\theta)}\right) \qquad \mbox{for $\tc/\te \to +\infty$}\,.
\feq
From the above relations we infer, especially, that for large times the scale factor diverges,
the scalar field behaves as a cosmological constant and becomes the dominant contribution
(since $\Omf = 1 - \Omm \to 1$ for $t \to +\infty$; see Eq.\,\eqref{OmmOmf}).
All these features are attained with exponential speed.

\subsubsection{Quantitative analysis of one of the previous cases}
\label{Models}
Hereafter we reconsider the general model described at the beginning
of the present subsection \ref{piedipal} and show how to fix all the
(so far unspecified) associated parameters $\ds,\te,\Oms,V_1,V_2,A,B,C,D$
so as to provide a physically plausible scenario.
\parn
To this purpose, we restrict the attention to the case of space dimension
and spatial curvature respectively given by (see Eq.\,\eqref{k0})
\beq
\ds = 3~, \qquad k = 0~.
\feq
Furthermore, we require that a Big Bang singularity occurs at $t = 0$; correspondingly,
we assume matter to be dominant near the Big Bang, i.e., $\Omm(t) \to 1$ for $t \to 0^{+}$.
The analysis of subsection \ref{subBigPP} (see, especially, case (iii) therein)
indicates that the above conditions can be realized only if
\beq
A = 0 \qquad \mbox{and} \qquad C = 0 ~. \label{AC0}
\feq
To proceed let us remark that, on account of the gauge invariance $\f \mapsto \f + \mbox{const.}$,
without any loss of generality we can assume that (see Eq.\,\eqref{VV1V2})
\beq V_1 = V_2 = V > 0 ~. \feq
Then, the potential \eqref{vclass1sol} reduces to
\beq \VV(\f) = 2\, V\,\cosh \f ~, \label{VPP} \feq
and the associated solution \eqref{uvsol} reads (see also Eq.\,\eqref{BD})
\beq
u(t) = B\, \sinh\big(\sqrt{V}\, t\big)~, \qquad
v(t) = D\, \sin\big(\sqrt{V}\, t\big) \qquad~
\mbox{with~ $B > |D| \geqs 0$} ~. \label{uvQuant}
\feq
Furthermore, note that the zero-energy constraint \eqref{E0Oms} becomes
\beq
\Oms = {V \over 9}\,\big(B^2 - D^2\big) > 0 ~. \label{zeroE}
\feq
\vfill\eject\noindent
$\phantom{a}$\vspace{-0.8cm}\\
Next, let us introduce a reference time $\ts > 0$ that we identify
with the present epoch; we prescribe
\beq \left\{\!\! \begin{array}{l}
\dd{a(\ts) = 1 ~,} \vspace{0.05cm}\\
\dd{\f(\ts) = \fs ~,} \vspace{0.05cm} \\
\dd{H(\ts) = 1/\te}
\end{array}\right. \label{conafH} \feq
where $\te$ is the usual time constant (see Eq.\,\eqref{tctlin})
and $\fs \in \reali$ is an arbitrary parameter; in the sequel we shall discuss
as examples a couple of sensible choices of $\fs$.
Expressing $a \equiv e^{\AA/3}$, $\f$ and $H \equiv \dot{\AA}/(3\,\te)$
(see Eqs.\,\eqref{aeaa}\eqref{H} and recall that here $\BB \equiv 0$)
in terms of the Lagrangian variables $u,v$ (see Eq.\,\eqref{Afuv}),
the above conditions \eqref{conafH} read
\beq \left\{\!\! \begin{array}{l}
\dd{u^2(\ts) - v^2(\ts) = 4 ~,} \vspace{0.05cm}\\
\dd{{u(\ts) - v(\ts) \over u(\ts) + v(\ts)} = e^{\fs} ,} \vspace{0.1cm} \\
\dd{{u(\ts)\, \dot{u}(\ts) - v(\ts)\, \dot{v}(\ts) \over u^2(\ts) - v^2(\ts)} = {3 \over 2}~.}
\end{array}\right. \feq
By simple algebraic manipulations (recalling the constraint $u > |v| \geqs 0$),
we infer
\beq \left\{\!\! \begin{array}{l}
\dd{u(\ts) = 2\,\cosh(\fs/2) ~,} \vspace{0.05cm}\\
\dd{v(\ts) = - \,2\,\sinh(\fs/2) ~,} \vspace{0.05cm}\\
\dd{\cosh(\fs/2)\, \dot{u}(\ts) + \sinh(\fs/2)\, \dot{v}(\ts) = 3 ~.}
\end{array}\right. \label{equv}\feq
Taking into account the explicit expressions for $u(\ts)$ and $v(\ts)$ (see Eq.\,\eqref{uvQuant}),
the first two relations in Eq.\,\eqref{equv} can be trivially solved in terms of the two unknown
parameters $B,D$; more precisely, introducing the short-hand notation
\beq \st := \sqrt{V}\,\ts~, \label{st} \feq
we get
\beq B = {2\,\cosh(\fs/2) \over \sinh(\st)} ~, \qquad\quad
D = - \,{2\,\sinh(\fs/2) \over \sin(\st)} ~. \label{BDsol} \feq
Substituting the above expressions for $B,D$ in the zero-energy
constraint \eqref{zeroE} and solving for $V$, we obtain
\beq V = {9\,\Oms\; \sinh^2(\st)\,\sin^2(\st) \over
4\,\big(\cosh^2(\fs/2)\,\sin^2(\st) - \sinh^2(\fs/2)\,\sinh^2(\st)\big)} ~.
\label{Vsol} \feq
Finally, the above relations \eqref{BDsol} \eqref{Vsol} and the last
identity in Eq.\,\eqref{equv} give
\begin{align}
& \sqrt{\cosh^2(\fs/2)\,\sin^2(\st) - \sinh^2(\fs/2)\,\sinh^2(\st)} \nonumber \\
& = \sqrt{\Oms}\; \Big[\cosh^2(\fs/2)\, \cosh(\st)\,\sin(\st) - \sinh^2(\fs/2)\,\sinh(\st)\, \cos(\st)\Big]
\label{num}
\end{align}
For assigned values of $\fs$ and $\Oms$, one can look for a solution $\st$
of the above equation by numerical methods (assuming that the said solution exists).
In the following we analyze in more detail a couple of examples corresponding,
respectively, to the choices $\fs = 0$ and $\fs = 1/2$.
\vspace{-0.3cm}

\paragraph{The case $\boma{\fs = 0}$.}
First of all let us remark that this particular choice of $\fs$ corresponds
to the (unique) minimum of the potential \eqref{VPP}.
Note that Eq.\,\eqref{num} reduces to
\beq \cosh(\st) = {1 \over \sqrt{\Oms}}~. \feq
Assuming $\Oms \!<\! 1$, the above equation can be solved analytically, which yields
\beq \st = \mbox{arccosh}\!\left({1 \over \sqrt{\Oms}}\right) . \label{stfi0} \feq
Substituting this solution in the expressions \eqref{BDsol}\eqref{Vsol} for $B,D$ and $V$ we obtain
\beq
B = 2 \,\sqrt{{\Oms \over 1- \Oms}} ~, \qquad ~~
D = 0 ~, \qquad~~ V = {9 \over 4}\, \big(1- \Oms\big) ~; \label{BDfi0}
\feq
besides, from Eq.\,\eqref{st} we infer
\beq
\ts = {2 \over 3\sqrt{1- \Oms}}\;\mbox{arccosh}\!\left({1 \over \sqrt{\Oms}}\right) .\label{tsfiszero}
\feq
On account of the relation $D = 0$ in Eq.\,\eqref{BDfi0}
and of Eqs.\,\eqref{uvsol}\eqref{AC0}, we have
\begin{equation}
v(t) = 0 \qquad \mbox{for all $t \in (0,\infty)$} ~.
\end{equation}
Recalling as well the expression for $\f$ in terms of the Lagrangian coordinates $u,v$
(see Eq.\,\eqref{Afuv}), this implies
\beq
\f(t) = \mbox{const.} = 0 = \fs ~. \label{f00}
\feq
Therefore, making reference to Eqs.\,\eqref{cosm1}-\eqref{wficos} and to the related comments,
we can say that in the present setting the field $\f$ plays the role of a cosmological constant.
Correspondingly, it can be checked by direct computations that
(see Eqs.\,\eqref{wfipal}\eqref{Ommpal}; cf. also Eq.\,\eqref{wficos})
\begin{gather}
\wF(t) = \mbox{const.} = -1 ~, \label{wF0}\\
\Omm(t) = {9\, \Oms \over \dot{u}^{2}}
= \cosh^{-2}\!\left({3 \over 2}\,\sqrt{1\!-\! \Oms}\; t\right) . \label{Om0}
\end{gather}
To say more, notice that the previous relations \eqref{stfi0}-\eqref{tsfiszero}
allow to express each of the parameters $\st,B,D,V,\ts$ in terms of $\Oms$;
this is the matter density at present time $t = \ts$ (see Eq.\,\eqref{OmsId}),
and choosing for the latter the accepted value
(see, e.g., \cite[p.\,128]{PDG18}, \cite{PLANK18})
\beq \Oms = 0.308 ~, \label{OmsNum} \feq
the said relations give
\beq \st = 1.194\,... ~,\qquad B = 1.334\,... ~, \qquad D = 0\,, \qquad
V = 1.557\,, \qquad \ts = 0.957\,... ~. \feq
Notably, setting (see, e.g., \cite[p.\,128]{PDG18}, \cite{PLANK18})
\beq H(\ts) \equiv \Hs \,=\, 67.89\, \frac{km}{s \cdot Mpc}
\,\simeq\, 2.20017 \times 10^{-18} \, s^{-1} \label{Hsval} \feq
one finds that the age of the universe in the cosmological model under analysis is
\beq \tc_* := \te \, \ts = \frac{\ts}{\Hs}
\simeq 4.34975 \times 10^{17} s
\simeq 13.793 \times 10^{9} \, \mbox{years} ~, \feq
in agreement with the accepted value for this quantity
(cf., e.g.,\,\cite[p.\,129]{PDG18}).
\vspace{-0.4cm}

\paragraph{The case $\boma{\fs = 1/2}$.}
In this case, once we have fixed $\Oms = 0.308 $ as in Eq.\,\eqref{OmsNum},
solving Eq.\,\eqref{num} for $\st$ yields
\beq \st = 1.09829\,... ~. \feq
Substituting the above value in the explicit expressions \eqref{BDsol}\eqref{Vsol}
we obtain
\beq B = 1.54775\,... ~, \qquad D = -\,0.56739\,... ~,  \qquad
V = 1.33682\,... ~, \feq
and from Eq.\,\eqref{st} it follows that
\beq \ts = 0.949905\,... ~. \feq
Fixing $ \Hs $ as in Eq.\,\eqref{Hsval}, the latter value for $\ts$
corresponds to an age of the universe of about
\beq \tc_* := \te \, \ts = \frac{\ts}{\Hs}
\simeq 4.31743 \times 10^{17} s
\simeq 13.6905 \times 10^{9} \, \mbox{years} ~. \label{tcsdef}
\vspace{-0.1cm}\feq

\paragraph{Further considerations on the previous cases with $\boma{\fs = 0}$ and $\boma{\fs = 1/2}$.}
Figs.\,1 and 2 give the plot of $a(\tc)$ as a function of the dimensionless ratio
$\tc/\te$ on two different intervals (namely, for $\tc/\te \in (0,1)$ and $\tc/\te \in (0,10)$),
in the two cases with $\fs = 0$ and $\fs = 1/2$ analyzed previously.
By close inspection of these figures it appears that after the Big Bang at $\tc = 0$
($a(\tc) \to 0^+$ for $\tc/\te \to 0^+$) the universe experiences an initial phase of decelerated
expansion (see Fig.\,1), followed by an endless accelerated expansion (see Fig.\,2).
\begin{figure}[t!]
    \centering
        \begin{subfigure}[b]{0.475\textwidth}\label{fig:1}
                \includegraphics[width=\textwidth]{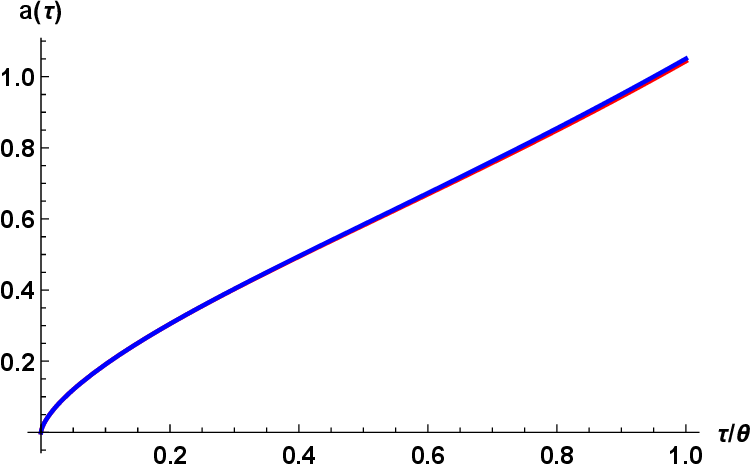}\vspace{-0.cm}
                \caption*{\textsc{Figure\! 1.}
                $a(\tc)$ as a function of $\tc/\te$,
                with $\fs \!= 0$ (in red) and $\fs \!= 1/2$ (in blue).}
        \end{subfigure}
        \hspace{0.3cm}
        \begin{subfigure}[b]{0.475\textwidth}
                \includegraphics[width=\textwidth]{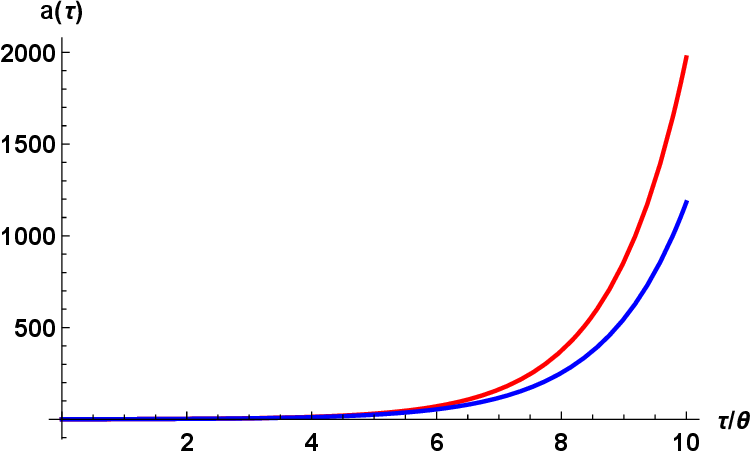}\label{fig:2}
                \caption*{\textsc{Figure\! 2.}
                $a(\tc)$ as a function of $\tc/\te$,
                with $\fs \!= 0$ (in red) and $\fs \!= 1/2$ (in blue).}
        \end{subfigure}
\end{figure}
\parn
Fig.\,3 gives plots of the (rescaled) scalar field $\f(\tc)$
for $\tc/\te \in (0,10)$ in the two cases with $\fs = 0$ and $\fs = 1/2$.
Let us recall that for $\fs = 0$ we get $\f = \mbox{const.} \equiv 0$
by purely analytic means (see Eq.\,\eqref{f00}).
On the other hand when $\fs = 1/2$ we have, in particular,
$\f(\tc) \to \log\big({B-D \over B+D}\big) = 0.76896\,... $ for $\tc/\te \to 0^+$
and $\f \to 0 $ for $ \tc/\te \to \infty $.
\parn
Fig.\,4 gives plots of the equation of state coefficient $\wF(\tc)$
for the field (see, especially, Eq.\,\eqref{wfipal}) for $\tc/\te \in (0,10)$
(with $\fs = 0$ and $\fs = 1/2$).
We already mentioned that $\wF = \mbox{const.} \equiv -1$ when $\fs = 0$
(see Eq.\,\eqref{wF0}). In the case with $\fs = 1/2$, we have the following features:
$\wF \to -1 $ for $\tc/\te \to 0^{+}$; $\wF(\tc_*) = -\,0.93632\,... $ at the present
cosmic time $\tc_*$ given in Eq.\,\eqref{tcsdef};
$\wF \to -1 $ for $\tc/\te \to \infty$.
So, in both the previous limits the scalar field behaves like a cosmological constant
(see the discussion after Eq.\,\eqref{defwfi}).
\begin{figure}[t!]
\vspace{-0.7cm}
    \centering
        \begin{subfigure}[b]{0.475\textwidth}\label{fig:3}
                \includegraphics[width=\textwidth]{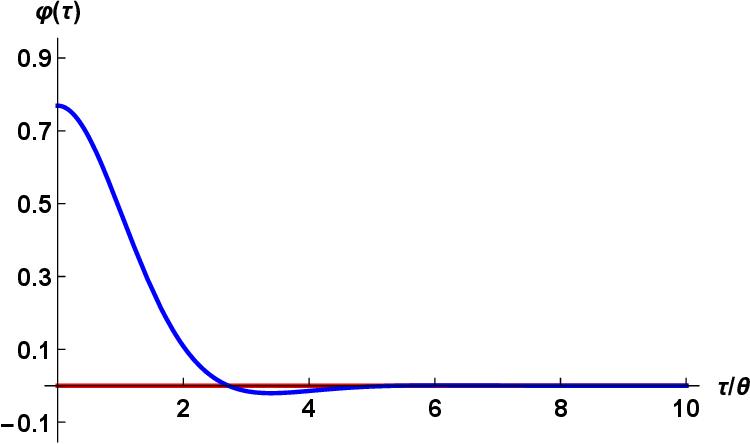}\vspace{-0.cm}
                \caption*{\textsc{Figure\! 3.}
                $\f(\tc)$ as a function of $\tc/\te$,
                with $\fs \!= 0$ (in red) and $\fs \!= 1/2$ (in blue).}
        \end{subfigure}
        \hspace{0.3cm}
        \begin{subfigure}[b]{0.475\textwidth}
                \includegraphics[width=\textwidth]{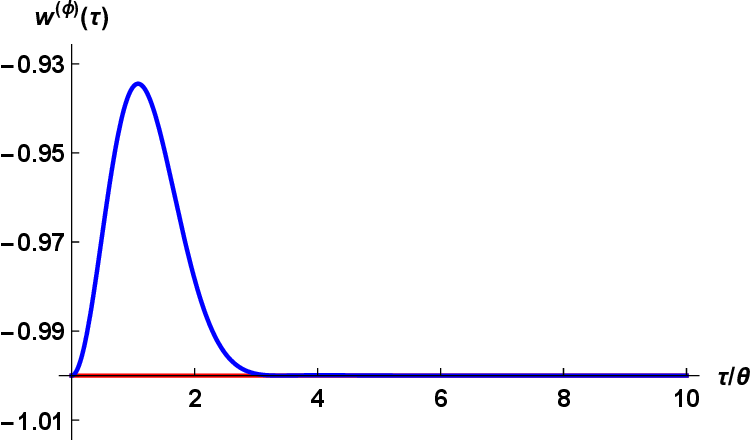}\label{fig:4}
                \caption*{\textsc{Figure\! 4.}
                $\wF(\tc)$ as a function of $\tc/\te$,
                with $\fs \!= 0$ (in red) and $\fs \!= 1/2$ (in blue).}
        \end{subfigure}
\vspace{-0.1cm}
\end{figure}
\begin{figure}[t!]
\vspace{-0.5cm}
    \centering
        \begin{subfigure}[b]{0.475\textwidth}\label{fig:5}
                \includegraphics[width=\textwidth]{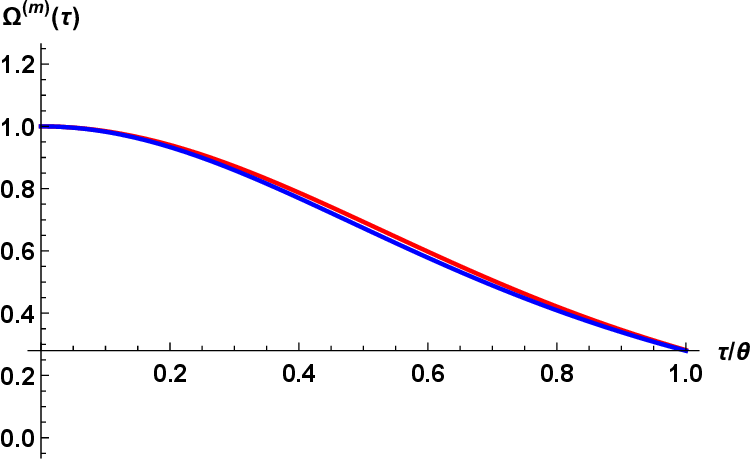}\vspace{-0.cm}
                \caption*{\textsc{Figure\! 5.}
                $\Omm(\tc)$ as a function of $\tc/\te$,
                with $\fs \!= 0$ (in red) and $\fs \!= 1/2$ (in blue).}
        \end{subfigure}
        \hspace{0.3cm}
        \begin{subfigure}[b]{0.475\textwidth}
                \includegraphics[width=\textwidth]{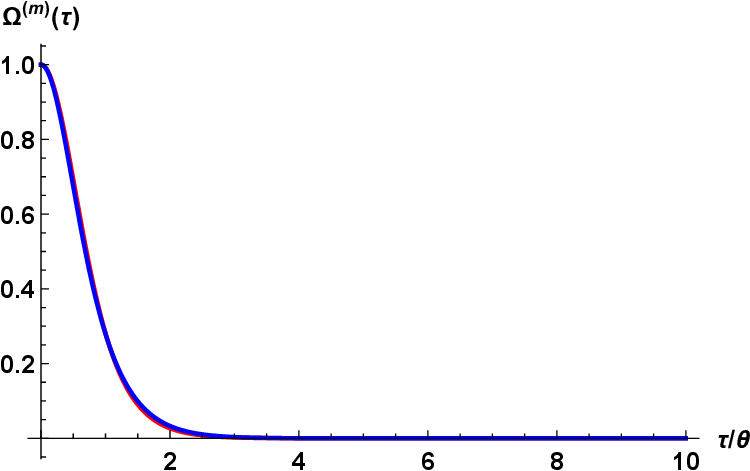}\label{fig:6}
                \caption*{\textsc{Figure\! 6.}
                $\Omm(\tc)$ as a function of $\tc/\te$,
                with $\fs \!= 0$ (in red) and $\fs \!= 1/2$ (in blue).}
        \end{subfigure}
\vspace{-0.3cm}
\end{figure}
\parn
Figs.\,5 and 6 give plots of the matter density parameter
$\Omm(\tc)$ (see Eq.\,\eqref{Ommpal}), respectively for $\tc/\te \in (0,1)$ and $\tc/\te \in (0,10)$
(with $\fs = 0$ and $\fs = 1/2$). From these figures we infer that the universe
is initially filled almost exclusively with matter ($\Omm(\tc) \to  1^{-}$, $\Omf(\tc) \to 0^{+}$
for $\tc/\te \to 0^{+}$). Afterwards, matter continues to dominate over the scalar field
(which is supposed to model dark energy) until the cosmic time $\bar{\tc}$
implicitly defined by the equality
\beq \Omm\big(\bar{\tc}\big) = \Omf\big(\bar{\tc}\big) = 1/2 ~;\feq
more precisely, from Eqs.\,\eqref{tctlin}\eqref{conafH}\eqref{Om0} we deduce
\beq \bar{\tc} = {2\,\mbox{arccosh}\sqrt{2} \over 3 \Hs \sqrt{1 - \Oms}} \simeq 10.18 \times 10^{9}\,\mbox{years}
\qquad\quad \mbox{for $\fs = 0$} ~, \feq
while by numerical methods from Eqs.\,\eqref{tctlin}\eqref{conafH}\eqref{Ommpal} we get
\beq \bar{\tc} \simeq 9.90505 \times 10^{9}\,\mbox{years}
\qquad\quad \mbox{for $\fs = 1/2$} ~. \feq
Of course, since $\Oms\! = 0.308$, at present time we have $\Omf(\tc_*) = 1 - \Omm(\tc_*)
= 1 - \Oms \!= 0.692$ (see Eq.\,\eqref{OmmOmf}), so the scalar field (namely, dark energy)
is the dominant contribution at the present time. In the future, the field continues to be dominant and eventually
fills the whole universe ($\Omm(\tc) \to 0^{+}, \Omf(\tc) \to 1^{-}$ for $\tc/\te \to \infty$).

\subsection{Solutions for class 2 potentials with a matter fluid}\label{class2}
In this subsection we analyze cosmological models corresponding to the integrable case (i)
of subsection \ref{subsec2} for potentials of class 2.
So, spacetime is $(\ds+1)$-dimensional, it is spatially flat ($k = 0$), the matter fluid
has an arbitrary equation of state parameter $w$, and the scalar field has self-interaction potential
\beq \VV(\f) = V_1\, e^{2\, w\, \f} + V_{2}\, e^{(1 + w)\, \f}
\qquad \big(V_1, V_2 \in \reali\big)~. \label{vclass2sol}\feq
The cited case (i) of subsection \ref{subsec2} prescribes $w \neq \pm 1$; 
here we make the more restrictive assumption
\beq
- 1 < w < 1 \,, \label{wpm1}
\feq
which generates two benefits. First of all, this restriction ensures that
all the hypergeometric functions appearing in the sequel are non-singular
({\footnote{Recall that ${}_2 F_1(a,b,c;z)$ is singular if $c = 0,-1,-2,\,...$.
The values of $c$ considered in this subsection depend on $w$, and our assumption
\eqref{wpm1} ensures that $c \neq 0,-1,-2,\,...$\,.\vspace{-0.3cm}}});
secondly, it simplifies the qualitative discussion of the solutions.
\parn
Making reference to the analysis of subsection \ref{subsec2},
we fix the gauge function $\BBB(\AA,\f)$ as in Eq.\,\eqref{gau2}
and introduce the pair of Lagragian coordinates $x,y$ related to $\AA,\f$
via Eq.\,\eqref{coor2}; let us remark that these coordinates must fulfill the condition
\beq
x, y >0 ~. \label{xypos}
\feq
With the above positions the Lagrangian function has the harmonic
triangular structure \eqref{triax}, and the related Lagrange equations \eqref{l11} become
\begin{gather}
\xtt -\,\sgn\, (V_1)\;\om^2\, x = 0 ~, \label{ex}\\
\ytt -\,\sgn\, (V_1)\;\om^2\, y = (1 - w)\, V_2\, x^{1 - w \over 1 + w}
- {w(1 - w)\, \ds^2\,\Oms \over 2}\; x^{-{1 + 3 w \over 1 + w}} ~, \label{ey}
\end{gather}
where for notational convenience we have put
\beq
\om := \sqrt{{(1-w^2)\, |V_1| \over 2}}\;.
\label{defom}
\feq
The corresponding solutions can be determined explicitly, treating separately the cases
$V_1 \!>\! 0$, $V_1 = 0$ and $V_1 \!<\! 0$.
Before providing a detailed analysis of these cases, let us point out that
Eqs.\,\eqref{rofi} \eqref{defwfi} \eqref{Omega} \eqref{gau2} \eqref{coor2} \eqref{vclass2sol}
yield the following expression for the coefficient $\wF$ and for the matter density
parameter $\Omm$:
\begin{gather}
\wF = {e^{2 \,w\, \f}\, \dot{\f}^2 - 2\VV(\f) \over e^{2 \,w\, \f}\, \dot{\f}^2 + 2\VV(\f)}
= {\big((1\!-\!w)\,\dot{x}\, y - (1\!+\!w)\,x\, \dot{y}\big)^2
- 2\,(1\!-\!w^2)^2 \big(V_1\, x^2 y^2 \!+\! V_2\, x^{{3+w \over 1+w}} y\big)
\over \big((1\!-\!w)\,\dot{x}\, y - (1\!+\!w)\,x\, \dot{y}\big)^2
+ 2\, (1\!-\!w^2)^2 \big(V_1\, x^2 y^2 \!+\! V_2\, x^{{3+w \over 1+w}} y\big)}~, \label{wfipa2} \\
\Omm = {\ds^2\, \Oms \, e^{-2\,w\, \f - (1+w) \AA} \over \dot{\AA}^{2}}
= {(1\!-\!w^2)^2\,\ds^2\,\Oms\;x^{\frac{1-w}{1+w}}\, y
\over \big((1\!-\!w)\,\dot{x}\,y + (1\!+\!w)\,x\,\dot{y}\big)^2} ~. \label{Ommpa2}
\end{gather}
In the forthcoming paragraphs we present analytic expressions for the solutions
$x(t),y(t)$ of the Lagrange equations \eqref{ex}\eqref{ey}. Afterwards,
in subsections \ref{subBig2}, \ref{subBigC2} we investigate the presence
of a Big Bang and the long time behavior in an example, to be specified below
(see Eq.\,\eqref{exemplary}).\vspace{-0.4cm}

\paragraph{i) $\boma{V_1 > 0}$.} In this case Eqs.\,\eqref{ex}\eqref{ey} read
\begin{gather}
\xtt - \om^2\, x = 0~, \label{ex1a} \\
\ytt - \om^2\, y =
(1-w)\,V_{2}\; x^{{1-w \over 1+w}} - {w(1-w)\, \ds^2\,\Oms \over 2}\; x^{-{1+3w \over 1+w}} ~.
\label{ey1a}
\end{gather}
After possibly a time translation $t \to t + \mbox{const.}$ and a time reflection $t \to -t$,
any solution of Eqs.\,\eqref{ex1a}\eqref{ey1a} compatible with Eq.\,\eqref{xypos}
can be written in one of the following forms
(see Appendix \ref{appsol} for the derivation of the following expressions):
\begin{gather}
x(t) = A \sinh(\om\, t) \qquad (A > 0)~,  \label{primasol} \\
y(t) = C \cosh(\om\, t) + D \sinh(\om\, t) \nonumber \\
+ \,{V_2 \over V_1}\, A^{{1 - w \over 1 + w}}
\sinh^{3 + w \over 1 + w}(\om\, t)\! \left[ 1  - {2 \over 3 + w}\,\cosh(\om\, t)
{~}_2 F_1\! \left( {1 \over 2}\,,  {3 + w \over 2+2w}\, , {5 + 3 w \over 2+2w}\,;
- \sinh^2(\om\, t)\! \right) \right] \nonumber \\
+ \,{\ds^2\,\Oms \over 2\,V_1}\, A^{-{1 + 3 w \over 1 + w}}
\sinh^{1 - w \over 1 + w}(\om\, t)\! \left[ 1  + {2\, w \over 1 - w}\,\cosh(\om\, t)
{~}_2 F_1\! \left( {1 \over 2}\,,  {1 - w \over 2+2w}\, , {3 +  w \over 2+2w}\,;
- \sinh^2(\om\, t) \!\right) \right] ;\nonumber 
\end{gather}
\begin{gather}
x(t) = A \cosh(\om\, t) \qquad (A > 0)~, \label{secondasol} \\
y(t) = C \cosh(\om\, t) + D \sinh(\om\, t) \nonumber \\
+\, {V_2 \over V_1}\, A^{{1 - w \over 1 + w}}\!
\left[\cosh(\om\,t) \Big(1 - \cosh^{{2 \over 1 + w}}(\om\,t)\Big)
+ {2 \over 1 \!+\! w}\, \sinh^2(\om\, t) {~}_2 F_1\! \left({1 \over 2}\,,  - \,{1 - w \over 2+2w}\, , {3 \over 2}\,;
- \sinh^2(\om\, t)\! \right) \right] \nonumber \\
+\, {\ds^2\, \Oms \over 2\,V_1}\, A^{-{1 + 3 w \over 1 + w}}\!
\left[ \cosh(\om\, t)\Big(1 - \cosh^{-{2 w \over 1 + w}}(\om\, t)\Big)
- {2\, w \over 1 \!+\! w}\, \sinh^2(\om\, t)
{~}_2 F_1\! \left( {1 \over 2}\,, {1 + 3 w \over 2+2w}\,, {3 \over 2}\,;
- \sinh^2(\om\, t)\! \right)\right] ;\nonumber 
\end{gather}
\begin{gather}
x(t) = A\, e^{\om\, t}\qquad (A > 0)~,  \label{terzasol} \\
y(t) = C \cosh(\om\, t) + D \sinh(\om t) \nonumber \\
+\, {V_2 \over V_1}\, A^{{1 - w \over 1 + w}}\,{1+w \over 2\,w}\!
\left[\cosh(\om\, t) + {1-w \over 1+w}\, \sinh(\om\, t)-e^{{1-w \over 1+w}\,\om\,t }\right] \nonumber \\
+\, {\ds^2\, \Oms \over 4\,V_1}\, A^{-{1+3w \over 1 + w}}\, {1+w \over 1+2w}\!
\left[\cosh(\om\, t) -{1+3w \over 1+w}\, \sinh(\om\, t) -e^{-{1+3w \over 1+w}\,\om\,t } \right] . \nonumber
\end{gather}
In the above Eqs.\,\eqref{primasol}\eqref{secondasol}\eqref{terzasol},
$\om$ is defined as in Eq.\,\eqref{defom}, $A,C,D$ are real integration constants,
${}_2 F_1$ is the ordinary, Gaussian hypergeometric function and it is assumed
\begin{equation}
t \in I~,
\end{equation}
where $I$ is a maximal interval such that
\begin{equation}\begin{array}{c}
\dd{I \subset (0,+\infty) \quad \mbox{in the case \eqref{primasol}}\,, \qquad
I \subset \reali \quad \mbox{in the cases \eqref{secondasol}\eqref{terzasol}}\,,}\vspace{0.15cm}\\
\dd{\mbox{and}\qquad \mbox{$y(t) > 0$~ for all $t \in I$}\,.} \label{IIypos}
\end{array}\end{equation}
(Note that the assumption $A > 0$ and the above conditions on $I$ grant $x(t) > 0$.)
\parn
Let us also remark that Eq.\,\eqref{terzasol} must be intended in a limit sense
for $w = 0$ and $w = -1/2$; more precisely, we understand that
\begin{align}
& \left[{1+w \over 2\,w}
\left(\cosh(\om\, t) + {1-w \over 1+w}\, \sinh(\om\, t)-e^{{1-w \over 1+w}\,\om\,t }\right)\right]_{w \,=\, 0}
:= \nonumber \\
& \lim_{w \to 0} \left[{1+w \over 2\,w}
\left(\cosh(\om\, t) + {1-w \over 1+w}\, \sinh(\om\, t)-e^{{1-w \over 1+w}\,\om\,t }\right)\right]
= \om\,t\,e^{\om\, t} - \sinh(\om\,t)\,, \label{lim1}
\end{align}
\begin{align}
& \left[{1+w \over 1+2w}\!
\left(\cosh(\om\, t) -{1+3w \over 1+w}\, \sinh(\om\, t) -e^{-{1+3w \over 1+w}\,\om\,t }\right) \right]_{w \,=\, -1/2} := \nonumber \\
& \lim_{w \to -1/2} \left[{1+w \over 1+2w}\!
\left(\cosh(\om\, t) -{1+3w \over 1+w}\, \sinh(\om\, t) -e^{-{1+3w \over 1+w}\,\om\,t }\right) \right]
= 2 \left(\om\,t\,e^{\om\, t} - \sinh(\om\,t)\right) \label{lim2}
\end{align}
To go on, let us recall the expression \eqref{E21} for the energy $\Ec$;
from here and from Eqs.\,\eqref{primasol} \eqref{secondasol} \eqref{terzasol} we obtain, respectively,
({{\footnote{For this computation, it is convenient to keep in mind that $\Ec$ is a constant
of motion. Therefore, it suffices to compute $\Ec \equiv \Ec(x(t),\xt(t),y(t),\yt(t))$
for any given $t$ or in a suitable limit, e.g. for $t \to 0^{+}$.
The facts mentioned in the present footnote apply to all the subsequent
energy computations in this work, but they will never be repeated.}})
\begin{gather}
\Ec = -\,V_{1}\,A\, D~, \\
\Ec = A \left({\ds^2\, \Oms \over 2}\,A^{-{1+3w \over 1+w }} + V_{2}\,A^{1-w \over 1+w} + V_{1}\,C\right) \,, \\
\Ec = A \left({\ds^2\, \Oms \over 2}\,A^{-{ 1+3w \over 1+w }}
+ V_{2}\, A^{1-w \over 1+w } + V_{1}\,(C - D) \right) \,;
\end{gather}
so, to fulfill the energy constraint $\Ec = 0$ we must put in
Eqs.\,\eqref{primasol} \eqref{secondasol} \eqref{terzasol}, respectively,
\begin{gather}
D = 0 ~, \label{energia1} \\
C = - \,{V_{2} \over V_{1}}\, A^{{1-w \over 1+w}} - \,{\ds^2\,\Oms \over 2\,V_{1}}\, A^{- { 1+3w \over 1+w }}~, \label{energia1bis} \\
C = D - {V_{2} \over V_{1}}\, A^{{ 1-w \over 1+w }} - \,{\ds^2\,\Oms \over 2\,V_{1}}\, A^{- {1+3w \over 1+w }}~. \label{energia1tris}
\end{gather}

\paragraph{ii) $\boma{V_1 = 0}$.} Eqs.\,\eqref{ex}\eqref{ey} reduce to
\begin{gather}
\xtt = 0 ~, \label{ex1b}\\
\ytt = (1 - w)\, V_2\; x^{{1 - w \over 1 + w}}
- {w(1 - w)\, \ds^2\,\Oms \over 2}\; x^{-{1 + 3 w \over 1 + w}} ~.\label{ey1b}
\end{gather}
After a time translation $t \to t + \mbox{const.}$ and possibly a time reflection $t \to -t$,
any solution of Eqs.\,\eqref{ex1b} \eqref{ey1b} compatible with Eq.\,\eqref{xypos}
can be written in one of the following forms (see Appendix \ref{appsol} for more details):
\begin{gather}
x(t) = A\, t \qquad (A > 0)~, \label{primasol1b} \\
y(t) = C + D\, t + {V_{2}\,(1+w)^2\, (1-w) \over 2\,(3+w)}\, A^{{1-w \over 1+w}}\, t^{{3+w \over 1+w}}
+ {(1+w)^2\,\ds^2\,\Oms \over 4}\; A^{-{1+3w \over 1+w}}\,  t^{{1-w \over 1+w}}\;; \nonumber 
\end{gather}
\begin{gather}
x(t) = A\qquad (A > 0) ~, \label{secondasol1b} \\
y(t) = C + D\, t + {t^2 \over 2} \left(V_2\,(1-w)\, A^{{1-w \over 1+w}}
- {w(1-w)\,\ds^2\,\Oms \over 2}\; A^{-{1+3w \over 1+w}} \right) . \nonumber 
\end{gather}
In both Eqs.\,\eqref{primasol1b}\eqref{secondasol1b}, $\om$ is defined as in Eq.\,\eqref{defom},
$A,C,D$ are real integration constants and it is assumed
\begin{equation}
t \in I~,
\end{equation}
where $I$ is a maximal interval such that
\begin{equation}\begin{array}{c}
\dd{I \subset (0,+\infty) \quad \mbox{in the case \eqref{primasol1b}}\,, \qquad
I \subset \reali \quad \mbox{in the case \eqref{secondasol1b}}\,,}\vspace{0.15cm}\\
\dd{\mbox{and}\qquad \mbox{$y(t) > 0$~ for all $t \in I$}\,.}
\end{array}\end{equation}
(Note that the assumption $A > 0$ and the above conditions on $I$ grant $x(t) > 0$.)
\parn
Recalling once more Eq.\,\eqref{E21} for the energy $\Ec$, from Eqs.\,\eqref{primasol1b}\eqref{secondasol1b}
we obtain, respectively,
\begin{gather}
\Ec = -\, {2 \over 1-w^2}\;A\,D ~; \\
\Ec = A^{-{2w \over 1+w}} \left(V_2\,A^2 + {\ds^2\,\Oms \over 2} \right)\,.
\end{gather}
Thus, to fulfill the zero-energy constraint $\Ec = 0$ we must require, respectively,
\begin{gather}
D = 0~; \label{energia1b} \\
V_2 < 0 ~~\mbox{and}~~ A = \sqrt{{\ds^2\,\Oms \over 2\,|V_2|}}~. \label{V2nz}
\end{gather}
(Let us recall that throughout this subsection we are assuming $\Oms \!> 0$.)
\vspace{-0.4cm}

\paragraph{iii) $\boma{V_1 < 0}$.} In this case, Eqs.\,\eqref{ex} \eqref{ey} become
\begin{gather}
\xtt + \om^2 x = 0~, \label{ex1c} \\
\ytt + \om^2 y = (1-w)\,V_2\; x^{{1-w \over 1+w}} -\, {w(1-w)\,\ds^2\,\Oms \over 2}\; x^{-{1+3w \over 1+w}}~.
\label{ey1c}
\end{gather}
After a time translation $t \to t + \mbox{const.}$, any solution of Eqs.\,\eqref{ex1c}\eqref{ey1c}
compatible with Eq.\,\eqref{xypos} can be written as follows (see Appendix \ref{appsol}):
\begin{gather}
x(t) = A\, \sin(\om\, t) \qquad (A>0)~, \label{primasol1c} \\
y(t) = C \cos(\om\, t) + D \sin(\om\, t) \nonumber \\
+ \,{V_2 \over V_1}\, A^{{1 - w \over 1 + w}}
\sin^{3 + w \over 1 + w}(\om\, t)\! \left[ 1  - {2 \cos(\om\, t) \over 3 + w}
{~}_2 F_1\! \left( {1 \over 2}\,,  {3 + w \over 2+2w}\, , {5 + 3 w \over 2+2w}\,;
\sin^2(\om\, t)\! \right) \right] \nonumber \\
\!\!\!- \,{\ds^2\,\Oms \over 2\,V_1}\, A^{-{1 + 3 w \over 1 + w}}
\sin^{1 - w \over 1 + w}(\om\, t)\! \left[ 1  + {2\, w \cos(\om\, t) \over 1 - w}
{~}_2 F_1\! \left( {1 \over 2}\,,  {1 - w \over 2+2w}\, , {3 +  w \over 2+2w}\,;
\sin^2(\om\, t) \!\right) \right]. \nonumber
\end{gather}
Also in this case, $\om$ is defined as in Eq.\,\eqref{defom}, $A,C,D$ are real integration constants,
${}_2 F_1$ denotes the Gaussian hypergeometric function and it is assumed
\begin{equation}
t \in I~,
\end{equation}
where $I$ is a maximal interval such that
\begin{equation}\begin{array}{c}
\dd{I \subset (0,\pi/\om) \qquad \mbox{and}\qquad \mbox{$y(t) > 0$~ for all $t \in I$}\,.}
\end{array}\end{equation}
(Note that the assumption $A > 0$ and the first condition on $I$ grant $x(t) > 0$.)
\parn
To proceed, let us mention that Eqs.\,\eqref{E21} \eqref{primasol1c} imply
\beq \Ec = V_1\,A\,D~; \feq
therefore, to fulfill the energy constraint $\Ec = 0$ we must require
\beq D = 0 ~. \label{enprimasol1c} \feq

\subsubsection{Big Bang analysis}\label{subBig2}
Recalling the general assumptions $\Oms > 0$ and $-1<w<1$ (see Eqs.\,\eqref{Omspos}\eqref{wpm1}),
in the present subsection we proceed to investigate the presence of an initial Big Bang singularity
and the asymptotic behavior close to it for one of the previous solutions,
taken as an example. More precisely, we assume
\beq V_1 > 0~. \label{exemplary} \feq
and we restrict the attention to the corresponding solution described in Eqs.\,\eqref{primasol}\eqref{energia1}
and proceed to examine the circumstances in which a Big Bang occurs at $t=0$ (cf. Eq.\,\eqref{BigBangt0}).
To this purpose, let us first remark that the expressions for $x(t)$ and $y(t)$ in
Eq.\,\eqref{primasol} and the constraint in Eq.\,\eqref{energia1} ensure the following
asymptotics, for $t \to 0^+$
({\footnote{To derive the expansions \eqref{primasolasx}\eqref{primasolasy}
it is useful to recall that the Gauss series for ${}_2 F_1$
(see, e.g., \cite[Eq.\,15.2.1]{NIST}) yields
${}_2 F_1(a,b,c;z) = 1 + O(z)$ for $z \!\to\! 0$ and any $a,b \!\in\! \reali$, $c \!\in \!\reali \backslash\{0,-1,-2,...\}$.}}):
\begin{gather}
x(t) = A\,\om \,t + {1 \over 6}\,A\,(\om\,t)^3 + O(t^5)~, \label{primasolasx} \\
y(t) = C + {1 \over 2}\,C\,(\om \,t)^2
+ {(1+w)\,\ds^2\,\Oms \over 2\,(1-w)\,V_1}\; A^{-{1 + 3 w \over 1 + w}} (\om\,t)^{{1 - w \over 1 + w}}
+O\!\left(t^{\min\left\{4,{3 + w \over 1 + w}\right\}}\right) .
\label{primasolasy}
\end{gather}
While the hypothesis $A > 0$ in Eq.\,\eqref{primasol} grants that $x(t) \!>\!0$
for all $t \in (0,\infty)$, the above expansion for $y(t)$ makes evident that,
under the assumptions \eqref{exemplary}, the analogous condition $y(t)\!>\!0$
(cf. Eq.\,\eqref{xypos}) can be fulfilled in a right neighborhood of $t = 0$ if and only if
\beq C \geqs 0\;. \label{COm}\feq
Hereafter we analyze separately the cases $C > 0$ and $C = 0$.
\vspace{-0.3cm}

\paragraph{i) $\boma{C>0}$.}
From Eqs.\,\eqref{coor2}\eqref{primasolasx}\eqref{primasolasy} we infer, for $t \to 0^{+}$,
\begin{gather}
\AA(t) = {1 \over 1 \!+\! w}\,\log t
\,+\, \log\!\left((A\,\om)^{{1 \over 1 + w}}\, C^{{1 \over 1 - w}}\right)
+ \, O\!\left(t^{\min\{2,{1-w \over 1+w}\}}\right) , \label{A2asy} \\
\f(t) = {1 \over 1\!+\! w}\,\log t
\,+\, \log\!\left((A\,\om)^{{1 \over 1+w}}\, C^{-{1 \over 1 - w}} \right)
+ \, O\!\left(t^{\min\{2,{1-w \over 1+w}\}}\right) . \label{fi2asy}
\end{gather}
In view of Eqs.\,\eqref{aeaa}\eqref{gau2} and of the above expansions, we further obtain
\begin{gather}
a(t) = (A\,\om)^{1 \over \ds(1 + w)}\, C^{1 \over \ds(1 - w)}\; t^{1 \over \ds(1 + w)}
\, +\, O\!\left(t^{\min\left\{{2\ds(1+w) + 1 \over \ds(1+w)},{\ds(1-w)
+ 1 \over \ds(1+w)}\right\}}\right) , \label{4a} \\
e^{\BB(t)} = (A\,\om)^{-{w \over 1 + w}}\, C^{{w \over 1 - w}}\; t^{-{w \over 1 + w}}
\,+\, O\!\left(t^{\min\{{2 + w \over 1 + w}, {1-2w \over 1+w} \}}\right) . \label{4b}
\end{gather}
Recalling again that $-1\!<\!w\!<\!1$,
Eq.\,\eqref{4a} shows that $a(t) \to 0$ for $t \to 0^+$, while Eq.\,\eqref{4b}
proves that $e^{\BB(t)}$ is integrable in a right neighborhood of $t = 0$.
Therefore, making reference to the arguments related to Eq.\,\eqref{BIGBANG},
we can properly speak of a Big Bang at $t = 0$.
In this connection, let us further remark that Eqs.\,\eqref{A2asy} and \eqref{4b} imply, for $t \to 0^+$,
\begin{gather}
e^{\BB(t) - \AA(t)/\ds} = (A\,\om)^{-{1 + \ds w \over n(1+w)}}\, C^{-{1 - \ds w \over \ds(1-w)}}\;
t^{-{1 + \ds w \over n(1+w)}} +
O\!\left(t^{\min\left\{{\ds(2+w) - 1 \over \ds(1+w)}, {\ds(1-2w) - 1 \over \ds(1 + w)}\right\}}\right).
\end{gather}
On account of Eq.\,\eqref{horizon} and of the fact that ${1+\ds\, w \over \ds(1+w)} < 1$
(in our case with $\ds \geq 2 $ and $ -1 < w < 1 $), the latter relation ensures
the presence of a
particle horizon.
\parn
Next, let us note that Eqs.\,\eqref{wfipa2}\eqref{Ommpa2} and the asymptotic expansions
in Eqs.\,\eqref{primasolasx}\eqref{primasolasy} (see also Eq.\,\eqref{defom}) give
\begin{gather}
\wF(t) = 1 - 8 \left({1\!+\!w \over 1\!-\!w}\right) (\om\, t)^2
+ O\!\left(t^{\min\left\{4, {3 + w \over 1 + w}\right\}}\right) \,, \label{4wf} \\
\Omm(t) = {(1\!+\!w)^2\, \ds^2\,\Oms \over C\,(A\,\om)^{\frac{1+3w}{1+w}}}\; t^{\frac{1-w}{1+w}}
+ O\!\left(t^{\min\left\{\frac{3+w}{1+w}, {2(1 - w) \over 1 + w}\right\}}\right) \,. \label{4om}
\end{gather}
Eq.\,\eqref{4wf} suggests that the scalar field behaves as stiff matter close to the Big Bang.
On the other hand, Eq.\,\eqref{4om} indicates that $\Omm(t) \to 0$ for $t \to 0^{+}$;
together with Eq.\,\eqref{OmmOmf}, this implies $\Omf(t) \to 1$ for $t \to 0^{+}$,
thus showing that the scalar field is dominant at the Big Bang.
\parn
Regarding the cosmic time $\tc \equiv \tc(t)$, let us notice that
Eqs.\,\eqref{BIGBANG}\eqref{4b} imply
\beq
\tc(t)/\te = {(1\!+\!w)\; C^{{w \over 1 - w}} \over (A\,\om)^{{w \over 1 + w}}}\; t^{{1 \over 1 + w}}
\,+\, O\!\left(t^{\min\{{3 + 2 w \over 1 + w}, {2 - w \over 1+w} \}}\right) \qquad
\mbox{for $t \to 0^+$}\,, \label{asytc2zero}
\feq
which can be locally inverted to give
\beq
t(\tc) = {(A\,\om)^{w} \over (1\!+\!w)^{1+w}\,C^{{w(1+w) \over 1 - w}}}\; (\tc/\theta)^{1+ w}
\,+\, O\!\left((\tc/\te)^{\min\{2,\, 3(1+w)\}}\right) \qquad
\mbox{for $\tc/\te \to 0^+$} . \label{espt4}
\feq
Using Eq.\,\eqref{espt4} for $t = t(\tc)$, the expansions \eqref{4a}-\eqref{4om}
can be reformulated in terms of the cosmic time $\tc$;
for example, Eq.\,\eqref{4a} yields
\beq
a(\tc) = \left({C A\,\om \over 1\!+\!w}\right)^{\!\!1/\ds}\! (\tc/\theta)^{1/\ds}
+\, O\!\left((\tc/\te)^{\min\left\{{2\ds(1+w) + 1 \over \ds},\,{\ds(1-w) + 1 \over \ds}\right\}}\right)
\qquad \mbox{for $\tc/\te \to 0^+$} . \vspace{-0.3cm} \label{aasytc0}\feq

\paragraph{ii) $\boma{C=0}$.}
First notice that Eqs.\,\eqref{coor2}\eqref{primasolasx}\eqref{primasolasy} imply, for $t \to 0^{+}$,
\begin{gather}
\AA(t) = {2 \over 1 + w}\,\log t\,
+ {1 \over 1-w}\,\log\!\left({(1\!+\!w)^2\,\ds^2\,\Oms \over 4\,(A\,\om)^{4w \over 1 + w}} \right) + O(t^2) ~, \label{A2asy2} \\
\f(t) = - \,{1 \over 1 - w}\, \log\!\left({(1\!+\!w)^2\,\ds^2\,\Oms \over 4\,(A\,\om)^{2}} \right) + O(t^2) ~. \label{fi2asy2}
\end{gather}
On account of Eqs.\,\eqref{aeaa}\eqref{gau2}, the above expansions allow us to infer that
\begin{gather}
a(t) = \left({\ds^2\,\Oms (1\!+\!w)^2 \over 4}\right)^{\!\!{1 \over \ds(1-w)}}
(A\,\om)^{-{4 w \over \ds(1 - w^2)}}\;
t^{{2 \over \ds(1 + w)}} + O\!\left(t^{{2 + 2\ds(1 + w) \over \ds(1 + w)}}\right), \label{41a} \\
e^{\BB(t)} = \left({(1\!+\!w)^2\,\ds^2\,\Oms \over 4\, (A\,\om)^2}\right)^{\!\!{w \over 1 - w}} \!+ O(t^2) ~. \label{41b}
\end{gather}
Similarly to the case with $C \!>\! 0$ discussed in the previous paragraph,
the above relations show that $a(t) \to 0$ for $t \to 0^+$ and even imply the
integrability of $e^{\BB(t)}$ in a right neighborhood of $t = 0$ (for $-1<w<1$);
so, we have a Big Bang at $t = 0$. To say more, Eqs.\,\eqref{A2asy2}\eqref{41b} give,
for $t \to 0^+$,
\begin{gather}
e^{\BB(t) - \AA(t)/\ds} \!=\!
\left({(1\!+\!w)^2\,\ds^2\,\Oms \over 4}\right)^{\!\!{\ds w - 1 \over \ds(1 - w)}}\!
(A\,\om)^{{2w(2-\ds (1+w)) \over \ds(1 - w^2)}}
t^{-{2 \over \ds(1 + w)}}+ O\!\left(\!t^{{2\ds(1+w)-2 \over \ds(1 + w)}}\!\right),
\end{gather}
which, together with Eq.\,\eqref{horizon}, indicates that a particle horizon
occurs whenever $ {\frac{2}{\ds(1+w)}} < 1 $. In our case with $\ds \geq 2 $ and $ -1 < w < 1 $,
this is equivalent to $w > (2/\ds) - 1$ (cf. Eq.\,\eqref{stEC} and the related comments);
especially, let us point out that the latter condition is fulfilled
in the case of radiation where $w = 1/\ds$.
\parn
Concerning the coefficient in the field equation of state and the matter density
parameter, from Eqs.\,\eqref{wfipa2}\eqref{Ommpa2}\eqref{primasolasx}\eqref{primasolasy}
we obtain
\begin{gather}
\wF(t) = -\,1 + {2 \left(\!{1 + w \over 3 + w}\!\right)^{\!2}\!
\left({2\,w \over 1 - w} + {2\,V_2\,A^2 \over \ds^2\,\Oms}\right)^{\!2}
\over \left({1 + w \over 1 - w} + {2\,V_2\,A^2 \over \ds^2\,\Oms}\right)}\;
(\om\,t)^{2} + O(t^4) ~, \label{41wf} \\
\Omm(t) = 1 - \left({1 \!+\! w \over 1 \!-\! w} + {2\,V_2\, A^2 \over \ds^2\,\Oms}\right)
(\om\,t)^{2} + O(t^{4}) ~. \label{41om}
\end{gather}
The above relations indicate, respectively, that close to the Big Bang
the scalar field $\f$ behaves as a cosmological constant whereas the dominant
contribution comes from the matter fluid.
\parn
To conclude, from Eqs.\,\eqref{BIGBANG} and \eqref{41b} we readily infer
\beq
\tc(t)/\te = \left({(1\!+\!w)^2\, \ds^2\,\Oms \over 4\, (A\,\om)^2}\right)^{\!\!{w \over 1 - w}} t \,+\, O(t^2)
\qquad \mbox{for $t \to 0^+$}\,,
\feq
which entails, by inversion,
\beq
t(\tc) = \left({(1\!+\!w)^2\,\ds^2\,\Oms \over 4\, (A\,\om)^2}\right)^{\!\!- {w \over 1 - w}} (\tc/\theta)
\,+\,O\big((\tc/\te)^2\big)
\qquad \mbox{for $\tc/\te \to 0^+$}\,, \label{espt14}
\feq
Of course, the previous expansions \eqref{41a}-\eqref{41om} can be rephrased in terms
of the cosmic time $\tc$, using Eq.\,\eqref{espt14}; for example, we have
\beq
a(\tc) = \left({(1\!+\!w)^2\,\ds^2\,\Oms \over 4}\right)^{\!\!{1 \over \ds(1+w)}}
(\tc/\te)^{{2 \over \ds(1 + w)}} + O\!\left((\tc/\te)^{{2 + \ds(1 + w) \over \ds(1 + w)}}\right) \quad
\mbox{for $\tc/\te \to 0^{+}$}. \vspace{0.cm}\feq

\subsubsection{Far future analysis}\label{subBigC2}
The qualitative behavior on large time scales of the model under analysis depends
sensibly on the choice of the parameters which characterize the solutions $x(t),y(t)$
of the Lagrange equations \eqref{ex}\eqref{ey}.
In the sequel we account (at least partially) for this rather predictable fact,
referring once more to the exemplary case whose Big Bang phenomenology was examined
in the previous paragraph.
\parn
Correspondingly, we assume again that $\Oms > 0$, $-1\!<\!w\!<\!1$ and $V_1 \!>\! 0$
(see Eqs.\,\eqref{Omspos}\eqref{wpm1}\eqref{exemplary}), and consider the solution given in
Eqs.\,\eqref{primasol}\eqref{energia1} with the condition $C \geqs 0$ (see Eq.\,\eqref{COm}),
which grants the occurrence of a Big Bang $t = 0$.
\vspace{-0.4cm} 

\paragraph{A discussion of the maximal domain.}
Let us recall that the domain of definition for the said solutions $x(t),y(t)$
is a maximal interval $I \subset (0,+\infty)$ such that $y(t) > 0$
for all $t \in I$ (see Eq.\,\eqref{IIypos}). 
In general, the maximal interval where the expression for $y(t)$ in Eq.\,\eqref{primasol}
gives $y(t) > 0$ cannot be determined by purely analytical means and one must perform a
numerical evaluation.
On the other hand, let us notice that for $t \to +\infty$ we have
({\footnote{To derive the expansions in Eq.\,\eqref{yasy2} one should
recall that for $z \to -\infty$ (see, e.g., \cite[Eqs.\,(15.2.2)(15.8.2)]{NIST})
$$ {~}_2 F_1(a,b,c;z) = {\Ga(c)\,\Ga(b-a)\,\Ga(1-(b-a)) \over  \Ga(b)\,\Ga(c-a)\,\Ga(a-b+1)}\;(-z)^{-a}
- {\Ga(c)\,\Ga(b-a)\,\Ga(1-(b-a)) \over \Ga(a)\,\Ga(c-b)\,\Ga(b-a+1)}\;(-z)^{-b}
+ O\!\left((-z)^{-\min\{a+1,b+1\}}\right). $$}})
\begin{gather}
y(t) = \label{yasy2} \\
\left\{\!\begin{array}{ll}
\dd{\left[ {C \over 2} + {A^{{1 - w \over 1 + w}} \over 4\, \sqrt{\pi}}\,
\Ga\!\left(\!{1\!+\!2w \over 1\!+\!w}\!\right) \Ga\!\left(\!{1\!-\!w \over 2(1\!+\!w)}\!\right)\!
\left({1\!-\!w \over w}\,V_2 + {\ds^2\, \Oms \over A^2}\right) \right]
e^{\om\,t} + O\!\left(e^{{1 - w \over 1 + w}\;\om\,t}\right)}
& \dd{\mbox{for $0\!<\!w \!<\! 1$}\,,} \vspace{0.2cm}\\
\dd{{(1\!-\!w^2)\,A\,V_2 \over 2\, \om^2}\;t\,e^{\om\, t} + O(e^{\om\,t})} & \dd{\mbox{for $w = 0$}\,,} \vspace{0.1cm}\\
\dd{-\,{(1\!-\!w)(1\!+\!w)^2\,A^{{1 - w \over 1 + w}}\,V_2 \over 4\,w\, \om^2}\;
e^{{1 - w \over 1 + w}\;\om\,t} + O\!\left(e^{({1 - w \over 1 + w}-1)\,\om\,t}\right)}
& \dd{\mbox{for $-1\!<\!w \!<\! 0$}\,.}
\end{array}\right. \nonumber
\end{gather}
Keeping in mind the previous assumptions on the parameters, the above asymptotics
show that the inequality $y(t) > 0$ holds only if
({\footnote{In this connection, note that $\Ga\!\left(\!{1\!+\!2w \over 1\!+\!w}\!\right)
\Ga\!\left(\!{1\!-\!w \over 2(1\!+\!w)}\!\right) > 0$ for all $0<w<1$.}})
\beq \left\{\!\begin{array}{ll}
\dd{V_2 > - \,{w \over 1\!-\!w} \left[
{2\, \sqrt{\pi} \over \Ga\big({1 + 2w \over 1 + w}\big)
\Ga\big({1 - w \over 2(1 + w)}\big)}\; {C \over A^{{1 - w \over 1 + w}}}
+ {\ds^2\, \Oms \over A^2}\right]}
& \dd{\mbox{for $0 < w < 1$}\,,} \vspace{0.15cm}\\
\dd{V_2 > 0}
& \dd{\mbox{for $-1 < w \leq 0$}\,.}
\end{array} \right. \label{yposinf}\feq
\vfill\eject\noindent
Recalling that $C \geqs 0$, we see that the above condition on $V_2$
for $0\!<\!w\!<\!1$ is certainly fulfilled if
\beq
V_2 > 0 ~.
\feq
Whenever the conditions in Eq.\,\eqref{yposinf} are violated, $y(t)$ eventually
becomes negative; since $y(t)$ is positive close to the Big Bang
(for $t \to 0^+$), it follows that $y(t)$ must vanish at some finite time, namely,
\beq \exists \;\ts \in (0,+\infty) \quad \mbox{such that}\quad y(\ts) = 0 ~. \feq
In this case the maximal admissible interval $I$ is a
subset of $(0,\ts)$. Besides, given that $x(t) = A\, \sinh(\om\,t)$
is strictly positive and finite for all $t \in (0,\ts)$,
from Eqs.\,\eqref{aeaa}\eqref{coor2} we see that
\beq a(t) = x(t)^{1 \over \ds(1 + w)}\, y(t)^{1 \over \ds(1 - w)} \to 0
\qquad \mbox{for $t \to \ts^-$}\,. \feq
The above relation suggests that a Big Crunch could occur at $t = \ts$;
in this regard, it should be recalled that the very definition of Big Crunch also
requires $e^{\BB(t)}$ to be integrable in a left neighborhood of $\ts$.
Since Eqs.\,\eqref{gau2}\eqref{coor2} give
\beq e^{\BB(t)} = e^{-w \,\f(t)} = x(t)^{-{w \over 1-w}}\, y(t)^{w \over 1-w}\,, \feq
it follows that $e^{\BB(t)}$ is certainly integrable for $t \to \ts^{-}$ if $0 \!\leqs\! w < 1$,
while a finer analysis is needed to ascertain the integrability of $e^{\BB(t)}$
when $-1 \!<\! w \!<\! 0$.
\parn
On the other side, let us stress that the fulfillment of the conditions in Eq.\,\eqref{yposinf}
is certainly not sufficient to ensure $y(t) > 0$ for all $t \in (0,+\infty)$. Notwithstanding,
in the upcoming subsection \ref{Models2} we are going to show that this condition is actually attained
at least for a specific choice of the parameters; for such a choice, the maximal admissible domain is in fact
\beq I = (0,+\infty) ~. \label{IINF}\feq
By continuity arguments, the same happens when the parameters are close to the above mentioned choice.
\vspace{-0.4cm}

\paragraph{Asymptotic expansions.}
Assuming Eq.\,\eqref{IINF} and restricting the attention to the case
\beq
0 < w < 1 ~,
\feq
of interest for the subsequent applications, for $t \to +\infty$ we obtain
\begin{align}
& \AA(t) =\;
{2 \over 1 \!-\! w^2}\;\om\,t \label{AInf2}\\
& + {1 \over 1 \!-\! w} \log\! \left[{A^{{1-w \over 1+w}} \over 2^{{2 \over 1+w}}}\!
\left(C + {A^{{1 - w \over 1 + w}} \over 2\, \sqrt{\pi}}\,
\Ga\!\left(\!{1\!+\!2w \over 1\!+\!w}\!\right)\! \Ga\!\left(\!{1\!-\!w \over 2(1\!+\!w)}\!\right)\!
\left({1\!-\!w \over w}\,V_2 + {\ds^2\, \Oms \over A^2}\right)\! \right)\!\right]
\!+ O\!\left(e^{-{2 w \over 1 + w}\,\om\,t}\right), \nonumber
\end{align}
\begin{align}
& \f(t) =\,
-{2\,w \over 1 \!-\! w^2}\;\om\,t \label{fInf2}\\
& - {1 \over 1\! -\! w} \log\! \left[{A^{-{1-w \over 1+w}} \over 2^{{2w \over 1+w}}}\!
\left(C + {A^{{1-w \over 1+w}} \over 2 \sqrt{\pi}}\,
\Ga\!\left(\!{1\!+\!2w \over 1\!+\!w}\!\right)\! \Ga\!\left(\!{1\!-\!w \over 2(1\!+\!w)}\!\right)\!
\left({1\!-\!w \over w}\,V_2 + {\ds^2\, \Oms \over A^2}\right)\!\right)\! \right]\!
+ O\!\left(e^{-{2 w \over 1 + w}\;\om\,t}\right) , \nonumber
\end{align}
\begin{gather}
\wF(t) =
-\,1 + 2 \,w^2 +\, O\!\left(e^{-{2 \,w \over 1 + w}\;\om\,t}\right) \label{wfInf2} \\
\Omm(t) = \label{OmmInf2} \\
{2\,(1 \!-\! w^2)^2\,\ds^2\, \Oms\; (A/2)^{{1-w \over 1 + w}} \over
(A\,\om)^2 \left(\!C \!+\! {A^{{1-w \over 1+w}} \over 2 \sqrt{\pi}}\,
\Ga\!\left(\!{1+2w \over 1+w}\!\right)\! \Ga\!\left(\!{1-w \over 2(1+w)}\!\right)\!
\left({1-w \over w}\,V_2 \!+\! {\ds^2\, \Oms \over A^2}\right)\!\right)}\;
e^{- {2 (1+2w) \over 1 + w}\;\om\,t}
+ O\!\left(e^{-{2 (1 + 3 w) \over 1 + w}\;\om\,t}\right) . \nonumber
\end{gather}
In particular, note that the field $\f$ is the dominant contribution for $t \to +\infty$
($\Omf = 1- \Omm \to 1$ for $t \to \infty$; see Eq.\,\eqref{OmmOmf}).
\vspace{-0.2cm}
\vfill\eject\noindent
$\phantom{a}$\vspace{-1.cm}\\
Also in this case, the previous asymptotic expansions  can be rephrased using
of the cosmic time $\tc$; to this purpose, one should first notice that
Eqs.\,\eqref{taut}\eqref{gau2}\eqref{fInf2} entail, for $t \to +\infty$,
({\footnote{Let us give some hints about the derivation of the asymptotic
expansion \eqref{asytc2inf}, understanding that the maximal domain of
definition of the solution is $I = (0,+\infty)$ as in Eq.\,\eqref{IINF}.
Firstly notice that Eqs.\,\eqref{taut}\eqref{gau2} imply, for any fixed
$t_* \!>\! 0$ and for all $t \!>\! 0$,
$$ \tc(t)/\te = \tc(t_*)/\te + \!\int_{t_*}^{t}\! d t'\; e^{-\,w\,\f(t')}\,. $$
Then, Eq.\,\eqref{asytc2inf} can be derived substituting the expansion \eqref{fInf2}
for $\f(t')$ in the above identity and integrating each term separately;
in this connection, it is worth noting that $\tc(t_*)/\te = O(1)$ and
$\int_{t_*}^{t}\! d t'\, O\Big(e^{{2 w(2w-1) \over 1 - w^2}\,\om\,t'}\Big)
= O\Big(e^{{2 w(2w-1) \over 1 - w^2}\,\om\,t}\Big) + O(1)
\equiv O\Big((e^{\om\,t})^{\max\left\{0,{2\,w (2w-1) \over 1 - w^2}\!\right\}}\Big)$ for $t \to +\infty$.\vspace{-0.3cm}}})
\begin{gather}
\tc(t)/\te = \label{asytc2inf} \\
{1 \!-\! w^2 \over 2^{{1 + w^2 \over 1-w^2}} w^2\! A^{{w \over 1+w}} \om\!}\!
\left(\!C \!+\! {A^{{1-w \over 1+w}} \over 2 \sqrt{\pi}}\,
\Ga\!\left(\!{1\!+\!2w \over 1\!+\!w}\!\right)\! \Ga\!\left(\!{1\!-\!w \over 2(1\!+\!w)}\!\right)\!\!
\left(\!{1\!-\!w \over w}\,V_2 \!+\! {\ds^2 \Oms \over A^2}\right)\!\!\right)^{\!\!{w \over 1 - w}}\!\!\!
e^{{2\,w^2 \over 1 - w^2}\,\om\,t}\!\! +\! O\!\!\left(\!\!(e^{\om\,t})^{\max\left\{0,{2\,w (2w-1) \over 1 - w^2}\!\right\}}\!\right) \nonumber
\end{gather}
As an example, let us mention that by inverting the above relation,
from Eqs.\,\eqref{aeaa}\eqref{AInf2} we get the following,
for $\tc/\te \to +\infty$:
\begin{gather}
a(\tc) = \label{atcInf}\\
\left({A \left({2\,w^2 \,\om \over 1 - w^2} \right)^{\!1/w} \over
C \!+\! {A^{{1 - w \over 1 + w}} \over 2\, \sqrt{\pi}}\,
\Ga\!\left(\!{1+2w \over 1+w}\!\right)\! \Ga\!\left(\!{1-w \over 2(1+w)}\!\right)\!
\left({1-w \over w}\,V_2 \!+\! {\ds^2\, \Oms \over A^2}\right)}
\right)^{\!\!{1 \over \ds\,w}}\!\!
(\tc/\te)^{1 \over \ds\,w^2} + O\!\left((\tc/\te)^{\max\left\{{1 - \ds\,w^2 \over \ds\,w^2},{1 - \ds\,w(1-w) \over \ds\,w^2}\right\}}\right) . \nonumber \vspace{0.cm}
\end{gather}

\subsubsection{Qualitative analysis of one of the previous cases. A model for inflation}\label{Models2}
We now proceed to examine in more detail a particular case of the cosmological
model analyzed before in the present subsection \ref{class2}, selecting specific
values for the associated free parameters.
Let us anticipate that the rationale behind the said choice of parameters is to
realize an inflationary scenario, in a very early stage of the universe.
This scenario would allow, among else, to resolve
the flatness, horizon and monopole problems. The above considerations and the arguments
to be presented in the sequel are largely inspired by the model portrayed in
\cite[Sec.\,11.4]{Ryd}, where inflation is triggered by a true cosmological constant;
on the contrary, here we plan to mimic this cosmological constant contribution
using the scalar field $\f$ with a self-interaction potential of the form \eqref{vclass2sol}.
\parn
To begin with, let us fix the space dimension and the spatial curvature as (see Eq.\,\eqref{k0})
\beq
\ds = 3~, \qquad k = 0~. \label{dsk}
\feq
We further suppose that the ordinary matter content of the universe can be described
by means of a perfect fluid of radiation type, i.e., we posit
\beq
w = 1/3 ~. \label{w13}
\feq
To proceed, let us refer to the considerations related to Eqs.\,\eqref{cosm1}-\eqref{wficos};
these indicate that the field $\f$ can effectively reproduce a cosmological constant contribution
whenever the self-interaction potential $\VV(\f)$ possesses a stationary point
(see Eq.\,\eqref{cosm2}), which can assumed to be zero after a translation $\f \mapsto \f + \mbox{const.}$.
More precisely, we require the potential in Eq.\,\eqref{vclass2sol} to attain a maximum at $\f = 0$,
which happens if
\beq
V_1 = 2\,V\,, \qquad V_2 = -\,V ~\qquad \mbox{for some $V> 0$}~. \label{V1V2}
\feq
With the above choices, the potential \eqref{vclass2sol} reduces to
\beq \VV(\f) = V \left(2\, e^{{2 \over 3}\, \f} -\, e^{{4 \over 3}\, \f}\right)
\label{vclass2solNum}\feq
(see Fig.\,7 for the plot of the map $\f \in \reali \mapsto \VV(\f)/V$).
\begin{figure}[t!]
\vspace{-0.5cm}
    \centering
        \begin{subfigure}[b]{0.5\textwidth}
                \includegraphics[width=\textwidth]{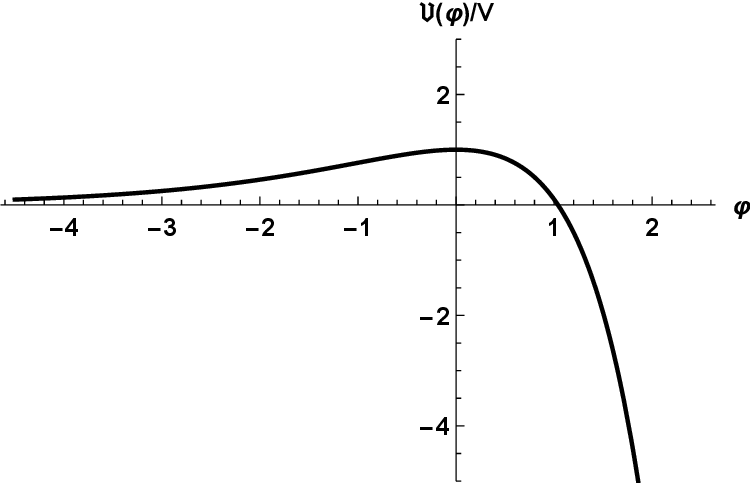}\label{fig:7}
        \end{subfigure}
        \caption*{\textsc{Figure\! 7.} Plot of the map $\f \mapsto \VV(\f)/V$, for $\VV(\f)$ as in Eq.\,\eqref{vclass2solNum}.}
\vspace{0.cm}
\end{figure}
\parn
Since the condition $V_1 > 0$ is certainly fulfilled in the case under analysis,
we can refer to the Lagrange equations \eqref{ex1a}\eqref{ey1a} and to
the corresponding solutions $x(t),y(t)$ written in Eq.\,\eqref{primasol};
taking also into account the associated zero-energy constraint \eqref{energia1}
and using some known relations for the hypergemetric functions ${}_2 F_1$
appearing in Eq.\,\eqref{primasol}
({\footnote{In particular, the derivation of the expression for $y(t)$ in Eq.\,\eqref{xyexp}
relies on the identity
$$ {~}_2 F_1\! \left( {1 \over 2}\,,  {5 \over 4}\, , {9 \over 4}\,; z \right) =
{5 \over 3\,z} \left[ \!{~}_2 F_1\! \left( {1 \over 2}\,,  {1 \over 4}\, , {5 \over 4}\,;z \!\right)
-\sqrt{1-z}\, \right] \qquad \mbox{for $z \in \reali$}~. $$
This identity can be derived with some elementary computations starting from the
Gauss series representation of ${}_2 F_1$ (see, e.g., \cite[Eq.\,15.2.1]{NIST})
and using some known relations for the Euler gamma functions $\Ga$ appearing therein.
The same identity could also be derived (with some more effort) from the
relations for contiguous hypergeometric functions (see, e.g., \cite[\S\,15.5(ii)]{NIST}).\vspace{-0.3cm}}}),
we obtain
\begin{gather}
x(t) = A \sinh(\om\, t)~, \nonumber \\
\begin{aligned}
y(t) =\; & C \cosh(\om\, t) + {\sqrt{A} \over 2}\left(1 + \,{9\,\Oms \over 2\,A^2\, V}\right) \sqrt{\sinh(\om\, t)} \\
& - {\sqrt{A} \over 2} \left(1 - {9\,\Oms \over 2\,A^2\, V}\right) \sqrt{\sinh(\om\, t)}\;\cosh(\om\, t)
{~}_2 F_1\! \left( {1 \over 2}\,,  {1 \over 4}\, , {5 \over 4}\,; - \sinh^2(\om\, t) \!\right) ,
\end{aligned} \label{xyexp}
\end{gather}
where, according to Eq.\,\eqref{defom},
\beq
\om = {2\, \sqrt{2\,V} \over 3}~. \label{omVexp}\vspace{-0.2cm}
\feq
\paragraph{Choice of a special solution.}
Quite understandably, the analysis of the model at issue becomes significantly simpler
if the parameter $A$ (left unspecified until now) is fixed so as to get rid
of the hypergeometric function ${}_2 F_1$ appearing in the expression for $y(t)$ in
Eq.\,\eqref{xyexp}; to this purpose, we set
\beq \Oms > 0~, \qquad~ A = \sqrt{{9\,\Oms \over 2\, V}} \equiv {2 \sqrt{\Oms} \over \om} ~. \label{Aexp}\feq
Next, let us require a Big Bang to occur at $t = 0$ and recall that this can happen
only if $C \geqs 0$ (see Eq.\,\eqref{COm}); accordingly,
for later convenience we put
\beq C = \sqrt{A}\;\eep \equiv \left({4\, \Oms \over \om^2}\right)^{\!\!1/4} \eep \qquad
\mbox{for ~$\eep \geqs 0$}~. \label{Cexp}\feq
With the above choices \eqref{Aexp}\eqref{Cexp}, Eq.\,\eqref{xyexp} reduces to
\begin{gather}
x(t) = {2 \sqrt{\Oms} \over \om} \; \sinh(\om\, t)~, ~\qquad
y(t) = \left({4\, \Oms \over \om^2}\right)^{\!\!1/4} \sqrt{\sinh(\om\, t)}\,
\left[ 1 + \eep\; {\cosh(\om\, t) \over \sqrt{\sinh(\om\, t)}}\right] . \label{xyexp2}
\end{gather}
It is evident that the above expressions fulfill $x(t) \!>\! 0$ and $y(t) \!>\! 0$ for all $t \!>\! 0$
(cf. Eq.\,\eqref{xypos}); so, we can understand the solution \eqref{xyexp2} to be defined
on the maximal admissible domain
\beq
I = (0,+\infty) ~. \label{Idefexp}
\feq
To go on, let us note that Eqs.\,\eqref{taut}\eqref{gau2}\eqref{coor2}
and Eq.\,\eqref{xyexp2} give the following expression for the cosmic time:
\beq
\tc(t)/\te \;= \int_0^t dt'\;x^{-{1/4}}(t')\; y^{{1/2}}(t')\,
= \int_0^t dt'\, \sqrt{1 + \eep\; {\cosh(\om\, t') \over \sqrt{\sinh(\om\, t')}}} ~.
\label{tcInt}
\feq
Concerning the scale factor and the field, from Eq.\,\eqref{aeaa}, Eq.\,\eqref{coor2} (with $\ga=w=1/3)$
and Eqs.\,\eqref{omVexp} \eqref{xyexp2} we infer
\begin{gather}
a(t) = x^{1/4}(t)\, y^{1/2}(t) =
\left({4\,\Oms \over \om^2}\right)^{\!\!1/4}\! \sqrt{\sinh(\om\, t)}\;
\sqrt{1 + \eep\; {\cosh(\om\, t) \over \sqrt{\sinh(\om\, t)}}}\; , \label{aexp} \\
\f(t) = \log\!\big(x^{3/4}(t)\, y^{-{3/2}}(t)\big) =
-\, {3 \over 2} \log\! \left[ 1 + \eep\; {\cosh(\om\, t) \over \sqrt{\sinh(\om\, t)}}\right]. \label{fildesp}
\end{gather}
Finally, the equation of state coefficient for the scalar field and the density
parameter of radiation can be determined using Eqs.\,\eqref{wfipa2}\eqref{Ommpa2}
and \eqref{xyexp2}, which gives
\begin{gather}
\wF(t) \,=\, -1 + {2\, \eep^2 \big(1\!-\!\sinh^2(\om\, t)\big)^2 \over
4\, \sinh^3(\om\, t)
+ 12\, \eep \cosh(\om\, t) \sinh^{5/2}(\om\, t)
+ \eep^2 \big(1\!+\!3 \sinh^2(\om\, t)\big)^2}~, \label{wfiexp2}\\
\Omm(t) \,=\, {\sinh(\om\, t) + \eep \cosh(\om\, t)\, \sqrt{\sinh(\om\, t)} \over
\big(\cosh(\om\, t) \sqrt{\sinh(\om\, t)} + {\eep \over 2} \big(1\!+\! 3 \sinh^2(\om\, t)\big)\big)^2}~. \label{Ommexp2}
\end{gather}
In the sequel we first discuss the exceptional case $\eep = 0$,
and then proceed to analyze the more generic configuration with $\eep > 0$.
\vspace{-0.2cm}

\paragraph{The case $\boma{\eep = 0}$.}
This case deserves a special mention because it corresponds to a scenario
where the field $\f$ behaves exactly as a cosmological constant.
Eqs.\,\eqref{tcInt}-\eqref{Ommexp2} with $\eep=0$ give the following results, for
$t \in (0,+\infty)$:
\beq \tc(t)/\te = t~, \feq
and $a(t) = (4\, \Oms/\om^2)^{1/4}\, \sqrt{\sinh(\om\, t)}$,
$\f(t) = \mbox{const.} = 0$, $\wF(t) = \mbox{const.} = -1$, $\Omm(t) = 1/\cosh^2(\om\, t)$.
Equivalently, viewing these observables as function of the cosmic time $\tc \in (0,+\infty)$:
\beq
\begin{array}{c}
\dd{a(\tc) = \left({4\, \Oms \over \om^2}\right)^{\!\!1/4}\! \sqrt{\sinh(\om\, \tc/\te)}~,
\qquad \f(\tc) = \mbox{const.} = 0~,} 
\vspace{0.1cm}\\
\dd{\wF(\tc) = \mbox{const.} = -1 ~, \qquad\quad \Omm(\tc) = {1 \over \cosh^2(\om\, \tc/\te)}~.}
\end{array}
\feq
In particular, from the above explicit expression for $a(\tc)$ we infer
({\footnote{One cannot naively refer to the asymptotic expansions in Eqs.\,\eqref{AInf2}-\eqref{atcInf}
for $\eep = 0$, since in this case the dominant contribution written in Eq.\,\eqref{yasy2} for $0\!<\!w\!<\!1$
vanishes identically.}})
\beq
a(\tc) = \left\{\!\!\begin{array}{ll}
\dd{\big(4\,\Oms\big)^{1/4}\, (\tc/\te)^{1/2} + O\big(\tc^{5/2}\big)}
&   \mbox{for\, $\tc/\te \to 0^+$}\,, \vspace{0.1cm}\\
\dd{(\Oms/\om^2)^{1/4}\, e^{{\om \over 2}\, (\tc/\te)} + O\big(e^{-{3\om \over 2}\,\tc/\te}\big)}
&   \mbox{for\, $\tc/\te \to +\infty$}\,. \label{atcasy0}
\end{array}\right.
\feq
\vfill\eject\noindent

\paragraph{The case $\boma{\eep > 0}$.}
Let us return to Eqs.\,\eqref{tcInt}-\eqref{Ommexp2}, that we now use with $\eep > 0$.
We first derive the asymptotic expansions of $\tc(t)/\te$, $a(t)$, $\wF(t)$, $\Omm(t)$
in the limit of small and large $t$. \parn
The behaviour of $\tc(t)/\te$ in these limits can be derived
from the general asymptotic expansions
\eqref{asytc2zero} \eqref{asytc2inf}, which in the present setting reduce to
({\footnote{More precisely, the asymptotics in Eq.\,\eqref{tcasy} follow from
the cited Eqs.\,\eqref{asytc2zero} and \eqref{asytc2inf} fixing
$\ds = 3$, $w = 1/3$, $V_1 = 2\,V$ and $V_2 = -\,V$ (with $V\!>\! 0$),
$A = \big(9\,\Oms\!/2 V\big)^{1/2}$ and $C = \big(9\,\Oms\!/2 V\big)^{1/4} \eep$ (with $\Oms \!>\! 0$),
in agreement with Eqs.\,\eqref{dsk}\eqref{w13}\eqref{V1V2}\eqref{Aexp}\eqref{Cexp}.}})
\begin{equation}
\tc(t)/\te = \left\{\!\begin{array}{ll}
\dd{{4 \over 3}\! \left({\eep^2 \over \om}\right)^{\!\!1/4} t^{3/4} + O\big(t^{5/4}\big)}
& \dd{\mbox{for $t \to 0^+$}\,,} \vspace{0.15cm}\\
\dd{{2^{7/4}\,\eep^{1/2} \over \om\!}\; e^{{1 \over 4}\,\om\,t} + O(1)}
& \dd{\mbox{for $t \to +\infty$} ~.}
\end{array}\right. \label{tcasy}
\end{equation}
The behavior of $a(t)$, $\wF(t)$, $\Omm(t)$ for small and large $t$
is obtained by direct inspection of Eqs.\,\eqref{aexp} \eqref{wfiexp2}\eqref{Ommexp2}
which give, respectively:
\begin{equation}
a(t) = \left\{\!\!\begin{array}{ll}
\dd{\sqrt{\eep} \left({4\,\Oms \over \om}\right)^{\!\!1/4} t^{1/4} + O\big(t^{3/4}\big)}
& \dd{\mbox{for $t \to 0^+$}\,,} \vspace{0.15cm}\\
\dd{\sqrt{\eep} \left({\Oms \over 2\, \om^2}\right)^{\!\!1/4} e^{{3 \over 4}\,\om\,t} + O\!\left(e^{{1 \over 4}\,\om\,t}\right)}
& \dd{\mbox{for $t \to +\infty$} ~;}
\end{array}\right. \label{aasyt}
\end{equation}
\beq
\wF(t) = \left\{\!\!\begin{array}{ll}
\dd{1 - 16\, \om^2\, t^2 + O(t^{5/2})}
& \dd{\mbox{for $t \to 0^+$}\,,} \vspace{0.1cm}\\
\dd{-\,{7 \over 9} + O\big(e^{-{1 \over 2}\,\om\,t}\big)}
& \dd{\mbox{for $t \to +\infty$}\,;} \\
\end{array}
\right. \label{wfiexp2asy}
\feq
\beq
\Omm(t) = \left\{\!\!\begin{array}{ll}
\dd{{4\, \sqrt{\om} \over \eep}\;t^{1/2} + O(t)}
& \dd{\mbox{for $t \to 0^+$}\,,} \vspace{0.2cm} \\
\dd{{32 \over 9\sqrt{2}\;\eep}\;e^{-\,{5 \over 2}\,\om\,t}+ O\big(e^{-3\,\om\,t}\big)}
& \dd{\mbox{for $t \to +\infty$}\,.}
\end{array}
\right.  \label{Ommexp2asy}
\feq
For our purposes it is also important to consider the behavior of the above observables when
$t$ ranges in a compact interval, and $\eep$ is small.
Indeed, let us fix any two times $0 < t_1 < t_2$; then, from
Eqs.\,\eqref{tcInt}\eqref{aexp}\eqref{wfiexp2}\eqref{Ommexp2}
it readily follows that
\beq\begin{array}{c}
\dd{\tc(t)/\te \,=\, t + O(\eep)~, \qquad\quad
a(t) = \left({4\,\Oms \over \om^2}\right)^{\!\!1/4}\! \sqrt{\sinh(\om\, t)}\, + O(\eep)~,}\vspace{0.1cm}\\
\dd{\wF(t) = -1 + O(\eep^2)~, \qquad\quad \Omm(t) = {1 \over \cosh^2(\om\,t)} + O(\eep)~,}\vspace{0.2cm}\\
\dd{\mbox{for $\eep \to 0^{+}$,\; uniformly in $t \in [t_1,t_2]$}.} \label{tccomp}
\end{array}\feq
Let us rephrase the previous results viewing the above observables as functions of
cosmic time. From Eqs.\,\eqref{tcasy} and \eqref{aasyt}-\eqref{Ommexp2asy} we deduce
\begin{equation}
a(\tc) = \left\{\!\!\begin{array}{ll}
\dd{\left({3\,\eep \over \sqrt{2\,\om}}\right)^{\!\!1/3} \big(\Oms\big)^{1/4}\; (\tc/\theta)^{1/3} +\, O(\tc/\te)}
& \dd{\mbox{for $\tc/\te \to 0^+$}\,,} \vspace{0.15cm}\\
\dd{{1 \over \eep}\, \sqrt{\om^{5} \over 2048}\; \big(\Oms\big)^{1/4}\; (\tc/\te)^{3} + O\big((\tc/\te)^2\big)}
& \dd{\mbox{for $\tc/\te \to +\infty$} ~;}
\end{array}\right. \label{aasytc}
\end{equation}
\begin{gather}
\wF(\tc) = \left\{\!\!\begin{array}{ll}
\dd{1 - \left({9\,\om^2 \over 2\,\eep}\right)^{\!\!4/3}\! (\tc/\te)^{8/3} + O\big( (\tc/\te)^{10/3}\big)}
& \dd{\mbox{for $\tc/\te \to 0^+$}\,,} \vspace{0.1cm}\\
\dd{-\,{7 \over 9} + O\big((\tc/\te)^{-2}\big)}
& \dd{\mbox{for $\tc/\te \to +\infty$}\,;} \\
\end{array}
\right. \label{wfiexp2asytc}
\end{gather}
\vfill\eject\noindent
\begin{gather}
\Omm(\tc) = \left\{\!\!\begin{array}{ll}
\dd{\left({6\,\om \over \eep^2}\right)^{\!\!2/3} (\tc/\te)^{2/3} + O\big((\tc/\te)^{4/3}\big)}
& \dd{\mbox{for $\tc/\te \to 0^+$}\,,} \vspace{0.2cm} \\
\dd{{4\,\eep^{3/2} \over 9\,(\om/4)^{10}}\; (\tc/\te)^{-\,10} + O\big((\tc/\te)^{-11} \big)}
& \dd{\mbox{for $\tc/\te \to +\infty$}\,.}
\end{array}
\right.  \label{Ommexp2asytc}
\end{gather}
Eq.\,\eqref{tccomp} implies the following, for any pair $0 < \tc_1 < \tc_2$ of cosmic time instants:
\begin{gather}
a(\tc) = \left({4\,\Oms \over \om^2}\right)^{\!\!1/4}\!\! \sqrt{\sinh(\om\, \tc/\te)} + O(\eep)\,, \nonumber \\
\wF(\tc) = -1 + O(\eep^2)\,, \quad \Omm(\tc) = {1 \over \cosh^2(\om\,\tc/\te)} + O(\eep)\,, \label{Belexptc} \\
\mbox{for $\eep \to 0^{+}$,\; uniformly in $\tc \in [\tc_1,\tc_2]$}. \nonumber
\end{gather}
Let us briefly comment the above results. According to Eq.\,\eqref{Belexptc}, on each
compact interval $[\tc_1, \tc_2]$, \textsl{with $\tc_1 > 0$ so as to ensure
a strict separation from the Big Bang}, for $\eep$ sufficiently small
the scale factor $a(\tc)$ grows exponentially
and $\wF(\tc)$ is close to $-1$, indicating that the field plays the role
a cosmological constant. As a consequence, the behavior of the system for $\tc \in [\tc_1, \tc_2]$ and small
$\eep$ is similar to that described in the previous paragraph for $\eep=0$ and all $\tc \in (0,+\infty)$.
The situation is completely different if we approach the Big Bang, or if we consider the very far
future; for example, Eq.\,\eqref{aasytc} shows that $a(\tc)$ has a power law dependence
on $\tc/\te$ with exponents $1/3$ and $3$, respectively, for $\tc/\te \to 0^{+}$
and $\tc/\te \to + \infty$. \parn
The presence of an epoch of exponential growth for $a(\tc)$,
preceded and followed by periods with slower expansion rates, is typical of inflationary models.
In the sequel we will show that one can adjust
the parameters of the system so as to obtain a rather realistic
model for inflation, even from a quantitative point of view.
\vspace{-0.4cm}

\paragraph{An interlude on the quantitative determination of cosmic time.}
Let us recall that $\tc(t)/\te$ is expressed via Eq.\,\eqref{tcInt}
as a nontrivial integral over the interval $(0, t]$.
The numerical computation of this integral
(for specified values of all parameters) is problematic,
especially in the situation of greatest interest for us.
In fact, for $t' \to 0^{+}$ the integrand function in Eq.\,\eqref{tcInt}
behaves like $\sqrt{\eep}\, (\om t')^{-1/4}$, the product
of the divergent factor $(\om t')^{-1/4}$ by the parameter
$\sqrt{\eep}$. To make things worse,
in the sequel we are mostly interested in a
case where $\eep$ is very small. \parn
Fortunately, the problem that we have just outlined can be overcome.
In fact, starting from the integral representation
\eqref{tcInt} it is possible to determine analytically
two elementary functions $T^{\pm}_{\eep}$ such that
$T^{-}_{\eep}(t) \leqs \tc(t)/\te \leqs T^{+}_{\eep}(t)$
for arbitrary $\eep, t > 0$; we refer to Appendix
\ref{appEsttc} for a detailed description
of such functions. The same Appendix shows that,
in the application with small $\eep$ considered
in next paragraph,
$T^{+}_{\eep}(t)$ and $T^{-}_{\eep}(t)$
are very close for
all the considered values of $t$, so that the
mean $(1/2) (T^{-}_{\eep}\! + T^{+}_{\eep})(t)$
is a very accurate approximant for $\tc(t)/\te$.
In the calculations mentioned in the next paragraph,
$\tc(t)/\te$ has always been approximated with
the previous mean.
\vspace{-0.4cm}

\paragraph{A plausible scenario with inflation.}
Let us now present a reasonable choice of the parameters not yet specified
for the model under analysis, which can in fact lead to a physically plausible inflationary scenario.
The key idea that we are going to pursue is that the scale factor
grows exponentially in a compact interval of cosmic time, at least for very small values of $\eep$
(a fact made evident by the asymptotic expansion written in Eq.\,\eqref{aexp}).
\parn
Inspired by standard arguments (see, e.g., \cite[Sec.\,11.4]{Ryd}),
we set the time parameter $\te$ equal to the alleged time of Grand Unified Theory (GUT), namely,
\beq
\te = 10^{-36}\, \sec ~,
\feq
and presume that the universe undergoes an inflationary expansion at least during the interval
of cosmic time approximately comprised between $\tc \simeq \te$ and $\tc \simeq N\,\te$, for some
given $N$ large enough. In the sequel we will refer to the case where
\beq
N = 100 ~;
\feq
analogous results could be derived for other values of $N$, with the same order of magnitude.
Correspondingly, let us assume that the dimensionless parameter $\eep$ introduced in
Eq.\,\eqref{Cexp} is exponentially small with respect to $N$; more precisely, we put
\beq
\eep \,=\, e^{-N} \,\simeq\, 3.72008\,... \times 10^{-44} ~.
\feq
On the contrary, let $\Oms$ and $V$ be independent of $N$ and comparable to unity, so that
the same holds true for $\om$ (due to Eq.\,\eqref{omVexp}); as an example, let us fix
\beq
\Oms = 0.308 ~, \qquad V = 1 ~, \qquad \om = {2\, \sqrt{2\,V} \over 3} \simeq 0.9428\,...~.
\feq
\begin{figure}[t!]
\vspace{-0.6cm}
    \centering
        \begin{subfigure}[b]{0.475\textwidth}\label{fig:8}
                \includegraphics[width=\textwidth]{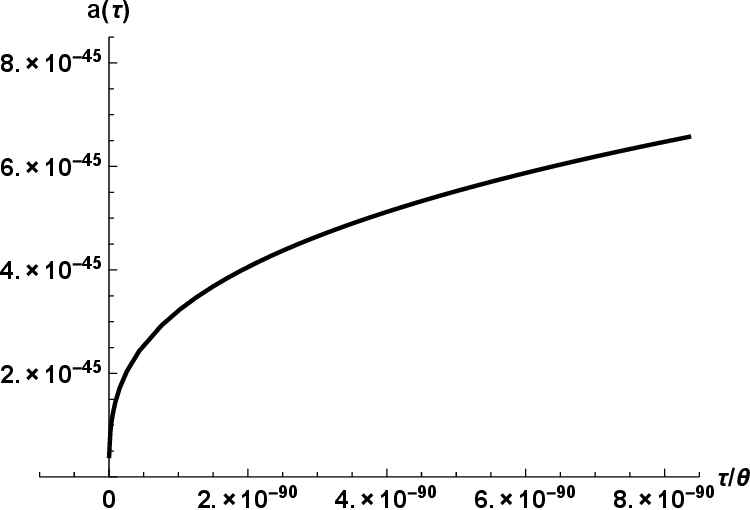}\vspace{-0.cm}
                \caption*{\textsc{Figure\! 8.} $a(\tc)$ as a function of $\tc/\te$.}
        \end{subfigure}
        \hspace{0.3cm}
        \begin{subfigure}[b]{0.475\textwidth}\vspace{-0.2cm}
                \includegraphics[width=\textwidth]{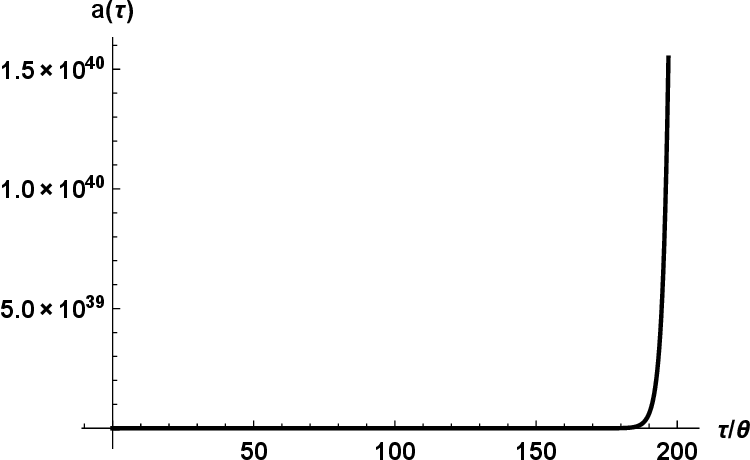}\label{fig:9}\vspace{0.2cm}
                \caption*{\textsc{Figure\! 9.} $a(\tc)$ as a function of $\tc/\te$.}
        \end{subfigure}
\end{figure}
\begin{figure}[t!]
\vspace{-0.4cm}
    \centering
        \begin{subfigure}[b]{0.475\textwidth}\label{fig:10}
                $\phantom{a}$\vspace{-1.3cm}\\
                \includegraphics[width=\textwidth]{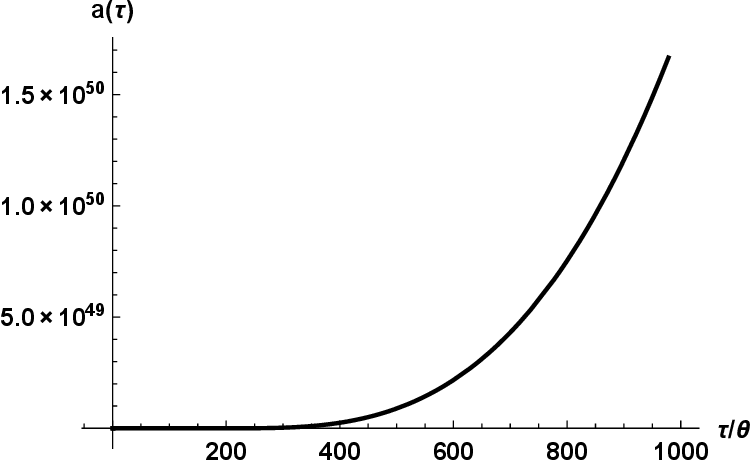}\vspace{0.1cm}
                \caption*{\textsc{Figure\! 10.} $a(\tc)$ as a function of $\tc/\te$.}
        \end{subfigure}
        \hspace{0.3cm}
        \begin{subfigure}[b]{0.475\textwidth}
                $\phantom{a}$\vspace{-0.cm}\\
                \includegraphics[width=\textwidth]{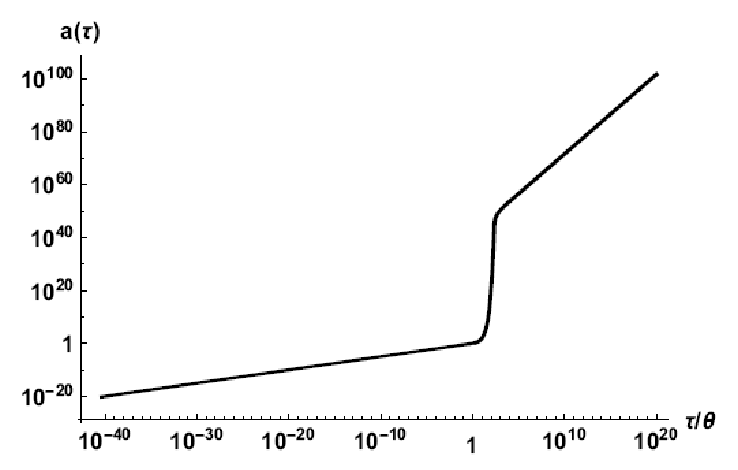}\label{fig:11}
                $\phantom{a}$\vspace{-1.0cm}\\
                \caption*{\textsc{Figure\! 11.} $a(\tc)$ as a function of $\tc/\te$. Logarithmic scales are used on both axes.}
                \vspace{-0.cm}
        \end{subfigure}
\end{figure}
\noindent \vspace{-0.2cm}\\
Having fixed all the involved parameters, for any given $t$
we can calculate the numerical values of the quantities
$\tc(t)/\te$, $a(t)$, $\wF(t)$, $\Omm(t)$. For $\tc(t)/\te$
we use, in place of the integral representation \eqref{tcInt},
the very accurate approximation method mentioned in
the previous paragraph and described in Appendix
\ref{appEsttc}; for $a(t)$, $\wF(t)$ and $\Omm(t)$
we use Eqs.\,\eqref{aexp}\eqref{wfiexp2}\eqref{Ommexp2}. \parn
Figs.\,8-11 give $a(\tc)$ as a function of the dimensionless quantity $\tc/\te$,
for different ranges of the latter; these figures have been obtained
drawing the curve $t \!\mapsto\! \big(\tc(t)/\te,a(\tc(t))\big)$
for $t$ within different intervals
(namely, for $t \in (0,10^{-90})$, $t \in (0,200)$, $t \in (0,240)$ and $t \in (0,400)$).
Logarithmic scales are used for both $\tc/\te$ and $a(\tc)$ in Fig.\,11;
this makes evident that there is a comparatively short interval of cosmic time where
the scale factor increases abruptly. Just to give an idea of the orders of magnitude
involved in these arguments, let us mention that
\begin{equation}
\tc_1 = 50\, \te\,, ~~\tc_2= 150\, \te
\qquad \Rightarrow \qquad
{a(\tc_2) \over a(\tc_1)} = 2.97056\,... \times 10^{20}\,.
\end{equation}
\begin{figure}[t!]
\vspace{-0.7cm}
    \centering
        \begin{subfigure}[b]{0.475\textwidth}\label{fig:12}
                \includegraphics[width=\textwidth]{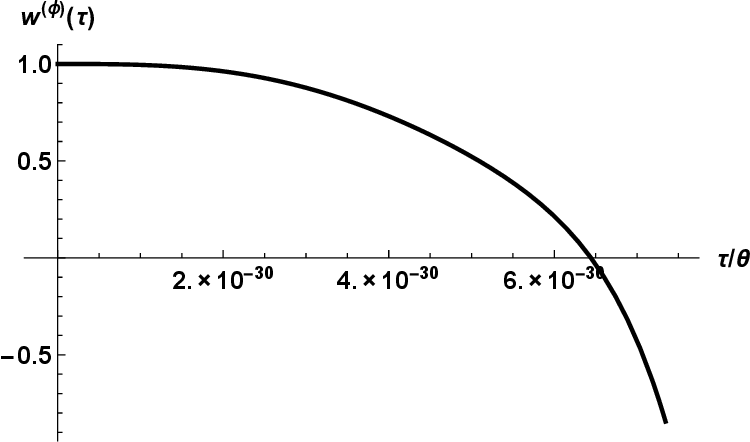}\vspace{-0.cm}
                \caption*{\textsc{Figure\! 12.} $\wF(\tc)$ as a function of $\tc/\te$.}
        \end{subfigure}
        \hspace{0.3cm}
        \begin{subfigure}[b]{0.475\textwidth}
                \includegraphics[width=\textwidth]{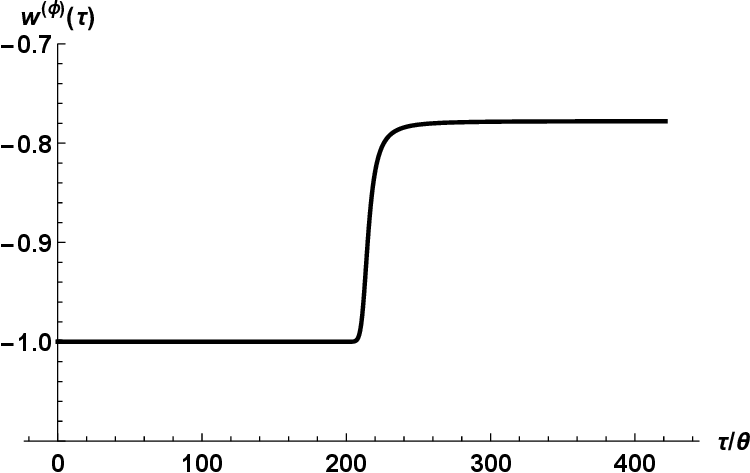}\label{fig:13}
                \caption*{\textsc{Figure\! 13.} $\wF(\tc)$ as a function of $\tc/\te$.}
        \end{subfigure}
\vspace{-0.15cm}
\end{figure}
\noindent \vspace{-0.3cm}\\
Fig.\,12 and Fig.\,13 refer to the equation of state parameter for the field;
more precisely they give $\wF(\tc)$ as a function of $\tc/\te$, for the latter variable
ranging in two different intervals. Again, these graphs have been obtained
as parametric plots  (the curve $t \mapsto \big(\tc(t)/\te,\wF(\tc(t))\big)$
has been plotted for $t \in (0,10^{-29})$ and $t \in (0,230)$, respectively). In particular, Fig.\,12 suggests
$\wF(\tc) \to 1^{-}$ for $\tc/\te \to 0^{+}$, in agreement with Eq.\,\eqref{wfiexp2asytc}.
On the other hand, Fig.\,13 exhibits a sharp transition from $\wF = -1$ to $\wF = -7/9 \simeq -0.777,...$\! ;
this behavior corresponds to the general features previously pointed out in Eqs.\,\eqref{tccomp}\eqref{wfiexp2asytc}.
\begin{figure}[t!]
    \centering
        \begin{subfigure}[b]{0.475\textwidth}\label{fig:14}
                \includegraphics[width=\textwidth]{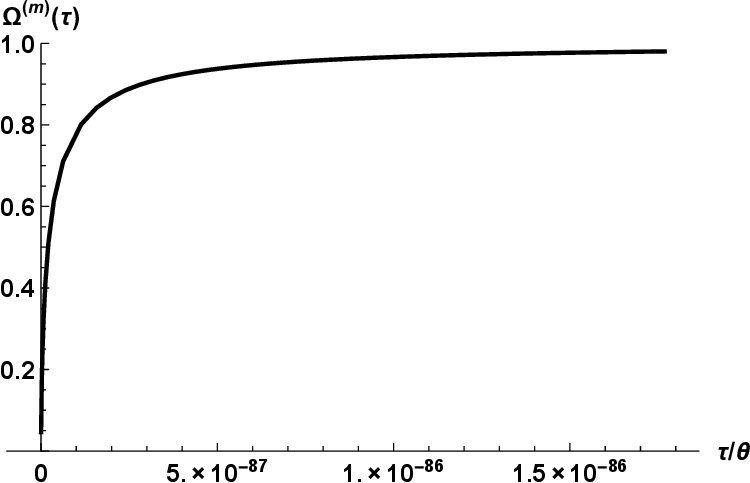}\vspace{-0.cm}
                \caption*{\textsc{Figure\! 14.} $\Omm(\tc)$ as a function of $\tc/\te$.}
        \end{subfigure}
        \hspace{0.3cm}
        \begin{subfigure}[b]{0.475\textwidth}
                \includegraphics[width=\textwidth]{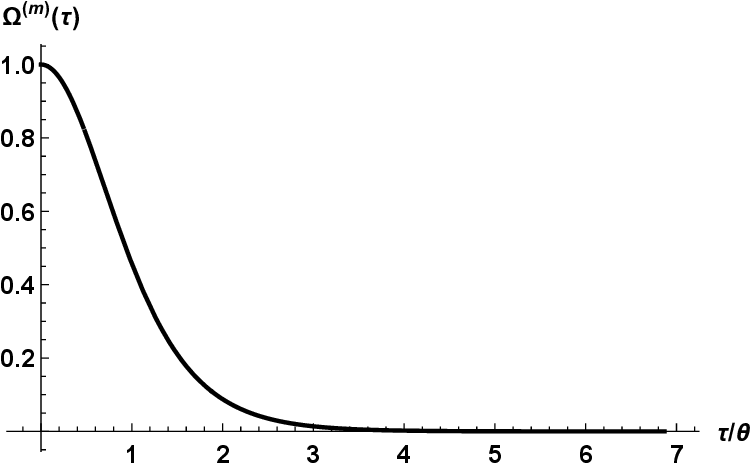}\label{fig:15}
                \caption*{\textsc{Figure\! 15.}
                $\Omm(\tc)$ as a function of $\tc/\te$.}
        \end{subfigure}
\vspace{-0.2cm}
\end{figure}
\noindent \vspace{-0.cm}\\
Finally, let us consider the density parameter $\Omm$ for the radiation content
of the universe. Fig.\,14 and Fig.\,15 represent $\Omm(\tc)$ as a function of $\tc/\te$
(and have been obtained plotting the curve
$t \mapsto \big(\tc(t)/\te,\Omm(\tc(t))\big)$ for $t \in (0,10^{-86})$ and $t \in (0,7)$, respectively).
Fig.\,14 makes evident that $\Omm(\tc) \to 0^{+}$ for $\tc/\te \to 0^{+}$
(see Eq.\,\eqref{Ommexp2asytc} for the leading order in the corresponding asymptotic expansion).
Fig.\,15 shows instead that $\Omm(\tc) \simeq 1$ for small (though, not too small) values of $\tc/\te$
and $\Omm(\tc)$ rapidly vanishes for larger values of the cosmic time, in accordance with the general
features mentioned in Eqs.\,\eqref{tccomp}\eqref{wfiexp2asytc}.
\vspace{-0.2cm}

\subsection{Solutions for class 7 potentials with a matter fluid. The nonlinear repulsor/oscillator model}\label{subclass7}
\vspace{-0.1cm}
In this section, we refer to the integrable subcase (i) in the analysis of
class 7 potentials (see page \pageref{L7i}), with an additional prescription: the exponent $2/\ga -2$
in the potential $\mathscr{V}$ of Eq.\,\eqref{VCase7} is required to be an even integer $2 \ell \geqs 4$.
Here are some motivations for this additional requirement: \parn
i) The map $\xi \mapsto \xi^{2/\ga - 2}$, appearing in Eqs.\,\eqref{VCase7}\eqref{LCase7}\eqref{ECase7},
is well defined, smooth and bounded from below for $\xi$ ranging throughout the whole
real axis if and only if $2/\ga - 2 = 2 \ell \in \{0,2,4,...\}$. \parn
ii) For $2/\ga - 2 = 0$ (i.e., $\ga\!=\!1$), the potential $\VV$ of Eq.\,\eqref{VCase7}
is constant. For $2/\ga - 2=2$ (i.e. $\ga \!= \!1/2$),
Eq.\,\eqref{VCase7} gives $\VV(\f) = (1/2)(V_1 - V_2) + (1/2) (V_1 + V_2) \cosh \f$;
thus, $\VV$ becomes a class 1 potential (cf. Eq.\,\eqref{vclass1}),
to be treated with the simpler methods already described for that class.
Summing up, the cases $2/\ga - 2 = 0, 2$ can be regarded as trivial; excluding them
from the considerations in the previous item (i),
we are left with the condition $2/\ga - 2 = 2 \ell \in \{4,6,8,...\}$.
\vfill\eject\noindent
$\phantom{a}$\vspace{-1.cm}\\
To proceed, let us recall that the subcase (i) of page \pageref{L7i} requires
$k=0$, $\ga = (1-w)/2$; on top of that, we assume $V_1$ and $V_2$ to be positive.
Thus, the complete list of our choices is the following:
\beq V_1 > 0~, \quad~ V_2 > 0~, \quad~ k = 0~, \quad~
\ga = {1 \over \ell + 1}~, \quad~ w = {\ell - 1 \over \ell + 1}~, \qquad \ell\in \{2,3,4,...\}~.
\label{lischoi} \feq
In particular, note that $1/3 \leqs w < 1$ for all $\ell \in \{2,3,4,...\}$
and $w=1/3$ if and only if $\ell = 2$. \parn
With the above choices \eqref{lischoi},
Eqs.\,\eqref{VCase7}\eqref{gau7} and \eqref{cor7bis} for the potential $\VV(\f)$,
the gauge function $\mathscr{B} \equiv \BB$
and the coordinate change $(x,y) \mapsto (\AA, \f)$ respectively reduce to:
\begin{gather}
\VV(\f) \,=\, V \!\left[\left(\cosh\!\Big({\f \over \ell\!+\!1}\Big)\!\right)^{\!\!2\ell}
+ \left(\sinh\!\Big({\f \over \ell\!+\!1}\Big)\!\right)^{\!\!2\ell}\right], \qquad
\f \in I_{\ga, V_2} \equiv (-\infty,+\infty) ~; \label{VVCase7} \\
\BB = {\ell - 1 \over 2}\,\log\big(x^2\! - y^2\big) ~; \label{ggau7}\\
\AA = {\ell + 1 \over 2}\,\log\big(x^2\! - y^2\big) \,, \qquad
\f = {\ell + 1 \over 2}\,\log\!\left({x+y \over x-y}\right) ~ ; \label{ccor7}\\
 (x,y) \,\in\, \mathcal{D}_{\gamma,V_2} \equiv \mathcal{D} := \big\{ (x,y) \in \reali^2 ~\big|~ x > 0\,,~ - x < y < x \big\}~. \label{ccor7bis}
\end{gather}
Note that, for $\AA$ as above one has $\AA \!\to -\infty$ when $x^2 \!- y^2 \to 0^{+}$.
Correspondingly, noting that the scale factor for the cosmology under analysis
is given by (see Eq.\,\eqref{aeaa})
\beq a = \big(x^2 - y^2\big)^{{\ell + 1 \over 2 \ds}}~, \label{scalex7} \feq
we have $a \!\to\! 0^{+}$ for $x^2 - y^2 \!\to\! 0^{+}$; this fact is especially
relevant for the presence of a Big Bang.
To say more, assuming that a Big Bang does actually occur at $t = 0$ in agreement with Eq.\,\eqref{BigBangt0}
({\footnote{We will show in the sequel that this condition can in fact be attained.}})
and adding the conventional prescription $\tc(t) \to 0^{+}$ for $t \to 0^{+}$, from
Eqs.\,\eqref{tauti}\eqref{ggau7} we derive the following expression for the cosmic time coordinate:
\beq
\tc(t)/\theta \,= \int_{0}^t d t'\, \big(x^2(t') - y^2(t')\big)^{{\ell - 1 \over 2}} ~. \label{cosmicx7}
\feq
Next let us recall that, in the subcase (i) for the potentials of class 7,
the Lagrangian \eqref{LCase7} and the corresponding energy \eqref{ECase7}
take the separable forms \eqref{L7i}\eqref{E7i}.
With the prescriptions stated in Eq.\,\eqref{lischoi},
the cited Eqs.\,\eqref{L7i}\eqref{E7i} give:
\begin{equation}
\begin{array}{c}
\dd{\LL(x,y,\xt,\yt) = \LL_1(x,\xt) + \LL_2(y,\yt) - {\ds^2\,\Oms \over 2} ~,}\vspace{0.1cm}\\
\dd{\LL_1(x,\xt) = - \,{(\ell + 1)^2 \over 2} \,\xt^2 - V_1 \, x^{2 \ell}~, \qquad
\LL_2(y,\yt) = {(\ell + 1)^2 \over 2} \,\yt^2 - V_2 \, y^{2 \ell}~;} \label{L7j}
\end{array}
\end{equation}
\begin{equation}
\begin{array}{c}
\dd{\Ec(x,y,\xt,\yt) = \Ec_1(x,\xt) + \Ec_2(y,\yt) + {\ds^2\,\Oms \over 2}~,} \vspace{0.1cm} \\
\dd{\Ec_1(x,\xt) = - \,{(\ell + 1)^2 \over 2} \,\xt^2 + V_1 \, x^{2 \ell}~, \qquad
\Ec_2(y,\yt) = {(\ell + 1)^2 \over 2} \,\yt^2 + V_2 \, y^{2 \ell}~ .} \label{E7j}
\end{array}
\end{equation}
In the sequel we discuss the solutions of the Lagrange equations corresponding to
the above Lagrangians $\LL_1, \LL_2$, providing explicit quadrature formulas
for them and discussing their asymptotic behaviors in different regimes of interest
for the applications to be discussed in the sequel.
\parn
Before proceeding with this analysis, let us express in terms of the coordinates $x,y$
two other relevant observables: namely, the coefficient $\wF$ in the equation of state for the field
and the dimensionless density parameter for matter $\Omm$.
The general relations \eqref{defwfi}\eqref{Omegadef},
combined with Eqs.\,\eqref{lischoi}\eqref{ggau7}\eqref{ccor7}
for the case under analysis, yield the following results:
\begin{gather}
\wF = {(\ell + 1)^2\, (x\,\dot{y} - \dot{x}\, y)^2 - 2\,(x^2-y^2)(V_1\,x^{2\ell} + V_2\,y^{2\ell})
\over (\ell + 1)^2\, (x\, \dot{y} - \dot{x}\, y)^2 + 2\,(x^2-y^2)(V_1\,x^{2\ell} + V_2\,y^{2\ell})}~, \label{wfx7} \\
\Omm = {\ds^2\, \Oms\, (x^2 - y^2) \over (\ell + 1)^2\, (x\, \xt - y\, \yt)^2} ~. \label{Omx7}
\end{gather}
\vfill\eject\noindent
$\phantom{a}$\vspace{-1.5cm}\\

\subsubsection{Constants of motion and quadrature formulas} \label{subsubquad}
From here to the end of the present subsection \ref{subclass7},
we stick to the configuration described by Eq.\,\eqref{lischoi}
and refer to Eqs.\,\eqref{L7j}\eqref{E7j} for the Lagrangian
$\LL$ and the corresponding energy $\Ec$.
\parn
The Lagrangian system described by $\LL$ can be analyzed in terms
of the separate $1$-dimensional subsystems with Lagrangians $\LL_1, \LL_2$,
whose energies $\Ec_1, \Ec_2$ are constants of motion.
Of course, the Lagrangians $\LL_1, \LL_2$ (as well as the energies $\Ec_1, \Ec_2$)
are well defined and smooth for any real $x$ and $y$.
Keeping this in mind, in the sequel we shall first study separately the
subsystems with Lagrangians $\LL_1, \LL_2$ assuming $x, y \!\in\! (-\infty,+\infty)$,
and reserve to a second step the implementation of the condition $(x,y) \!\in\! \mathcal{D}_{\gamma,V_2}$
(see Eq.\,\eqref{ccor7bis}).
\parn
From the expression \eqref{E7j} for the total energy $\Ec$, we see that the constraint $\Ec=0$
is fulfilled if and only if $\Ec_1$ and $\Ec_2$ are expressed as follows:\vspace{-0.1cm}
\beq \Ec_1 = - \,\EE~, \qquad \Ec_2 = \FF~, \qquad \EE := \FF + {\ds^2\,\Oms \over 2}~. \label{ee} \feq
The expression for $\Ec_2$ in Eq.\,\eqref{E7j} makes evident
that $\FF \geqs 0$, and that $\FF=0$ only along motions with $y(t)=0$ and $\dot{y}(t)=0$
for all $t$; in the sequel we exclude such motions, fixing
\beq \FF \in (0,+\infty)~. \label{ee2} \feq
Since we are assuming $\Oms > 0$ (see Eq.\,\eqref{Omspos}), the above condition
also grants that $\EE \in (0,+\infty)$.
\parn
Let us now consider two motions $t \mapsto x(t)$ and $t \mapsto y(t)$ fulfilling the Lagrange
equations, with energies fixed according to the above prescriptions (and $t$ ranging within suitable intervals).
Then, from Eqs.\,\eqref{E7j}\eqref{ee} we infer
\beq {(\ell + 1)^2 \over 2} \;\xt^2 - V_1 \, x^{2 \ell} = \EE~, \qquad
{(\ell + 1)^2 \over 2} \;\yt^2 + V_2 \, y^{2 \ell} = \FF~. \label{exy} \feq
The above equations can be interpreted as the conservation laws for the energies of two
fictitious mechanical systems with kinetic energies ${(\ell + 1)^2 \over 2} \,\xt^2$,
${(\ell + 1)^2 \over 2} \,\yt^2$ and potential energies $- V_1 \, x^{2 \ell}$, $V_2\, y^{2 \ell}$,
that we can denominate, respectively, a \textsl{nonlinear repulsor} and a
\textsl{nonlinear oscillator}. \parn
From the second equality in Eq.\,\eqref{exy} we see that $t \mapsto y(t)$ is an oscillatory motion
such that
\beq y(t) \in \left[- \big(\FF/V_2\big)^{1/(2 \ell)},\big(\FF/V_2\big)^{1/(2 \ell)} \right]
\qquad \mbox{for all $t$}~\label{motiv} \feq
(the above interval is the set $\{ y \!\in\! \reali~|~V_2\, y^{2 \ell} \!\leqs\! \FF \}$
and the times $t$ such that $y(t) = \pm \big(\FF/V_2\big)^{1/(2 \ell)}$ are
inversion times for the motion). Since $V_1 \,x^{2 \ell} \!\geqs\! 0$ and $ \EE \!>\! 0$,
from the first equality in Eq.\,\eqref{exy} we infer $\xt^2(t) \!>\! 0$, i.e. $\xt(t) \neq 0$
for all $t$. Thus $\xt(t)$ has a constant sign, and the function $t \mapsto x(t)$ is strictly monotonic.
\parn
From Eq.\,\eqref{exy} we also infer quadrature formulas containing the
hypergeometric-type function
\beq F_{\ell}(z) \,:=\, {}_{2} F_{1}\!\left( {1 \over 2}\,,\, {1 \over 2 \ell}\,,\, {1 \over 2 \ell} + 1\,,\, z \right) .
\label{defell} \feq
More precisely, we have the following implications
({\footnote{Let us give a few details on the derivation of Eq.\,\eqref{quad1}.
To express the integral in Eq.\,\eqref{quad1}
in terms of $F_\ell$ write
$\int_{x(t_1)}^{x(t_2)} d x/\sqrt{\EE \!+\! V_1 x^{2 \ell}} \!= {1 \over \sqrt{\EE}}
\left(\int_{0}^{x(t_2)} \!-\! \int_{0}^{x(t_1)}\right) d x/\sqrt{1 \!+\! (V_1/\EE) x^{2 \ell}}$
and then use for $i=1,2$ the following relations, based on the change
of variable $x = x(t_i)\, s^{1/(2 \ell)}$ with $s \!\in\! [0,1]$:
$\int_{0}^{x(t_i)} d x/\sqrt{1 \!+\! (V_1/\EE)\, x^{2 \ell}}$
$= {x(t_i) \over 2 \ell}\!  \int_{0}^{1} d s \;
s^{{1 \over 2 \ell} - 1}/\sqrt{1 \!+\! (V_1/\EE)\, x^{2 \ell}(t_i)\, s}$
$ = x(t_i)\, F_{\ell}\big( - (V_1/\EE)\, x^{2 \ell}(t_i)\big)$.
One proceeds similarly to express in terms of $F_{\ell}$ the integral in Eq.\,\eqref{quad2}.\vspace{-0.4cm}}}):
\begin{gather}
\mbox{$x(t)$ is well defined for $t \in [t_1, t_2]$ \,and\, $\sgn \, \dot x(t) = \xi \in \{\pm 1\}$
for all $t \in (t_1, t_2)$} \label{quad1} \\
\Rightarrow\; \xi\, (t_2 - t_1)
= {\ell + 1 \over \sqrt{2\, \EE}} \left[ x(t_2)\, F_{\ell}\! \left(- (V_1/\EE)\,x^{2 \ell}(t_2) \right) -
x(t_1)\, F_{\ell}\! \left(- (V_1 /\EE)\, x^{2 \ell}(t_1) \right) \right] ; \nonumber\\
\mbox{$y(t)$ is well defined for $t \in [t_1, t_2]$ \,and\, $\sgn \, \dot y(t) = \sigma \in \{\pm 1\}$
for all $t \in (t_1, t_2)$} \label{quad2} \\
\Rightarrow\; \sigma\, (t_2 - t_1)
= {\ell + 1 \over  \sqrt{2 \, \FF} } \left[ y(t_2)\, F_{\ell}\! \left( (V_2/\FF)\, y^{2 \ell}(t_2) \right) -
y(t_1)\, F_{\ell}\! \left( (V_2/\FF)\, y^{2 \ell}(t_1) \right) \right] . \nonumber
\end{gather}
\vfill\eject\noindent
$\phantom{a}$\vspace{-1.2cm}\\
For future use, let us mention the asymptotic expansion
({\footnote{This expansion follows from the definition \eqref{defell} of $F_{\ell}$
and from some known identities about hypergeometric functions, namely:
a Kummer transformation (see, e.g., \cite[Eq.\,15.8.2]{NIST}) and the elementary relations
${}_{2} F_{1}\left(a, 0, c, \zeta \right)=1$ for all $\zeta$, ${}_{2} F_{1}\left(a, b, c, \zeta \right)=1 + O(\zeta)$
for $\zeta \to 0$.}})
\begin{gather}
F_{\ell}(-z) = {C_\ell \over z^{1/(2 \ell)}} - {1 \over (\ell-1) \sqrt{z}} + O\!\left({1 \over z^{3/2}}\right)
~~\mbox{for $z \to + \infty$}~, \label{efas} \\
C_\ell := {1 \over \sqrt{\pi}} \, \Gamma\!\left( {\ell - 1 \over 2 \ell}\right)
\Gamma\!\left({2\ell+1 \over 2 \ell}\right); \nonumber
\end{gather}
here the dominating term is $C_{\ell}/z^{1/(2 \ell)}$, since $0 \!<\! 1/(2 \ell) \!\leqs\! 1/4$.
Again for future use, let us also mention the special value
({\footnote{Eq.\,\eqref{effe1} follows from the definition \eqref{defell} of $F_{\ell}$ and
from a general result about ${}_{2} F_{1}\left(a, b, c, 1 \right)$
(see, e.g., \cite[Eq.\,15.4.20]{NIST}).\vspace{-0.5cm}}})
\beq F_{\ell}(1) = {\sqrt{\pi}~\Gamma\!\left({2\ell + 1 \over 2 \ell}\right) \over
\Gamma\!\left({\ell + 1 \over 2 \ell}\right)}~. \label{effe1} \vspace{0.2cm}\feq

\subsubsection{Choosing the initial data}\label{subdata}
We now fix the attention on the solutions $t \mapsto x(t)$ and $t \mapsto y(t)$
of the Lagrange equations for $\LL_1, \LL_2$ with energies as in Eqs.\,\eqref{ee}\eqref{ee2}
and with the following initial data, specified at time $t=0$ by convention:
\beq x(0) = y(0) = Y \in \left(0, (\FF/V_2)^{1/(2 \ell)}\right), \qquad \xt(0) = u > 0~,
\quad \yt(0) = v > 0~.
\label{idata} \feq
The upper bound $(\FF/V_2)^{1/(2 \ell)}$ prescribed here for $Y$ is motivated
by Eq.\,\eqref{motiv}. The equality $x(0)=y(0)$ will be employed in the sequel to infer
that the scale factor $a(t)$ vanishes for $t \to 0^{+}$
(see the considerations after Eqs.\,\eqref{ggau7}-\eqref{ccor7bis}), a fact
related to the occurrence of a Big Bang. Let us also
point out that Eq.\,\eqref{exy} for the energies (here employed at time $t = 0$)
and the assumptions in Eq.\,\eqref{idata} give
\beq u  =  {\sqrt{2} \over \ell + 1}\, \sqrt{\EE + V_1\, Y^{2 \ell}}~,\qquad
v =  {\sqrt{2} \over \ell + 1 }\, \sqrt{\FF - V_2\, Y^{2 \ell}}~. \label{uandv} \feq
Each one of the solutions $t \mapsto x(t)$ and $t \mapsto y(t)$ is intended
to be defined on the maximal admissible domain, that is on the largest interval
containing $t = 0$ on which the solution is well defined. \parn
The discussion of subsection \ref{subsubquad}, combined with the present assumptions,
ensures that $t \mapsto x(t)$ is a strictly increasing function, while $t \mapsto y(t)$ oscillates.
\parn
The map $t \mapsto x(t)$ has a bounded domain of the form
\beq (\tm, \tp) \qquad \mbox{with} \qquad  -\infty < \tm < 0 < \tp < + \infty~. \feq
The finite times $\tm, \tp$ are characterized by the fact that
\beq x(t) \to - \infty~~ \mbox{for} ~~t \to \tm^{+}~,
\qquad x(t) \to + \infty~ ~\mbox{for} ~~t \to \tp^{-}~. \feq
To determine $\tp$, it suffices to employ the quadrature formula
\eqref{quad1} with $\xi=+1$, $t_1 = 0$, $t_2=\tp$ and $x(t_1),
x(t_2)$ replaced by $Y,+\infty$, respectively; in this way we
obtain
\begin{align}
\tp & = {\ell + 1 \over \sqrt{2}} \int_{Y}^{+\infty} \!\!{d x
\over \sqrt{ \EE \!+\! V_1 x^{2 \ell}}} = {\ell + 1 \over \sqrt{2
\,\EE} } \left[ \lim_{x \to + \infty} x\, F_{\ell}\! \left(-
(V_1/\EE)\, x^{2 \ell} \right) -
Y\, F_{\ell}\! \left(- (V_1/\EE)\, Y^{2 \ell} \right) \right] \nonumber \\
& = {\ell + 1 \over \sqrt{2 \EE} } \left[ {C_{\ell} \over (V_1/\EE)^{1/(2\ell)}}
- Y\, F_{\ell} \!\left(- (V_1/\EE)\, Y^{2 \ell} \right) \right] \label{eqtp}
\end{align}
(the limit $x \to + \infty$ indicated above is computed using the
asymptotic expansion written in Eq.\,\eqref{efas}; the cited
equation also defines the constant $C_{\ell}$). The time $\tm$
could be determined by similar computations, but is irrelevant for
the subsequent applications. \parn Let us now pass to the function
$t \mapsto y(t)$, which oscillates in the range indicated by
Eq.\,\eqref{motiv}. This function is well defined for all $t \in
(-\infty,+\infty)$ and periodic:
\beq y(t+T) = y(t)~. \feq
The period $T$ is twice the time needed for $y(t)$ to pass from the
minimum to the maximum of the interval in Eq.\,\eqref{motiv}, and
this time can be computed using the quadrature formula
\eqref{quad2}; this gives
\beq T = 2 \;{\ell + 1 \over \sqrt{2}}
\int_{- (\FF/V_2)^{1/(2 \ell)}}^{(\FF/V_2)^{1/(2 \ell)}}\! {d y
\over \sqrt{ \FF \!-\! V_2\, y^{2 \ell}}} = {2 \sqrt{2}\, (\ell +
1)\, F_{\ell}(1) \over \sqrt{\FF}\; (V_2/\FF)^{1/(2\ell)}}~, \label{eqTper}\feq
with $F_{\ell}(1)$ given by Eq.\,\eqref{effe1}.
\parn
From here to the end of the present Section \ref{subclass7}, $x(t)$ and
$y(t)$ are the functions discussed above, i.e., the solutions
of maximal domain of the Lagrange equations with initial data \eqref{idata}
and energies \eqref{ee}\eqref{ee2}.
\vspace{-0.2cm}

\paragraph{Behavior of $\boma{x(t)}$ for positive times. The limits $\boma{t \to 0^{+}}$ and $\boma{t \to \tp^{-}}$.}
For $t \!\in\! (0, \tp)$ the function $x(t)$ increases,
starting from the initial value $x(0)=Y > 0$ and ultimately diverging.
The small $t$ behavior of $x(t), \xt(t)$ is determined by the smoothness of
these functions and by the initial data \eqref{idata}, which of course imply
\beq x(t) = Y + u \, t + O(t^2)~,~~\xt(t) = u + O(t)  \quad \mbox{for $t \to 0^{+}$}~. \label{xzero} \feq
On the other hand, the quadrature formula \eqref{quad1} with
$\xi=+1$, $t_1=0, t_2=t \in (0,\tp)$ and $x(t_1)=Y$ gives
\beq t = {\ell + 1 \over \sqrt{2}}\! \int_{Y}^{x(t)}\!\!\!
{d x \over \sqrt{ \EE \!+\! V_1 x^{2 \ell}}} = {\ell + 1 \over \sqrt{2\, \EE}}
\left[ x(t)\, F_{\ell}\! \left(- (V_1/\EE)\, x^{2 \ell}(t) \right)
- Y\, F_{\ell}\!\left(- (V_1/\EE)\, Y^{2 \ell} \right)\right] . \label{tx} \feq
From here and from Eq.\,\eqref{eqtp} for $\tp$, we obtain
\beq \tp\! - t = {\ell + 1 \over \sqrt{2\, \EE} }
\left[ {C_{\ell} \over (V_1/\EE)^{1/(2\ell)}} - x(t)\, F_{\ell}\!
\left(- (V_1/\EE)\, x^{2 \ell}(t) \right)\right] \qquad \mbox{for
all $t \in (0, \tp)$}\, . \feq
We know that $x(t) \to + \infty$ for $t \to \tp^{-}$; this fact,
together with the asymptotics \eqref{efas}, entails
\beq \tp\! - t = {\ell + 1 \over (\ell-1)\,
\sqrt{2 V_1}}\; {1 \over x^{\ell-1}(t)}
+ O\!\left({1 \over x^{3\ell-1}(t)}\!\right) , \feq
whence
({\footnote{To invert the asymptotic relation between $\tp - t$ and $x(t)$ one should note that,
since $\tp - t = {\mbox{const.} \over x^{\ell-1}(t)} \big(1+ O\big(x^{-2\ell}(t)\big)\big)$
for some nonzero constant, it follows $x(t) = {\mbox{const.} \over (\tp-t)^{1/(\ell-1)}}
\big(1 + O\big( (\tp\! - t)^{2 \ell/(\ell - 1)}\big)\big)$.\vspace{-0.3cm}}})
\beq x(t) = \left( { \ell + 1 \over (\ell-1)\, \sqrt{2 V_1}}\right)^{ \!{1 \over \ell - 1}}
{1 \over (\tp\! - t)^{1 \over \ell - 1}} + O\!\left((\tp\! - t)^{2 \ell-1 \over \ell - 1} \right)
\qquad \mbox{for $t \to \tp^{-}$}~. \label{xesp} \feq
The above result also allows to determine the behavior of
$\xt(t)$ as $t$ approaches $\tp$. In fact, from Eq.\,\eqref{exy}
and from the positivity of $\xt(t)$ we infer $\xt(t) = {\sqrt{2}
\over \ell + 1} \sqrt{\EE \!+\! V_1 x^{2 \ell}(t)}$, which
together with Eq.\,\eqref{xesp} gives
\beq \xt(t) = \left( { \ell + 1 \over (\ell - 1)^{\ell} \sqrt{2 V_1}} \right)^{\! {1 \over \ell
- 1} } {1 \over  (\tp\! - t)^{\ell \over \ell - 1}}
+ O\!\left((\tp\! - t)^{\ell \over \ell - 1} \right) \qquad
\mbox{for $t \to \tp^{-}$}~. \label{xtesp} \feq
\vfill\eject\noindent

\paragraph{Behavior of $\boma{y(t)}$ for positive times. The limit $\boma{t \to 0^{+}}$.}
Since $\yt(0) > 0$ (see Eq.\,\eqref{idata}),
$\dot{y}(t)$ will be positive from $t = 0$ up to the first
positive inversion time
\beq t_{*} := \min \big\{ t \!\in\!
(0,+\infty)~|~\yt(t) = 0 \big\} \label{eqts}\feq
at which $y$ attains its maximum value $y(t_{*}) = (\FF/V_2)^{1/(2 \ell)}$.
The small $t$ behavior of $y(t), \yt(t)$ is determined by the same smoothness considerations
that, combined with the initial data \eqref{idata}, give \beq y(t)
= Y + v \, t + O(t^2)~,~~\yt(t) = v + O(t) \quad \mbox{for $t \to
0^{+}$}\,. \label{yzero} \feq
On the other hand, using the
quadrature formula \eqref{quad2} with $\sigma=+1$, $t_1=0$, $t_2 =
t \in (0,t_{*})$ and $y(t_1)=Y$, we get
\beq t = {\ell + 1 \over \sqrt{2}} \int_{Y}^{y(t)}\!\!
{d y \over \sqrt{ \FF \!-\! V_2\, y^{2\ell}}}
= {\ell + 1 \over \sqrt{2 \,\FF} } \left[ y(t)\,
F_{\ell}\! \left( (V_2/\FF)\, y^{2 \ell}(t) \right) - Y\,
F_{\ell}\! \left( (V_2/\FF)\, Y^{2 \ell} \right) \right] .
\label{ty} \feq
Taking the limit $t \to t_{*}$ in the last equation we get
\beq t_{*} = {\ell + 1 \over
\sqrt{2}} \int_{Y}^{(\FF/V_2)^{1/(2 \ell)}}\!\! {d y \over \sqrt{
\FF \!-\! V_2\, y^{2 \ell}}} = {\ell + 1 \over \sqrt{2\, \FF}}
\left[ (\FF/V_2)^{1/2\ell}\, F_{\ell}(1) - Y\, F_{\ell}\! \left(
(V_2/\FF)\, Y^{2 \ell} \right) \right] . \label{eqtss} \feq
After the inversion time $t_{*}$, $y(t)$ decreases until it reaches the
subsequent inversion time, and so on.
Note that, depending on the choices of the parameters,
it can be $t_{*}< \tp$, or $\tp > t_{*}$, or even $t_{*} = \tp$;
in the sequel we will give explicit examples of the first two alternatives
(see Eqs.\,\eqref{tpT1}\eqref{tpT2} and the related comments).
When $t_{*}< \tp$, $y(t)$ has enough time to invert its motion (at least once) before
the explosion of $x(t)$.
\parn
In any case, the smoothness of $y(t)$ for all $t \in (-\infty,+\infty)$ ensures the finiteness
of $y(\tp), \yt(\tp)$ and yields the obvious relations
\beq y(t) = y(\tp) + O(t - \tp)~,
\quad \yt(t) = \yt(\tp) + O(t - \tp) \qquad \mbox{for $t \to
\tp$}\,. \label{yesp} \feq

\paragraph{The condition $\boma{(x(t),y(t)) \in \mathcal{D}}$.}
We now claim that the functions $t \mapsto x(t), y(t)$ fulfill
\beq x(t) > 0
~~\mbox{and}~-x(t) < y(t) < x(t) \quad \mbox{for all ~$t \in
(0,\tp)$}~. \label{eprove} \feq This is means that,
$\big(x(t),y(t)\big) \!\in \!\mathcal{D}$ (see
Eq.\,\eqref{ccor7bis}) for all $t \in (0,\tp)$, a mandatory
requirement for associating to the motion $t \mapsto (x(t),y(t))$
a cosmological model via the transformation
\eqref{ggau7}-\eqref{ccor7}. For the proof of Eq.\,\eqref{eprove},
we refer to Appendix \ref{appclass7}.

\subsubsection{Big Bang analysis}
From now on we consider the cosmology corresponding to the motion
$t \in (0,\tp) \mapsto (x(t),y(t)) \in \mathcal{D}$ described in
the preceding subsection.
\parn
The asymptotic behavior of $x(t)$ and $y(t)$ for $t \to 0^{+}$ is obviously determined by
Eqs.\,\eqref{xzero}\eqref{yzero}, which involve the parameters $Y, u, v$
introduced in Eqs.\,\eqref{idata}\eqref{uandv};
in the sequel we will frequently use the related quantity
\beq Z := (u - v)\, Y > 0~. \label{eqdefz} \feq
From Eqs.\,\eqref{scalex7}\eqref{wfx7}\eqref{Omx7} we deduce the following, for $t \to 0^{+}$:
\begin{gather}
a(t) = (2\, Z)^{\ell + 1 \over \ds} \,t^{\ell + 1 \over \ds}
+ O\!\left(t^{{\ds + \ell + 1 \over \ds}}\right) ; \label{scaleasz} \\
\wF(t) = 1 - {8\,(V_1\!+\!V_2)\,Y^{2 \ell} \over (\ell+1)^2\, Z}\;t + O(t^2)~; \label{wfias}\\
\Omm(t) = {2\, \ds^2\, \Oms  \over (\ell + 1)^2\,Z} \; t + O(t^2) ~. \label{ommasz}
\end{gather}
It is convenient to describe the limit $t \to 0^{+}$ in terms of
the cosmic time $\tc$. This can be done starting from Eq.\,\eqref{cosmicx7},
that gives an integral representation for $\tc(t)$; from here we obtain
\beq
\tc(t)/\te = \!\int_{0}^t\! d t' \left[ (2 Z)^{{\ell - 1 \over 2}}\,(t')^{{\ell - 1 \over 2}}\!
+ O\!\left((t')^{{\ell + 1 \over 2}}\right)\right]
= {(2 Z)^{\ell + 1 \over 2} \over (\ell + 1)\, Z}\;t^{\ell + 1 \over 2}
+ O\!\left(t^{\ell + 3 \over 2}\right)
\quad\; \mbox{for $t \to 0^{+}$}. \label{tcasz} \feq
The above relation shows, in particular, that $\tc(t) \to 0^{+}$ for $t \to 0^{+}$.
Considering the inverse function $t \mapsto t(\tc)$, one readily checks that Eq.\,\eqref{tcasz}
implies
\beq t(\tc) = {(\ell + 1)^{2 \over \ell + 1} \over 2\, Z^{\ell - 1 \over \ell + 1}}\;
\big(\tc/\te\big)^{2 \over \ell + 1}
+ O\!\left(\big(\tc/\te\big)^{4 \over \ell + 1} \right)
\qquad \mbox{for $\tc \to 0^{+}$}\,. \label{tasz} \feq
To go on, let us view the observables of the model as functions of the cosmic time $\tc$;
inserting the asymptotic relation \eqref{tasz} into Eqs.\eqref{scaleasz}-\eqref{ommasz},
we find the following for $\tc \to 0^{+}$:
\begin{gather}
a(\tc) = \big( (\ell + 1)\, Z\big)^{2 \over \ds} \,(\tc/\te)^{2 \over \ds}
+ O\!\left((\tc/\te)^{2(\ds + \ell + 1) \over \ds(\ell + 1)} \right); \label{scaleasez} \\
\wF(\tc) = 1 + O\!\left((\tc/\te)^{2 \over \ell + 1}\right) ; \label{wfiasez} \\
\Omm(\tc) = {\ds^2\, \Oms \over \big((\ell + 1)\, Z\big)^{2\ell \over \ell + 1}}\;
(\tc/\te)^{2 \over (\ell + 1)}
+ O\left((\tc/\te)^{4 \over \ell + 1}\right) . \label{ommasez}
\end{gather}
From the above expansions it appears that, for $\tc \to 0^{+}$:
the scale factor $a(\tc)$ vanishes, which indicates the occurrence of a Big Bang;
the reciprocal $1/a(\tc)$ diverges in a non-integrable way if $n = 2$, indicating the
absence of a particle horizon in the case of a $(2+1)$-dimensional spacetime,
while it diverges in an integrable way if $n > 2$,
showing that a particle horizon occurs when the space dimension
is equal to or greater than $3$;
$\Omm(\tc) \to 0$, which on account of Eq.\,\eqref{OmmOmf} proves that the field
energy density dominates on matter density close to the Big Bang.

\subsubsection{Far future analysis}\vspace{-0.15cm}
From Eqs.\,\eqref{scalex7}\eqref{wfx7}\eqref{Omx7} and
\eqref{xesp}\eqref{yesp} we infer that the scale factor, the
equation of state coefficient for the field and the matter density
parameter behave, respectively, as follows for $ t \to \tp^{-}$:
\begin{gather}
a(t) = \left( { \ell + 1 \over (\ell-1) \sqrt{2 V_1}}
\right)^{\!\! {\ell + 1 \over \ds (\ell-1)}}
(\tp\! - t)^{-\,{\ell + 1 \over \ds (\ell-1)}}\,
+ O\!\left( (\tp\! - t)^{2\ds - (\ell + 1) \over \ds(\ell - 1)}\right) ; \label{scaleas} \\
\wF(t) = - 1 + O\!\left( (\tp\! - t)^{2 \over \ell - 1}\right) ; \label{wfiasse} \\
\Omm(t) = \ds^2\, \Oms\,(2\,V_1)^{1 \over \ell-1} \left({ \ell\!-\!1
\over \ell \!+\! 1}\right)^{\!\!{2\ell \over \ell-1}}
(\tp\! - t)^{2\ell \over \ell-1}
+ O\!\left((\tp\! - t)^{2(\ell + 1)\over \ell-1} \right) . \label{ommas}
\end{gather}
It is convenient to describe the limit $t \to \tp^{-}$ in terms of
the cosmic time. To this purpose, we must first
determine the asymptotics of $\tc(t)$ in this limit
starting from the integral representation \eqref{cosmicx7};
since it is not so obvious how to proceed, we have given
some detail on this computation in Appendix \ref{appclass7}. Here we only report
the final result, which reads
\begin{gather}
\tc(t)/\te = {\ell + 1 \over (\ell -1) \sqrt{2 V_1}}\, \log\!
\left({\tp \over \tp\! - t} \right)
+ P + O\!\left( (\tp\! - t)^{2 \over \ell-1} \right) \qquad \mbox{for $t \to \tp^{-}$} , \label{tcas} \\
P := \int_{0}^{\tp}\! d t' \left[ \big(x^2(t') -
y^2(t')\big)^{{\ell - 1 \over 2}} - { \ell + 1 \over (\ell-1)
\sqrt{2 V_1}}\;(\tp - t')^{-1} \right] . \nonumber
\end{gather}
Thus $\tc(t) \to + \infty$ for $t \to \tp^{-}$. Considering the
inverse function $t \mapsto t(\tc)$, it can be checked
that Eq.\,\eqref{tcas} implies the following, for $\tc \to +
\infty$:
\beq \tp\! - t(\tc) = Q \; e^{- \,{(\ell - 1) \sqrt{2 V_1} \over \ell + 1}\, (\tc/\te)}
+ O \!\left(e^{- \,\sqrt{2 V_1}\, (\tc/\te)}\right), \qquad Q :=
\tp\, e^{{{(\ell - 1) \sqrt{2 V_1} \over \ell + 1}\, P }}~.
\label{dtas}\feq
Inserting the above relation into Eq.s
\eqref{scaleas}-\eqref{ommas}, we find the following for $\tc \to +
\infty$:
\begin{gather}
a(\tc) = \left( {\ell + 1 \over (\ell - 1) \sqrt{2 V_1}\,Q} \right)^{\!\!{\ell + 1 \over \ds (\ell-1)}}
e^{{ \sqrt{2 V_1} \over \ds}\,(\tc/\te)}
+ O\!\left( e^{{(\ell + 1 - 2\ds) \sqrt{2 V_1} \over \ds(\ell +1)}\,(\tc/\te)}\right) ; \label{scalease} \\
\wF(\tc) = - 1 + O\!\left( e^{{ -{2 \sqrt{2 V_1} \over \ell + 1}\, (\tc/\te)} }\right) ; \label{wfiase}\\
\Omm(\tc) = \ds^2\, \Oms\, (2 V_1)^{1 \over \ell-1} \left( {(\ell-1)\, Q \over \ell+1} \right)^{\!\!{2 \ell \over \ell-1}}
e^{-\,{ 2 \ell \sqrt{2 V_1} \over \ell + 1}\,(\tc/\te)}
+ O\!\left(e^{-\,2 \sqrt{2 V_1}\,(\tc/\te)} \right) . \label{ommase}
\end{gather}
Thus, for $\tc \to + \infty$ the following phenomena occur with
exponential speed: the scale factor diverges, the field behaves like a cosmological constant
($\wF(\tc) \to -1$) and the field energy density dominates on matter density.

\subsubsection{Some numerical examples}\vspace{-0.1cm}
Let us now restrict the attention to realistic, $3$-dimensional scenarios;
to this purpose, we put (cf. Eq.\,\eqref{OmsNum})
\begin{equation}
n = 3~, \qquad \Oms = 0.308~. \label{chon}
\end{equation}
In the sequel we consider two exemplary configurations corresponding,
respectively, to the following choices of the parameters $\ell,V_1,V_2$
which characterize the potential $\VV(\f)$ of Eq.\,\eqref{VVCase7}:
\begin{gather}
l = 2~, \qquad V_1 = 1~, \qquad V_2 = 0.1~; \label{cho1}\\
l = 2~, \qquad V_1 = 1~, \qquad V_2 = 10^3~. \label{cho2}
\end{gather}
Making reference to Eq.\,\eqref{ee}, we specify the energies
of the two Lagrangian subsystems setting
\begin{equation}
\FF = 10 ~, \qquad \EE = \FF + {n^2\,\Oms \over 2} = 11.386 ~. \label{choEF}
\end{equation}
Furthermore, keeping in mind the upper bound in Eq.\,\eqref{idata} we choose
\begin{equation}
Y = 0.1 ~. \label{choY}
\end{equation}
Let us remark that on account of Eqs.\,\eqref{idata}\eqref{uandv},
the above choices \eqref{chon}-\eqref{choY} completely determine
the initial data:
\begin{gather}
x(0) = y(0) = Y = 0.1~, \\
\dot{x}(0) \!= {\sqrt{2} \over \ell \!+\! 1}\, \sqrt{\EE \!+\! V_1\, Y^{2 \ell}} = 1.5906\,...\,, \;\quad
\dot{y}(0) \!= {\sqrt{2} \over \ell \!+\! 1 }\, \sqrt{\FF \!-\! V_2\, Y^{2 \ell}} =
\left\{\!\!\begin{array}{ll}
\dd{1.4907\,...}\!  &   \dd{\mbox{for $V_2 = 0.1$},} \vspace{0.1cm}\\
\dd{1.4832\,...}\!  &   \dd{\mbox{for $V_2 = 10^3$}.}
\end{array}\right.\!\! \nonumber
\end{gather}
Concerning the final time $\tp$ and the inversion time $t_*$ described,
respectively, by Eqs.\,\eqref{eqtp} and \eqref{eqts}\eqref{eqtss}, let us point out that the two
cases \eqref{cho1}\eqref{cho2} (together with the other choices specified above)
describe qualitatively different scenarios. In fact, $\tp = 2.0782\,...$ for $V_1 = 1$ (and any $V_2$),
while $t_* = 2.7140\,...$ for $V_2 = 0.1$ and $t_* = 0.2109\,...$ for $V_2 = 10^3$ (independently of $V_1$).
Thus, we have
\begin{gather}
\tp < t_* \qquad \mbox{for~ $V_1 = 1$,~ $V_2 = 0.1$}\,, \label{tpT1}\\
\tp > t_* \qquad \mbox{for~ $V_1 = 1$,~ $V_2 = 10^3$}\,. \label{tpT2}
\end{gather}
\begin{figure}[t!]
\vspace{0.2cm}
    \centering
        \begin{subfigure}[b]{0.465\textwidth}\label{fig:16}
                \includegraphics[width=\textwidth]{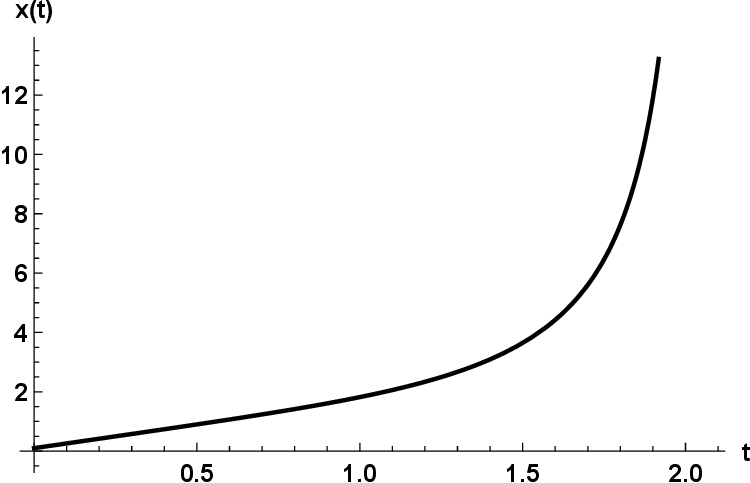}\vspace{-0.cm}
                \caption*{\textsc{Figure\! 16.}
                Plot of $x(t)$,\! with parameters fixed as in Eqs.\,\eqref{chon}\eqref{choEF}\eqref{choY}
                and either Eq.\eqref{cho1} or Eq.\eqref{cho2}. Recall that $\tp= 2.0782\,...$\,.}
        \end{subfigure}
        \hspace{0.3cm}
        \begin{subfigure}[b]{0.465\textwidth}
                \includegraphics[width=\textwidth]{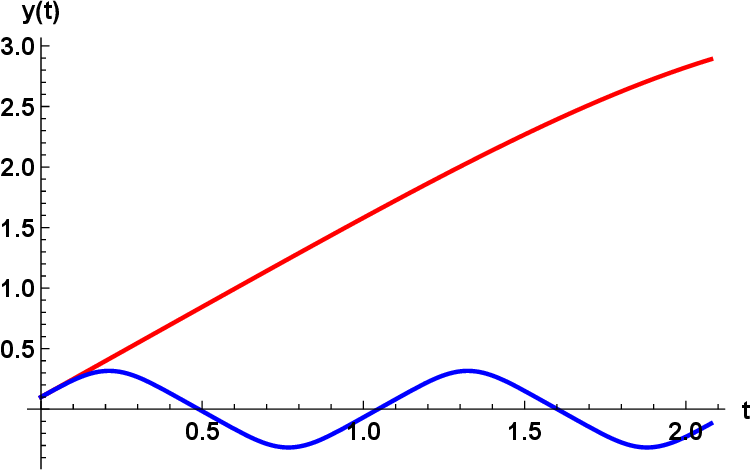}\label{fig:17}
                \caption*{\textsc{Figure\! 17.}
                Plot of $y(t)$,
                with parameters fixed as in Eqs.\,\eqref{chon}\eqref{choEF}\eqref{choY}
                and \eqref{cho1} (in red) or \eqref{cho2} (in blue).\! Recall that $\tp= 2.0782\,...$\,.}
        \end{subfigure}
\end{figure}
\noindent\vspace{-0.3cm}\\
Figs.\,16 and 17 give plots of the Lagrangian coordinates $x(t),y(t)$ for $t \in (0,\tp)$.
Especially, Fig.\,17 makes evident that $y(t)$ is strictly increasing for $t \in (0,\tp)$
if $\ell,V_1,V_2$ are fixed as in Eq.\,\eqref{cho1}, while $y(t)$ oscillates if $\ell,V_1,V_2$
are as in Eq.\,\eqref{cho2}. This fact is in agreement with the previous considerations related
to Eqs.\,\eqref{tpT1}\eqref{tpT2}.
\begin{figure}[t!]
    \centering
        \begin{subfigure}[b]{0.475\textwidth}\label{fig:18}
                \includegraphics[width=\textwidth]{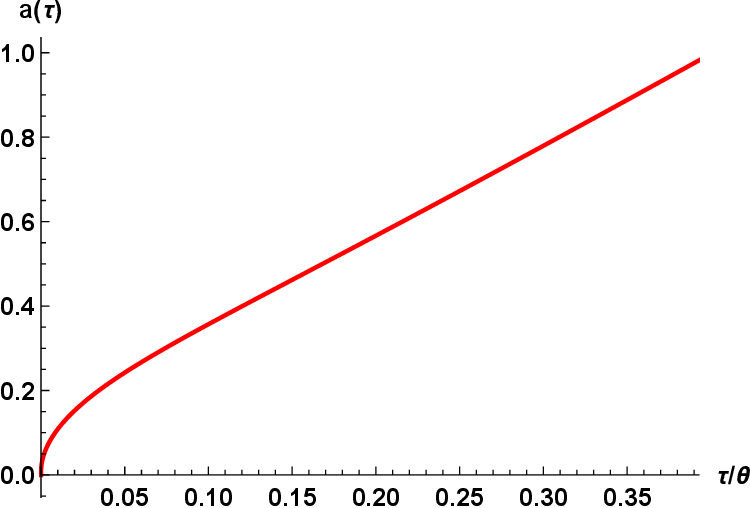}\vspace{-0.cm}
                \caption*{\textsc{Figure\! 18.}
                $a(\tc)$ as a function of $\tc/\te$,
                with parameters fixed as in Eqs.\,\eqref{chon}\eqref{choEF}\eqref{choY}
                and \eqref{cho1}.}
        \end{subfigure}
        \hspace{0.3cm}
        \begin{subfigure}[b]{0.475\textwidth}
                \includegraphics[width=\textwidth]{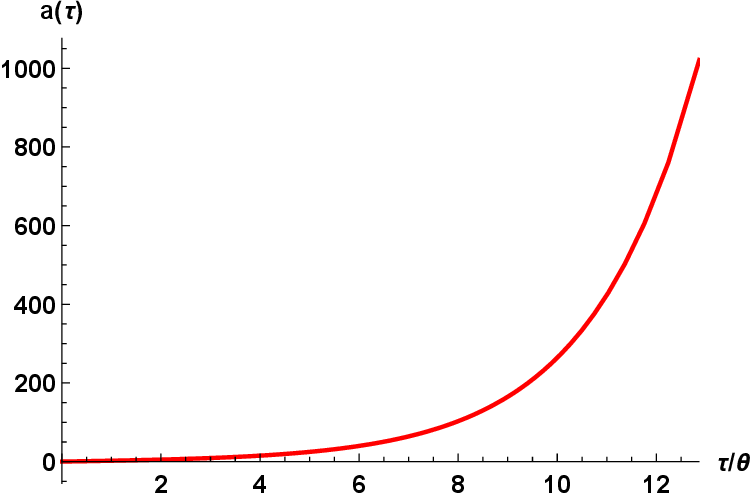}\label{fig:19}
                \caption*{\textsc{Figure\! 19.}
                $a(\tc)$ as a function of $\tc/\te$,
                with parameters fixed as in Eqs.\,\eqref{chon}\eqref{choEF}\eqref{choY}
                and \eqref{cho1}.}
        \end{subfigure}
\vspace{-0.05cm}
\end{figure}
\begin{figure}[t!]
    \centering
        \begin{subfigure}[b]{0.475\textwidth}\label{fig:20}
                \includegraphics[width=\textwidth]{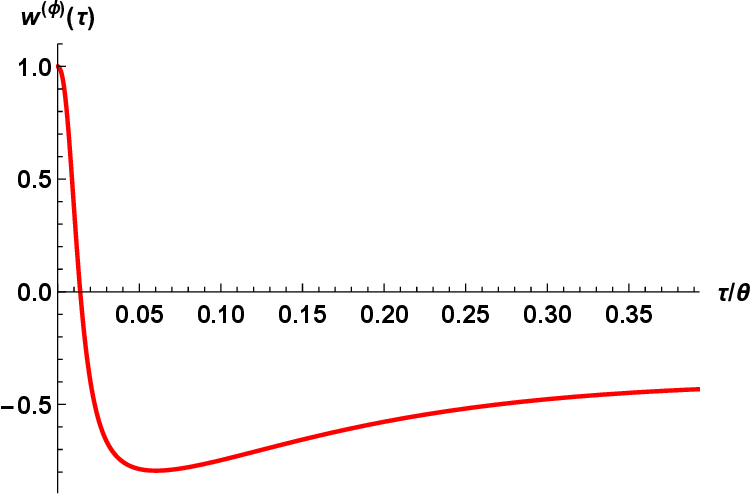}\vspace{-0.cm}
                \caption*{\textsc{Figure\! 20.}
                $\wF(\tc)$ as a function of $\tc/\te$,
                with parameters fixed as in Eqs.\,\eqref{chon}\eqref{choEF}\eqref{choY}
                and \eqref{cho1}.}
        \end{subfigure}
        \hspace{0.3cm}
        \begin{subfigure}[b]{0.475\textwidth}
                \includegraphics[width=\textwidth]{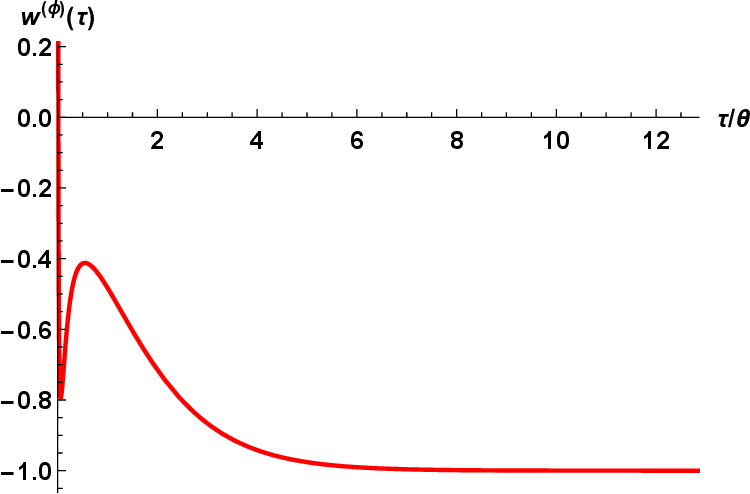}\label{fig:21}
                \caption*{\textsc{Figure\! 21.}
                $\wF(\tc)$ as a function of $\tc/\te$,
                with parameters fixed as in Eqs.\,\eqref{chon}\eqref{choEF}\eqref{choY}
                and \eqref{cho1}.}
        \end{subfigure}
\vspace{-0.05cm}
\end{figure}
\begin{figure}[t!]
    \centering
        \begin{subfigure}[b]{0.475\textwidth}\label{fig:22}
                \includegraphics[width=\textwidth]{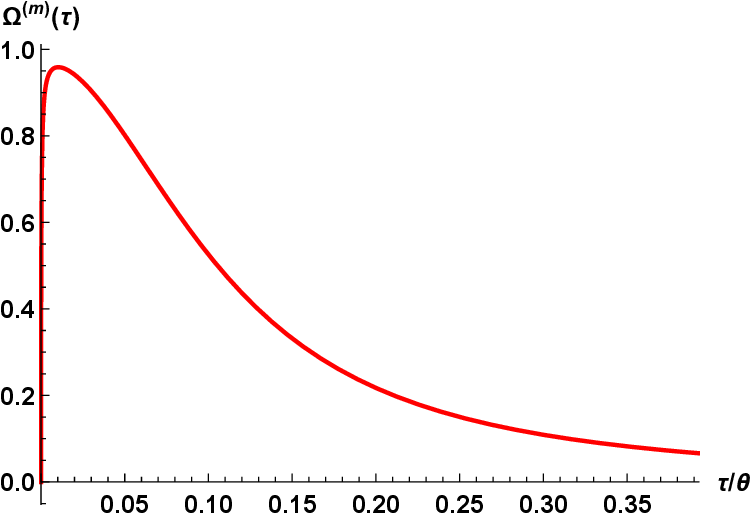}\vspace{-0.cm}
                \caption*{\textsc{Figure\! 22.}
                $\Omm(\tc)$ as a function of $\tc/\te$,
                with parameters fixed as in Eqs.\,\eqref{chon}\eqref{choEF}\eqref{choY}
                and \eqref{cho1}.}
        \end{subfigure}
        \hspace{0.3cm}
        \begin{subfigure}[b]{0.475\textwidth}
                \includegraphics[width=\textwidth]{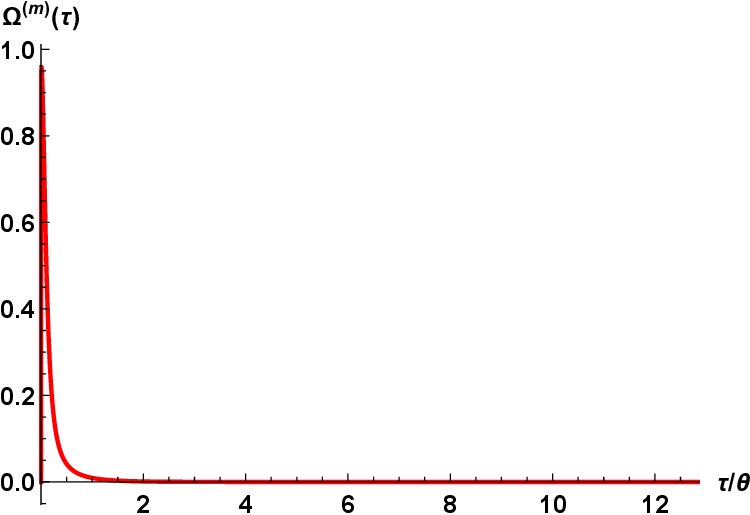}\label{fig:23}
                \caption*{\textsc{Figure\! 23.}
                $\Omm(\tc)$ as a function of $\tc/\te$,
                with parameters fixed as in Eqs.\,\eqref{chon}\eqref{choEF}\eqref{choY}
                and \eqref{cho1}.}
        \end{subfigure}
\vspace{-0.2cm}
\end{figure}
\vskip 0.1cm \noindent
Figs.\,18-29 represent the observables $a(\tc), \wF(\tc), \Omm(\tc)$ as functions
of $\tc/\te$. For each one of these observables we consider the choices
\eqref{chon}\eqref{choEF}\eqref{choY} and both choices \eqref{cho1}\eqref{cho2}
for the involved parameters; in addition, for each one of the previous choices
we consider two possible ranges for $\tc/\te$, corresponding in terms of the
coordinate time $t$ to the intervals $(0,\tp/2)$ (figures with an odd numbering)
and $(0, \vartheta\, \tp)$ (figures with an even numbering),
with $\vartheta = 0.999$ so that $\tc/\te \in (0,12.8673\,...)$ in the case \eqref{cho1}
and $\tc/\te \in (0,14.0804\,...)$ in the case \eqref{cho2}.
As a matter of fact, all these figures were obtained plotting the curves
$t \mapsto (\tc(t)/\te, a(t))$, $(\tc(t)/\te, \wF(t))$,
$(\tc(t)/\te, \Omm(t))$ for $t \in (0,\tp/2)$ or $t \in \big(0,\vartheta \tp\,\big)$.\vspace{-0.6cm}
\begin{figure}[t!]
\vspace{-0.7cm}
    \centering
        \begin{subfigure}[b]{0.475\textwidth}\label{fig:24}
                \includegraphics[width=\textwidth]{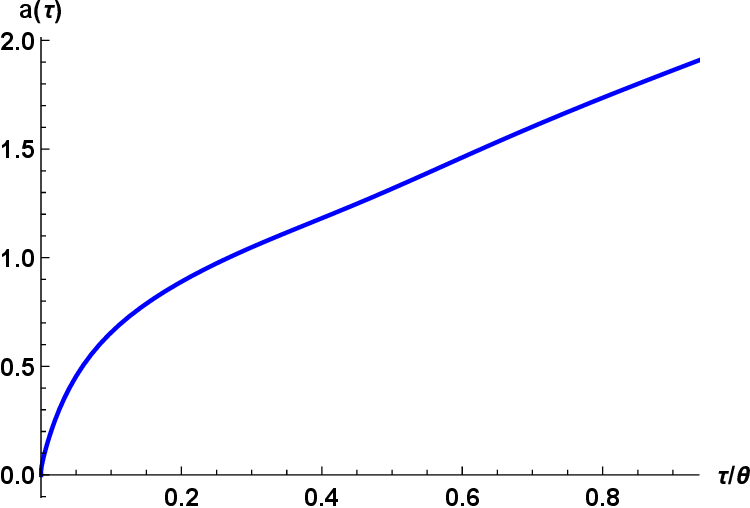}\vspace{-0.cm}
                \caption*{\textsc{Figure\! 24.}
                $a(\tc)$ as a function of $\tc/\te$,
                with parameters fixed as in Eqs.\,\eqref{chon}\eqref{choEF}\eqref{choY}
                and \eqref{cho2}.}
        \end{subfigure}
        \hspace{0.3cm}
        \begin{subfigure}[b]{0.475\textwidth}
                \includegraphics[width=\textwidth]{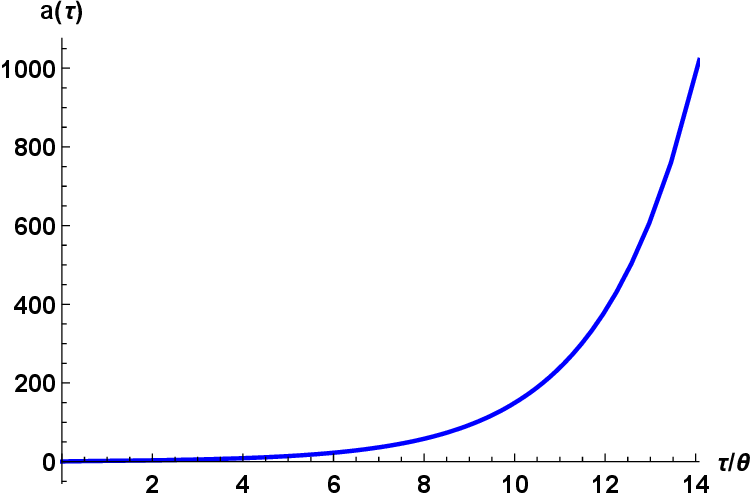}\label{fig:25}
                \caption*{\textsc{Figure\! 25.}
                $a(\tc)$ as a function of $\tc/\te$,
                with parameters fixed as in Eqs.\,\eqref{chon}\eqref{choEF}\eqref{choY}
                and \eqref{cho2}.}
        \end{subfigure}
\vspace{0.05cm}
\end{figure}
\begin{figure}[t!]
    \centering
        \begin{subfigure}[b]{0.475\textwidth}\label{fig:26}
                \includegraphics[width=\textwidth]{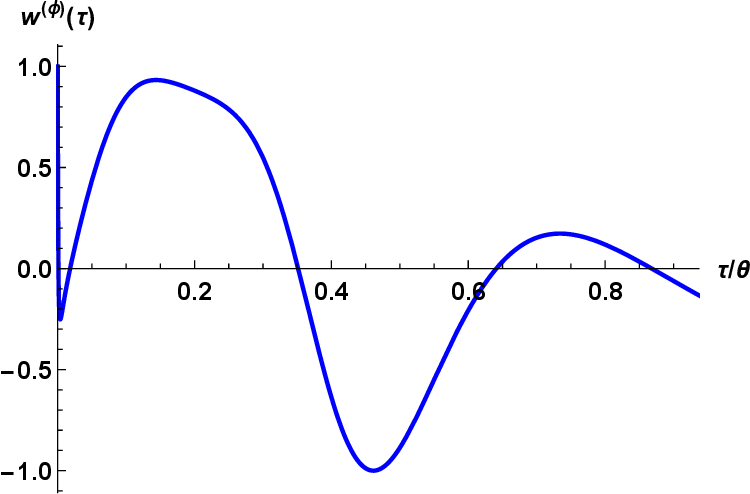}\vspace{-0.cm}
                \caption*{\textsc{Figure\! 26.}
                $\wF(\tc)$ as a function of $\tc/\te$,
                with parameters fixed as in Eqs.\,\eqref{chon}\eqref{choEF}\eqref{choY}
                and \eqref{cho2}.}
        \end{subfigure}
        \hspace{0.3cm}
        \begin{subfigure}[b]{0.475\textwidth}
                \includegraphics[width=\textwidth]{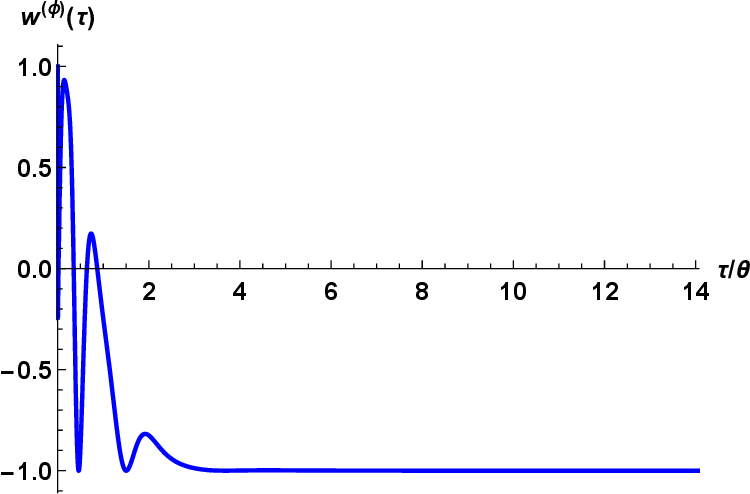}\label{fig:27}
                \caption*{\textsc{Figure\! 27.}
                $\wF(\tc)$ as a function of $\tc/\te$,
                with parameters fixed as in Eqs.\,\eqref{chon}\eqref{choEF}\eqref{choY}
                and \eqref{cho2}.}
        \end{subfigure}
\vspace{0.05cm}
\end{figure}
\vfill\eject\noindent
\begin{figure}[t!]
    \centering
        \begin{subfigure}[b]{0.475\textwidth}\label{fig:28}
                \includegraphics[width=\textwidth]{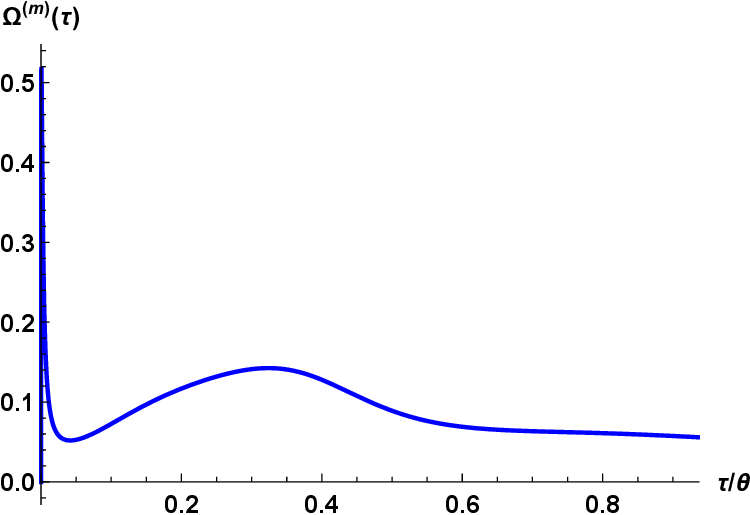}\vspace{-0.cm}
                \caption*{\textsc{Figure\! 28.}
                $\Omm(\tc)$ as a function of $\tc/\te$,
                with parameters fixed as in Eqs.\,\eqref{chon}\eqref{choEF}\eqref{choY}
                and \eqref{cho2}.}
        \end{subfigure}
        \hspace{0.3cm}
        \begin{subfigure}[b]{0.475\textwidth}
                \includegraphics[width=\textwidth]{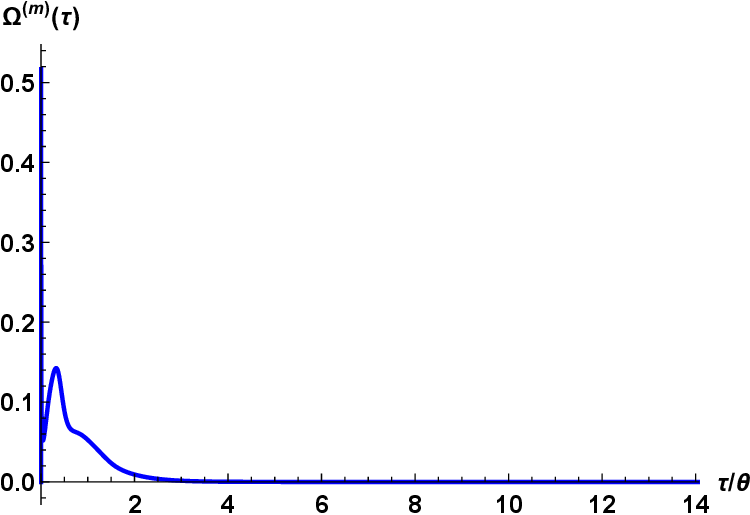}\label{fig:29}
                \caption*{\textsc{Figure\! 29.}
                $\Omm(\tc)$ as a function of $\tc/\te$,
                with parameters fixed as in Eqs.\,\eqref{chon}\eqref{choEF}\eqref{choY}
                and \eqref{cho2}.}
        \end{subfigure}
\end{figure}

\subsection*{Acknowledgments}
\vspace{-0.1cm}
This work was supported by:
Universit\`{a} degli Studi di Milano;
INFN, Istituto Nazionale di Fisica Nucleare;
MIUR, PRIN 2010 Research Project ``Geometric and analytic theory of Hamiltonian systems in finite and infinite dimensions'';
INdAM, Gruppo Nazio\-nale per la Fisica Matematica.

\vfill\eject\noindent
\vfill\eject\noindent

\appendix
\numberwithin{equation}{section}

\section{Appendix. On the setting of Section \ref{bafre}}
\label{appa}
We consider the spacetime metric \eqref{robw} and the coordinate system \eqref{sc0i}
(recalling that Greek indexes range from $0$ to $\ds$, while Latin indexes range from $1$ to $\ds$).
Moreover, we make all the assumptions stated in Section \ref{bafre}
about the scalar field and the matter fluid.
\vspace{-0.4cm}

\paragraph{The metric $\boma{g_{\mu\nu}}$.} From Eq.\,\eqref{robw} we infer (for $i,j=1,...,\ds$)
\beq g_{0 0} = - \,\te^2\, e^{2 \BB}~,\qquad
g_{0 i} = g_{i 0} = 0~, \qquad
g_{i j} = e^{2 \AA/\ds}\, h_{i j} ~. \label{gimunu} \feq
Recall that $\AA,\BB$ are functions of $x^0 \equiv t$, while the coefficients $h_{i j}$
are functions of the space coordinates $(x^i) \equiv \bx$.
For future use, let us mention that $g := \det(g_{\mu \nu})$
can be expressed as follows in terms of $h:= \det(h_{i j}) \!>\! 0$:
\beq g = -\,\te^2 e^{2 \AA + 2 \BB}\, h~. \label{detg}
\vspace{-0.1cm}\feq

\paragraph{Ricci tensor $\boma{R_{\mu \nu}}$ and scalar curvature $\boma{R}$.}
Given the metric \eqref{gimunu}, these are ($i,j \!=\! 1,...,\ds$):
\beq \begin{array}{c}
\dd{R_{0 0} = -\, \ddot{\AA} + \dot{\AA}\, \dot{\BB} - {1 \over \ds}\,\dot{\AA}^2 \,, \qquad
R_{i j} = \left[{1 \over \ds}\;e^{2 \AA/\ds - 2 \BB}\,
(\ddot{\AA} + \!\dot{\AA}^2\! - \dot{\AA}\, \dot{\BB}) + (\ds-1)\,k\right]\!{h_{i j} \over \te^2} ~,} \vspace{0.1cm}\\
\dd{R_{0 i} = R_{i 0} = 0~; \qquad
R = {e^{- 2 \BB} \over \te^2} \left(2 \ddot{\AA} - 2 \dot{\AA} \dot{\BB} + {\ds + 1 \over \ds}\, \dot{\AA}^2 \right)\!
+ {\ds (\ds-1)\, k \over \te^2}\; e^{- 2 \AA/\ds}}\,. \label{Rij1}
\end{array} \feq
Note that $(\ds-1)\, k\, h_{i j}/\te^2 = (\ds-1)\, \cu\, h_{i j}$
is the Ricci tensor of the Riemannian manifold $(\MK, h_{i j})$.
\vspace{-0.4cm}

\paragraph{Stress-energy tensor for the scalar field.}
Eqs.\,\eqref{tifield0}\eqref{fiad} imply
\beq \Tf_{\mu \nu} = {\ds-1 \over \ds\, \kappa^2_\ds}
\left[\partial_{\mu} \f\, \partial_{\nu} \f
- {1 \over 2}\,g_{\mu \nu}\,\partial_{\alpha}\f \,\partial^{\alpha} \f
- {1 \over \te^2}\,g_{\mu \nu} \VV(\f) \right] . \label{tifield} \feq
Since $\varphi$ depends only on $t$ as indicated in Eq.\,\eqref{firomt} and
$g_{\mu \nu}$ is as in Eq.\,\eqref{gimunu}, for $i,j=1,...,\ds$ we have
\beq \begin{array}{c}
\dd{\Tf_{0 0} = {\ds-1 \over \ds\, \kappa^2_\ds} \left({1 \over 2}\,\ft^2  + e^{2 \BB} \, \VV(\f)\! \right) ,
\quad\qquad
\Tf_{0 i} = \Tf_{i 0} = 0 ~, }\vspace{0.1cm}\\
\dd{\Tf_{i j} = {\ds-1 \over \ds\, \kappa^2_\ds \te^2} \left({1 \over 2}\,\ft^2 - e^{2 \BB} \, \VV(\f) \right)
e^{2 \AA/\ds - 2 \BB} \,h_{i j} ~.} \label{tfield1}
\end{array}\feq
Comparing this result with Eq.\,\eqref{umurob} for $U_\mu$
and Eq.\,\eqref{gimunu} for $g_{\mu\nu}$, it follows that
$\Tf_{\mu \nu}$ can be written in the fluid-like form
\beq
\dd{\Tf_{\mu \nu} = \big(\pfi +\rofi\big)\, U_\mu\, U_\nu + \pfi\, g_{\mu \nu}~,} \label{tirofi} \feq
with $\pfi$, $\rofi$ as in Eq.\,\eqref{rofi}.
\vspace{-0.4cm}

\paragraph{Stress-energy tensor for the matter fluid.}
This has the form \eqref{Tm}, where the pressure and density are related by the equation
of state $\ppm = w\,\rom$ (see Eq.\,\eqref{pwro}). Using Eq.\,\eqref{umurob} for $U_\mu$
and Eq.\,\eqref{gimunu} for $g_{\mu\nu}$, we obtain
\beq
\Tm_{0 0} = \te^2\, e^{2 \BB} \rom \, ,\qquad
\Tm_{0 i} = \Tm_{i 0}  = 0 \,, \qquad
\Tm_{i j} = w\, e^{2 \AA/\ds} \rom\, h_{i j}~; \label{tfim}
\feq
recall that, according to \eqref{firomt}, $\rom$ depends only on $t$.
\vspace{-0.4cm}

\paragraph{Conservation law for the stress-energy tensor of the matter fluid.}
Let $\nabla_{\mu}$ be the covariant derivative associated to the metric \eqref{gimunu};
then, from Eq.\,\eqref{tfim} for $\Tm_{\mu \nu}$ we get the following:
\beq \nabla_{\!\mu} \Tm^{\mu}_{\;\;0} = -\, \conm \,,
\quad \nabla_{\!\mu} \Tm^{\mu}_{\;\;i} = 0 ~;
\qquad \conm := \drom\! + (w + 1) \, \dot{\AA} \, \rom\,.
\label{due2} \feq
So, the conservation law $\nabla_{\!\mu} \Tm^{\mu}_{\;\;\nu} \!=\! 0$ is equivalent to $\conm\!=\!0$;
clearly, the latter relation holds if and only if $\rom(t) \!=\! \roms e^{-(w+1) \AA(t)}$
for some constant $\roms$, which can be written in the form \eqref{pmrom1}.
\vspace{-0.4cm}

\paragraph{Einstein equations.} Let us consider the Eistein equations \eqref{eqs1}, i.e.,
\beq E_{\mu \nu} := 0~, \qquad
E_{\mu \nu} := R_{\mu \nu} - {1 \over 2}\, g_{\mu \nu} R -
\kappa^2_\ds \big(\Tf_{\!\mu \nu} + \Tm_{\!\mu \nu}\big)~. \label{eqs1e} \feq
From Eqs.\,\eqref{Rij1}\eqref{tfield1}\eqref{tfim}
for $R_{\mu \nu}$, $R$, $\Tf_{\!\mu \nu}$, $\Tm_{\!\mu \nu}$
and \eqref{pmrom1} for $\rom$, we infer ($i,j=1,...,\ds$):
\beq E_{0 0} = {\ds-1 \over \ds}\; \einsd~, \qquad
E_{0 i} = E_{i 0} = 0 ~, \qquad
E_{i j} = - \,{\ds-1 \over \ds\, \te^2} \; e^{2 \AA/\ds - 2 \BB} \; \einsu \, h_{i j} ~, \label{eqeins} \feq
with $\einsu, \einsd$ as in Eqs.\,\eqref{EE1}\eqref{EE2}. Thus, if we assume
Eq.\,\eqref{pmrom1} for $\rom$ (in agreement with the conservation
law for $\Tm_{\mu \nu}$), the Einstein equations are equivalent to
$\einsu=0$, $\einsd=0$.
\vspace{-0.4cm}

\paragraph{Evolution equation for the scalar field.} According to Eq.\,\eqref{eqsX},
this is $\Box \phi - V'(\phi)=0$. Let us recall that
$\Box \phi = {1 \over \sqrt{|g|}}\,
\partial_\mu \big(\sqrt{|g|}\, g^{\mu\nu}\, \partial_\nu \phi\big)$ with
$g := \det (g_{\mu\nu})$.
Expressing the metric and the corresponding determinant as in Eqs.\,\eqref{gimunu}\eqref{detg},
the field $\phi$ and its potential $V$ as in Eq.\,\eqref{fiad}
(involving the dimensionless equivalents $\f, \VV$)
and finally recalling that $\f$ depends only on $t$, we obtain
\beq \Box \phi - V'(\phi) = - \,\sqrt{{\ds-1 \over \ds}}\; {e^{-2 \BB} \over \kappa_\ds\,\te^2} \, \einsf~, \feq
with $\einsf$ as in Eq.\,\eqref{ensca}. Thus, the field equation \eqref{eqsX} is equivalent to
$\einsf=0$.
\vspace{-0.4cm}

\paragraph{Evaluating the action functional on a history as above.}
Let us consider the action functional $\mathcal{S}$ defined by Eq.\,\eqref{Functional},
and evaluate it on a history of the type considered in the previous
paragraphs, so that: the spacetime metric $g_{\mu\nu}$ is given by Eq.\,\eqref{gimunu}
(implying Eqs.\,\eqref{detg} \eqref{Rij1} for the determinant $g$ and the scalar curvature $R$);
the scalar field $\phi$ depends only on $t$ and is represented with the related
potential as in Eq.\,\eqref{fiad}; the matter density is given by Eq.\,\eqref{pmrom1}.
In this way we obtain (with $h := \det(h_{i j})$)
\beq \begin{array}{c}
\hspace{-0.5cm}
\dd{\mathcal{S}  = {1 \over \kappa^2_\ds \, \te} \int\! d t \, d^\ds \bx\, \sqrt{h(\bx)} \,
\Bigg[ e^{\AA - \BB} \left(\ddot{\AA} - \dot{\AA}\, \dot{\BB} + {\ds + 1 \over 2 \ds}\;\dot{\AA}^2
+ {\ds -1 \over 2 \ds}\; \dot{\f}^2 \right)} \vspace{0.2cm}\\
\hspace{2.5cm}
\dd{-\, {\ds - 1 \over \ds}\, e^{\AA + \BB} \, \VV(\f)
- {\ds (\ds - 1)\, \Oms \over 2}\; e^{-w \AA + \BB}
+ {\ds (\ds - 1)\,k \over 2}\; e^{{\ds - 2 \over \ds}\, \AA + \BB} \Bigg]\; .}
\end{array} \feq
It is straightforward to put the above result in the form \eqref{eLL}.

\numberwithin{equation}{subsection}
\section{Appendix. On the explicit solutions of subsection \ref{class2}\label{appsol}}
This appendix contains some details on the derivation of the explicit expressions
reported in subsection \ref{class2} for the solutions $x(t),y(t)$ of the Lagrange
equations \eqref{ex}\eqref{ey} fulfilling the condition $x(t),y(t) > 0$ stated in Eq.\,\eqref{xypos}.
Let us recall that $-1\!<\!w\!<\!1$ (see Eq.\,\eqref{wpm1}) and treat separately the cases
$V_1 \!>\! 0$, $V_1 = 0$, $V_1 \!<\! 0$.

\subsection[The case $V_1 > 0$]{The case $\boma{V_1 > 0}$}\label{subepm1}
To begin with, let us recall that in this case Eqs.\,\eqref{ex}\eqref{ey} reduce, respectively,
to Eqs.\,\eqref{ex1a} \eqref{ey1a}.
By elementary arguments one infers that any positive solution of Eq.\,\eqref{ex1a}
can be written in one of the following forms, after possibly a time translation
$t \to t + \mbox{const.}$ and a time reflection $t \to -t$:
\begin{align}
x(t) = A\, \sinh(\om\, t) \qquad & \mbox{for ~$A> 0$~ and ~$t \in (0,+\infty)$}~; \label{form11a} \\
x(t) = A\, \cosh(\om\, t) \qquad & \mbox{for ~$A> 0$~ and ~$t \in \reali$}~; \label{form21a} \\
x(t) = A\, e^{\om\, t}    \qquad & \mbox{for ~$A> 0$~ and ~$t \in \reali$}~. \label{form31a}
\end{align}
Supposing that the time coordinate $t$ has been fixed so that the solution of Eq.\,\eqref{ex1a}
takes one of the forms \eqref{form11a}\eqref{form21a}\eqref{form31a}, we proceed to determine
the corresponding solution of Eq.\,\eqref{ey1a} (on the intervals mentioned above) starting
from the familiar representation
\begin{align}
y(t) & = C \cosh(\om\,t) + D \sinh(\om\, t) \nonumber \\
& \quad +\, {1 \over \om} \int_{0}^ t d s\; \sinh\big(\om(t-s)\big)
\Big[ (1-w)\,V_{2}\, x(s)^{{1-w \over 1+w}} - {w\,(1-w)\,\ds^2\,\Oms \over 2}\; x(s)^{-{1+3w \over 1+w}} \Big]~,
\label{espy}
\end{align}
where $C,D \in \reali$ are arbitrary constants. This representation is understood to hold for all values
of $w \in (-1,1)$ granting the convergence of the integral over $s \in (0,t)$;
in cases \eqref{form21a}\eqref{form31a} there are no convergence problems, while in case \eqref{form11a}
we must require $ {1-w \over 1+w} > -1 $ and $ -\,{1+3w \over 1+w} > -1$, which happens if and only if $ -1 < w < 0 $.
Nevertheless, we shall see later that even in the case \eqref{form11a}
the final result can be extended to all values $w \in (-1,1)$
by analytic continuation (see, especially, the remark at the end of this subsection).
\parn
Of course, the evaluation of the integral in Eq.\,\eqref{espy} with $x$ given by one of
Eqs.\,\eqref{form11a}\eqref{form21a}\eqref{form31a}
can be reduced to the computation of the following integrals,
for $\lp = {1-w \over 1+w}$ or $\lp = -\,{1+3w \over 1+w}$\,:
\begin{gather}
{1 \over \om} \int_{0}^t d s\; \sinh\!\big(\om\,(t-s)\big) \sinh^\lp(\om\, s)\,, \qquad
{1 \over \om} \int_{0}^t d s\; \sinh\big(\om\,(t-s)\big) \cosh^\lp (\om\, s)\,, \nonumber \\
{1 \over \om} \int_{0}^t d s\; \sinh\big(\om\,(t-s)\big)\, e^{\lp\, \om\, s}\,.
\end{gather}
On one hand, by basic trigonometric identities, for any $\om, t > 0$ we have the following relations,
holding under the condition $\lp > -1$ that ensures the convergence of the forthcoming integrals
({\footnote{The last equality in \eqref{sinhInt} follows making the
change of variable $s = {1 \over \om}\,\mbox{arcsinh}\big(\sinh(\om\,t)\sqrt{\tu}\big)$
in the second integral of the preceding expression.
The same change of variable is used to derive the last identity in Eq.\,\eqref{coshInt}.
}}):
\begin{align}
& {1 \over \om} \int_{0}^t d s\; \sinh\big(\om\,(t-s)\big)\, \sinh^\lp(\om\, s) \nonumber \\
& = {\sinh(\om\,t) \over \om} \int_{0}^t d s\;\cosh(\om\,s)\, \sinh^\lp(\om\,s)
 -{\cosh(\om\,t) \over \om}\! \int_{0}^t d s\; \sinh^{\lp+1}(\om\, s) \nonumber \\
& = {\sinh^{\lp + 2}(\om\,t) \over \om^2\,(\lp + 1)}
- {\cosh(\om\,t)\,\sinh^{\lp + 2}(\om\,t) \over 2\,\om^2}
\int_{0}^{1}\! d \tu\; {\tu^{\lp/2} \over \sqrt{1 + \sinh^2(\om\, t)\, \tu}}\;.
\label{sinhInt}
\end{align}
Similarly, for any $\om,t > 0$ and $\lp \in \reali$ we get
\begin{align}
& {1 \over \om} \int_{0}^t\! d s\; \sinh\big(\om\,(t-s)\big)\; \cosh^\lp(\om\,s) \nonumber \\
& = -\,{\cosh(\om\,t) \over \om}\! \int_{0}^t\! d s\; \sinh(\om\, s)\, \cosh^{\lp}(\om\, s)
+ {\sinh(\om\, t) \over \om}\! \int_{0}^t\! d s\; \cosh^{\lp+1}(\om\, s) \nonumber \\
& = {\cosh(\om\,t)\, \big(1 - \cosh^{\lp+1}(\om t) \big) \over \om^2\, (\lp + 1)}
+ {\sinh^2(\om\, t) \over 2\,\om^2}\! \int_{0}^{1}\! d \tu\;
{\big(1 + \sinh^2(\om\,t)\,\tu\big)^{\lp/2} \over \sqrt{\tu}}\;.
\label{coshInt}
\end{align}
The integrals over $v \in (0,1)$ appearing in Eqs.\,\eqref{sinhInt}\eqref{coshInt} can be
expressed in terms of hypergeometric functions ${}_2 F_1(\al,\be,\ga;z)$,
recalling the integral identity (see, e.g., \cite[Eqs.\,15.1.2 and 15.6.1]{NIST})
\beq \int_0^1\! dv\;{v^{\be-1} (1\!-\!v)^{\ga-\be-1} \over (1-z\,v)^{\al}} =
{\Ga(\be)\,\Ga(\ga\!-\!\be) \over \Ga(\ga)}\,{}_{2} F_1(\al,\be,\ga;z)
\quad~ \mbox{for\, $\al,\be,\ga,z \!\in\! \reali$\, with\, $\ga \!>\! \be \!>\! 0$}\,. \label{Inf2F1} \feq
\vfill\eject\noindent
Using the above identity with $\al = 1/2$, $\be = 1 + \lp/2$, $\ga = 2 + \lp/2$
and $z = -\,\sinh^2(\om\, t)$ (along with the basic relations
$\Ga(1) = 1$, $\Ga(2+\lp/2) = (1+\lp/2)\,\Ga(1+\lp/2)$\,),
from Eq.\,\eqref{sinhInt} we infer
\begin{align}
& {1 \over \om} \int_{0}^t d s\; \sinh\big(\om\,(t-s)\big)\, \sinh^\lp(\om\, s) \nonumber\\
& = {\sinh^{\lp + 2}(\om\,t) \over \om^2} \left[{1 \over \lp + 1}
-\, {\cosh(\om\,t) \over \lp + 2}\;{}_{2} F_1\!\left({1 \over 2}\,,1+{\lp \over 2}\,,
2+{\lp \over 2}\,;\,-\,\sinh^2(\om\,t)\right)\right] . \label{prima}
\end{align}
Likewise, using the identity \eqref{Inf2F1} with
$\al = -\lp/2$, $\be = 1/2$, $\ga = 3/2$ and $z = -\,\sinh^2(\om\, t)$
(along with the relations $\Ga(1/2)/\Ga(3/2) = 2$ and ${}_2 F_1(\be,\al,\ga;z) = {}_2 F_1(\al,\be,\ga;z)$\,),
from Eq.\,\eqref{coshInt} we infer
\begin{align}
& {1 \over \om} \int_{0}^t\! d s\; \sinh\big(\om\,(t-s)\big)\; \cosh^\lp(\om\,s) \nonumber \\
& = {1 \over \om^2} \left[ {\cosh(\om\,t)\, \big(1 - \cosh^{\lp+1}(\om t) \big) \over \lp + 1}
+ \sinh^2(\om\, t)\;
{}_2 F_1\!\left({1 \over 2}\,, -{\lp \over 2}\,, {3 \over 2}\,; -\,\sinh^2(\om\,t)\right) \right] .
\label{seconda}
\end{align}
On the other hand, for any $\om,t > 0$ and $\lp \in \reali$, by direct computations we get
\beq {1 \over \om} \int_{0}^t d s\; \sinh\big(\om\,(t-s)\big)\, e^{\lp\,\om\, s}
= {\cosh(\om\,t) + \eta \sinh(\om\,t) - e^{\eta\,\om\,t} \over (1 - \eta^2)\, \om^2} ~. \label{terza}\feq
Let us remark that the right-hand sides of Eqs.\,\eqref{seconda}\eqref{terza}
must be intended in a natural limit sense for $\lp = \pm 1$;
more precisely, we understand that
\begin{align}
{1 - \cosh^{\lp+1}(\om\, t) \over 1 + \lp}\bigg\rvert_{\lp = -1}
& := \lim_{\lp \rightarrow -1} {1 - \cosh^{\lp+1}(\om\, t) \over 1 + \lp} = -\, \log\big(\cosh(\om\, t)\big) ~, \label{Lim1} \\
{\cosh(\om\, t) + \lp\, \sinh(\om\,t) - e^{\lp\, \om\, t} \over 1-\lp^2}\bigg\rvert_{\lp = \pm 1}
& := \lim_{\lp \rightarrow \pm 1} {\cosh(\om\, t) + \lp \sinh(\om\, t) - e^{\lp\, \om \,t} \over 1-\lp^2} \nonumber\\
&\; = \pm\, {\om\,t \, e^{\pm\,\om \,t} -\, \sinh(\om\, t) \over 2} ~. \label{Lim2}
\end{align}
Summing up, Eqs.\,\eqref{form11a}\eqref{form21a}\eqref{form31a} for $x(t)$
and the corresponding expressions for $y(t)$ descending from Eqs.\,\eqref{espy} and
\eqref{prima} \eqref{seconda} \eqref{terza}
give rise to the explicit solutions \eqref{primasol} \eqref{secondasol} \eqref{terzasol}
reported in the main text.
Correspondingly, Eqs.\,\eqref{Lim1}\eqref{Lim2} account for Eqs.\,\eqref{lim1}\eqref{lim2}.
\vspace{-0.2cm}

\paragraph{A final remark on the solution \eqref{primasol}.}
Let us recall that in all the previous manipulations yielding the expression \eqref{primasol}
for $y(t)$ to grant the convergence of the involved integrals we have assumed
$ {1-w \over 1+w} > -1 $ and $ -\,{1+3w \over 1+w} > -1$, which happens if and only if $ -1 < w < 0 $.
However, after proving that the expression \eqref{primasol} for $y(t)$ gives a solution
of Eq.\,\eqref{ey1a} for $ -1 < w < 0 $, by elementary consideration based on
analytic continuation we can infer that the same holds on the entire region of analyticity
w.r.t. $w$.
Regarding this region, let us recall that for any fixed $z \in (-\infty, 1)$, ${}_2 F_1(a,b,c\,;z)$
is analytic w.r.t. the parameters $a,b \in \reali$ and $c \in \reali \backslash \{0, -1, -2, ... \}$
(see, e.g., \cite{NIST}). In Eq.\,\eqref{primasol} there appear two hypergeometric terms with
$c = \frac{3+w}{2+2w}$ and $c = \frac{5+3w}{2+2w} = \frac{3+w}{2+2w} + 1$, which are both
different from $ 0, -1, -2, ...$ if and only if
\beq w \neq - \frac{3+2 h}{1+2 h} \qquad \mbox{for all $h \in \{0, 1, 2, ... \}$~;} \label{whcond} \feq
so, Eq.\,\eqref{primasol} gives the general solution of Eq.\,\eqref{ey1a} for all 
$w$ as in Eq.\,\eqref{whcond} and, in particular, for all $-1 < w < 1$.
\vfill\eject\noindent
$\phantom{a}$\vspace{-1.2cm}

\subsection[The case $V_1 = 0$]{The case $\boma{V_1 = 0}$}
In this case Eqs.\,\eqref{ex}\eqref{ey} respectively reduce
to Eqs.\,\eqref{ex1b} \eqref{ey1b}.
It appears that, after a time translation $t \to t + \mbox{const.}$
and possibly a time reflection $t \to -t$, any positive solution of Eq.\,\eqref{ex1b}
can be written in one of the following ways:
\begin{align}
x(t) = A\, t    \qquad & \mbox{for ~$A> 0$~ and ~$t \in (0,+\infty)$}~; \label{form11b} \\
x(t) = A        \qquad & \mbox{for ~$A> 0$~ and ~$t \in (-\infty,+\infty)$}~. \label{form21b}
\end{align}
The related solutions of Eq.\,\eqref{ey1b} can be easily derived via the following integral representation,
evaluating the basic integrals which result upon substitution of the expressions
\eqref{form11b}\eqref{form21b} for $x(t)$:
\beq y(t) = C_0 + D_0\,t \,+ \int_{t_0}^{t}\! ds \int_{t_0}^{s}\! ds' \left[(1 - w)\, V_2\,x(s')^{{1 - w \over 1 + w}}
- {w(1 - w)\,\ds^2\,\Oms \over 2}\; x(s')^{-{1 + 3 w \over 1 + w}}\right] , \label{Intrep}\feq
where $C_0,D_0 \in \reali$ are integration constants and $t_0 \in (0,+\infty)$ is fixed arbitrarily.\parn
On one hand, from Eqs.\,\eqref{form11b}\eqref{Intrep} we infer
\begin{gather}
y(t) = C + D\, t + {V_{2}\,(1 + w)^2\, (1 - w) \over 2\,(3+w)}\; A^{{1-w \over 1+w}}\, t^{{3+w \over 1+w}}
+ {\ds^2\,(1 + w)^2\, \Oms\, \over 4}\;A^{-{1+3w \over 1+w}}\, t^{{1-w \over 1+w}}\;, \nonumber \\
C := C_0 + {V_{2}\,(1 - w^2) \over 3+w}\; A^{{1-w \over 1+w}}\, t_0^{{3+w \over 1+w}}
- {\ds^2\,(1 + w)\,w\, \Oms\, \over 2}\;A^{-{1+3w \over 1+w}}\, t_0^{{1-w \over 1+w}}\,, \nonumber \\
D := D_0 - {V_{2}\,(1 - w^2) \over 2}\; A^{{1-w \over 1+w}}\, t_0^{{2 \over 1+w}}
- {\ds^2\,(1 - w^2)\, \Oms\, \over 4}\;A^{-{1+3w \over 1+w}}\, t_0^{-{2w \over 1+w}}\,. \label{form11by}
\end{gather}
On the other hand, Eqs.\,\eqref{form21b}\eqref{Intrep} imply
\begin{gather}
y(t) = C + D\, t + {t^2 \over 2} \left((1 - w)\, V_2\,A^{{1 - w \over 1 + w}}
- {\ds^2 \over 2}\, (1 - w)\, w\, \Oms\, A^{-{1 + 3 w \over 1 + w}}\right) , \nonumber \\
C := C_0 + {t_0^2 \over 2} \left((1 - w)\, V_2\,A^{{1 - w \over 1 + w}}
- {\ds^2 \over 2}\, (1 - w)\, w\, \Oms\, A^{-{1 + 3 w \over 1 + w}}\right), \nonumber \\
D := D_0 - t_0 \left((1 - w)\, V_2\,A^{{1 - w \over 1 + w}}
- {\ds^2 \over 2}\, (1 - w)\, w\, \Oms\, A^{-{1 + 3 w \over 1 + w}}\right). \label{form21by}
\end{gather}
The expressions \eqref{form11b}\eqref{form21b} for $x(t)$ and \eqref{form11by}\eqref{form21by}
for $y(t)$ are patently equivalent to the explicit solutions \eqref{primasol1b}\eqref{secondasol1b}
reported in the main text.

\subsection[The case $V_1 < 0$]{The case $\boma{V_1 < 0}$}
In this case Eqs.\,\eqref{ex}\eqref{ey} reduce, respectively,
to Eqs.\,\eqref{ex1c}\eqref{ey1c}.
By direct inspection it appears that any positive solution of Eq.\,\eqref{ex1c}
can be written as follows, after a time translation $t \to t + \mbox{const.}$:
\beq x(t) = A\,\sin(\om\, t) \qquad \mbox{for ~$A> 0$~ and ~$t \in (0,\pi/\om)$}~. \label{form21c} \feq
For the general solution of Eq.\,\eqref{ey1c} (on the interval mentioned above),
we have the familiar representation
\begin{align}
y(t) & = C\, \cos(\om\, t) + D\, \sin(\om\, t) \nonumber \\
& \quad +\, {1 \over \om} \int_{0}^ t\! d s\; \sin\big(\om\,(t-s)\big) \left[ V_2\,(1-w)\, x(s)^{{1-w \over 1+w}}
- \, {w (1-w)\,\ds^2\,\Oms \over 2}\; x(s)^{-{1+3w \over 1+w}} \right]\, , \label{espy1c}
\end{align}
where $C,D \!\in\! \reali$ are integration constants. This representation is understood to hold
for all values of $w \!\in\! (-1,1)$ granting the convergence of the integral
over $s \!\in\! (0,t)$. As in a similar situation occurring in subsection \ref{subepm1},
convergence holds if and only if $ {1-w \over 1+w} > -1 $ and $ -\,{1+3w \over 1+w} > -1$,
which is equivalent to $ -1 < w < 0 $; however, as in the cited subsection
the final result will be extendable to all $-1<w<1$.
\vfill\eject\noindent
The calculation of the integral in Eq.\,\eqref{espy1c} with $x$ as in Eq.\,\eqref{form21c} is reduced
to the evaluation of the subsequent integral, for $\lp = {1-w \over 1+w}$ or $\lp = -\,{1+3w \over 1+w}$\,:
\beq {1 \over \om} \int_{0}^t\! d s\; \sin\!\big(\om\,(t-s)\big)\, \sin^\lp(\om\, s)~. \label{sinInt} \feq
By arguments similar to those mentioned in subsection \ref{subepm1},
for any $\om \!>\! 0$, $0\!<\! t \!<\! \pi/\om$ and $\lp \!>\! -1$ we have
({\footnote{The condition $\lp > -1$ is required to ensure the convergence of the integral
in Eq.\,\eqref{sinInt}. Correspondingly, let us note that the last identity in the cited equation
can be derived making the change of variable $s = {1 \over \om}\,\mbox{arcos}\big(\sqrt{1 - \sin^2(\om\,t)\,\tu}\big)$.}})
\begin{align}
& {1 \over \om} \int_{0}^t d s\; \sin\big(\om\,(t-s)\big)\, \sin^\lp(\om\, s) \nonumber \\
& = {\sin(\om\,t) \over \om} \int_{0}^t d s\;\cos(\om\,s)\, \sin^\lp(\om\,s)
 -{\cos(\om\,t) \over \om}\! \int_{0}^t d s\; \sin^{\lp+1}(\om\, s) \nonumber \\
& = {\sin^{\lp + 2}(\om\,t) \over \om^2\,(\lp + 1)}
- {\cos(\om\,t)\,\sin^{\lp + 2}(\om\,t) \over 2\,\om^2}
\int_{0}^{1}\! d \tu\; {\tu^{\lp/2} \over \sqrt{1 -  \sin^2(\om\, t)\, \tu}}\;.
\end{align}
Also in this case, the remaining integral can be expressed in terms of the hypergeometric function ${}_2 F_1$,
resorting to the identity \eqref{Inf2F1}. More precisely, employing the cited identity
with $\al = 1/2$, $\be = 1 + \lp/2$, $\ga = 2 + \lp/2$ and $z = \sin^2(\om\, t)$
(along with the basic relations $\Ga(1) = 1$, $\Ga(2+\lp/2) = (1+\lp/2)\,\Ga(1+\lp/2)$\,),
we obtain
\begin{align}
& {1 \over \om} \int_{0}^t d s\; \sin\big(\om\,(t-s)\big)\, \sin^\lp(\om\, s) \nonumber \\
& = {\sin^{\lp + 2}(\om\,t) \over \om^2} \left[{1 \over \lp + 1}
-\, {\cos(\om\,t) \over \lp + 2}\;{}_{2} F_1\!\left({1 \over 2}\,,1+{\lp \over 2}\,,
2+{\lp \over 2}\,;\,\sin^2(\om\,t)\right)\right]\,. \label{sinintex}
\end{align}
Eq.\,\eqref{form21c} for $x(t)$ and the corresponding expression for $y(t)$
deduced from Eqs.\,\eqref{espy1c}\eqref{sinintex} give rise to the explicit
solution \eqref{primasol1c} reported in the main text.
\parn
Considerations of analyticity analogous to those reported in the concluding remark of
subsection \ref{subepm1} allow us to infer that, although in principle the expression \eqref{primasol1c}
for $y(t)$ would hold only for $ -1 < w < 0 $, a posteriori it holds
for any $ w \neq - \frac{3+2h}{1+2h}$ $(h= 0, 1, 2, \ldots)$ (cf. Eq.\,\eqref{whcond}),
and in particular for all $-1<w<1$\,.

\numberwithin{equation}{section}
\section{Appendix. Upper and lower bounds for the integral \textbf{\boma{\eqref{tcInt}}}}\label{appEsttc}
Let us refer to the framework of subsection \ref{Models2} and consider the expression
for cosmic time given in Eq.\,\eqref{tcInt}. With an obvious change of
integration variable, this can be written as
\beq
\tc(t)/\te \;=\, {1 \over \om}\! \int_0^{\,\om\,t}\! ds\, \sqrt{1 + \eep\; {\cosh s \over \sqrt{\sinh s}}}
\qquad \mbox{for $t \in (0,+\infty)$}~.
\label{tcInt2}
\feq
From here to the end of this Appendix we assume $\eep > 0$. Our goal is to derive from
Eq.\,\eqref{tcInt2} upper and lower bounds $T^{\pm}_{\eep}(t)$ for $\tc(t)/\te$,
expressed via elementary functions. To this purpose
let us introduce the following pair of functions, for $z \in (0,+\infty)$\,:
\begin{gather}
\Bm(z) := \sqrt{\sqrt{z} + z} \left({1 \over 2} + \sqrt{z}\right)
- {1 \over 4} \log\!\left(1 + 2\, \sqrt{z} + 2\, \sqrt{\sqrt{z} + z}\right) ; \label{Pdef} \\
\Bp(z) := 2\, \sqrt{z+1} + \log\!\left({\sqrt{z+1}-1 \over \sqrt{z+1}+1}\right) . \label{Qdef}
\end{gather}
\vfill\eject\noindent
$\phantom{a}$\vspace{-1.3cm}\\
To go on, let us fix two real parameters $\elm,\elp$ such that
\beq 0 < \elm < \log\big(1\!+\!\sqrt{2}\,\big) < \elp < \infty ~, \label{elmelpdef} \feq
and set
\beq
\Mel := \max\!\left\{{\cosh \elm \over \sqrt{\sinh \elm}}\,,\,{\cosh \elp \over \sqrt{\sinh \elp}} \right\} .
\label{Mel}
\feq
Finally let us define two continuous, piecewise smooth functions $T^{\pm}_{\eep}$ on $(0,+\infty)$, setting:
\begin{equation}
T^{-}_{\eep}(t) := \left\{\!\!\begin{array}{ll}
\dd{{\eep^2 \over \om}\; \Bm\!\left({\om \over \eep^2}\;t\right)}
&   \quad\dd{\mbox{for\, $0 < t \leqs \elm/\om$}\,,} \vspace{0.15cm}\\
\dd{{\eep^2 \over \om}\; \Bm\!\left({\elm \over \eep^2}\!\right)
+ \sqrt{1 \!+\! \eep \sqrt{2}}\,\left(t - {\elm \over \om} \right)}
&   \quad\dd{\mbox{for\, $\elm/\om < t \leqs \elp/\om$}\,,} \vspace{0.15cm}\\
\dd{{\eep^2 \over \om}\; \Bm\!\left({\elm \over \eep^2}\!\right)
+ \sqrt{1 \!+\! \eep \sqrt{2}}\;\,{\elp - \elm \over \om} } & \vspace{-0.07cm}\\
\dd{\hspace{2cm} +\; {2 \over \om} \left[\Bp\!\left({\eep \over \sqrt{2}}\;e^{\om\,t/2}\right)
- \Bp\!\left({\eep \over \sqrt{2}}\;e^{\elp/2}\right) \right]}
&   \quad\dd{\mbox{for\, $\elp/\om < t < +\infty$}\,;}
\end{array} \right. \label{Tmdef}
\end{equation}
\begin{equation}
T^{+}_{\eep}(t) := \left\{\!\!\begin{array}{ll}
\dd{{\eep^2\;\elm \cosh^2\! \elm \over \om\;\sinh \elm}\;
\Bm\!\left({\om\; \sinh \elm \over \eep^2\,\elm \cosh^2\! \elm}\;t\right) }
&   \quad\!\!\dd{\mbox{for\, $0 < t \leqs \elm/\om$}\,,} \vspace{0.15cm}\\
\dd{{\eep^2\, \elm \cosh^2\! \elm \over \om\,\sinh \elm}\;
\Bm\!\left({\sinh \elm \over \eep^2 \cosh^2\! \elm}\right)
\!+\! \sqrt{1 \!+\! \eep \Mel}\left(t - {\elm \over \om} \right)}
&   \quad\!\!\dd{\mbox{for\, $\elm/\om < t \leqs \elp/\om$}\,,} \vspace{0.15cm}\\
\dd{{\eep^2\, \elm \cosh^2\! \elm \over \om\,\sinh \elm}\;
\Bm\!\left({\sinh \elm \over \eep^2 \cosh^2\! \elm}\right)
\!+\! \sqrt{1 \!+\! \eep \Mel}\;{\elp - \elm \over \om}} & \vspace{0.07cm}\\
\dd{\hspace{1.2cm}+\; {2 \over \om} \left[\Bp\!\left(\!{\eep\,\cosh \elp \over \sqrt{\sinh \elp}}\,e^{(\om\,t\,-\,\elp)/2}\!\right)
\!-\! \Bp\!\left({\eep\,\cosh \elp \over \sqrt{\sinh \elp}}\right)\! \right]}
&   \quad\!\!\dd{\mbox{for\, $\elp/\om < t < +\infty$}\,.}
\end{array} \right. \label{Tpdef}
\end{equation}
We now claim that
\begin{equation}
T^{-}_{\eep}(t) \,\leqs\, \tc(t)/\te \,\leqs\, T^{+}_{\eep}(t) \qquad \mbox{for $t \in (0,+\infty)$}\,.
\label{TmtcTp}
\end{equation}
Most of this Appendix is devoted to the proof of this statement. After the end
of the proof, in the last two paragraphs of the Appendix we discuss
the asymptotics of $T^{\pm}_{\eep}(t)$ for small and large $t$, and present
a numerical appreciation of these upper and lower bounds.
\vspace{-0.4cm}

\paragraph{Preliminaries to the proof of Eq.\,\eqref{TmtcTp}.} Let us consider the function
appearing under the square root in Eq.\,\eqref{tcInt2}, namely,
\beq
J : (0,+\infty) \to (0,+\infty)~, \qquad J(s) := {\cosh s \over \sqrt{\sinh s}} ~. \label{Jdef}
\feq
It can be easily checked that $J$ is a convex function, attaining its global minimum
at a point $s_* \in (0,+\infty)$; more precisely, we have
($\mbox{asinh}$ is the inverse hyperbolic sine function)
\beq
s_* := \mbox{asinh} (1) = \log\big(1\!+\!\sqrt{2}\,\big)~, \qquad
J(s_*) =\! \min_{s \,\in\, (0,+\infty)}\! J(s) = \sqrt{2}~.
\feq
Let us note that $s_{*}$ appears in the inequalities
\eqref{elmelpdef} regarding the parameters $\ell, L$.
The following bounds can be deduced by elementary arguments
({\footnote{Let us give a few more details on the derivation
of Eqs.\,\eqref{0elm}\eqref{elmelp}\eqref{elpinf}.
To prove Eq.\,\eqref{0elm} it suffices to note that the map
$s \in (0,+\infty) \mapsto \sqrt{s}\;J(s)$ is strictly increasing
and further fulfills $\sqrt{s}\;J(s) \to 1^{+}$ for $s \to 0^+$.
Eq.\,\eqref{elmelp} follows straightforwardly from the features of $J(s)$
mentioned in the main text. Finally, to infer Eq.\,\eqref{elpinf}
just notice that the map $s \in (0,+\infty) \mapsto e^{-s/2}\,J(s)$
is strictly decreasing and such that $e^{-s/2}\,J(s) \to (1/\sqrt{2})^{+}$
for $s \to +\infty$.}}):
\begin{align}
{1 \over \sqrt{s}}
& \,\leqs\, J(s) \,\leqs\, \sqrt{\elm\,\cosh^2\! \elm \over \sinh \elm}\; {1 \over \sqrt{s}} \hspace{-2cm}
& \mbox{for~ $0<s \leqs \elm$} ~; \label{0elm}\\
\sqrt{2}
& \,\leqs\, J(s) \,\leqs\, \Mel \hspace{-2cm}
& \mbox{for~ $\elm \leqs s \leqs \elp$} ~; \hspace{-0.07cm} \label{elmelp}\\
{1 \over \sqrt{2}}\;e^{s/2}
& \,\leqs\, J(s) \,\leqs\, {\cosh \elp \over e^{\elp/2} \sqrt{\sinh \elp}}\;e^{s/2} \hspace{-2cm}
& \mbox{for~ $\elp \leqs s < + \infty$} ~. \hspace{-0.6cm} \label{elpinf}
\end{align}
\vfill\eject\noindent
We now proceed to prove Eq.\,\eqref{TmtcTp}, analyzing separately the cases
$0\!<\!t\!\leqs\!\elm/\om$, $\elm/\om\!<\!t\!\leqs\!\elp/\om$ and $\elp\!<\!t\!\leqs\!+\infty$.\vspace{-0.4cm}

\paragraph{Proof of Eq.\,\eqref{TmtcTp} for $\boma{0\!<\!t\!\leqs\!\elm/\om}$.}
For the said values of $t$, from Eqs.\,\eqref{tcInt2}\eqref{0elm} we infer
\beq
{1 \over \om} \int_0^{\,\om\,t}\! ds\; \sqrt{1 + {\eep \over \sqrt{s}}}
\;\leqs\; \tc(t)/\te \;\leqs\;
{1 \over \om} \int_0^{\,\om\,t}\! ds\; \sqrt{1 + \sqrt{\elm\,\cosh^2\! \elm \over \sinh \elm}\;
{\eep \over \sqrt{s}}} ~,
\feq
which by obvious changes of the integration variables can be rephrased as
\beq
{\eep^2 \over \om}\! \int_0^{\om\,t/\eep^2}\!\!\! d\sigma\; \sqrt{1 + {1 \over \sqrt{\sigma}}}
\,\leqs\, \tc(t)/\te \,\leqs
{\eep^2\;\elm \cosh^2\! \elm \over \om\;\sinh \elm} \!
\int_0^{\,(\om\,t \sinh \elm)/(\eep^2 \elm \cosh^2\! \elm)}\!  d\sigma\; \sqrt{1 +
{1 \over \sqrt{\sigma}}} ~.
\feq
Then, noting the basic identity
\beq
\int_0^{z}\! d\sigma\; \sqrt{1 + {1 \over \sqrt{\sigma}}} =
\sqrt{\sqrt{z} + z} \left({1 \over 2} + \sqrt{z}\right)
- {1 \over 4} \log\!\left(1 + 2\, \sqrt{z} + 2\, \sqrt{\sqrt{z} + z}\right)
\quad \mbox{for any $z > 0$}\,,
\feq
and recalling the definition of $\Bm$ given in Eq.\,\eqref{Pdef}, we obtain
\begin{equation}
{\eep^2 \over \om}\, \Bm\!\left({\om \over \eep^2}\;t\right)
\leqs\, \tc(t)/\te \,\leqs\,
{\eep^2\;\elm \cosh^2\! \elm \over \om\;\sinh \elm}\;
\Bm\!\left({\om\; \sinh \elm \over \eep^2\,\elm \cosh^2\! \elm}\;t\right)
\qquad~ \mbox{for\, $0 < t \leqs \elm/\om$}~.
\label{B0elm}
\end{equation}
The upper and lower bounds in Eq.\,\eqref{B0elm} are just $T^{\pm}_{\eep}(t)$ (for the specified values of $t$).
\vspace{-0.4cm}

\paragraph{Proof of Eq.\,\eqref{TmtcTp} for $\boma{\elm/\om\!< \!t\!\leqs\!\elp/\om}$.}
Splitting the integral in the representation \eqref{tcInt2} for $\tc(t)/\te$ at $s = \elm$
and using the bounds \eqref{0elm}\eqref{elmelp},
for the values of $t$ under analysis we obtain
\beq
\begin{aligned}
& {1 \over \om} \int_0^{\,\elm}\! ds\; \sqrt{1 + {\eep \over \sqrt{s}}}
+ {1 \over \om} \int_{\elm}^{\,\om\,t}\! ds\; \sqrt{1 + \eep\, \sqrt{2}} \\
& \leqs\; \tc(t)/\te \;\leqs\;
{1 \over \om} \int_0^{\,\elm}\! ds\; \sqrt{1 + \sqrt{\elm\,\cosh^2\! \elm \over \sinh \elm}\; {\eep \over \sqrt{s}}}
+ {1 \over \om} \int_{\elm}^{\,\om\,t}\! ds\; \sqrt{1 + \eep\,\Mel} ~.
\end{aligned}
\feq
Now, evaluating the integrals for $s \!\in\! (0,\elm)$ with
the same methods presented in the previous paragraph, and computing explicitly
the trivial integrals for $s \!\in\! (\elm,\om\,t)$ we readily get
\begin{gather}
{\eep^2 \over \om}\, \Bm\!\left({\elm \over \eep^2}\!\right)
\!+\! \sqrt{1 \!+\! \eep \sqrt{2}}\left(t - {\elm \over \om} \right)
\leqs\, \tc(t)/\te \,\leqs\,
{\eep^2\, \elm \cosh^2\! \elm \over \om\,\sinh \elm}\,
\Bm\!\left({\sinh \elm \over \eep^2 \cosh^2\! \elm}\right)
\!+\! \sqrt{1 \!+\! \eep \Mel}\left(t - {\elm \over \om} \right)  \nonumber \\
\mbox{for\, $\elm/\om \leqs t \leqs \elp/\om$} ~.
\label{Belmelp}
\end{gather}
The upper and lower bounds in Eq.\,\eqref{Belmelp} are just $T^{\pm}_{\eep}(t)$ (for the considered values of $t$).
\vspace{-0.4cm}

\paragraph{Proof of Eq.\,\eqref{TmtcTp} for $\boma{\elp/\om \! < \! t \!<\! + \infty}$.}
Let us separate the integral in Eq.\,\eqref{tcInt2} at $s = \elm$
and at $s = L$; then, recalling the bounds \eqref{0elm}\eqref{elmelp}\eqref{elpinf},
for the considered values of $t$ we obtain
\begin{align}
& {1 \over \om}\! \int_0^{\,\elm}\!\! ds\, \sqrt{1 + {\eep \over \sqrt{s}}}
+ {1 \over \om}\! \int_{\elm}^{\,\elp}\!\! ds\, \sqrt{1 + \eep\, \sqrt{2}}
+ {1 \over \om}\! \int_{\elp}^{\,\om\,t}\!\! ds\, \sqrt{1 + {\eep \over \sqrt{2}}\;e^{s/2}} \nonumber \\
& \leqs\; \tc(t)/\te \;\leqs \\
& {1 \over \om}\! \int_0^{\,\elm}\!\! ds\, \sqrt{1 + \sqrt{\elm\,\cosh^2\! \elm \over \sinh \elm}\; {\eep \over \sqrt{s}}}
+ {1 \over \om}\! \int_{\elm}^{\,\elp}\!\! ds\, \sqrt{1 + \eep\,\Mel}
+ {1 \over \om}\! \int_{\elp}^{\,\om\,t}\!\! ds\,
\sqrt{1 + \eep\; {\cosh \elp \over e^{\elp/2} \sqrt{\sinh \elp}}\;e^{s/2}} ~. \nonumber
\end{align}
The integrals for $s \!\in\! (0,\elm)$ and for $s \!\in\! (\elm,\elp)$
can be treated as described in the previous paragraph.
On the other hand, the integrals for $s \!\in\!(\elp,\om\,t)$ can both be recast
in the following form, performing the change of the integration variable $\sigma := \eta\,e^{s/2}$
with $\eta = {\eep \over \sqrt{2}}$ and $\eta = {\eep\,\cosh \elp \over e^{\elp/2} \sqrt{\sinh \elp}}$, respectively:
\beq
\int_{L}^{\,\om\,t}\!\! ds\, \sqrt{1 + \eta\;e^{s/2}}
= 2 \int_{\eta\,e^{\elp/2}}^{\eta\,e^{\om\,t/2}}\!\!\! d\sigma\; {\sqrt{1 + \sigma} \over \sigma}
= 2 \left[\Bp\big(\eta\,e^{\om\,t/2}\big) - \Bp\big(\eta\,e^{\elp/2}\big) \right],
\feq
where $\Bp$ is defined as in Eq.\,\eqref{Qdef}.
Summing up, the above arguments allow us to infer that
\begin{align}
& {\eep^2 \over \om}\, \Bm\!\left({\elm \over \eep^2}\!\right)
\!+\! \sqrt{1 \!+\! \eep \sqrt{2}}\;\,{\elp - \elm \over \om}
+ {2 \over \om} \left[\Bp\!\left({\eep \over \sqrt{2}}\;e^{\om\,t/2}\right)
- \Bp\!\left({\eep \over \sqrt{2}}\;e^{\elp/2}\right) \right] \nonumber \\
& \leqs \tc(t)/\te \,\leqs \label{Belpinf}\\
& {\eep^2\, \elm \cosh^2\! \elm \over \om\,\sinh \elm}\,
\Bm\!\left({\sinh \elm \over \eep^2 \cosh^2\! \elm}\right)
\!+\! \sqrt{1 \!+\! \eep \Mel}\;{\elp - \elm \over \om}
+ {2 \over \om}\! \left[\Bp\!\left(\!{\eep\,\cosh \elp \over \sqrt{\sinh \elp}}\,e^{(\om\,t\,-\,\elp)/2}\!\right)
\!-\! \Bp\!\left({\eep\,\cosh \elp \over \sqrt{\sinh \elp}}\right)\! \right]\!,  \nonumber
\end{align}
for\, $\elp/\om < t < +\infty$, where the obtained upper and lower bounds coincide
with $T^{\pm}_{\eep}(t)$. \parn
The arguments described in the previous two paragraphs prove Eq.\,\eqref{TmtcTp}
for all $t \in (0,+\infty)$.
\vspace{-0.1cm}

\paragraph{Asymptotics of $T^{\pm}_{\eep}(t)$ for small and large $t$.}
The asymptotic behavior of $\Bm(z),\Bp(z)$ for $z \to 0^{+}$ and $z \to + \infty$
is readily derived from the definitions \eqref{Pdef}\eqref{Qdef}.
Especially, note that
\begin{gather}
\Bm(z) = {4 \over 3}\;z^{3/4} + O\big(z^{5/4}\big) \quad \mbox{for\, $z \to 0^+$},
\qquad\;
\Bp(z) = 2\,\sqrt{z} + O\big(z^{-1/2}\big) \quad \mbox{for\, $z \to +\infty$}\,.
\end{gather}
From here and from Eqs.\,\eqref{Tmdef} \eqref{Tpdef} one infers
\begin{equation}
T^{-}_{\eep}(t)  = \left\{\!\begin{array}{ll}
\dd{{4 \over 3}\! \left({\eep^2 \over \om}\right)^{\!\!1/4} t^{3/4} + O\big(t^{5/4}\big)}
& \dd{\mbox{for $t \to 0^+$}\,,} \vspace{0.15cm}\\
\dd{{2^{7/4}\,\eep^{1/2} \over \om\!}\; e^{{1 \over 4}\,\om\,t} + O(1)}
& \dd{\mbox{for $t \to +\infty$} ~;}
\end{array}\right. \label{tcasym}
\end{equation}
\begin{equation}
T^{+}_{\eep}(t)  = \left\{\!\begin{array}{ll}
\dd{{4 \over 3}\! \left({\eep^2\;\elm \cosh^2\! \elm \over \om\;\sinh \elm}\right)^{\!\!1/4} t^{3/4} + O\big(t^{5/4}\big)}
& \dd{\mbox{for $t \to 0^+$}\,,} \vspace{0.15cm}\\
\dd{{4 \over \om}\!\left(\!{\eep\,\cosh \elp \over e^{\elp/2}\,\sqrt{\sinh \elp}}\right)^{\!\!1/2}\; e^{{1 \over 4}\,\om\,t} + O(1)}
& \dd{\mbox{for $t \to +\infty$} ~.}
\end{array}\right. \label{tcasyp}
\end{equation}
By comparison with Eq.\,\eqref{tcasy} for $\tc(t)/\te$,
we see that $T^{-}_{\eep}(t)$ has just the same asymptotics as $\tc(t)/\te$
both for small and large $t$; on the other hand, the asymptotics of $T^{+}_{\eep}(t)$ and $\tc(t)/\te$
are very similar in both limits.
\vspace{-0.1cm}

\paragraph{A numerical test.} Let us consider the following
prescriptions for the parameters of the model, which are used
in the final paragraph of subsection \ref{Models2} to get a realistic
model of inflation:
\beq
\eep = e^{-100} \simeq 3.72008\,... \times 10^{-44} \,,
\qquad \Oms = 0.9 \,, \qquad V = 1 \,, \qquad \om = {2\, \sqrt{2\,V} \over 3} \simeq 0.9428\,...~.
\feq
To determine the upper and lower bounds $T^{\pm}_{\eep}$ of Eqs.\,\eqref{Tmdef}\eqref{Tpdef}, let us fix
as follows the parameters $\elm, \elp$ appearing therein (and in Eq.\,\eqref{elmelpdef}):
\beq
\elm = {1 \over 2}\,\log\big(1\!+\!\sqrt{2}\big) \simeq 0.4406\,... ~, \qquad
\elp = 2 \,\log\big(1\!+\!\sqrt{2}\big) \simeq 1.7627\,... ~. \label{fixelmp} \vspace{-0.5cm}
\feq
\vfill\eject\noindent
\begin{figure}[t!]
    \centering
        \begin{subfigure}[b]{0.475\textwidth}\label{fig:30}
                \includegraphics[width=\textwidth]{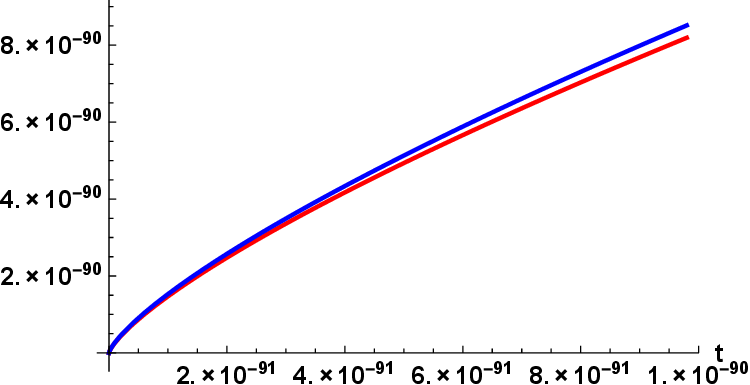}\vspace{-0.cm}
                \caption*{\textsc{Figure 30.} Plot of $T_{\eep}^{-}(t)$ (in red) and $T_{\eep}^{+}(t)$ (in blue).}
        \end{subfigure}
        \hspace{0.3cm}
        \begin{subfigure}[b]{0.475\textwidth}
                \includegraphics[width=\textwidth]{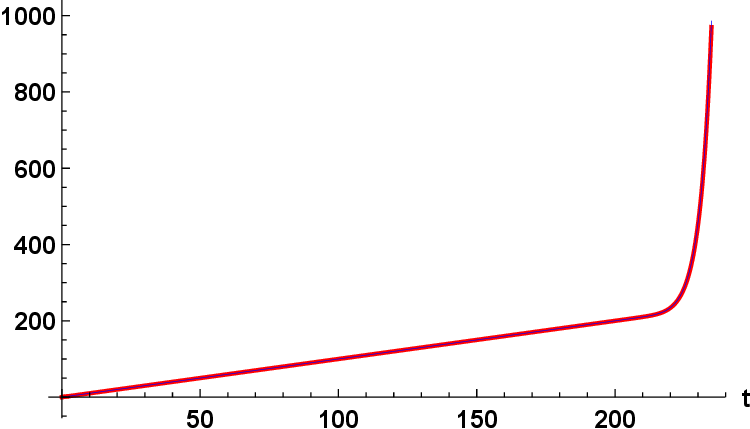}\label{fig:31}
                \caption*{\textsc{Figure 31.} Plot of $T_{\eep}^{-}(t)$ (in red) and $T_{\eep}^{+}(t)$ (in blue).}
        \end{subfigure}
\vspace{-0.5cm}
\end{figure}
\noindent\\
Fig.\,30 and 31 are plots of the functions $T_{\eep}^{\pm}$ over
different time intervals (namely: for $t \in (0,10^{-90})$ and for $t \in (0,240)$, respectively).
The graphs of $T_{\eep}^{\pm}$ are very close in Fig.\,30, and practically coincide in Fig.\,31.
Let us recall that, according to Eq.\,\eqref{TmtcTp}, we
have $T^{-}_{\eep}(t) \leqs \tc(t)/\te \leqs T^{+}_{\eep}(t)$
for all $t >0$. For $t$ in the ranges of the two figures these
bounds, being very close, determine $\tc(t)/\te$ up to a very
small uncertainty; in particular, $\tc(t)/\te$ is approximated
with excellent accuracy by the mean value $(1/2) (T^{-}_{\eep} + T^{+}_{\eep})(t)$.
We have used this approximation of $\tc(t)/\te$ (with $\ell,L$ as in Eq.\,\eqref{fixelmp})
for all the computations in the final paragraph of subsection \ref{Models2}
(and in particular, to construct Figs.\,8-15).
To conclude, let us remark that Fig.\,31 exhibits the approximate linear dependence $\tc(t)/\te \simeq t$
for positive (not too large) values of the coordinate time $t$,
in accordance with the expansion given in Eq.\,\eqref{tccomp}.

\numberwithin{equation}{subsection}
\section{Appendix. On the model of subsection \ref{subclass7}}
\label{appclass7}
Let us keep all the assumptions and notations of the cited subsection; in particular,
we consider a motion $t \mapsto x(t),y(t)$ with initial conditions as in
\eqref{idata} \eqref{uandv}. Hereafter we justify some technical statements appearing
without proof in subsection \ref{subclass7}.

\subsection[Proof of Eq.\,\eqref{eprove}: $x(t) > 0$ and $x(t) < y(t) < x(t)$ for all $t \in (0,\tp)$]{Proof of Eq.\,\eqref{eprove}: $\boma{x(t) > 0}$ and $\boma{x(t) < y(t) < x(t)}$ for all $\boma{t \in (0,\tp)}$}
The statement in Eq.\,\eqref{eprove} about $x(t)$ is obvious,
since $x(0) = Y > 0$ and $t \mapsto x(t)$ is a strictly increasing
function. In the rest of this subsection we show how to derive the
inequalities $-x(t)< y(t) < x(t)$ for $t \in (0, \tp)$. Our
arguments also involve the inversion time $t_{*}$ of Eqs.
\eqref{eqts} \eqref{eqtss}; we will treat separately the cases $\tp \leqs t_{*}$
and $\tp > t_{*}$.
\parn \textbf{i) $\boma{\tp \leqs t_{*}}$\,.} For any $t \!\in\! (0, \tp)$,
we have $Y \!<\! x(t)$, $Y \!<\! y(t) \!<\! (\FF/V)^{1/(2\ell)}$ and
Eqs.\,\eqref{tx}\eqref{ty} give
\beq
\int_{Y}^{x(t)}\!\! {d x \over \sqrt{ \EE \!+\! V_1 x^{2 \ell}}} = {\sqrt{2} \over \ell +
1}\; t = \!\int_{Y}^{y(t)}\!\! {d y \over \sqrt{ \FF \!-\! V_2\, y^{2 \ell}}}~.
\feq
Clearly, the above chain of identities holds true if and only if
\beq
\int_{Y}^{x(t)}\!\!{d z \over \sqrt{\EE \!+\! V_1 z^{2 \ell}}}
= \int_{Y}^{y(t)}\!\! {d z \over \sqrt{\FF \!-\! V_2\, z^{2 \ell}}} \qquad
\mbox{for all}~ t \in (0,\tp)~. \label{inteuguali} \feq
On the other hand, since $1/\sqrt{\EE \!+\! V_1 z^{2 \ell}} <  1/\sqrt{ \FF \!-\! V_2\, z^{2 \ell}}$
for all $z \!\in\! [Y, (\FF/V_2)^{1 \over {2 \ell}})$, the above
identity \eqref{inteuguali} is certainly violated if $x(t) \leqs
y(t)$. This suffices to infer that
\beq x(t) > y(t) > Y \qquad \mbox{for all}~~ t \in (0, \tp)~. \label{quasiprov} \feq
This also implies $y(t) > Y > 0 > - x(t)$. Therefore, summing up we have $x(t)
> y(t) > - x(t)$, as required.
\vfill\eject\noindent
\textbf{ii) $\boma{\tp > t_{*}}$}\,. In this case, the relations
written in Eqs.\,\eqref{tx}\eqref{ty}\eqref{inteuguali} hold true
for $t \in (0, t_{*})$ and, by continuity, even for $t = t_{*}$.
Then, repeating the considerations which led to
Eq.\,\eqref{quasiprov}, we deduce
\beq x(t) > y(t) > Y \qquad \mbox{for all}~~ t \in (0, t_{*}]~. \label{quasiprovvv} \feq
The above inequalities also imply $y(t) > Y > 0 > - x(t)$, which
allows us to infer
\beq - x(t) < y(t) < x(t) \qquad \mbox{for all}~~ t \in (0, t_{*}]~. \label{quasii} \feq
Finally, let $t \in (t_{*}, \tp)$. Recalling that $x(t)$ is a strictly increasing
function, and using Eq.\,\eqref{quasii} at $t = t_{*}$ we infer
$x(t) > x(t_{*}) > y(t_{*}) = (\FF/V_2)^{1/(2 \ell)}$; on the
other hand, from Eq.\,\eqref{motiv} we know that in any case
$-(\FF/V_2)^{1/(2 \ell)} \leqs y(t) \leqs (\FF/V_2)^{1 /(2
\ell)}$. Summing up, we obtain
\beq -x(t) < y(t) < x(t) \quad \mbox{for all}~~ t \in (t_{*}, \tp)~ \label{quasjj} \feq
and this fact, along with Eq.\,\eqref{quasii}, ensures $-x(t) < y(t) < x(t)$ for all $t \in (0,\tp)$.

\subsection[Proof of Eq.\,\eqref{tcas}: asymptotics of $\tc(t)$ for $t \to \tp^{-}$]{Proof of Eq.\,\eqref{tcas}: asymptotics behaviour of $\boma{\tc(t)}$ for $\boma{t \to \tp^{-}}$}
Let us consider the integral representation of $\tc(t)$
given in Eq.\,\eqref{cosmicx7}. Due to Eqs.\eqref{xesp}\eqref{yesp}, the integrand
function therein has the asymptotic expansion
\beq \big(x^2(t') -
y^2(t')\big)^{{\ell - 1 \over 2}} = { \ell + 1 \over (\ell-1)
\sqrt{2 V_1}}\; (\tp\! - t')^{-1} \left( 1 + O\big((\tp\! - t')^{2
\over \ell - 1} \big) \right) \quad \mbox{for $t' \to \tp^{-}$}~.
\label{integrand} \feq
We now re-write Eq.\,\eqref{cosmicx7} isolating the dominant
contribution for $t \to \tp^{-}$, which gives
\begin{align}
& \tc(t)/\te = \int_{0}^t\! d t'\, \big(x^2(t') - y^2(t')\big)^{{\ell - 1 \over 2}}  \\
& = \int_{0}^t\! d t'\, { \ell + 1 \over (\ell-1) \sqrt{2
V_1}}\;(\tp\! - t')^{-1} + \int_{0}^t\! d t' \left[ \big(x^2(t') -
y^2(t')\big)^{{\ell - 1 \over 2}} - { \ell + 1 \over (\ell-1)
\sqrt{2 V_1}}\,(\tp\! - t')^{-1} \right]. \nonumber
\end{align}
Computing the first integral above and splitting the second one in two parts, we obtain
\begin{align}
\tc(t)/\te & = {\ell + 1 \over (\ell -1) \sqrt{2 V_1}} \, \log \!\left({\tp \over \tp - t} \right) \nonumber \\
& \qquad + \left(\int_{0}^{\tp} - \int_{t}^{\tp} \right) d t'
\left[ \big(x^2(t') - y^2(t')\big)^{{\ell - 1 \over 2}} - { \ell +
1 \over (\ell-1) \sqrt{2 V_1}}\; (\tp\! - t')^{-1} \right] .
\end{align}
Let us write $[...]$ for the expression appearing above between
square brackets; according to Eq.\,\eqref{integrand} we have
$[...] =  O\big( (\tp - t')^{-1 + {2 \over \ell - 1}}\big)$ for
$t' \!\to\! \tp^{-}$, thus $\int_{0}^{\tp} d t' [...]$ is
convergent and $\int_{t}^{\tp} d t' [...]$ $=\int_{t}^{\tp} d t'
O\big( (\tp - t')^{-1 + {2 \over \ell - 1}}\big)$ $ = O\big( (\tp
- t)^{2 \over \ell-1} \big)$ for $t \!\to\! \tp^{-}$. In
conclusion,
\begin{align}
\tc(t)/\te & = {\ell + 1 \over (\ell -1) \sqrt{2 V_1}}\, \log\!
\left({\tp \over \tp\! - t} \right) \\
& \qquad + \int_{0}^{\tp}\!\! d t' \left[ \big(x^2(t')\! -
y^2(t')\big)^{{\ell - 1 \over 2}}\! - { \ell + 1 \over (\ell-1)
\sqrt{2 V_1}}\;(\tp\! - t')^{-1} \right]\!
+ O\big( (\tp\! - t)^{2 \over \ell-1} \big)\,,  \nonumber
\end{align}
as stated in Eq.\,\eqref{tcas}.

\vfill \eject 

\noindent
\begin{thebibliography}{99}

\bibitem{Barr2} J. D. Barrow,
\textsl{Cosmic no-hair theorems and inflation},
Phys. Lett. B \textbf{187}(1-2) (1987), 12--16.

\bibitem{Bar90} J. D. Barrow,
\textsl{Graduated inflationary universes},
Phys. Lett. B \textbf{235}(1-2) (1990), 40--43.

\bibitem{Barr} J. D. Barrow, A. Paliathanasis,
\textsl{Observational constraints on new exact inflationary scalar-field solutions},
Phys. Rev. D \textbf{94} (2016), 083518 [17 pp].

\bibitem{Burd} A. B. Burd, J. D. Barrow,
\textsl{Inflationary models with exponential potentials},
Nucl. Phys. B \textbf{308}(4) (1988), 929--945.

\bibitem{CFP} C. Cacciapuoti, D. Fermi, A. Posilicano,
\textsl{Relative-Zeta and Casimir energy for a semitrasparent hyperplane selecting transverse modes},
pp. 71--97 in G.F. Dell’Antonio, A. Michelangeli (Eds.), ``Advances in Quantum Mechanics: contemporary trends and open problems'', Springer INdAM Series, Springer (2017).

\bibitem{Cald} R. R. Caldwell, R. Dave, P.J. Steinhardt,
\textsl{Cosmological imprint of an energy component with general equation of state},
Phys. Rev. Lett. \textbf{80}(8) (1998), 1582--1585.

\bibitem{Chim} L. P. Chimento,
\textsl{General solution to two-scalar field cosmologies with exponential potentials},
Class. Quant. Grav. \textbf{15}(4) (1998), 965--974.

\bibitem{Mar1} R. de Ritis, G. Marmo, G. Platania, C. Rubano, P. Scudellaro, C. Stornaiolo,
\textsl{New approach to find exact solutions for cosmological models with a scalar field},
Phys. Rev. D \textbf{42}(4) (1990), 1091--1097.

\bibitem{Mar2} R. de Ritis, G. Marmo, G. Platania, C. Rubano, P. Scudellaro, C. Stornaiolo,
\textsl{Scalar field, nonminimal coupling, and cosmology},
Phys. Rev. D \textbf{44}(10) (1991), 3136--3146.

\bibitem{Dima} N. Dimakis, A. Karagiorgos, A. Zampeli, A. Paliathanasis, T. Christodoulakis, P. A. Terzis,
\textsl{General analytic solutions of scalar field cosmology with arbitrary potential},
Phys. Rev. D \textbf{93} (2016), 123518 [16 pp].

\bibitem{East} R. Easther,
\textsl{Exact superstring motivated cosmological models},
Class. Quant. Grav. \textbf{10}(11) (1993), 2203--2215.

\bibitem{Ema} G.F.R. Ellis, M.S. Madsen, \textsl{Exact scalar field cosmologies},
Class. Quant. Grav. \textbf{8} (1991), 667--676.

\bibitem{phantom} D. Fermi, M. Gengo, L. Pizzocchero,
\textsl{On the necessity of phantom fields for solving the horizon problem in scalar cosmologies},
Universe 2019 \textbf{5}(3), 76 [20 pp].

\bibitem{FPWS} D. Fermi, L. Pizzocchero,
``Local Zeta Regularization and the Scalar Casimir Effect. A General Approach based on Integral Kernels'',
World Scientific Publishing Co., Singapore (2017).

\bibitem{FPpoint} D. Fermi, L. Pizzocchero, \textsl{Local Casimir Effect for a Scalar Field in Presence of a Point Impurity},
Symmetry \textbf{2018}, 10(2) (2018).

\bibitem{Fre} P. Fr\'e, A. Sagnotti, A. S. Sorin,
\textsl{Integrable scalar cosmologies, I. Foundations and links with string theory},
Nucl. Phys. B \textbf{877}(3) (2013), 1028--1106.

\bibitem{Fut} T. Futamase, K. Maeda,
\textsl{Chaotic inflationary scenario of the Universe with a nonminimally coupled ``inflaton'' field},
Phys. Rev. D \textbf{39}(2) (1989), 399--404.

\bibitem{thesis} M. Gengo,
\textsl{Integrable multidimensional cosmologies with matter and a scalar field},
PhD Thesis, Doctoral Program in Mathematical Sciences, Universit\`a degli Studi di Milano (2019).

\bibitem{Hawk} S. W. Hawking, G. F. R. Ellis,
``The large scale structure of spacetime'',
Cambridge University Press, Cambridge (1973).

\bibitem{KasZh1} U. Kasper, M. Rainer, A. Zhuk,
\textsl{Integrable multicomponent perfect fluid multidimensional cosmology II: scalar fields},
Gen. Rel. Grav. \textbf{29}(9) (1997), 1123--1162.

\bibitem{KasZh2} U. Kasper, A. Zhuk,
\textsl{Integrable multicomponent perfect fluid multidimensional cosmology. I},
Gen. Rel. Grav. \textbf{28}(10) (1996), 1269--1292.

\bibitem{Lind} A. D. Linde,
\textsl{Chaotic inflation},
Phys. Lett. B \textbf{129}(3-4) (1983), 177--181.

\bibitem{Luc} F. Lucchin, S. Matarrese,
\textsl{Power-law inflation},
Phys. Rev. D \textbf{32}(6) (1985), 1316--1322.

\bibitem{Mads} M. S. Madsen, P. Coles,
\textsl{Chaotic inflation},
Nucl. Phys. B \textbf{298}(4) (1988), 701--725.

\bibitem{Maed} K. Maeda,
\textsl{Towards the Einstein-Hilbert action via conformal transformation},
Phys. Rev. D \textbf{39}(10) (1989), 3159--3162.

\bibitem{Ma18} H. Maeda, C. Martinez,
\textsl{Energy conditions in arbitrary dimensions},
arXiv:1810.02487 [gr-qc] (2018).

\bibitem{NIST} F. W. J. Olver, D. W. Lozier, R. F. Boisvert, C. W. Clark,
``NIST Handbook of Mathematical Functions'',
Cambridge University Press, Cambridge (2010).

\bibitem{Perl} S. Perlmutter \textsl{et al.},
\textsl{Measurements of Omega and Lambda from 42 high redshift supernovae},
Astroph. J. \textbf{517}(2) (1999), 565--86.

\bibitem{PLC} E. Piedipalumbo, M. De Laurentis, S. Capozziello,
\textsl{Noether symmetries in Interacting Quintessence
Cosmology}, arXiv:1912.08089 [gr-qc] (2019).

\bibitem{Piedipalumbo} E. Piedipalumbo, P. Scudellaro, G. Esposito, C. Rubano,
\textsl{On quintessential cosmological models and exponential potentials},
Gen. Rel. Grav. \textbf{44} (2012), 2611--2643.

\bibitem{PLANK18} Planck Collaboration (N. Aghanim (Orsay, IAS) et al.),
\textsl{Planck 2018 results. VI. Cosmological parameters},
arXiv:1807.06209 [astro-ph.CO] (2018).

\bibitem{Peeb} P. Ratra, L. Peebles,
\textsl{Cosmological consequences of a rolling homogeneous scalar field},
Phys. Rev. D \textbf{37}(12) (1988), 3406--3427.

\bibitem{Riess} A. G. Riess \textsl{et al.},
\textsl{Observational evidence from supernovae for an accelerating universe and a cosmological constant},
Astron. J. \textbf{116}(3) (1998), 1009--38.

\bibitem{Rub} C. Rubano, P. Scudellaro,
\textsl{On some exponential potentials for a cosmological scalar field as quintessence},
Gen. Rel. Grav. \textbf{34}(2) (2002), 307--328.

\bibitem{Ryd} B. Ryden,
``Introduction to Cosmology'',
Addison Wesley, San Francisco (2002).

\bibitem{Starob}  T. D. Saini, S. Raychaudhury, V. Sahni, A. A. Starobinsky,
\textsl{Reconstructing the cosmic equation of state from supernova distances},
Phys. Rev. Lett. \textbf{85}(6) (2000), 1162--1165.

\bibitem{SokSor} V. V. Sokolov, A. S. Sorin,
\textsl{Integrable cosmological potentials},
Lett. Math. Phys. \textbf{107} (2017), 1741–-1768.

\bibitem{PDG18} M. Tanabashi \emph{et al}. (Particle Data Group),
\textsl{Review of Particle Physics},
Phys. Rev. D \textbf{98}(3) (2018), 030001 and 2019 update.

\bibitem{Wei} S. Weinberg,
``Cosmology'',
Oxford University Press, Oxford (2008).

\bibitem{Zhuk} A. Zhuk,
\textsl{Integrable scalar field multi-dimensional cosmologies},
Class. Quant. Grav. \textbf{13} (1996), 2163-–2178.

\end{thebibliography}
\end{document}